\documentclass[11pt]{article}

\usepackage[english]{babel}
\usepackage[a4paper,margin=1in]{geometry}
\usepackage{setspace}
\usepackage{ragged2e}
\usepackage{csquotes}

\usepackage{amsmath,amssymb}
\usepackage{graphicx}
\usepackage{booktabs}
\usepackage{multirow}
\usepackage{tabularx}
\usepackage{array}
\usepackage{adjustbox}
\usepackage{rotating}
\usepackage{siunitx}
\usepackage[final]{microtype}
\usepackage{placeins}
\usepackage{float}
\usepackage{enumerate}
\usepackage{algorithm}
\usepackage{algorithmic}
\usepackage{tocloft}

\cftsetindents{section}{0em}{2.4em}
\cftsetindents{subsection}{2.4em}{3.2em}
\cftsetindents{subsubsection}{5.6em}{4.1em}
\usepackage{xurl}
\usepackage[no-math]{fontspec}
\newfontfamily\prismuifont{texgyreheros-regular.otf}[
  BoldFont=texgyreheros-bold.otf,
  ItalicFont=texgyreheros-italic.otf,
  BoldItalicFont=texgyreheros-bolditalic.otf]
\usepackage[defaultmathsizes]{mathastext}

\usepackage[dvipsnames]{xcolor}
\definecolor{sectioncolor}{HTML}{9c4847} 
\definecolor{linkcolor}{HTML}{008ebe}    
\definecolor{black}{HTML}{000000}
\definecolor{llmcore}{RGB}{220, 53, 69}
\definecolor{skillblue}{RGB}{13, 110, 253}
\definecolor{datagreen}{RGB}{25, 135, 84}
\definecolor{userpurple}{RGB}{111, 66, 193}
\definecolor{mcpyellow}{RGB}{255, 193, 7}
\definecolor{deeponet}{RGB}{253, 126, 20}
\definecolor{langgraph}{RGB}{52, 152, 219}
\definecolor{physicsgreen}{RGB}{46, 204, 113}
\definecolor{stategray}{RGB}{149, 165, 166}
\definecolor{webui}{RGB}{155, 89, 182}

\usepackage{tikz}
\usetikzlibrary{shapes,arrows,shapes.geometric,arrows.meta,positioning,fit,backgrounds,calc,shadows,decorations.pathreplacing,chains,matrix,patterns,shapes.arrows}
\DeclareMathOperator{\csch}{csch}

\tikzset{
  Arr/.style={-{Stealth[length=5pt]}, thick},
  ArrDash/.style={-{Stealth[length=5pt]}, thick, dashed},
  Ortho/.style={rounded corners=4pt},
  loop/.style={-{Stealth[length=5pt]}, thick, gray!70},
}

\usepackage[colorlinks=true, linkcolor=linkcolor, citecolor=linkcolor, urlcolor=linkcolor, breaklinks=true]{hyperref}

\usepackage[capitalise, noabbrev, nameinlink]{cleveref}
\crefdefaultlabelformat{#2#1#3}
\Crefname{figure}{Figure}{Figures}
\Crefname{table}{Table}{Tables}
\Crefname{section}{\textbf{Section}}{\textbf{Sections}}

\usepackage{titlesec}
\titleformat{\section}
{\color{sectioncolor}\normalfont\Large\bfseries\raggedright}
{\thesection}{1em}{}

\titleformat{\subsection}
{\color{sectioncolor}\normalfont\large\bfseries\raggedright}
{\thesubsection}{1em}{}

\titleformat{\subsubsection}
{\color{sectioncolor}\normalfont\normalsize\bfseries\raggedright}
{\thesubsubsection}{1em}{}

\usepackage[labelsep=period]{caption}
\usepackage{subcaption}
\DeclareCaptionLabelFormat{bold}{{(#2)}}
\DeclareCaptionFont{figuretitlecolor}{\color{sectioncolor}}
\DeclareCaptionFont{figurecaptioncolor}{\color{black}\footnotesize}
\DeclareCaptionFont{figurecaptionsmall}{\color{black}\scriptsize}
\DeclareCaptionFont{tabletitlecolor}{\color{sectioncolor}}
\DeclareCaptionFont{tablecaptioncolor}{\color{black}\footnotesize}
\graphicspath{{figs/}}
\DeclareGraphicsExtensions{.pdf,.jpeg,.JPG,.png,.PNG,.eps,.tiff}

\usepackage[backend=biber, citestyle=numeric-comp, bibstyle=authoryear, sorting=none, natbib=true, url=false, minbibnames=10, maxbibnames=10, uniquename=false, uniquelist=false, giveninits=true]{biblatex}

\defbibenvironment{bibliography}
  {\list
    {\printtext[labelnumberwidth]{%
       \printfield{labelprefix}%
       \printfield{labelnumber}}}
    {\setlength{\labelwidth}{\labelnumberwidth}%
     \setlength{\leftmargin}{\labelwidth}%
     \setlength{\labelsep}{\biblabelsep}%
     \addtolength{\leftmargin}{\labelsep}%
     \setlength{\itemsep}{\bibitemsep}%
     \setlength{\parsep}{\bibparsep}}%
     }
  {\endlist}
  {\item}

\DeclareFieldFormat{labelnumberwidth}{#1.}
\DeclareNameAlias{author}{family-given}
\AtEveryBibitem{\clearfield{month}}
\AtEveryBibitem{\clearfield{number}}
\AtEveryBibitem{\clearfield{issn}}
\DeclareFieldFormat[article]{title}{#1}
\DeclareFieldFormat{journaltitle}{#1\isdot}
\renewbibmacro{in:}{}
\renewbibmacro*{volume+number+eid}{%
  \printfield{volume}%
}
\DeclareFieldFormat[article]{volume}{\mkbibitalic{#1}}
\DeclareFieldFormat{doi}{\mkbibbrackets{\url{https://doi.org/#1}}}
\DefineBibliographyStrings{english}{%
  page              = {},
  pages             = {},
}
\usepackage{authblk}
\DeclareRobustCommand{\prismAffilMark}[1]{\raisebox{0.5ex}{\normalfont\scriptsize #1}}
\makeatletter
\renewcommand{\AB@authnote}[1]{\protect\prismAffilMark{#1}}
\renewcommand{\AB@affilnote}[1]{\protect\prismAffilMark{#1}}
\makeatother

\makeatletter
\renewcommand\maketitle{\par
  \begingroup
    \flushleft
    \Large{\textbf{\@title}}\par
  \endgroup
  \vspace{0.8em}
  \begin{flushleft}
    \@author
  \end{flushleft}
}
\makeatother

\renewenvironment{abstract}{
  \begin{flushleft}
    \textbf{\abstractname}
  \end{flushleft}
  \vspace{-1.2em}
  \par
  \RaggedRight\rightskip=0pt plus 1fil\relax
  \noindent
}{
  \par 
}

\newcommand{\makeAbstract}[1]{
  \renewcommand{\abstractname}{\color{sectioncolor}SUMMARY}
  \begin{abstract}
    \noindent #1
  \end{abstract}
}

\newcommand{\siref}[1]{\textcolor{linkcolor}{(#1)}}

\newcommand{\bNF}{b_{\mathrm{NF}}}
\newcommand{\NFcell}[2]{\begin{tabular}[t]{@{}c@{}}#1\\ #2\end{tabular}}

\newcommand{\SInumbering}{%
  \setcounter{page}{1}%
  \setcounter{section}{0}%
  \setcounter{subsection}{0}%
  \setcounter{subsubsection}{0}%
  \setcounter{equation}{0}%
  \setcounter{figure}{0}%
  \setcounter{table}{0}%
  \renewcommand{\thepage}{S\arabic{page}}%
  \renewcommand{\thesection}{S\arabic{section}}%
  \renewcommand{\thesubsection}{S\arabic{section}.\arabic{subsection}}%
  \renewcommand{\thesubsubsection}{S\arabic{section}.\arabic{subsection}.\arabic{subsubsection}}%
  \renewcommand{\theequation}{S\arabic{equation}}%
  \renewcommand{\thefigure}{S\arabic{figure}}%
  \renewcommand{\thetable}{S\arabic{table}}%
  \renewcommand{\theHsection}{SI.section.\arabic{section}}%
  \renewcommand{\theHsubsection}{SI.subsection.\arabic{section}.\arabic{subsection}}%
  \renewcommand{\theHsubsubsection}{SI.subsubsection.\arabic{section}.\arabic{subsection}.\arabic{subsubsection}}%
  \renewcommand{\theHequation}{SI.equation.\arabic{equation}}%
  \renewcommand{\theHfigure}{SI.figure.\arabic{figure}}%
  \renewcommand{\theHtable}{SI.table.\arabic{table}}%
}

\begin{document}
\addtocontents{toc}{\protect\setcounter{tocdepth}{-2}}
\setstretch{1.15}

\title{PRISM: Unifying multiphysics optothermal transport and identifiability-guided inference for nanoscale thermal metrology}

\author[1]{Yicong~QIU}
\author[1,2,3]{Ziyang~ZHONG}
\author[1]{Ziyan~QIAN}
\author[1]{Yang~LU}
\author[1]{Yong~Jian~HONG}
\author[4]{Chenhan~LIU}
\author[1,5,*]{Qiye~ZHENG}

\affil[1]{Department of Mechanical and Aerospace Engineering, The Hong Kong University of Science and Technology, Clear Water Bay, Hong Kong SAR, China 999077}
\affil[2]{Department of Computer Science and Engineering, The Hong Kong University of Science and Technology, Clear Water Bay, Hong Kong SAR, China 999077}
\affil[3]{Department of Electronic and Computer Engineering, The Hong Kong University of Science and Technology, Clear Water Bay, Hong Kong SAR, China 999077}
\affil[4]{School of Energy and Mechanical Engineering, Nanjing Normal University, Nanjing 210023, China}
\affil[5]{Lead contact}
\affil[*]{Correspondence: [qiyezheng@ust.hk]}
\affil[ ]{ }
\date{}

\maketitle

\makeAbstract{

Quantifying thermal transport across micro- and nanoscale heterostructures is critical for solid-state devices and relies on fitting heat-transfer models to thermal-wave spectra. When probing scales approach optical absorption depths, electron--phonon equilibration lengths, or phonon mean free paths, classical surface-heating Fourier models can match measured phase spectra while returning biased, measurement-dependent parameters. Moreover, because subsurface optical, hot-carrier, and quasi-ballistic processes couple non-additively across interfaces, single-mechanism extensions leave cross-regime biases unresolved, whereas enabling all mechanisms simultaneously induces severe multi-parameter ill-posedness. Here we introduce PRISM (Physics-Resolved Inference and Screening Meta-framework), an agentic framework coupling multilayer optical transfer-matrix electrodynamics, two-temperature electron--phonon relaxation, and nonlocal Guyer--Krumhansl phonon hydrodynamics within a unified admittance formalism, paired with singular-value-decomposition identifiability screening and knowledge-guided routing. Demonstrated with frequency-domain thermoreflectance across stacks relevant to microelectronic contacts and dielectric barriers, PRISM corrects a 24--28\% interfacial conductance underestimation in $\text{Au/Si}$, recovers intrinsic silica conductivity in $\text{Au/Ti/SiO}_2$, and eliminates a 40\% cross-plane conductivity underestimation in $\text{Au/MoS}_2$. In bimetallic $\text{Au/Ti/Si}$ stacks, where all three non-classical mechanisms act simultaneously, nonlocal modeling reconciles the $116\text{--}\SI{137}{\watt\per\metre\per\kelvin}$ local-fit scatter across spot sizes and frequency windows into a configuration-independent intrinsic silicon conductivity of $\SI{141}{\watt\per\metre\per\kelvin}$. Furthermore, for ion-irradiated and graded media with continuous conductivity profiles, we show that PRISM can autonomously constructs and verifies a deep operator network solver to reconstruct steep damage gradients. By uniting non-equilibrium transport physics and identifiability screening with autonomous model synthesis, PRISM establishes a generalizable paradigm readily extensible to broader optothermal and thermal-wave metrology modalities.

}

{\RaggedRight \vspace{-\baselineskip}\paragraph{Keywords:} nanoscale thermal metrology, frequency-domain thermoreflectance, non-Fourier phonon transport, electron--phonon nonequilibrium, inverse problem identifiability, multiphysics modeling, neural operator learning, autonomous scientific inference.\par}

\clearpage

\section{Introduction}
\label{sec:introduction}

Thermal transport across micro- and nanoscale dimensions governs the power density, performance, and reliability of solid-state technologies, ranging from three-dimensional integrated circuits (3D~ICs) and back-end-of-line (BEOL) interconnects~\citep{moore1998cramming,pop2010energy,moore2014emerging,warzoha2021applications,warzoha2024buried,zhang2025thermal,sharma2026thermal,shen2026beol,lu2026process} to van der Waals optoelectronics~\citep{yuan2022thermoreflectance,guo2026light} and radiation-hardened semiconductors~\citep{huynh2023depth,xiang2023machine}. Across these architectures, device integration and microfabrication introduce three intertwined thermal transport complexities: thin-film and substrate conductivities ($k$) deviate from classical bulk behavior due to boundary confinement, structural anisotropy ($k_r, k_z$), and size effects; multilayer deposition, contact metallization, and hybrid wafer bonding create dense stacks of internal interfaces governed by boundary conductances ($G$)~\citep{warzoha2024buried,shen2026beol,lu2026process}; and energetic processes such as ion implantation and plasma etching---alongside intentional compositional grading---generate continuous depth-dependent conductivity profiles, $k(z)$~\citep{pfeifer2022measuring,huynh2023depth,xiang2023machine}. While first-principles density functional theory coupled to the Boltzmann transport equation (DFT-BTE) and molecular dynamics (MD) provide fundamental insights into pristine crystals and idealized boundaries~\citep{sellan2010size,lindsay2019perspective,mcgaughey2019phonon,chen2021nonfourier}, they cannot a priori parameterize the process-dependent disorder, interfacial intermixing, and defect gradients of fabricated device stacks. Direct experimental characterization is therefore indispensable~\citep{jiang2018tutorial}, demanding metrologies that combine multi-regime physical fidelity with well-constrained, computationally efficient parameter extraction.

Over the past four decades, transient thermal-wave metrologies---notably pump--probe time-domain (TDTR)~\citep{cahill2004analysis,mohan2025tdtr}, frequency-domain (FDTR)~\citep{schmidt2009frequency,malen2011optical,regner2013broadband,kirsch2024}, steady-state~\citep{braun2019steady}, spatial-domain~\citep{jiang2022spatial}, and square-pulsed-source thermoreflectance~\citep{song2025sps}, alongside photothermal radiometry~\citep{kusiak2013using,chatterjee2024photothermal} and electrothermal $3\omega$ methods~\citep{cahill1990thermal,kim1999application,borcatasciuc2001data}---have served as the primary experimental paradigms for probing micro- and nanoscale heat transport~\citep{cahill2014nanoscale,jiang2018tutorial}. Across these techniques, target transport properties---including cross-plane and in-plane thermal conductivities ($k_z, k_r$) and interfacial thermal conductances ($G$)---are extracted by fitting a forward heat-transfer model to measured phase or amplitude spectra. Standard planar stacks traditionally employ the semi-analytical Feldman transfer-matrix formulation~\citep{feldman1999algorithm} in Hankel or Fourier space~\citep{cahill2004analysis,schmidt2009frequency}, while three-dimensional nanostructures require numerical solvers~\citep{warzoha2024buried,lu2026process}. A common baseline across both regimes is the classical surface-heating, single-temperature model (SH-STM), which rests on four constitutive assumptions: (i)~absorbed pump power is deposited strictly as an idealized surface heat flux; (ii)~energy carriers reside in instantaneous local thermodynamic equilibrium; (iii)~heat conduction obeys local Fourier diffusion with a scalar interface $G$; and (iv)~the specimen consists of discrete, piecewise-homogeneous planar sublayers.

These foundational assumptions break down when characteristic probing scales---laser spot radius $w_r$, layer thickness $d$, or thermal penetration depth $\delta_p \approx \sqrt{k/(\pi f C)}$ (at modulation frequency $f$ and volumetric heat capacity $C$)---approach microscopic carrier relaxation lengths, optical absorption depths, or spatial defect gradients. When this occurs, SH-STM regressions compensate for omitted physics by distorting free parameters, yielding measurement-dependent ($f, w_r$) apparent values of $G$ and $k$ despite matching experimental spectra within noise tolerances~\citep{yang2016analysis,jiang2017anisotropic,sood2019quasiballistic}. First, the surface-flux assumption fails when transducer thickness $d$ approaches the optical absorption depth $d_{\mathrm{opt}} = \lambda/(4\pi\kappa)$ (at wavelength $\lambda$ and extinction coefficient $\kappa$), allowing pump and probe light to penetrate into underlying layers. Multi-interface interference volumetrically distributes pump absorption and probe reflection across the stack~\citep{yeh1988optical,yang2016analysis,guo2026light,he2026interference}; omitting these depth profiles introduces high-frequency phase lags that SH regressions misattribute to thermal resistance, depressing $k_z$ or $G$ by up to 30\% in FDTR studies of semiconductors~\citep{yang2016analysis,jiang2017anisotropic,guo2026light}---a challenge mirrored by subsurface infrared emission in photothermal radiometry~\citep{nordal1979photothermal,kusiak2013using,breitenstein2018lockin}. Second, local equilibrium fails in noble-metal transducers where optical absorption selectively heats electrons, while weak electron--phonon coupling $g_{\mathrm{ep}}$ and dominant electronic conduction ($k_{\mathrm{el}} \gg k_{\mathrm{ph}}$) sustain carrier nonequilibrium across a characteristic equilibration length $L_{\mathrm{ep}} \approx \sqrt{k_{\mathrm{ph}}/g_{\mathrm{ep}}}$ comparable to $d$~\citep{majumdar2004role,lin2008electron,wang2012limits,li2015thermal,zhang2023epcoupling}. Across metallic boundaries with coupled electronic and phononic channels~\citep{chen2022interfacial,tong2022augan,chen2024algan,desanfeliciano2025metalsilicon}, STM regressions lump the series equilibration resistance ($L_{\mathrm{ep}}/k_{\mathrm{ph}}$) into an apparent $G$, underestimating intrinsic conductances by 20--60\%~\citep{majumdar2004role,wangruanroy2012ttmnemd,wuluo2015nonmonotonic,regner2015interpretation}. Third, local Fourier diffusion fails in crystalline semiconductors (e.g., Si, GaAs, SiC) when the Knudsen number $\mathrm{Kn} = \lambda_{\mathrm{ph}}/\min(w_r, \delta_p) \gtrsim 0.05\text{--}0.1$, as phonons with long mean free paths (MFPs, $\lambda_{\mathrm{ph}}$) traverse the heated region ballistically without thermalizing collisions~\citep{chen1996nonlocal,koh2007frequency,sellan2010cross,siemens2010quasi,minnich2011thermal,minnich2012determining,regner2013broadband,regner2013mfpfdtr,hua2014transport,wilson2014anisotropic,chen2021nonfourier}. Fourier regressions misinterpret the elevated hotspot temperature as a suppressed apparent $k$  that varies artifactually with $w_r$ and $f$~\citep{regner2014analytical,yang2015heating,li2025complexk}. Fourth, piecewise-homogeneous layering fails in ion-irradiated or functionally graded media with continuous conductivity profiles $k(z)$~\citep{huynh2023depth,xiang2023machine}; fitting the conductivities of many fine sublayers independently increases the inverse dimension, inducing severe cross-correlations and non-uniqueness.

Reliable parameter extraction in heterogeneous micro- and nanostructures requires a forward model that combines the relevant optothermal transport mechanisms and remains computationally efficient for iterative inversion~\citep{pfeifer2025limitations}. Mathematically coupling these regimes is non-trivial: subsurface optical standing waves inject spatially distributed heat into the electronic subsystem, which relaxes into the lattice across nanometer-scale nonequilibrium boundary layers that directly feed wavevector-dependent nonlocal phonon modes in the underlying substrate. Because these mechanisms modify overlapping high-frequency phase bands in opposite or correlated directions, correcting any single physical axis in isolation leaves residual compensation errors across the remaining parameters. Although prior model extensions have targeted individual SH-STM breakdowns, existing formulations remain fragmented (\Cref{fig:arch}A--D). Optical transfer-matrix (OTh) models resolve multi-interface interference and subsurface absorption in semitransparent van der Waals crystals and irradiated multilayers~\citep{guo2026light,he2026interference}, yet remain restricted to Fourier diffusion and local equilibrium. Conversely, two-temperature models (TTMs) capture electron--phonon nonequilibrium across noble-metal/semiconductor interfaces in high-frequency FDTR and ultrafast TDTR~\citep{regner2015interpretation,wuluo2015nonmonotonic}, but rely on simplified Beer--Lambert heating and omit quasi-ballistic phonon suppression. For non-Fourier transport, deterministic and synthetic BTE solvers such as OpenBTE~\citep{romano2021openbte} and GiftBTE~\citep{hu2024giftbte} resolve non-gray phonon spectra in porous nano-silicon and 3D device microstructures, but nesting microscopic BTE iterations inside multi-parameter regressions remains computationally prohibitive. Alternatively, mesoscopic non-Fourier (NF) continuum models based on the Guyer--Krumhansl (GK) equation~\citep{guyer1966solution,beardo2020phonon,sendra2021derivation,alajlouni2021geometrical,xiang2022time,kovacs2024nonfourier} project linearized BTE dynamics onto a nonlocal length $\ell$ and flux relaxation time $\tau$~\citep{beardo2020phonon,beardo2026theory} to reconcile quasi-ballistic FDTR and TDTR measurements in crystalline $\mathrm{Si}$ and $\mathrm{Ge}$---isolating bulk flux-curvature damping and nonequilibrium boundary jumps from the intrinsic conductivity $k$ while preserving the fitting degrees of freedom of Fourier's law at minimal computational cost.  Meanwhile, for spatially graded media, continuous functional representations avoid independently fitting sublayer conductivities to infer depth-dependent $k(z)$ directly~\citep{huynh2023depth,xiang2023machine}, yet remain isolated from generalized multilayer solvers. Together, these complementary capabilities motivate a reconfigurable model hierarchy that selectively activates optical, nonequilibrium, nonlocal, and continuous-gradient physics as dictated by the transport regime, addressing model-form mismatch while preserving computational efficiency and screening fit targets before optimization.

Compounding this physical fragmentation is the mathematical ill-conditioning of multi-parameter thermal-wave measurement inversion. Because measured phase spectra integrate the multilayer's cumulative thermal response, optimization landscapes often feature shallow basins, high parameter correlation, and ill-posed Jacobian null-spaces~\citep{tu2023universal,tao2024unraveling,chen2024svd}. Conventional workflows require tuning bounds by trial and error, assessing parameter identifiability, and selecting physical fitting models that may mismatch or deviate for unknown, complex geometries. Although data-driven machine learning (ML) surrogates---such as feedforward neural networks (ANNs) for rapid regression~\citep{pang2021machine,mao2024deep} or kernel ridge regression (KRR) for depth profiling~\citep{xiang2023machine}---accelerate parameter estimation, their applicable transport regimes and profile representations are set by the solvers and parameter ranges used to construct their training libraries (Figure~\ref{fig:arch}E)~\citep{hinton2006reducing,kingma2013auto,raissi2019physics,karniadakis2021physics}. Networks for graded media likewise require a specified profile representation and training prior~\citep{ikeda2025}. Agentic frameworks such as Vibe-FDTR~\citep{yang2026vibefdtr} automate scripting, sensitivity analysis, and execution; PRISM integrates these workflow capabilities with explicit physical-branch selection and user-supervised solver extensions (Figure~\ref{fig:arch}E,F).

Here, we introduce PRISM (Physics-Resolved Inference and Screening Meta-framework)---an agentic scientific architecture that unifies an expandable hierarchy of optothermal transport solvers with automated model routing and identifiability screening to resolve ill-posed multi-parameter inversions in micro- and nanoscale thermal metrology (\Cref{fig:arch}). Implemented for frequency-domain thermoreflectance (PRISM-FDTR), the framework pairs a unified transfer-matrix and admittance formalism---coupling optical transfer-matrix (OTh) electrodynamics, two-temperature (TTM) electron--phonon relaxation, and nonlocal GK quasi-ballistic phonon transport---with a stateful execution layer that routes transport models via dimensionless physical scaling criteria, prunes unidentifiable parameters through noise-whitened singular value decomposition (SVD), and executes reduced-space global optimization via domain-transform differential evolution (DTDE)~\citep{zhang2023dtde}. When structured residuals or cross-configuration discrepancies indicate unmodeled physics, the agent diagnoses the regime breakdown and either escalates the physical branch or autonomously constructs, trains, and verifies new solver modules through a knowledge-graph-guided workflow (\Cref{fig:arch}F). Across experimental benchmarks representative of microelectronic contacts, dielectric barriers, and 2D heterostructures, PRISM-FDTR resolves pervasive cross-regime parameter distortions: it decouples the near-interface electron--phonon equilibration resistance in $\text{Au/Si}$ (correcting a 24--28\% underestimation in $G_{\mathrm{Au/Si}}$), recovers intrinsic $k_{\mathrm{SiO}_2}$ in $\text{Au/Ti/SiO}_2$, and eliminates a 40\% underestimation in cross-plane $k_{\mathrm{MoS}_2,z}$ caused by subsurface optical heating in thin-transducer $\text{Au/MoS}_2$. In bimetallic $\text{Au/Ti/Si}$ contacts, where subsurface optical absorption, electron--phonon nonequilibrium, and quasi-ballistic phonon transport act simultaneously, nonlocal modeling reconciles the severe spot-size and frequency-window scatter of local Fourier fits ($k_{\mathrm{app}} = 116\text{--}\SI{137}{\watt\per\metre\per\kelvin}$) into a configuration-independent intrinsic bulk silicon conductivity of $140.5\text{--}\SI{141.4}{\watt\per\metre\per\kelvin}$. Furthermore, for ion-irradiated and functionally graded media with continuous conductivity profiles $k(z)$, we show that PRISM autonomously constructs and verifies a deep operator network (DeepONet)~\citep{lu2021learning} solver across six gradient families, avoiding discrete-layer overfitting and reducing root-mean-square percentage error (RMSPE) from 3.3\% to 2.7\% ($R^2 = 0.9986$, an 18\% error reduction relative to kernel ridge regression). Extensible across transfer-matrix thermal-wave techniques (including FDTR, TDTR, $3\omega$, and photothermal radiometry), PRISM establishes a generalizable paradigm for reliable, configuration-invariant nanoscale thermal characterization.

\section{Results}
\label{sec:results}

\subsection{Physics-resolved optothermal inference and agentic workflow}
\label{sec:inference_framework}

\begin{figure}[!htbp]
\centering
\includegraphics[width=0.99\textwidth]{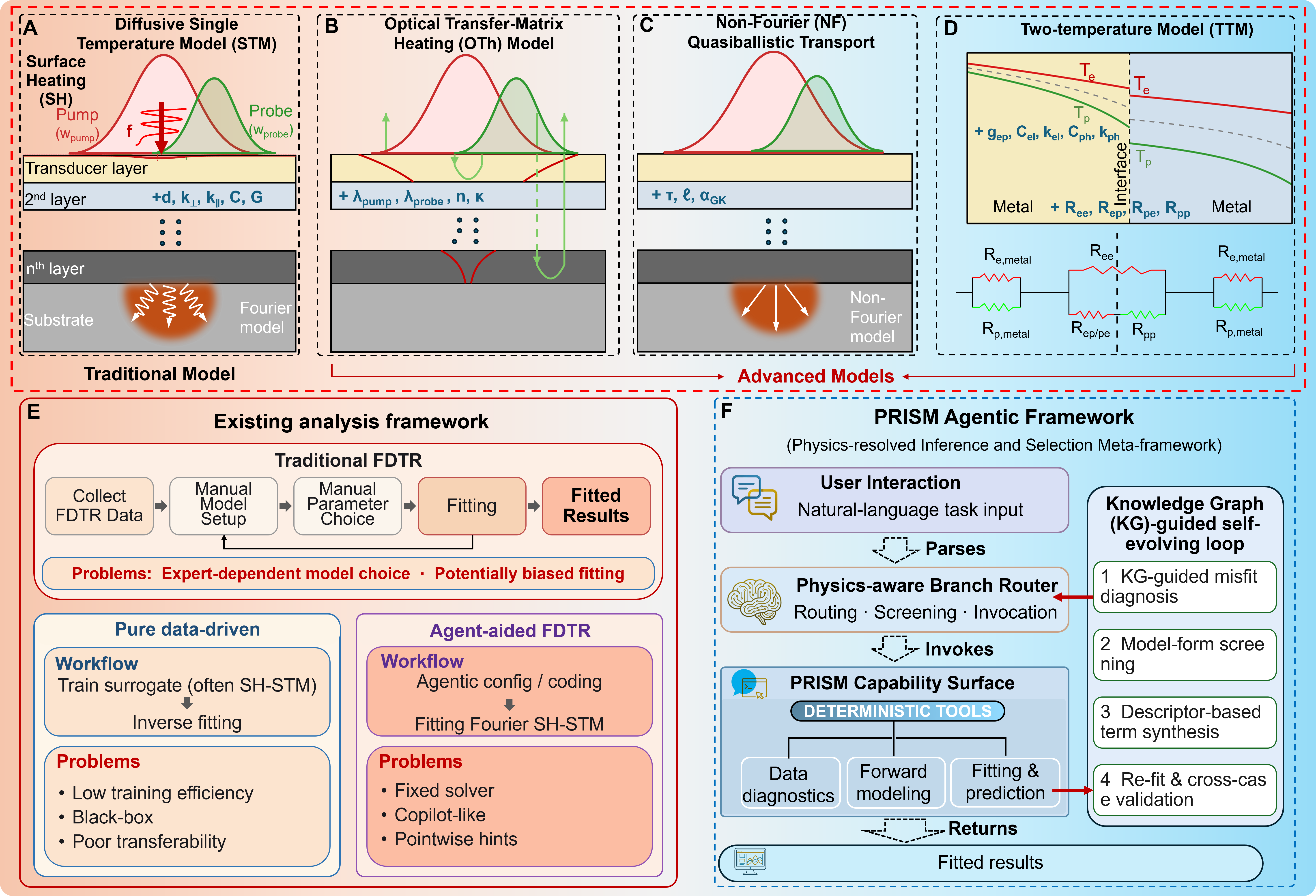}
\caption{PRISM physics-resolved model hierarchy and agentic architecture.
(A--D) Forward-model hierarchy spanning the traditional baseline (A) and advanced transport formulations (B--D).
(A) Diffusive surface-heating, single-temperature model (SH-STM): Gaussian pump and probe spots (radii $w_{\mathrm{pump}}, w_{\mathrm{probe}}$) illuminate a transducer-coated multilayer stack defined by layer properties (thickness $d$, conductivities $k_x,k_y,k_z$, volumetric heat capacity $C$) and interface conductance $G$ under pump modulation frequency $f$, assuming local Fourier diffusion under surface heat-flux boundary conditions.
(B) Optical transfer-matrix (OTh) model: pump and probe field propagation at wavelengths $\lambda^{\mathrm{pump}}$ and $\lambda^{\mathrm{probe}}$ using complex refractive indices ($n, \kappa$) resolves depth-resolved volumetric absorption, multi-interface interference, and subsurface probe reflection.
(C) Non-Fourier heat transfer (NF): mesoscopic nonlocal Guyer--Krumhansl (GK) transport incorporating relaxation time $\tau$, nonlocal length $\ell$, and weighting factor $\alpha_{\mathrm{GK}}$ to capture quasi-ballistic phonon suppression.
(D) Two-temperature model (TTM): independent electron ($T_e$) and phonon ($T_p$) depth relaxation in metallic films mediated by volumetric coupling $g_{\mathrm{ep}}$, decomposing boundary transport into multi-channel resistances ($R_{\mathrm{ee}}, R_{\mathrm{ep/pe}}, R_{\mathrm{pp}}$).
(E) Analysis workflows: expert-led fitting combines model setup, sensitivity analysis, and regression; data-driven surrogates learn phase-to-property mappings from a training library; agent-assisted workflows automate scripting and tool execution.
(F) PRISM agentic execution framework: a branch router parses natural-language tasks and coordinates data diagnosis, forward modeling, and SVD identifiability screening. Systematic discrepancies trigger a knowledge-graph-guided extension loop comprising physical diagnosis, model-form screening, module construction, and cross-case validation, with researcher review before a new module enters routine use.
Branch names default to Fourier conduction (e.g., OTh-TTM), with GK transport denoted by an (NF) suffix (e.g., OTh-TTM(NF)).}
\label{fig:arch}
\end{figure}

\Cref{fig:arch} presents the PRISM forward-model hierarchy (\Cref{fig:arch}A--D) and execution workflow (\Cref{fig:arch}F), alongside existing analysis workflows (\Cref{fig:arch}E). In the forward hierarchy, optical transfer-matrix electrodynamics (OTh), nonlocal GK phonon transport (NF), and nonequilibrium carrier relaxation (TTM) are formulated as compatible transfer-matrix and admittance blocks around the diffusive SH-STM baseline. The execution layer (\Cref{fig:arch}F) parses task schemas, evaluates physical gating criteria to assemble the required solver branch, screens identifiable targets via SVD, and coordinates global optimization. When structured residuals or cross-configuration discrepancies indicate omitted physics, the agent queries a cited domain knowledge graph to revise the governing transport model or construct a validated solver module for researcher review.

Transient thermal-wave metrologies infer transport properties through non-linear regression of a forward heat-transfer model against experimental signals recorded across modulation frequencies or delay times~\citep{cahill1990thermal,cahill2004analysis,schmidt2009frequency,jiang2018tutorial,chatterjee2024photothermal}. We formulate this inverse problem generically as:
\begin{equation}
\mathbf{y}_{\mathrm{obs}}  = \mathcal{F}_{\mathcal{M}}\!\left(\mathbf{p}_{\mathrm{fit}}, \mathbf{p}_{\mathrm{fixed}}\right) + \boldsymbol{\epsilon},
\label{eq:observation_model}
\end{equation}
where, taking FDTR as the primary platform, $\mathbf{y}_{\mathrm{obs}} \in \mathbb{R}^{N_f}$ is the experimental lock-in phase vector $\boldsymbol{\phi} = [\phi(f_1), \dots, \phi(f_{N_f})]^T$ across $N_f$ frequencies, $\mathbf{p}_{\mathrm{fit}}$ denotes the reduced set of identifiable targets, $\mathbf{p}_{\mathrm{fixed}}$ collects independently known inputs---layer thicknesses $d$, heat capacities $C$, conductivity tensors $\mathbf{k}$ (reducing to $\mathrm{diag}(k_x, k_y, k_z)$ along principal orthotropic axes, or to $k_r$ and $k_z$ for transversely isotropic layers), complex refractive indices $\tilde{n} = n + i\kappa$, laser wavelengths ($\lambda^{\mathrm{pump}}, \lambda^{\mathrm{probe}}$), calibrated beam radii ($w_{\mathrm{pump}}, w_{\mathrm{probe}}$), and lateral offset $\Delta$---and $\boldsymbol{\epsilon} \sim \mathcal{N}(\mathbf{0}, \boldsymbol{\Sigma}_\phi)$ is zero-mean phase noise with covariance $\boldsymbol{\Sigma}_\phi$~\citep{tu2023universal,chen2024svd}. Reliable inversion of \Cref{eq:observation_model} requires both configuring the physical operator $\mathcal{F}_{\mathcal{M}}$ so that unmodeled transport mechanisms do not bias the fit, and partitioning the parameters into $\mathbf{p}_{\mathrm{fit}}$ and $\mathbf{p}_{\mathrm{fixed}}$ so that the regression remains well-conditioned under noise $\boldsymbol{\epsilon}$.

To configure the forward model systematically and resolve the coupled transport breakdowns characteristic of real heterostructures, PRISM organizes $\mathcal{F}_{\mathcal{M}}$ over a four-dimensional discrete model space:
\begin{equation}
\mathcal{M} = \left(\mathcal{E}, \mathcal{X}, \mathcal{T}, \mathcal{R}\right) \in \mathcal{S}_{\mathcal{E}} \times \mathcal{S}_{\mathcal{X}} \times \mathcal{S}_{\mathcal{T}} \times \mathcal{S}_{\mathcal{R}},
\label{eq:model_space_compact}
\end{equation}
{\RaggedRight
where the four dimensions specify: (i)~optical heating and detection $\mathcal{S}_{\mathcal{E}} = \{\mathrm{SH}, \mathrm{VH}, \mathrm{OTh}\}$ (with $\mathrm{VH}$ denoting Beer--Lambert volumetric heating), (ii)~electron--phonon coupling and thermalization $\mathcal{S}_{\mathcal{X}} = \{\mathrm{STM}, \mathrm{TTM}\}$, (iii)~phonon transport $\mathcal{S}_{\mathcal{T}} = \{\mathrm{Fourier}, \mathrm{NF}\}$, and (iv)~spatial depth parameterization $\mathcal{S}_{\mathcal{R}} = \{\mathrm{discrete}, \mathrm{continuous}\}$. PRISM accounts for the simultaneous effects of optical penetration, electron--phonon nonequilibrium, and quasi-ballistic phonon transport in multilayer stacks---where omitting one mechanism biases the parameters extracted along another (as demonstrated in our $\mathrm{Au/Ti/SiO}_2$ and $\mathrm{Au/Ti/Si}$ benchmarks below). By formulating optical absorption, electron--phonon relaxation, and nonlocal phonon conduction within a unified transfer-matrix and admittance framework, PRISM integrates these modules into a composable solver library where any combined branch (e.g., $\mathrm{OTh}\text{-}\mathrm{TTM(NF)}$) can be assembled from the corresponding model components (\hyperref[sec:methods]{materials and methods}, sections~\ref{sec:tmm} and~\ref{sec:optical_tmm} of the supplemental information). Guided by dimensionless physical gating criteria, SVD identifiability screening, and post-fit cross-configuration consistency checks, the agent retains the baseline $\mathrm{SH}\text{-}\mathrm{STM}$ model where valid and escalates to advanced transport models only when sample dimensions or probing scales cross regime thresholds.
\par}

In optical excitation and detection modeling ($\mathcal{S}_{\mathcal{E}}$), classical surface heating ($\mathrm{SH}$, $q_0(x,y) \propto I_{\mathrm{pump}}(x,y)$) and Beer--Lambert volumetric heating ($\mathrm{VH}$) suffice for optically thick transducers ($d \gg d_{\mathrm{opt}}$). However, in thin or semitransparent metallic transducers ($d \sim 1\text{--}3\,d_{\mathrm{opt}}$) employed to enhance substrate thermal sensitivity, subsurface light penetration invalidates the surface-flux assumption. To resolve this, PRISM activates the optical transfer-matrix ($\mathrm{OTh}$) module~\citep{yeh1988optical,guo2026light,he2026interference}, solving Maxwell's boundary-value problem across the multilayer stack. In each layer $n$ (thickness $d_n$), the pump field establishes a depth-resolved absorption profile per unit incident irradiance ($\mathrm{m^{-1}}$): $Q_n(z) = Q_{n,\mathrm{fwd}} e^{-\alpha_{\mathrm{opt},n}^{\mathrm{pump}} z} + Q_{n,\mathrm{bwd}} e^{-\alpha_{\mathrm{opt},n}^{\mathrm{pump}}(d_n - z)} + Q_{n,\mathrm{int}} \cos(\beta_{\mathrm{opt},n}^{\mathrm{pump}} z + \varphi_n)$, where $\alpha_{\mathrm{opt},n}^{\mathrm{pump}} = 4\pi\kappa_n/\lambda^{\mathrm{pump}}$ is the optical attenuation coefficient, $\beta_{\mathrm{opt},n}^{\mathrm{pump}} = 4\pi n_n/\lambda^{\mathrm{pump}}$ is the standing-wave wavenumber, $\varphi_n$ is the interference phase shift, and $Q_{n,\mathrm{fwd}}$, $Q_{n,\mathrm{bwd}}$, and $Q_{n,\mathrm{int}}$ are the forward, backward, and interference absorption coefficients. Multiplying $Q_n(z)$ by the lateral Gaussian pump intensity $I_{\mathrm{pump}}(x,y) \propto \exp(-2x^2/w_{\mathrm{pump},x}^2 - 2y^2/w_{\mathrm{pump},y}^2)$ yields the three-dimensional volumetric heat source $S_n(x,y,z) = I_{\mathrm{pump}}(x,y) Q_n(z)$ that drives the heat conduction equations in each layer. Concurrently, subsurface probe sensing is modeled by integrating the oscillating temperature field over depth against the optical sensitivity kernel $\exp(g_{f,n} z)$, where $g_{f,n} = -\alpha_{\mathrm{opt},n}^{\mathrm{probe}} + i\beta_{\mathrm{opt},n}^{\mathrm{probe}}$ accounts for probe attenuation and optical phase accumulation at $\lambda^{\mathrm{probe}}$ (\hyperref[sec:methods]{materials and methods}, section~\ref{sec:volumetric_readout} of the supplemental information).

Along the electron--phonon coupling axis ($\mathcal{S}_{\mathcal{X}}$), the local thermal equilibrium $\mathrm{STM}$ baseline ($T_{\mathrm{el}} = T_{\mathrm{ph}}$) holds in transition metals with strong electron--phonon coupling (e.g., Ti, Pt, Cr; $g_{\mathrm{ep}} \sim 10^{18}~\SI{}{\watt\per\metre\cubed\per\kelvin}$~\citep{lin2008electron,tong2019comprehensive}), where electrons and phonons thermalize within a coupling length $L_{\mathrm{ep}} \sim \SI{1}{\nano\metre}$. Conversely, in noble metals with weak electron--phonon coupling (e.g., Au, Cu, Ag; $g_{\mathrm{ep}} \sim 10^{16}~\SI{}{\watt\per\metre\cubed\per\kelvin}$, $k_{\mathrm{el}} \gg k_{\mathrm{ph}}$) or bimetallic transducers where electron-dominated heat flux must transfer to the lattice before entering a dielectric or semiconductor substrate, PRISM activates the two-temperature model ($\mathrm{TTM}$; \hyperref[sec:methods]{materials and methods}, section~\ref{sec:ttm_model} of the supplemental information)~\citep{majumdar2004role,regner2015interpretation}:
\begin{align}
C_{\mathrm{el},n}\frac{\partial T_{\mathrm{el},n}}{\partial t} &= \nabla \cdot \left(\mathbf{k}_{\mathrm{el},n}\nabla T_{\mathrm{el},n}\right) - g_{\mathrm{ep},n}\left(T_{\mathrm{el},n} - T_{\mathrm{ph},n}\right) + I_{\mathrm{pump}}(x,y)Q_n(z), \label{eq:ttm_el_main} \\
C_{\mathrm{ph},n}\frac{\partial T_{\mathrm{ph},n}}{\partial t} &= \nabla \cdot \left(\mathbf{k}_{\mathrm{ph},n}\nabla T_{\mathrm{ph},n}\right) + g_{\mathrm{ep},n}\left(T_{\mathrm{el},n} - T_{\mathrm{ph},n}\right), \label{eq:ttm_ph_main}
\end{align}
for the coupled electron ($\mathrm{el}$) and phonon ($\mathrm{ph}$) subsystems. Interfacial heat transfer across metallic boundaries is formulated via a four-element conductance matrix $\mathbf{G}_j = \begin{pmatrix} G_{\mathrm{ee}} & G_{\mathrm{ep}} \\ G_{\mathrm{pe}} & G_{\mathrm{pp}} \end{pmatrix}$~\citep{chen2022interfacial,tong2022augan,chen2024algan}, where metal/metal contacts (e.g., $\mathrm{Au/Ti}$) are parameterized by parallel electronic ($G_{\mathrm{ee}}$) and phononic ($G_{\mathrm{pp}}$) priors, and metal/non-metal interfaces with blocked electronic transmission ($\widehat{q}_{\mathrm{el}} = 0$) reduce to the single fitted phonon-mediated conductance $G_{\mathrm{pp}}$. At such metal/non-metal interfaces, electronic heat converts into lattice vibrations across a near-interface nonequilibrium zone of width $L_{\mathrm{ep},z} \approx \sqrt{k_{\mathrm{ph},z}/g_{\mathrm{ep}}} \sim 10\text{--}15~\SI{}{\nano\metre}$ (for $k_{\mathrm{ph}} \approx 2\text{--}4~\SI{}{\watt\per\metre\per\kelvin}$ in Au~\citep{jain2016thermal,tong2019comprehensive}). In thin transducers ($d_{\mathrm{trans}} \sim 30\text{--}100~\SI{}{\nano\metre}$), PRISM explicitly decouples this volumetric electron--phonon equilibration resistance ($L_{\mathrm{ep},z}/k_{\mathrm{ph},z} \sim 4.5\times 10^{-9}~\SI{}{\metre\squared\kelvin\per\watt}$) from the intrinsic interfacial conductance (\hyperref[sec:methods]{materials and methods}, section~\ref{sec:interface_coupling} of the supplemental information). Because room-temperature phonon MFPs ($<\SI{10}{\nano\metre}$) and electron MFPs ($5\text{--}25~\SI{}{\nano\metre}$~\citep{tong2019comprehensive}, up to $\approx \SI{38}{\nano\metre}$ in Au~\citep{gall2016search}) remain shorter than typical transducer thicknesses, intra-film conduction within each subsystem remains diffusive.

For phonon transport ($\mathcal{S}_{\mathcal{T}}$), local $\mathrm{Fourier}$ diffusion holds in amorphous dielectrics and in crystalline layers when the Knudsen number $\mathrm{Kn} = \lambda_{\mathrm{ph}}/L_{\mathrm{char}} \ll 0.05$. Here, $L_{\mathrm{char}}$ is primarily set by the thermal gradient scales---spot radius $w_r$ and penetration depth $\delta_p$---acting in tandem with geometric confinement from the layer thickness $d$. When long-MFP phonons in crystalline semiconductors drive transport into the quasi-ballistic regime ($\mathrm{Kn} \gtrsim 0.05\text{--}0.1$), PRISM activates the non-Fourier module via the mesoscopic GK equation~\citep{guyer1966solution,beardo2020phonon}:
\begin{equation}
\mathbf{q} + \tau \frac{\partial \mathbf{q}}{\partial t} - \ell^2 \left[\nabla^2 \mathbf{q} + \alpha_{\mathrm{GK}} \nabla(\nabla \cdot \mathbf{q})\right] = -\mathbf{k} \nabla T,
\label{eq:gk_transport_main}
\end{equation}
with $\alpha_{\mathrm{GK}} = 1/3$ weighting the nonlocal divergence~\citep{sendra2021derivation,beardo2020phonon}. While Equation~\eqref{eq:gk_transport_main} applies generally to both finite thin films and bulk media, in our experimental benchmarks it is formulated as a semi-infinite substrate boundary module. Applying in-plane spatial transforms and Helmholtz decomposition decouples $\mathbf{q}$ into compressive and shear modes (\hyperref[sec:methods]{materials and methods}, section~\ref{sec:non_fourier_si} of the supplemental information). Enforcing normal flux continuity and diffuse tangential slip yields the bulk thermal impedance $Z_{\mathrm{bulk}}$ and nonequilibrium boundary temperature jump $T_{\mathrm{diffusive}} - T_{\mathrm{non-Fourier}} = Z_{\mathrm{jump}} q_n$, where $Z_{\mathrm{jump}}$ combines second-order nonlocal stress corrections with the Kapitza-like resistance $R = \frac{b_{\mathrm{NF}}}{2}(\gamma_0^{-1} + \gamma_{0\Gamma}^{-1})$ defined by the substrate and transducer modal energy-transfer coefficients $\gamma_0$ and $\gamma_{0\Gamma}$. Condensing this hydrodynamic response into the effective nonlocal substrate admittance $Y_{\mathrm{sub}}^{\mathrm{NF}} = (Z_{\mathrm{bulk}} + Z_{\mathrm{jump}})^{-1}$ provides the exact bottom boundary condition for the upward Schur-complement admittance recursion on the $4\times 1$ metallic TTM state vector $\boldsymbol{\Theta}_N = [\widehat{T}_{\mathrm{el}}, \widehat{T}_{\mathrm{ph}}, \widehat{q}_{\mathrm{el}}, \widehat{q}_{\mathrm{ph}}]^\mathsf{T}$, coupling the phonon heat-flux channel ($\widehat{q}_{\mathrm{ph}} = Y_{\mathrm{sub}}^{\mathrm{NF}} \widehat{T}_{\mathrm{ph}}$) across an insulated electronic boundary ($\widehat{q}_{\mathrm{el}} = 0$; Equation~\eqref{eq:nf_replacement_map}). With the Si bulk parameters at 311\,K ($\tau = \SI{42.0}{\pico\second}$, $\ell = \SI{185.0}{\nano\metre}$, $\alpha_{\mathrm{GK}} = 1/3$) and reference modal transfer coefficients ($\gamma_0, \gamma_{0\Gamma}$) fixed to the literature values of Ref.~\citep{beardo2020phonon} (Table~\ref{tab:archive_si_nf_params}), the dimensionless contact factor $b_{\mathrm{NF}}$ serves as the sole free interfacial parameter in the non-Fourier branch, directly replacing $G$ with equal degrees of freedom (where the extracted equivalent conductance $G_{\mathrm{equiv}} = 1/R$ is invariant to the reference prior chosen for $\gamma_{0\Gamma}$) and reducing analytically to Fourier diffusion as $\tau, \ell \to 0$.

\begin{figure}[!htbp]
  \centering
  \includegraphics[width=0.99\textwidth]{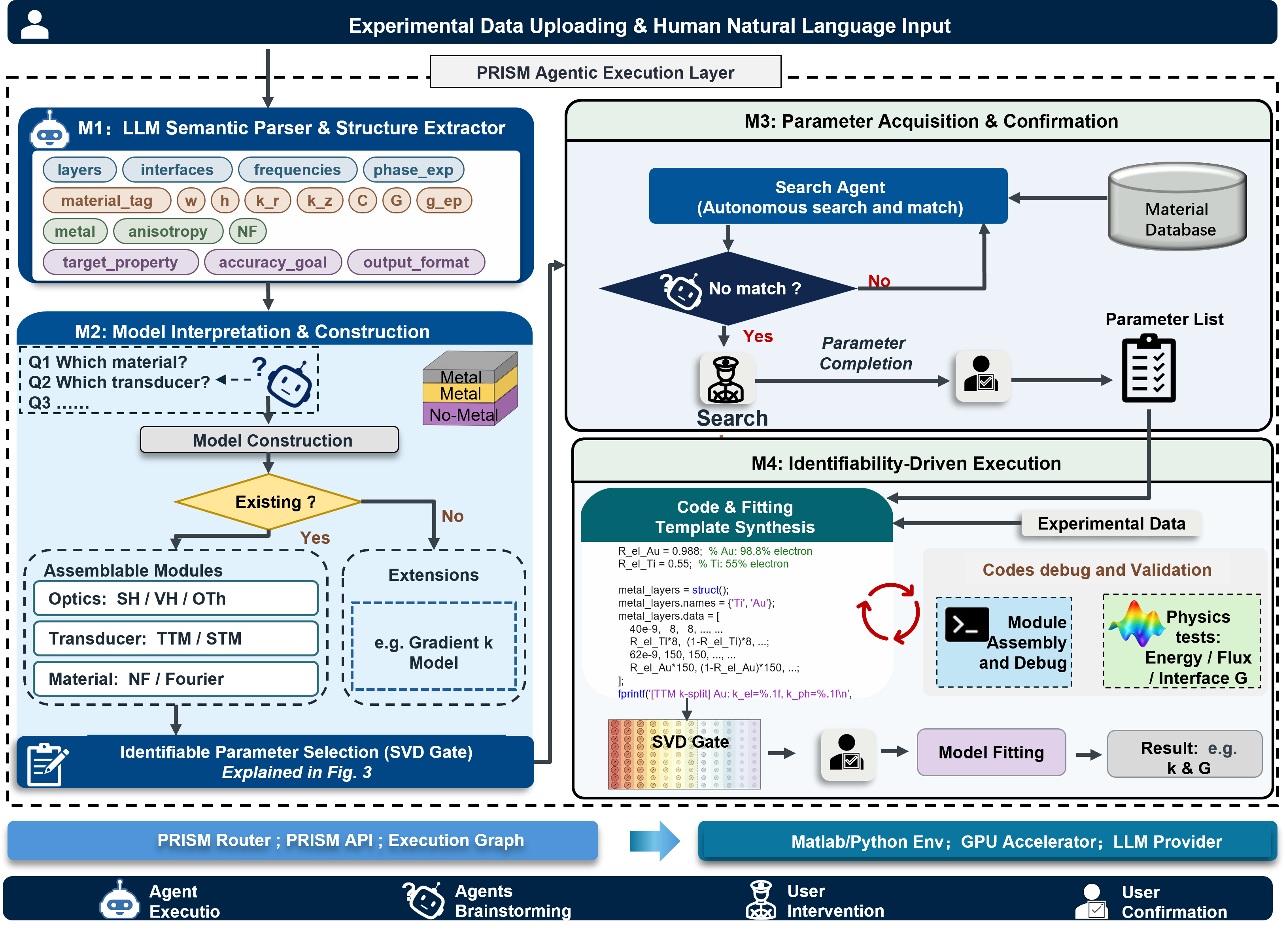}
  \caption{Architecture and operational workflow of the PRISM Agentic Execution Layer. Module 1 (LLM Semantic Parser \& Structure Extractor) ingests natural-language task requests and raw measurement data, mapping them to structured, typed stack schemas. Module 2 (Model Interpretation \& Construction) evaluates physical gating criteria (optical absorption depth, transducer nonequilibrium, ballistic Knudsen scaling) and asks whether the required branch already exists: if it does, the solver is assembled from the composable optics (SH / VH / OTh), transducer (STM / TTM), and substrate-transport (NF / Fourier) modules; if it does not, the branch is added through a user-supervised extension, for example the gradient-$k$ model of Figure~\ref{fig:deeponet}. The assembled model proceeds to parameter acquisition and subsequent SVD screening (Figure~\ref{fig:workflow}). Module 3 (Parameter Acquisition \& Confirmation) dispatches an autonomous search agent against a material database to populate initial parameter values, looping back whenever no match is returned and escalating to parameter completion and user confirmation when uncataloged properties occur, before emitting the confirmed parameter list. Module 4 (Identifiability-Driven Execution) takes that parameter list together with the experimental data, synthesizes the code and fitting template, and iterates module assembly and debugging against physical conservation tests (energy, heat flux, interface conductance) until the solver passes; the SVD gate then prunes unidentifiable targets and multi-algorithm reduced-space optimization (DTDE, CMA-ES) returns thermal properties with propagated uncertainties such as $k$ and $G$. The layer runs on the PRISM router, PRISM API, and execution graph, backed by MATLAB/Python environments, GPU acceleration, and a large language model (DeepSeek V4 Flash); the icons distinguish autonomous agent execution and agent brainstorming from the two points of human intervention and confirmation.}
  \label{fig:routing_framework}
\end{figure}

\begin{figure}[!htbp]
  \centering
  \includegraphics[width=0.99\textwidth]{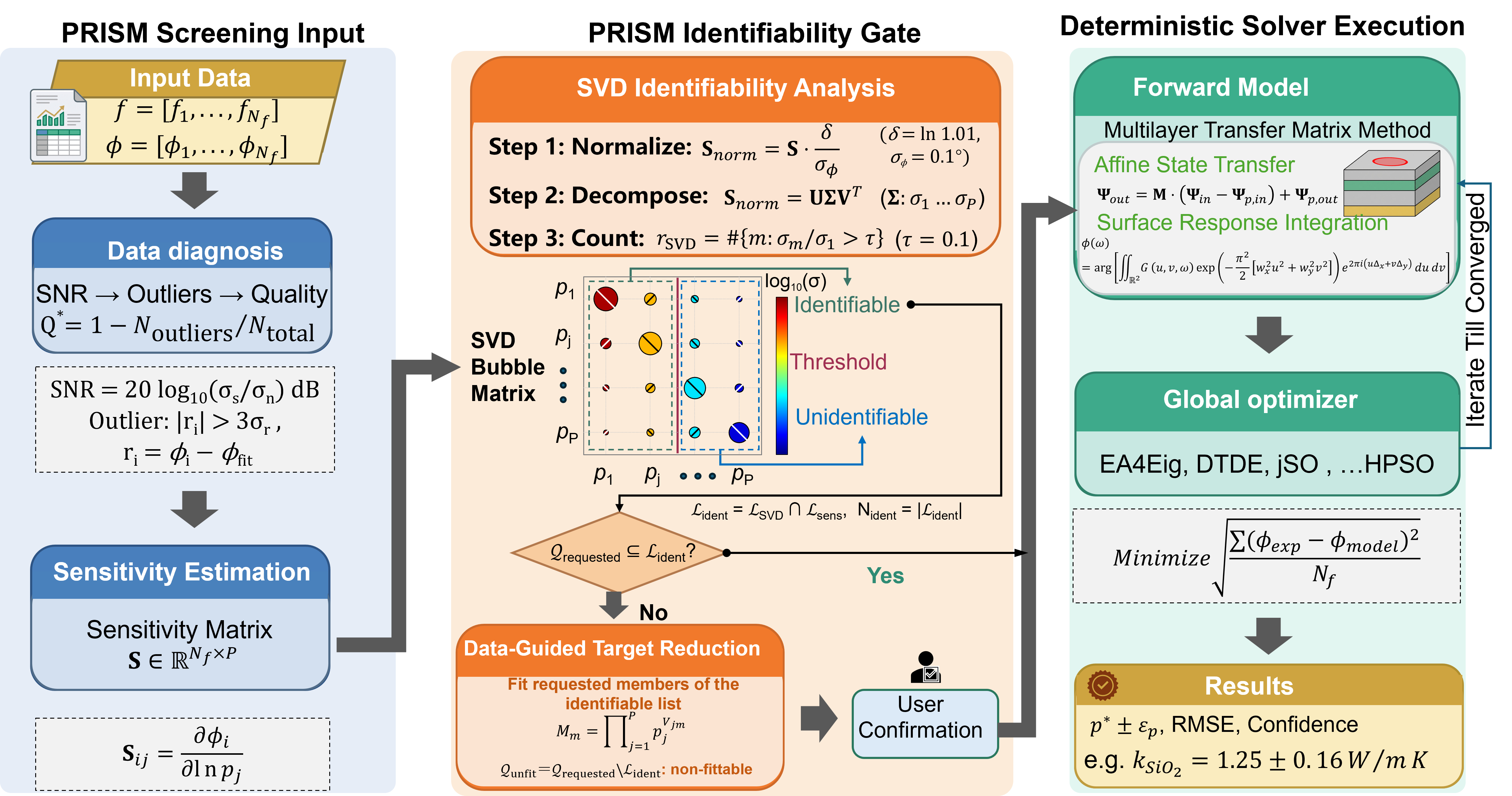}
  \caption{SVD-guided parameter screening and optimization workflow. Stage 1 (PRISM Screening Input) receives frequency and measured phase vectors $f=[f_1,\ldots,f_{N_f}]$ and $\boldsymbol\phi=[\phi_1,\ldots,\phi_{N_f}]$, diagnoses signal-to-noise ratio, residual outliers and data quality $Q^*$, and constructs $\mathbf S\in\mathbb R^{N_f\times P}$, with $S_{ij}=\partial\phi(f_i)/\partial\ln p_j$. Here $N_f$ counts frequencies used at the corresponding stage, and $P$ counts candidate parameters $p_j$; diagnostic scales and rejected-point counts are defined in section~\ref{sec:quality_gate} of the supplemental information. Stage 2 (PRISM Identifiability Gate) normalizes $\mathbf S_{\mathrm{norm}}=(\delta/\sigma_\phi)\mathbf S$, with $\delta=\ln1.01$ and $\sigma_\phi=0.1^\circ$, and decomposes it as $\mathbf U\boldsymbol\Sigma\mathbf V^T$. Columns represent right singular vectors (singular modes) ordered by decreasing singular value, and rows identify parameters. The parameter ranking below the matrix is independent of the mode-column order; its divider separates eligible from excluded parameters. Bubble area increases with $|V_{jm}|$, and color encodes $\log_{10}\sigma_m$ for the corresponding mode column (raw-Jacobian singular values in the numerical diagnostic charts). The diagonal marks $\backslash$ and $/$ denote positive and negative components, respectively, after fixing each column's largest-magnitude component positive. The retained mode index set is $\mathcal{I}_{\mathrm{SVD}}=\{m:\sigma_m/\sigma_1>\tau\}$, where the figure's dimensionless $\tau\equiv\tau_{\mathrm{SNR}}=0.1$ and $r_{\mathrm{SVD}}=|\mathcal{I}_{\mathrm{SVD}}|$. Parameter projections and retained sensitivity energy determine the ordered list $\mathcal L_{\mathrm{SVD}}$ (Algorithm~\ref{alg:svd} in the supplemental information; Equation~\eqref{eq:svd_parameter_score}). Filtering by raw RMS sensitivity $\sqrt{N_f^{-1}\sum_i|S_{ij}|^2}\ge S_{\min}=0.2^\circ$ gives $\mathcal L_{\mathrm{ident}} = \mathcal{L}_{\mathrm{SVD}} \cap \mathcal{L}_{\mathrm{sensitivity}}$, with $N_{\mathrm{ident}}=|\mathcal L_{\mathrm{ident}}|$, preserving the parameter order of $\mathcal L_{\mathrm{SVD}}$. The diagram's $\mathcal L_{\mathrm{sens}}$ abbreviates $\mathcal L_{\mathrm{sensitivity}}$. Requested members of this intersection are admitted to fitting under the declared model and fixed inputs; excluded requests form $\mathcal Q_{\mathrm{unfit}}=\mathcal Q_{\mathrm{requested}}\setminus\mathcal L_{\mathrm{ident}}$. The displayed $M_m=\prod_jp_j^{V_{jm}}$ describes a composite singular direction. The list of fitted parameters is presented for user confirmation. Stage 3 (Deterministic Solver Execution) propagates the affine thermal state and integrates the two-dimensional Fourier response using independent widths $w_x,w_y$ and offsets $\Delta_x,\Delta_y$. In Stage~3 of the diagram, $\boldsymbol\Psi \equiv \boldsymbol\Theta$ and $\boldsymbol\Psi_p \equiv \boldsymbol\Theta_p$ denote the four-component layer thermal state vector and its particular solution (section~\ref{sec:affine_transfer} of the supplemental information; distinct from the scalar GK shear streamfunction $\Psi$ in section~\ref{sec:nf_helmholtz}), and $G(u,v,\omega) \equiv \mathcal{K}(u,v,\omega)$ denotes the unit-pump-Fourier-amplitude linear response kernel defined in Equation~\eqref{eq:workflow_response_kernel} (distinct from the interfacial thermal conductance $G$). A configured global optimizer minimizes phase RMSE, with DTDE as the default (section~\ref{sec:optimizer_benchmark} of the supplemental information). The Results box reports best-fit $p^*$, propagated standard uncertainty $\epsilon_n$, and fit diagnostics; the illustrative $1.25\pm0.16$ conductivity is quoted at one standard deviation.}
  \label{fig:workflow}
\end{figure}

To automate model routing and prevent ill-conditioned regressions across $\mathcal{M}$, PRISM's execution layer coordinates a four-stage pipeline (\Cref{fig:routing_framework}; \hyperref[sec:methods]{materials and methods} and section~\ref{sec:agent_system_overview} of the supplemental information). Module~1 parses natural-language inputs and raw spectra into a structured stack schema while screening spectral signal-to-noise ratio and rejecting statistical outliers (Figure~\ref{fig:diagnose-data}). Module~2 evaluates dimensionless physical scaling criteria from the domain knowledge graph---the optical penetration ratio $d_{\mathrm{trans}}/d_{\mathrm{opt}}$, carrier equilibration ratio $d_{\mathrm{trans}}/L_{\mathrm{ep},z}$, and Knudsen number $\mathrm{Kn} = \lambda_{\mathrm{ph}}/\min(w_r, \delta_p)$---alongside forward-phase deviation gates. Prior to fitting, the agent simulates candidate branches at the nominal stack parameters and activates $\mathrm{OTh}$ or $\mathrm{TTM}$ whenever their maximum phase deviation from the baseline across the modulation band exceeds three noise standard deviations ($\max_f\lvert\phi_{\mathrm{OTh}}-\phi_{\mathrm{SH}}\rvert > 3\sigma_\phi$ or $\max_f\lvert\phi_{\mathrm{OTh\text{-}TTM}}-\phi_{\mathrm{OTh\text{-}STM}}\rvert > 3\sigma_\phi$, e.g., when $d_{\mathrm{trans}} \lesssim 3\,d_{\mathrm{opt}}$), whereas escalation to the non-Fourier ($\mathrm{NF}$) branch is triggered when local inversions exhibit systematic parameter drift across spot sizes, beam offsets, or frequency windows (\Cref{sec:ausi_nonlocal_result,sec:nonlocal_result}). Module~3 retrieves cataloged thermophysical priors from literature repositories for user confirmation, and Module~4 compiles the C++/CUDA and MATLAB solver blocks, verifies energy conservation and interfacial flux continuity, and applies the SVD identifiability gate (\Cref{fig:workflow}) before launching global optimization. When unmodeled regimes such as continuous conductivity gradients arise, Module~2 invokes a user-supervised extension route (\Cref{fig:deeponet}; Figure~\ref{fig:si_deeponet_log}).

To limit optimizer drift along weakly constrained directions and trade-offs among correlated parameters, PRISM applies an SVD identifiability gate before regression (\Cref{fig:workflow}; \hyperref[sec:methods]{materials and methods}, section~\ref{sec:sensitivity_svd} of the supplemental information). For $P$ candidate parameters across $N_f$ modulation frequencies, the agent evaluates the logarithmic sensitivity matrix $S_{ij} = \partial \phi(f_i)/\partial \ln p_j$ via central finite differences ($\delta = \ln 1.01$) and normalizes it by the phase noise floor $\sigma_\phi = 0.1^\circ$ (or a calibrated $\sigma_\phi(f)$) to form the dimensionless whitened Jacobian $\mathbf{S}_{\mathrm{norm}} = (\delta/\sigma_\phi)\mathbf{S}$~\citep{tu2023universal,chen2024svd}. Factorizing $\mathbf{S}_{\mathrm{norm}} = \mathbf{U}\boldsymbol{\Sigma}\mathbf{V}^T$ yields orthogonal frequency modes in $\mathbf{U}$, ordered singular values $\sigma_1 \ge \sigma_2 \ge \dots \ge \sigma_P$ in $\boldsymbol{\Sigma}$, and parameter-space right singular vectors $\mathbf{v}_m$ forming the columns of $\mathbf{V}$, where $V_{jm}$ is the component associated with parameter $p_j$ in mode $m$. In logarithmic parameter space, each vector $\mathbf{v}_m$ defines an orthogonal composite direction $M_m = \prod_{j=1}^P p_j^{V_{jm}}$ with local sensitivity strength $\sigma_m$~\citep{chen2024svd}. In multilayer models, these directions combine properties from different layers. PRISM uses the parameter projections for screening and retains the composite products as diagnostics. Specifically, the gate retains singular modes within the index set $\mathcal{I}_{\mathrm{SVD}} = \{m : \sigma_m/\sigma_1 > \tau_{\mathrm{SNR}}\}$ ($\tau_{\mathrm{SNR}} = 0.1$) and orders parameters satisfying the retained-subspace projection screen $\eta_j = \sum_{m\in\mathcal{I}_{\mathrm{SVD}}} |V_{jm}|^2 \ge 1/P$ by descending retained sensitivity energy $E_j$ to form the SVD pass list $\mathcal{L}_{\mathrm{SVD}}$ (Equation~\eqref{eq:svd_parameter_score}). Intersecting $\mathcal{L}_{\mathrm{SVD}}$ with the absolute sensitivity pass list $\mathcal{L}_{\mathrm{sensitivity}} = \{p_j : \bar{S}_{p_j,\mathrm{rms}} \ge 0.2^\circ\}$ defines the ordered parameter-eligibility list $\mathcal{L}_{\mathrm{ident}} = \mathcal{L}_{\mathrm{SVD}} \cap \mathcal{L}_{\mathrm{sensitivity}}$ (\hyperref[sec:methods]{materials and methods}, Equation~\eqref{eq:identifiable_list_main}). Parameters outside $\mathcal{L}_{\mathrm{ident}}$ are held at their specified prior values during global optimization (Table~\ref{tab:fit_ledger}).

For reduced-space non-linear parameter estimation on $\mathcal{L}_{\mathrm{ident}}$, we benchmarked seven derivative-free global evolutionary optimizers---CMA-ES, DTDE~\citep{zhang2023dtde}, jSO, EA4Eig, L-SHADE~\citep{brest2017single,bujok2022ea4eig,tanabe2014lshade}, canonical DE, and HPSO~\citep{xiao2025hybridpso}---on a canonical Au/Si FDTR inverse problem (\hyperref[sec:methods]{materials and methods}, section~\ref{sec:optimizer_benchmark} of the supplemental information, and Figure~\ref{fig:si_optimizer}). Because all seven algorithms converge to an identical terminal phase RMSE ($\approx 0.208^\circ$) while DTDE achieves the lowest evolutionary wall-clock runtime, PRISM adopts DTDE as its default optimizer, ensuring that parameter differences across model branches reflect physical model form rather than local-minimum trapping.

\subsection{PRISM validation and SVD-guided parameter identifiability in buried multilayers}
\label{sec:buried_results}

\begin{figure}[!htbp]
\centering
\includegraphics[width=0.99\textwidth]{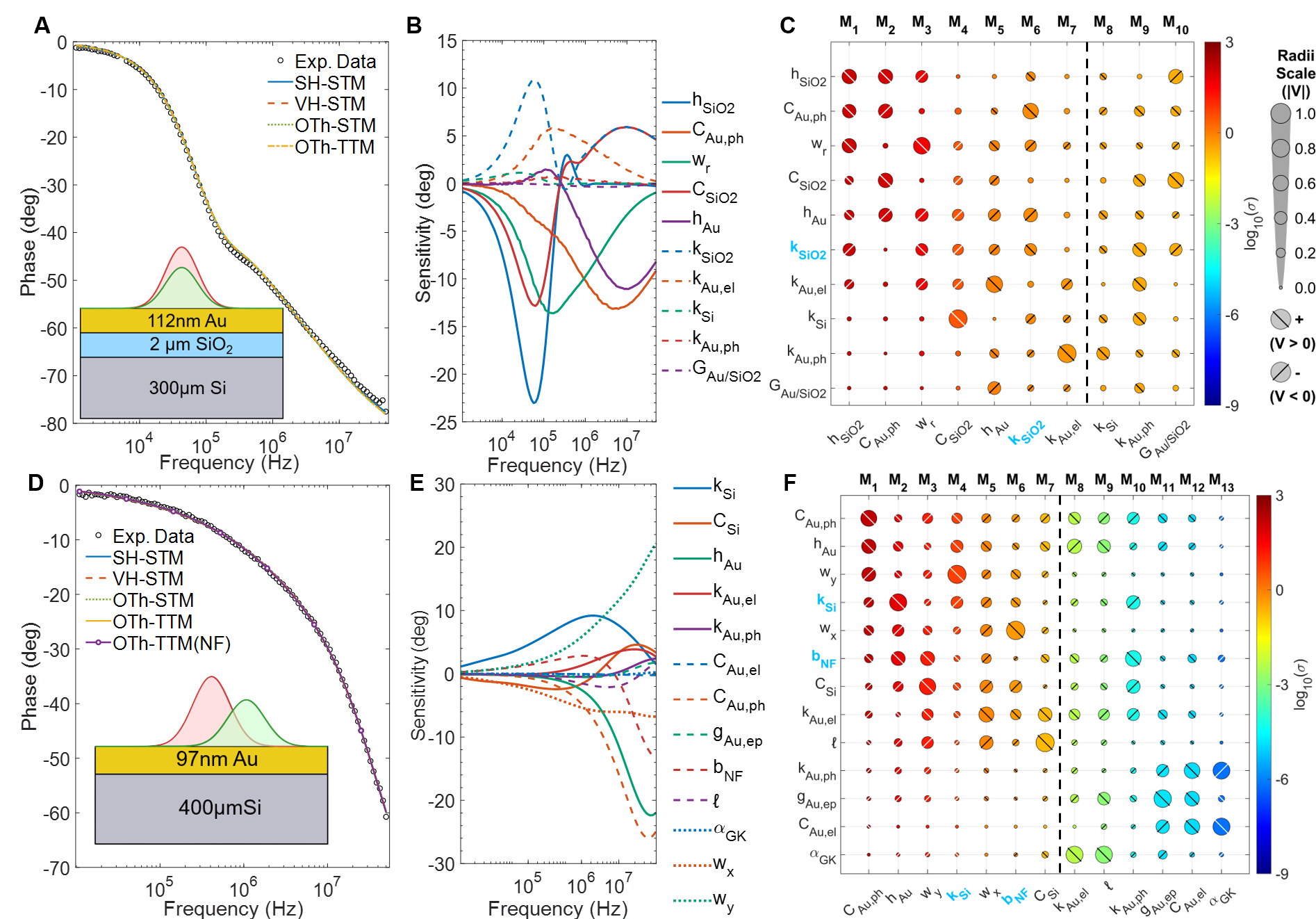}
\caption{Phase fitting and SVD-guided parameter identifiability in buried-oxide Au/SiO\textsubscript{2}/Si and offset-beam Au/Si. (A and B)~Coaxial Au/SiO\textsubscript{2}/Si with $w_r=\SI{1.86}{\micro\metre}$: (A)~measured phase and fits from the four local branches; (B)~logarithmic phase sensitivities. (C)~SVD bubble chart for the same stack in the large-spot configuration ($w_r\approx\SI{10.44}{\micro\metre}$). (D--F)~Offset Au/Si with $\Delta\approx\SI{2}{\micro\metre}$, $w_x=\SI{2.19}{\micro\metre}$ and $w_y=\SI{1.84}{\micro\metre}$: (D)~measured phase and model fits; (E)~sensitivities; (F)~SVD bubble chart. Final list membership follows Algorithm~\ref{alg:svd} and Equation~\eqref{eq:ordered_intersection_si}. Cyan labels mark the requested targets released to fitting under the fixed inputs. The keys in (C) also apply to (F).}
\label{fig:ausi}
\end{figure}

We first validate the PRISM model hierarchy and its pre-fit SVD identifiability gate on two baseline sample configurations shown in \Cref{fig:ausi} and summarized in \Cref{tab:silica}: a buried-oxide $\text{Au/SiO}_2/\text{Si}$ stack (\SI{112}{\nano\metre}~Au / \SI{2}{\micro\metre}~SiO\textsubscript{2} / \SI{300}{\micro\metre}~Si; \Cref{fig:ausi}A--C), where classical surface heating and local Fourier diffusion hold and SVD isolates a single identifiable parameter ($k_{\mathrm{SiO}_2}$), and an $\text{Au/Si}$ stack (\SI{97}{\nano\metre}~Au / \SI{400}{\micro\metre}~Si; \Cref{fig:ausi}D--F) measured in both coaxial and offset beam geometries, where SVD admits two coupled targets ($k_{\mathrm{Si}}$ and $G_{\mathrm{Au/Si}}$, or the nonlocal contact factor $\bNF$ in OTh-TTM(NF)). For comparison across silica stacks, \Cref{tab:silica}A also lists the $\text{Au/Ti/SiO}_2$ bilayer control analyzed in \Cref{sec:optical_result}; all remaining fixed inputs are collected in Tables~\ref{tab:archive_si_thermal_props}--\ref{tab:archive_si_nf_params}.

The $\text{Au/SiO}_2/\text{Si}$ stack serves as an ideal diffusive reference. In amorphous silica, heat is carried predominantly by non-propagating, wave-like vibrational modes (diffusons) and short-wavelength propagons with nanometer-scale relaxation lengths~\citep{simoncelli2019unified}, precluding quasi-ballistic transport, while the \SI{2}{\micro\metre} oxide thermal resistance ($h_{\mathrm{SiO}_2}/k_{\mathrm{SiO}_2} \approx \SI{1.5e-6}{\metre\squared\kelvin\per\watt}$) heavily dominates the stack. Moreover, electron--phonon nonequilibrium in the relatively thick \SI{112}{\nano\metre} Au transducer has negligible influence: the Au equilibration length ($L_{\mathrm{ep}} \sim 10\text{--}15~\SI{}{\nano\metre}$) is much shorter than the film thickness, and its series equilibration resistance ($R_{\mathrm{ep}} \approx \SI{4.5e-9}{\metre\squared\kelvin\per\watt}$) is over two orders of magnitude smaller than the oxide resistance. Consequently, neither volumetric optical absorption nor carrier nonequilibrium alters the inversion: all four local branches (SH-STM, VH-STM, OTh-STM, and OTh-TTM) reproduce the experimental phase spectrum from \SI{1.1}{\kilo\hertz} to \SI{50}{\mega\hertz} (\Cref{fig:ausi}A) and yield consistent $k_{\mathrm{SiO}_2}$ values (with $G_{\mathrm{Au/SiO}_2}$ fixed at its literature prior of \SI{50}{\mega\watt\per\metre\squared\per\kelvin}~\citep{regner2015interpretation,hopkins2013thermal} due to its negligible phase sensitivity). At $w_r \approx \SI{1.86}{\micro\metre}$, the recovered $k_{\mathrm{SiO}_2}$ increases by only 5\% across the hierarchy, from $1.25\pm\SI{0.16}{\watt\per\metre\per\kelvin}$ (SH-STM) to $1.32\pm\SI{0.17}{\watt\per\metre\per\kelvin}$ (OTh-TTM; \Cref{tab:silica}A). Enlarging the spot radius 5.6-fold to $w_r \approx \SI{10.44}{\micro\metre}$ on the same sample preserves this branch ordering and increases $k_{\mathrm{SiO}_2}$ by approximately 3\% within each model. At this larger spot, the model hierarchy gives $1.29\pm\SI{0.16}{\watt\per\metre\per\kelvin}$ (SH-STM) to $1.36\pm\SI{0.16}{\watt\per\metre\per\kelvin}$ (OTh-TTM), a difference of approximately 5\%, while fitting across restricted modulation-frequency windows yields no discernible variation, confirming the absence of quasi-ballistic artifacts and matching the bulk amorphous value of \SI{1.35}{\watt\per\metre\per\kelvin} measured by the $3\omega$ method~\citep{cahill1990thermal} within uncertainty (see \Cref{tab:silica} caption for the uncertainty propagation protocol).

\Cref{fig:ausi}B,C illustrates how PRISM determines parameter identifiability before fitting. Logarithmic sensitivities $S_p = \partial\phi/\partial\ln p$ (\Cref{fig:ausi}B) confirm that $k_{\mathrm{SiO}_2}$ is well resolved, peaking at $+10.5^\circ$ near \SI{7.3e4}{\hertz} where the thermal wave penetrates the full oxide layer. While the pre-calibrated layer thicknesses, heat capacities, and laser spot radius also exhibit strong phase sensitivities across distinct frequency windows, substrate conductivity $k_{\mathrm{Si}}$ ($<1.1^\circ$) and interface conductance $G_{\mathrm{Au/SiO}_2}$ ($<0.4^\circ$) remain small relative to the oxide response across the band; $G_{\mathrm{Au/SiO}_2}$ is excluded from the recorded target list by the combined sensitivity and SVD screen. Sensitivity curves alone, however, cannot detect whether the strongly responding parameters are mutually independent or correlated; the SVD in \Cref{fig:ausi}C resolves these parameter couplings.

The SVD bubble chart (\Cref{fig:ausi}C) shows ten right singular vectors (modes $M_1$--$M_{10}$) ordered by decreasing singular value. Prior sensitivity-matrix reductions for the classical Fourier SH-STM~\citep{tu2023universal,chen2024svd} rounded singular vectors to simple integer or fractional exponents to construct analytical parameter groups. PRISM evaluates parameter projections directly from the Jacobian of the selected model, including its distributed heat sources and multi-channel boundary conditions. In this diffusive reference stack, the leading modes couple oxide, transducer and beam-geometry parameters. They describe mixed responses involving oxide heat flow, lateral spreading over the laser spot and heat storage in the oxide and transducer. The substrate conductivity $k_{\mathrm{Si}}$ and interface conductance $G_{\mathrm{Au/SiO}_2}$ also have components in weak modes, including $M_4$ and $M_5$ in the displayed diagnostic. Fixing the pre-calibrated thicknesses $d$, volumetric heat capacities $C$, and spot radius $w_r$, while holding $k_{\mathrm{Si}}$ and $G_{\mathrm{Au/SiO}_2}$ at their specified values, leaves $k_{\mathrm{SiO}_2}$ (highlighted in cyan) as the fitted parameter.

\begin{table}[!htbp]
\centering
\caption{Recovered thermal properties of the baseline stacks across the model hierarchy. (A)~Silica stacks: silica thermal conductivity $k_{\mathrm{SiO}_2}$ of $\text{Au/SiO}_2/\text{Si}$ at two spot radii, measured on the same \SI{112}{\nano\metre}~Au / \SI{2}{\micro\metre}~SiO\textsubscript{2} / \SI{300}{\micro\metre}~Si sample, alongside $k_{\mathrm{SiO}_2}$ and Ti/SiO\textsubscript{2} interface conductance $G_{\mathrm{Ti/SiO}_2}$ of $\text{Au/Ti/SiO}_2$ on bulk silica. For $\text{Au/SiO}_2/\text{Si}$, $G_{\mathrm{Au/SiO}_2}$ is not an identifiable fit target: its phase sensitivity remains below $0.4^\circ$ across the modulation band (\Cref{fig:ausi}B), and it is excluded from the recorded fit-target list (\Cref{fig:ausi}C). It is fixed at its prior of \SI{50}{\mega\watt\per\metre\squared\per\kelvin} (Table~\ref{tab:archive_si_interface_G}), rendering each $\text{Au/SiO}_2/\text{Si}$ evaluation a single-parameter fit. (B)~$\text{Au/Si}$ in coaxial and offset beam geometries under isotropic substrate parameterization; the OTh-TTM(NF) entries report the equivalent interface conductance $G_{\mathrm{equiv}} = 1/R = \left[\frac{\bNF}{2}\left(\gamma_0^{-1} + \gamma_{0\Gamma}^{-1}\right)\right]^{-1} = \SI{1446}{\mega\watt\per\metre\squared\per\kelvin}/\bNF$ alongside the dimensionless contact factor $\bNF$. All uncertainties are quoted at one standard deviation, calculated by propagating a 5\% relative standard deviation across every fixed constitutive input through the linearized inverse problem, weighted by the experimental phase noise floor $\sigma_\phi = 0.1^\circ$; the resulting relative uncertainties span 6--18\% for fitted thermal conductivities and 11--65\% for fitted interface parameters.}
\label{tab:silica}\label{tab:ausi}
\footnotesize
\begin{tabular*}{\textwidth}{@{\extracolsep{\fill}}lcccc@{}}
\toprule
\multicolumn{5}{@{}l}{\textbf{(A)} Silica stacks}\\
\midrule
& \multicolumn{2}{c}{$\text{Au/SiO}_2/\text{Si}$, $k_{\mathrm{SiO}_2}$ (\si{\watt\per\metre\per\kelvin})} & \multicolumn{2}{c}{$\text{Au/Ti/SiO}_2$, $w_r\approx\SI{1.86}{\micro\metre}$}\\
\cmidrule(lr){2-3}\cmidrule(lr){4-5}
Model & $w_r\approx\SI{1.86}{\micro\metre}$ & $w_r\approx\SI{10.44}{\micro\metre}$ & $k_{\mathrm{SiO}_2}$ (\si{\watt\per\metre\per\kelvin}) & $G_{\mathrm{Ti/SiO}_2}$ (\si{\mega\watt\per\metre\squared\per\kelvin})\\
\midrule
SH-STM  & $1.25\pm0.16$ & $1.29\pm0.16$ & $1.22\pm0.14$ & $65\pm42$\\
VH-STM  & $1.28\pm0.16$ & $1.32\pm0.16$ & $1.25\pm0.15$ & $66\pm43$\\
OTh-STM & $1.30\pm0.17$ & $1.34\pm0.16$ & $1.33\pm0.15$ & $71\pm46$\\
OTh-TTM & $1.32\pm0.17$ & $1.36\pm0.16$ & $1.39\pm0.16$ & $75\pm48$\\
\midrule
\multicolumn{5}{@{}l}{\textbf{(B)} $\text{Au/Si}$}\\
\midrule
& \multicolumn{2}{c}{Coaxial, $\Delta=0$} & \multicolumn{2}{c}{Offset, $\Delta\approx\SI{2}{\micro\metre}$}\\
\cmidrule(lr){2-3}\cmidrule(lr){4-5}
Model & $k_{\mathrm{Si}}$ (\si{\watt\per\metre\per\kelvin}) & $G_{\mathrm{Au/Si}}$ (\si{\mega\watt\per\metre\squared\per\kelvin}) & $k_{\mathrm{Si}}$ (\si{\watt\per\metre\per\kelvin}) & $G_{\mathrm{Au/Si}}$ (\si{\mega\watt\per\metre\squared\per\kelvin})\\
\midrule
SH-STM      & $139\pm12$ & $58\pm6$ & $117\pm7$ & $54\pm6$\\
VH-STM      & $140\pm12$ & $59\pm7$ & $118\pm7$ & $54\pm6$\\
OTh-STM     & $143\pm13$ & $60\pm7$ & $121\pm8$ & $55\pm6$\\
OTh-TTM     & $142\pm13$ & $83\pm9$ & $119\pm8$ & $72\pm8$\\
OTh-TTM(NF) & $147\pm13$ & \NFcell{$87\pm9$}{($\bNF=16.7\pm1.8$)} & $132\pm8$ & \NFcell{$73\pm8$}{($\bNF=19.9\pm2.2$)}\\
\bottomrule
\end{tabular*}
\end{table}

Complementing single-target isolation in the diffusive stack (\Cref{fig:ausi}A--C), \Cref{fig:ausi}D--F tests the identifiability gate on coupled parameters in offset $\text{Au/Si}$ ($\Delta \approx \SI{2}{\micro\metre}$). In the SVD bubble chart (\Cref{fig:ausi}F), both $k_{\mathrm{Si}}$ and the interface parameter $\bNF$ (or $G_{\mathrm{Au/Si}}$) belong to the final intersection list under the confirmed inputs, so PRISM releases both as requested fit targets. Yet while all five branches reproduce the phase spectrum with low residuals (\Cref{fig:ausi}D), the Fourier SH-STM model suffers severe measurement-condition-dependent bias, depressing $k_{\mathrm{Si}}$ from $139\pm\SI{12}{\watt\per\metre\per\kelvin}$ (coaxial) to $117\pm\SI{7}{\watt\per\metre\per\kelvin}$ (offset; \Cref{tab:silica}B). Mathematical identifiability and low residuals alone therefore cannot prevent model-form bias, motivating PRISM's nonlocal physics escalation detailed in \Cref{sec:ausi_nonlocal_result}.

\subsection{PRISM optical-source and two-temperature validation in thin transducers}
\label{sec:optical_result}

\begin{figure}[!htbp]
\centering
\includegraphics[width=0.99\textwidth]{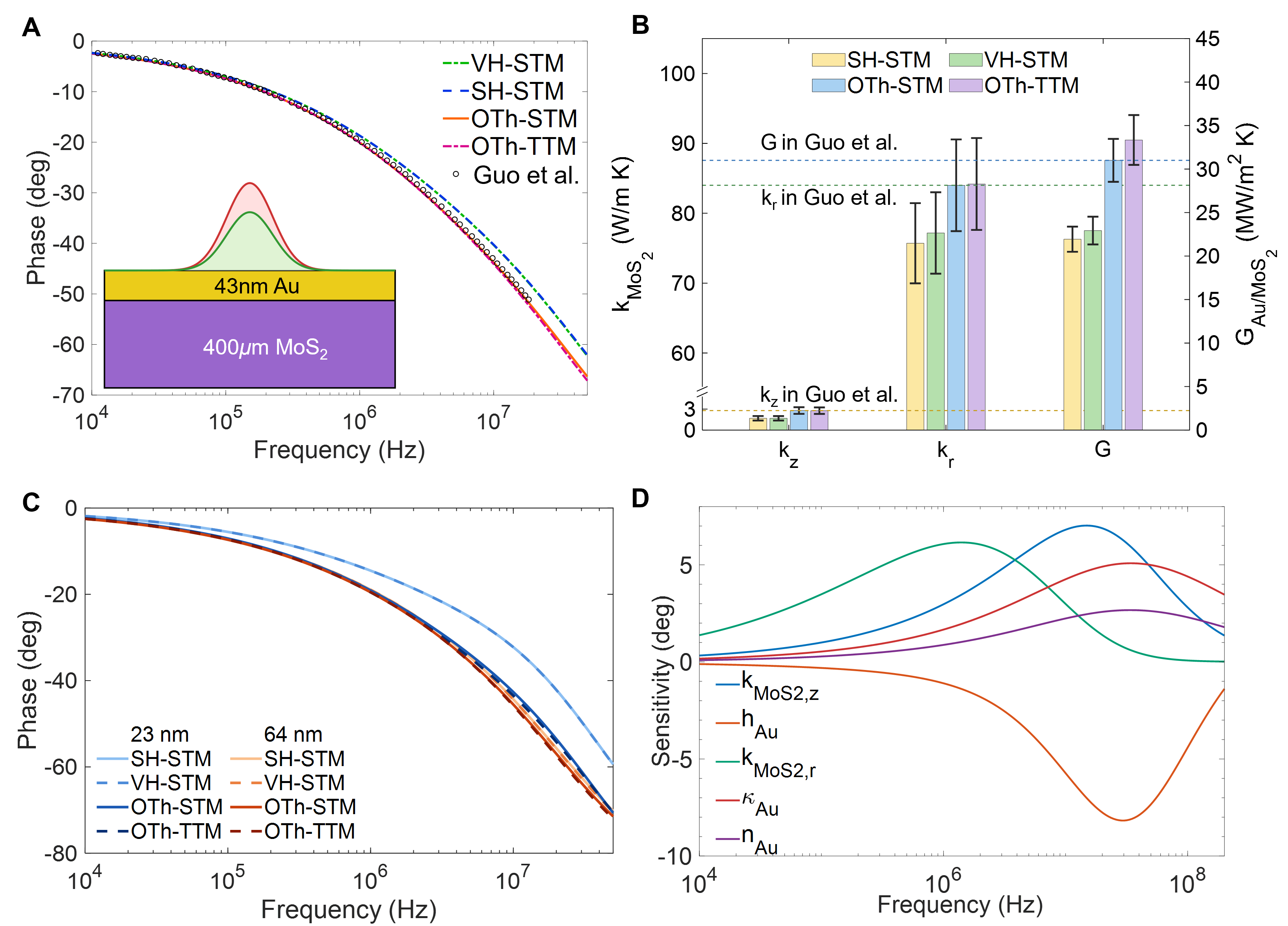}
\caption{Effects of depth-dependent optical absorption in a \SI{43}{\nano\metre} Au/MoS\textsubscript{2} stack~\citep{guo2026light}. (A)~Measured phase spectrum of Ref.~\citep{guo2026light} (open circles) overlaid with phase spectra simulated by the four model branches at the reference properties of that work. Above \SI{1}{\mega\hertz}, SH-STM and VH-STM underestimate the phase lag, whereas OTh-STM and OTh-TTM overlap, confirming that the inter-branch spread is optical in origin. (B)~Three-parameter inversions of the digitized spectrum in (A) using the stack and optical inputs of Ref.~\citep{guo2026light}. Dashed lines mark the reference values: $k_{\mathrm{MoS}_2,z}=\SI{2.8}{\watt\per\metre\per\kelvin}$ from independent TDTR, alongside $k_{\mathrm{MoS}_2,r}=84\pm\SI{9}{\watt\per\metre\per\kelvin}$ and $G=31\pm\SI{8}{\mega\watt\per\metre\squared\per\kelvin}$ from the reference optothermal fit. The fitted triples $(k_{\mathrm{MoS}_2,z},\;k_{\mathrm{MoS}_2,r},\;G)$ are $(1.7\pm0.3,\;76\pm6,\;21.9\pm1.4)$ for SH-STM, $(1.7\pm0.3,\;77\pm6,\;22.9\pm1.6)$ for VH-STM, $(2.8\pm0.5,\;84\pm7,\;31.0\pm2.5)$ for OTh-STM, and $(2.8\pm0.5,\;84\pm7,\;33.3\pm2.9)$ for OTh-TTM, with conductivities in $\mathrm{W\,m^{-1}\,K^{-1}}$ and $G$ in $\mathrm{MW\,m^{-2}\,K^{-1}}$. Error bars indicate $1\sigma$ uncertainties. (C)~Phase spectra simulated for \SI{23}{\nano\metre} and \SI{64}{\nano\metre} Au transducers on $\text{MoS}_2$, showing a $>3^\circ$ phase deviation of SH-STM and VH-STM relative to the OTh branches at \SI{23}{\nano\metre} and full convergence across all four models at \SI{64}{\nano\metre}. (D)~Logarithmic phase sensitivities $S_p$ for the \SI{43}{\nano\metre} stack across five selected parameters ($k_{\mathrm{MoS}_2,z}$, $k_{\mathrm{MoS}_2,r}$, $h_{\mathrm{Au}}$, $n_{\mathrm{Au}}$, and $\kappa_{\mathrm{Au}}$), showing that the Au optical constants carry sensitivities comparable to those of the substrate conductivities.}
\label{fig:optical}
\end{figure}

In thermal-wave metrology, thin metallic transducers or composite bilayers featuring a low-conductivity interlayer beneath the Au film are frequently adopted to suppress lateral thermal shunting within the metal transducer and enhance phase sensitivity to the underlying substrate (\Cref{fig:optical}D; Figure~\ref{fig:si_transducer_sensitivity}A)---particularly to in-plane conductivity $k_r$~\citep{guo2026light,islam2024evaluating}. However, reducing transducer thickness invalidates the classical SH approximation, as pump energy deposits volumetrically across the film depth rather than strictly at the surface~\citep{yang2016analysis,guo2026light}. At our pump wavelength of \SI{445}{\nano\metre}, for example, the optical absorption depth of Au ($d_{\mathrm{opt}}$) is approximately \SI{18.4}{\nano\metre}. For gold films around \SI{60}{\nano\metre} or thinner ($d \lesssim 3\,d_{\mathrm{opt}}$), this distributed internal heating---and, in ultrathin films, partial optical transmission into underlying layers---deviates substantially from a surface heat-flux boundary condition, introducing severe parameter bias in conventional SH analyses~\citep{yang2016analysis,guo2026light}.

We first examine volumetric absorption and electron--phonon decoupling within a metallic transducer using an $\text{Au}(\SI{62}{\nano\metre})/\text{Ti}(\SI{40}{\nano\metre})$ bilayer on bulk silica under \SI{445}{\nano\metre} pumping (\hyperref[sec:methods]{materials and methods}; Table~\ref{tab:archive_si_optical_constants}). Since Ti absorbs more strongly than Au and couples electrons to phonons much more rapidly, OTh-TTM explicitly resolves distributed bilayer absorption and phonon-mediated heat transfer into the silica substrate. Across the hierarchy, recovered $k_{\mathrm{SiO}_2}$ rises by 14\% from an artificially depressed $1.22\pm\SI{0.14}{\watt\per\metre\per\kelvin}$ (SH-STM) to $1.39\pm\SI{0.16}{\watt\per\metre\per\kelvin}$ (OTh-TTM), matching the bulk silica reference within uncertainty~\citep{cahill1990thermal}, with OTh absorption and TTM nonequilibrium contributing $\sim 9\%$ ($1.33\pm\SI{0.15}{\watt\per\metre\per\kelvin}$ in OTh-STM) and $\sim 5\%$ increases, respectively (\Cref{tab:silica}A). The TTM correction arises because heat within the metal bilayer is carried primarily by electrons but must enter insulating silica via Ti phonons. This diagnostic comparison co-fits $G_{\mathrm{Ti/SiO}_2}$ with $k_{\mathrm{SiO}_2}$ to examine the model dependence of the weakly constrained conductance (\Cref{tab:silica}A; Figure~\ref{fig:si_chat_ui}). Under OTh-TTM, $G_{\mathrm{Ti/SiO}_2}$ has a 64\% relative uncertainty, and traditional STM regressions lump the omitted spatial equilibration resistance partly into $G_{\mathrm{Ti/SiO}_2}$ ($65\pm 42$ vs.\ $75\pm\SI{48}{\mega\watt\per\metre\squared\per\kelvin}$) and partly into the substrate response, artificially depressing both parameters.

While optical absorption in $\text{Au/Ti/SiO}_2$ remains confined within the metallic bilayer above a transparent substrate, reducing the transducer thickness further on an absorbing semiconductor introduces direct subsurface substrate heating. This regime is demonstrated by the $\text{Au}(\SI{43}{\nano\metre})/\text{MoS}_2$ phase spectrum reported by Guo et al.~\citep{guo2026light}, digitized from their Figure~4B over \SI{11}{\kilo\hertz}\text{--}\SI{18}{\mega\hertz} with $w_r = \SI{2.8}{\micro\metre}$ and $\lambda^{\mathrm{pump}} = \SI{488}{\nano\metre}$ (\Cref{fig:optical}A). Since $\text{MoS}_2$ absorbs strongly at \SI{488}{\nano\metre} ($\tilde{n} = 3.98 + 1.17i$), pump light transmitted through the thin Au film deposits directly into the $\text{MoS}_2$ crystal immediately beneath the $\text{Au/MoS}_2$ interface. Using the reported stack parameters, we invert the spectrum simultaneously for cross-plane $k_{\mathrm{MoS}_2,z}$, in-plane $k_{\mathrm{MoS}_2,r}$, and interfacial $G$.

Simulated at the benchmark properties of Ref.~\citep{guo2026light} ($k_{\mathrm{MoS}_2,z} = \SI{2.8}{\watt\per\metre\per\kelvin}$; $k_{\mathrm{MoS}_2,r} = 84\pm\SI{9}{\watt\per\metre\per\kelvin}$ and $G = 31\pm\SI{8}{\mega\watt\per\metre\squared\per\kelvin}$), SH-STM and VH-STM underestimate the measured phase lag above \SI{1}{\mega\hertz} by up to $4^\circ$, whereas OTh-STM and OTh-TTM reproduce the experimental data (\Cref{fig:optical}A). Physically, pump light transmitted through the \SI{43}{\nano\metre} Au film deposits heat directly within the top $d_{\mathrm{opt,MoS}_2} \approx \SI{33}{\nano\metre}$ of the $\text{MoS}_2$ crystal beneath the interface; this subsurface heat must diffuse back across the low-$k_z$ $\text{MoS}_2$ basal planes and the $\text{Au/MoS}_2$ boundary before reaching the probe-sensed Au transducer, introducing an additional high-frequency phase lag. Inverting the digitized experimental spectrum with each model branch reveals how surface-heating assumptions distort the extracted properties (\Cref{fig:optical}B): restricting heat deposition to the Au surface or interior forces SH-STM and VH-STM to compensate for the missing subsurface back-diffusion delay by artificially lowering the subsurface conductances, returning $(k_{\mathrm{MoS}_2,z},\; k_{\mathrm{MoS}_2,r},\; G) = (1.7\pm 0.3~\si{\watt\per\metre\per\kelvin},\; 76\pm 6~\si{\watt\per\metre\per\kelvin},\; 21.9\pm 1.4~\si{\mega\watt\per\metre\squared\per\kelvin})$ for SH-STM and $(1.7\pm 0.3~\si{\watt\per\metre\per\kelvin},\; 77\pm 6~\si{\watt\per\metre\per\kelvin},\; 22.9\pm 1.6~\si{\mega\watt\per\metre\squared\per\kelvin})$ for VH-STM---underestimating $k_{\mathrm{MoS}_2,z}$ by $40\%$, $G$ by $26\text{--}29\%$, and $k_{\mathrm{MoS}_2,r}$ by $8\text{--}10\%$. Conversely, resolving multilayer optical absorption and volumetric probe sensing under OTh-STM yields $(2.8\pm 0.5~\si{\watt\per\metre\per\kelvin},\; 84\pm 7~\si{\watt\per\metre\per\kelvin},\; 31.0\pm 2.5~\si{\mega\watt\per\metre\squared\per\kelvin})$, matching both the independent TDTR $k_{\mathrm{MoS}_2,z}$ benchmark and the reference optothermal fit. Upgrading to OTh-TTM returns $(2.8\pm 0.5~\si{\watt\per\metre\per\kelvin},\; 84\pm 7~\si{\watt\per\metre\per\kelvin},\; 33.3\pm 2.9~\si{\mega\watt\per\metre\squared\per\kelvin})$, shifting $k_{\mathrm{MoS}_2,z}$ and $k_{\mathrm{MoS}_2,r}$ by $<1\%$ while raising $G$ by $8\%$ (well within the $1\sigma$ uncertainty). Unlike $\text{Au/Si}$ and $\text{Au/Ti/Si}$ ($G \approx 73\text{--}\SI{137}{\mega\watt\per\metre\squared\per\kelvin}$) or $\text{Au/Ti/SiO}_2$ (where pump absorption is confined to the metal bilayer), TTM produces only a minor shift here for two reasons: first, the van der Waals $\text{Au/MoS}_2$ boundary resistance ($1/G \approx \SI{3.2e-8}{\metre\squared\kelvin\per\watt}$) is roughly an order of magnitude larger than the Au electron--phonon equilibration resistance, and second, pump photons transmitted directly into the semiconducting $\text{MoS}_2$ substrate heat the lattice without traversing the Au electron--phonon coupling bottleneck (reducing the apparent interfacial resistance difference between OTh-STM and OTh-TTM to $\approx\SI{2.3e-9}{\metre\squared\kelvin\per\watt}$).

Phase spectra calculated across transducer thicknesses (\Cref{fig:optical}C) and sensitivity diagnostics (\Cref{fig:optical}D) further clarify this optical absorption and readout behavior. Thinning the Au transducer to \SI{23}{\nano\metre} ($\approx d_{\mathrm{opt}}$ at $\lambda^{\mathrm{pump}} = \SI{488}{\nano\metre}$) increases optical transmission into $\text{MoS}_2$, causing SH-STM and VH-STM to deviate from the OTh-STM and OTh-TTM phase curves by $>3^\circ$ above \SI{10}{\mega\hertz}, whereas thickening the Au film to \SI{64}{\nano\metre} ($\approx 3\,d_{\mathrm{opt}}$) confines pump absorption within the metal so that all four branches converge (\Cref{fig:optical}C), matching PRISM's pre-fit phase-deviation gate (\Cref{sec:inference_framework}). In \Cref{fig:optical}D, the logarithmic sensitivity spectra and supplemental SVD screening (Table~\ref{tab:fit_ledger}) document the selection of the three fitted targets under the fixed inputs: $S_{k_{\mathrm{MoS}_2,r}}$ peaks near \SI{1}{\mega\hertz} where radial heat spreading dominates, whereas $S_{k_{\mathrm{MoS}_2,z}}$ peaks a decade higher where shallow thermal penetration confines the thermal wave near the interface. Since unmodeled subsurface heating primarily perturbs the $>1\,\si{\mega\hertz}$ phase, our simultaneous three-parameter inversion absorbs the optical error chiefly into $k_{\mathrm{MoS}_2,z}$ and $G$; by contrast, fixing $k_{\mathrm{MoS}_2,z}$ at its TDTR value as in Ref.~\citep{guo2026light} forces that discrepancy into $k_{\mathrm{MoS}_2,r}$. Meanwhile, the Au optical constants ($n_{\mathrm{Au}}, \kappa_{\mathrm{Au}}$) exhibit high-frequency sensitivities comparable to $k_{\mathrm{MoS}_2,z}$ (\Cref{fig:optical}D), underscoring the necessity of accurate refractive-index inputs in thin-transducer inversions.

\subsection{PRISM nonlocal transport validation in coaxial and offset Au/Si}
\label{sec:ausi_nonlocal_result}
Having validated volumetric optical absorption and carrier nonequilibrium in thin transducers, we next examine coupled electron--phonon relaxation and quasi-ballistic phonon transport in the $\text{Au}(\SI{97}{\nano\metre})/\text{Si}(\SI{400}{\micro\metre})$ benchmark across coaxial ($\Delta = 0$) and laterally offset ($\Delta \approx \SI{2}{\micro\metre}$) beam geometries fitted over the full \SI{10.9}{\kilo\hertz}\text{--}\SI{50}{\mega\hertz} band (\Cref{fig:ausi}D--F; Figure~\ref{fig:si_spatial_overlap}A--C). In room-temperature Si, phonons with micrometer-scale MFPs ($\lambda_{\mathrm{ph}} > \SI{1}{\micro\metre}$) carry a substantial fraction of heat over distances comparable to the elliptical spot radii ($w_x = \SI{2.19}{\micro\metre}$, $w_y = \SI{1.84}{\micro\metre}$) and high-frequency thermal penetration depth~\citep{minnich2012determining,regner2013mfpfdtr,wilson2014anisotropic}. Because these long-MFP modes escape the heated volume before scattering locally, the hotspot temperature exceeds the local diffusion prediction, which Fourier inversions misinterpret as a suppressed apparent conductivity~\citep{regner2014analytical,yang2015heating,li2025complexk}. Displacing the probe beam by $\Delta \approx \SI{2}{\micro\metre}$ accentuates this quasi-ballistic signature because the offset probe samples the steep lateral temperature gradient at the Gaussian beam wing---weighting high-in-plane-wavevector modes where ballistic phonon escape and nonlocal flux curvature are most pronounced~\citep{feser2012probing,wilson2014anisotropic,beardo2020phonon}. Consequently, although all five model branches fit the full-band phase spectra within experimental noise (\Cref{fig:ausi}D), the four local Fourier models (SH-STM through OTh-TTM) recover $k_{\mathrm{Si}} = 139\text{--}\SI{143}{\watt\per\metre\per\kelvin}$ in the coaxial geometry but drop to $117\text{--}\SI{121}{\watt\per\metre\per\kelvin}$ under beam offset (\Cref{tab:ausi}B). Enforcing a single isotropic $k_{\mathrm{Si}}$ across both configurations thus exposes the breakdown of local diffusion through geometry-dependent parameter drift rather than fit residuals.

Escalating to the $\mathrm{OTh\text{-}TTM(NF)}$ branch expands the candidate parameter space to 13 dimensions (\Cref{fig:ausi}E,F): alongside $k_{\mathrm{Si}}$, $C_{\mathrm{Si}}$, $h_{\mathrm{Au}}$, and the spot radii ($w_x, w_y$), TTM introduces five metallic carrier parameters ($C_{\mathrm{Au,el}}, C_{\mathrm{Au,ph}}, k_{\mathrm{Au,el}}, k_{\mathrm{Au,ph}}, g_{\mathrm{Au,ep}}$), while the GK model adds the nonlocal length $\ell$, divergence factor $\alpha_{\mathrm{GK}}$, and dimensionless interfacial contact factor $\bNF$. Sensitivity magnitudes alone do not establish joint identifiability in this 13-parameter space: in \Cref{fig:ausi}E, internal parameters such as $k_{\mathrm{Au,el}}$ (peaking near $+4.5^\circ$), $\ell$ ($\pm 2^\circ$), $k_{\mathrm{Au,ph}}$ ($+2.5^\circ$), and $g_{\mathrm{Au,ep}}$ ($+2.5^\circ$) show appreciable high-frequency phase sensitivity; absolute-sensitivity admission uses the band-RMS criterion in Equation~\eqref{eq:absolute_sensitivity_list}. The SVD bubble chart (\Cref{fig:ausi}F) additionally displays parameter coupling. Singular-mode retention is evaluated separately using $\sigma_m/\sigma_1>0.1$ (Algorithm~\ref{alg:svd}). The parameters $k_{\mathrm{Au,ph}}$, $g_{\mathrm{Au,ep}}$, and $C_{\mathrm{Au,el}}$ have large components in trailing modes $M_{10}$--$M_{13}$, while $\alpha_{\mathrm{GK}}$ has substantial components in $M_8$ and $M_9$. The parameters $k_{\mathrm{Au,el}}$ and $\ell$ cross-couple with transducer and substrate components across $M_3$--$M_9$. Fixing the five pre-calibrated geometry and heat-capacity inputs ($C_{\mathrm{Au,ph}}, h_{\mathrm{Au}}, w_y, w_x, C_{\mathrm{Si}}$) and holding the six unidentifiable TTM and GK parameters at their literature priors isolates $k_{\mathrm{Si}}$ and $\bNF$ (highlighted in cyan) as the two fit targets.

Comparing the fitted parameters across the hierarchy (\Cref{tab:ausi}B) shows how OTh-TTM and the GK model separately correct interfacial and bulk biases. First, upgrading from OTh-STM to OTh-TTM isolates the electron--phonon equilibration resistance in the \SI{97}{\nano\metre} Au film from the interface, raising $G_{\mathrm{Au/Si}}$ by 38\% in the coaxial geometry (from $60\pm 7$ to $83\pm\SI{9}{\mega\watt\per\metre\squared\per\kelvin}$) and by 31\% in the offset geometry (from $55\pm 6$ to $72\pm\SI{8}{\mega\watt\per\metre\squared\per\kelvin}$). From $1/G_{\mathrm{STM}} = 1/G_{\mathrm{TTM}} + L_{\mathrm{ep}}/k_{\mathrm{ph}}$ for $k_{\mathrm{el}} \gg k_{\mathrm{ph}}$~\citep{majumdar2004role}, both geometries yield an equilibration resistance of $R_{\mathrm{ep}} \approx \SI{4.5e-9}{\metre\squared\kelvin\per\watt}$, corresponding to $k_{\mathrm{ph}} \approx \SI{2}{\watt\per\metre\per\kelvin}$ and $L_{\mathrm{ep}} \approx \SI{9}{\nano\metre}$ in agreement with first-principles values~\citep{jain2016thermal,tong2019comprehensive}. Second, activating the GK model in $\mathrm{OTh\text{-}TTM(NF)}$ replaces $G_{\mathrm{Au/Si}}$ with $\bNF$, returning $\bNF = 16.7\pm 1.8$ (coaxial) and $19.9\pm 2.2$ (offset). Converting $\bNF$ into the equivalent conductance $G_{\mathrm{equiv}} = 1/R = \SI{1446}{\mega\watt\per\metre\squared\per\kelvin}/\bNF$ (\Cref{tab:ausi}B) gives $87\pm\SI{9}{\mega\watt\per\metre\squared\per\kelvin}$ (coaxial) and $73\pm\SI{8}{\mega\watt\per\metre\squared\per\kelvin}$ (offset), consistent with the OTh-TTM conductances within their uncertainties. Simultaneously, the GK model captures quasi-ballistic phonon suppression in Si, raising $k_{\mathrm{Si}}$ from $142\pm 13$ to $147\pm\SI{13}{\watt\per\metre\per\kelvin}$ (coaxial, within the $142\text{--}\SI{148}{\watt\per\metre\per\kelvin}$ bulk reference band) and from $119\pm 8$ to $132\pm\SI{8}{\watt\per\metre\per\kelvin}$ (offset). While this brings the two geometries into agreement within their $1\sigma$ uncertainties, a residual offset difference remains under the isotropic slip condition, motivating the coaxial multi-spot and frequency-window validation on $\text{Au/Ti/Si}$ below.

\subsection{Coupled optical, electron--phonon, and quasi-ballistic transport in Au/Ti/Si}
\label{sec:nonlocal_result}

\begin{figure}[!htbp]
\centering
\includegraphics[width=0.82\textwidth]{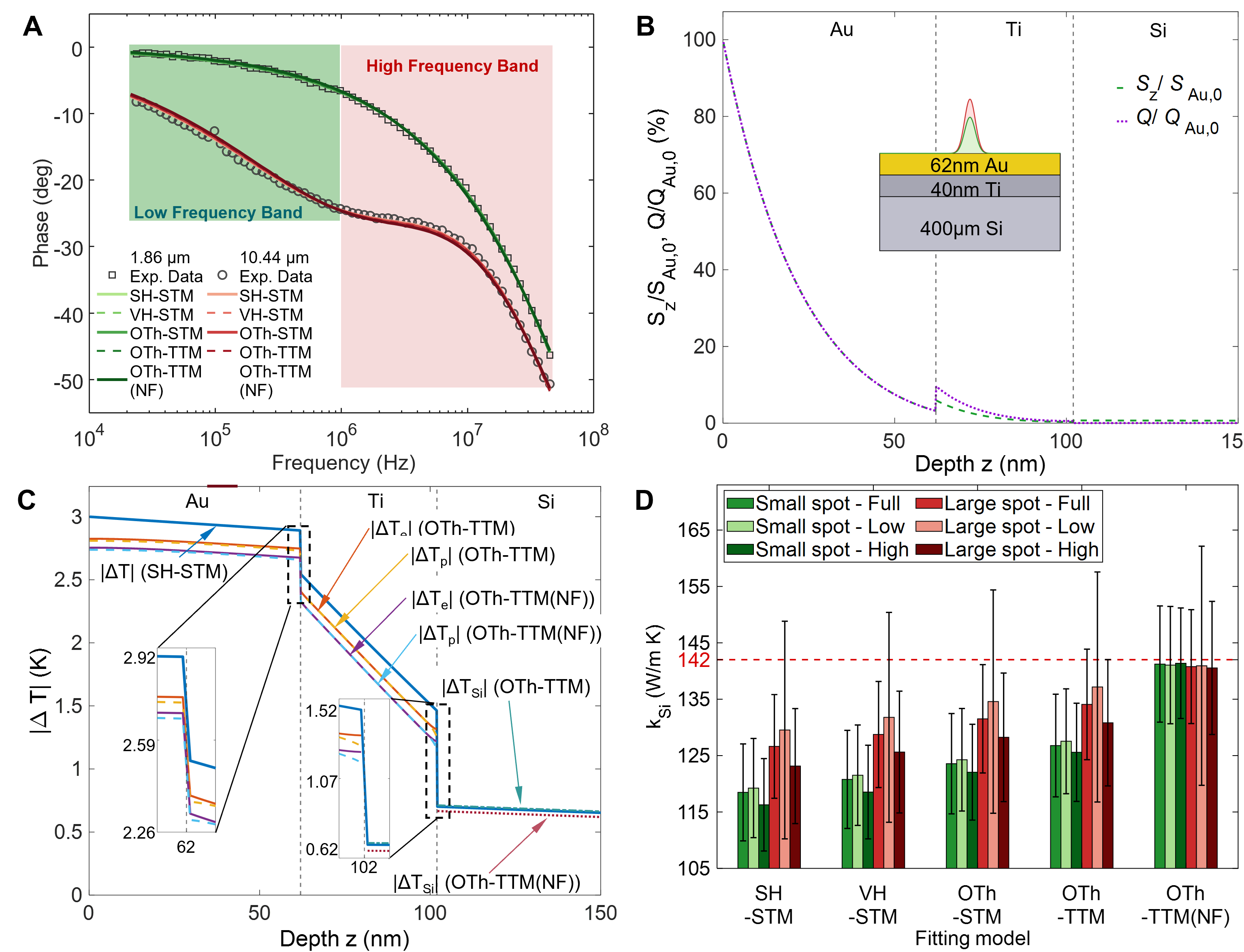}
\caption{Spot-size and frequency-window dependence of the Au/Ti/Si thermal response (\SI{62}{\nano\metre}~Au / \SI{40}{\nano\metre}~Ti / \SI{400}{\micro\metre}~Si stack). (A)~Measured phase spectra at small ($w_r \approx \SI{1.86}{\micro\metre}$) and large ($w_r \approx \SI{10.44}{\micro\metre}$) spot radii overlaid with model fits across the low- and high-frequency bands. (B)~Depth profiles of normalized Poynting flux $S_z/S_{\mathrm{Au},0}$ and volumetric absorbed power density $Q/Q_{\mathrm{Au},0}$ at $\lambda^{\mathrm{pump}} = \SI{445}{\nano\metre}$, showing a jump in absorbed power density upon entering the Ti layer and $<1\%$ transmission into the Si substrate; the inset illustrates the stack and Gaussian beams. (C)~Depth-resolved temperature oscillation amplitudes $|\Delta T(z)|$ at \SI{50}{\mega\hertz}, showing interfacial temperature drops at $z = \SI{62}{\nano\metre}$ ($\text{Au/Ti}$) and $z = \SI{102}{\nano\metre}$ ($\text{Ti/Si}$) and localized electron--phonon decoupling ($|\Delta T_e|$ vs.\ $|\Delta T_p|$) immediately above the $\text{Ti/Si}$ interface. (D)~Recovered silicon thermal conductivity $k_{\mathrm{Si}}$ across the five model branches, two spot radii, and three frequency windows (\Cref{tab:nonlocal_windows}); error bars denote $1\sigma$ propagated uncertainties, and the dashed line marks the \SI{142}{\watt\per\metre\per\kelvin} lower bound of bulk single-crystal Si. Whereas the 24 local Fourier evaluations scatter over $116\text{--}\SI{137}{\watt\per\metre\per\kelvin}$ and track the spot radius, the nonlocal OTh-TTM(NF) branch converges to $\SI{141}{\watt\per\metre\per\kelvin}$ across all six spot-size and window combinations.}
\label{fig:nonlocal}
\end{figure}

The Au/Ti/Si stack combines volumetric optical absorption, electron--phonon nonequilibrium in the metal layers, and quasi-ballistic phonon transport in Si. We examine these coupled effects in an $\text{Au}(\SI{62}{\nano\metre})/\text{Ti}(\SI{40}{\nano\metre})/\text{Si}(\SI{400}{\micro\metre})$ heterostructure (\Cref{fig:nonlocal}; Figure~\ref{fig:si_spatial_overlap}D--F) across two coaxial spot radii ($w_r \approx \SI{1.86}{\micro\metre}$ and $\SI{10.44}{\micro\metre}$, a 5.6-fold change in radial heating scale) and three spectral windows: the full band (\SI{20}{\kilo\hertz}\text{--}\SI{50}{\mega\hertz}), a low-frequency window (\SI{20}{\kilo\hertz}\text{--}\SI{1}{\mega\hertz}), and a high-frequency window (\SIrange{1}{50}{\mega\hertz}). Subdividing the modulation band systematically varies the axial thermal penetration depth $\delta_p$ alongside the radial spot scale $w_r$ within the same dataset. Although all five model branches reproduce the measured phase spectra nearly indistinguishably across both spot sizes (\Cref{fig:nonlocal}A), the inverted $k_{\mathrm{Si}}$ and $G_{\mathrm{Ti/Si}}$ diverge substantially across the local models and probing windows (\Cref{fig:nonlocal}D; \Cref{tab:nonlocal_windows}).

\begin{table}[!htbp]
\centering
\caption{Recovered silicon thermal conductivity $k_{\mathrm{Si}}$ and Ti/Si interface conductance $G_{\mathrm{Ti/Si}}$ for the Au/Ti/Si stack across both laser spot sizes and the three fitting windows plotted in \Cref{fig:nonlocal}D. The four local Fourier branches fit $G_{\mathrm{Ti/Si}}$ directly, whereas OTh-TTM(NF) replaces it with the dimensionless nonlocal contact factor $b_{\mathrm{NF}}$, maintaining equal degrees of freedom; its entries report the equivalent conductance $G_{\mathrm{Ti/Si}} = \SI{1446}{\mega\watt\per\metre\squared\per\kelvin}/b_{\mathrm{NF}}$. The fitted $b_{\mathrm{NF}}$ values, in order of full, low, and high windows, are $11\pm3$, $11\pm3$, and $11\pm3$ for the small spot, and $12\pm3$, $16\pm10$, and $12\pm3$ for the large spot. The three spectral windows span \SI{20}{\kilo\hertz}\text{--}\SI{50}{\mega\hertz} (full), \SI{20}{\kilo\hertz}\text{--}\SI{1}{\mega\hertz} (low), and \SIrange{1}{50}{\mega\hertz} (high). Uncertainties represent one standard deviation, propagated as in \Cref{tab:ausi}.}
\label{tab:nonlocal_windows}
\footnotesize
\begin{tabular*}{\textwidth}{@{\extracolsep{\fill}}llcccc@{}}
\toprule
& & \multicolumn{2}{c}{Small spot ($w_r \approx \SI{1.86}{\micro\metre}$)} & \multicolumn{2}{c}{Large spot ($w_r \approx \SI{10.44}{\micro\metre}$)}\\
\cmidrule(lr){3-4}\cmidrule(lr){5-6}
Window & Model & $k_{\mathrm{Si}}$ (\si{\watt\per\metre\per\kelvin}) & $G_{\mathrm{Ti/Si}}$ (\si{\mega\watt\per\metre\squared\per\kelvin}) & $k_{\mathrm{Si}}$ (\si{\watt\per\metre\per\kelvin}) & $G_{\mathrm{Ti/Si}}$ (\si{\mega\watt\per\metre\squared\per\kelvin})\\
\midrule
Full   & SH-STM        & $118\pm9$  & $108\pm28$ & $127\pm9$  & $102\pm26$\\
       & VH-STM        & $121\pm9$  & $113\pm29$ & $129\pm9$  & $104\pm26$\\
       & OTh-STM       & $124\pm9$  & $118\pm30$ & $132\pm10$ & $106\pm27$\\
       & OTh-TTM       & $127\pm9$  & $128\pm33$ & $134\pm10$ & $114\pm29$\\
       & OTh-TTM(NF)   & $141\pm10$ & $136\pm34$ & $141\pm10$ & $118\pm29$\\
\midrule
Low    & SH-STM        & $119\pm9$  & $102\pm30$ & $130\pm19$ & $93\pm57$\\
       & VH-STM        & $122\pm9$  & $106\pm30$ & $132\pm19$ & $85\pm53$\\
       & OTh-STM       & $124\pm9$  & $112\pm32$ & $135\pm20$ & $87\pm54$\\
       & OTh-TTM       & $128\pm9$  & $124\pm36$ & $137\pm20$ & $87\pm53$\\
       & OTh-TTM(NF)   & $141\pm10$ & $137\pm35$ & $141\pm21$ & $92\pm58$\\
\midrule
High   & SH-STM        & $116\pm8$  & $110\pm28$ & $123\pm10$ & $98\pm22$\\
       & VH-STM        & $119\pm8$  & $115\pm29$ & $126\pm11$ & $101\pm23$\\
       & OTh-STM       & $122\pm9$  & $120\pm31$ & $128\pm11$ & $103\pm23$\\
       & OTh-TTM       & $126\pm9$  & $130\pm33$ & $131\pm11$ & $111\pm25$\\
       & OTh-TTM(NF)   & $141\pm10$ & $135\pm33$ & $141\pm12$ & $116\pm26$\\
\bottomrule
\end{tabular*}
\end{table}

The depth-resolved optical and thermal profiles in \Cref{fig:nonlocal}B,C explain why models that fit the same phase spectrum (\Cref{fig:nonlocal}A) yield different transport properties. In \Cref{fig:nonlocal}B, the normalized Poynting flux $S_z/S_{\mathrm{Au},0}$ decays continuously through the \SI{62}{\nano\metre} Au film to $\sim 6\%$ at the $\text{Au/Ti}$ boundary ($z = \SI{62}{\nano\metre}$), whereas the absorbed power density $Q/Q_{\mathrm{Au},0}$ jumps from $\sim 4\%$ to $\sim 10\%$ upon entering the Ti layer because the imaginary permittivity of Ti is $2.5$ times that of Au at \SI{445}{\nano\metre} (Table~\ref{tab:archive_si_optical_constants}). The buried Ti film therefore receives direct optical heating that surface-heating models omit. Before the $\text{Ti/Si}$ interface ($z = \SI{102}{\nano\metre}$), both $S_z$ and $Q$ fall below $1\%$ of their surface values, confirming that optical absorption is confined to the metal bilayer and that direct photocarrier generation in Si is negligible~\citep{galovic2025carrier,park2025surfacerecomb}. In the \SI{50}{\mega\hertz} temperature profiles (\Cref{fig:nonlocal}C), high electronic conductivity keeps $|\Delta T|$ nearly uniform across the Au film, followed by a $\sim\SI{0.4}{\kelvin}$ drop across the electron-dominated $\text{Au/Ti}$ interface and a $\sim\SI{1.0}{\kelvin}$ drop across the \SI{40}{\nano\metre} Ti layer. Here, $G_{\mathrm{Au/Ti}}$ is fixed at \SI{500}{\mega\watt\per\metre\squared\per\kelvin}, where its phase sensitivity remains negligible (Figure~\ref{fig:si_spatial_overlap}E). Within the Ti layer, strong electron--phonon coupling ($g_{\mathrm{ep,Ti}} \approx 74\,g_{\mathrm{ep,Au}}$; Table~\ref{tab:archive_si_ttm_params}) maintains $T_{\mathrm{el}} \approx T_{\mathrm{ph}}$ across the film interior, whereas the Ti/Si boundary condition blocks electronic transmission ($\widehat{q}_{\mathrm{el}} = 0$) and forces electron-to-phonon energy conversion within $\sim\SI{1}{\nano\metre}$ of the $\text{Ti/Si}$ boundary (insets of \Cref{fig:nonlocal}C) ahead of a $0.6\text{--}\SI{0.8}{\kelvin}$ interfacial temperature jump. Consequently, model branches with indistinguishable phase spectra (\Cref{fig:nonlocal}A) differ by $\sim 8\%$ in surface temperature amplitude ($\approx\SI{3.0}{\kelvin}$ in SH-STM vs.\ $\approx\SI{2.75}{\kelvin}$ in OTh-TTM(NF)) and partition thermal resistance differently between the $\text{Ti/Si}$ interface and the Si substrate.

This resistance mispartitioning and omitted quasi-ballistic transport drive the spread across the 24 local Fourier fits (\Cref{fig:nonlocal}D, \Cref{tab:nonlocal_windows}). Within each window, upgrading from SH-STM to OTh-STM and OTh-TTM removes the optical error due to the thin Au and Au/Ti interface and near-interface TTM resistance, raising the full-band small-spot $k_{\mathrm{Si}}$ from $118\pm 9$ (SH-STM) to $124\pm 9$ (OTh-STM) and $127\pm\SI{9}{\watt\per\metre\per\kelvin}$ (OTh-TTM), while $G_{\mathrm{Ti/Si}}$ rises from $108\pm 28$ to $128\pm\SI{33}{\mega\watt\per\metre\squared\per\kelvin}$. Even under OTh-TTM, however, $k_{\mathrm{Si}}$ remains below the single-crystal band ($142\text{--}\SI{148}{\watt\per\metre\per\kelvin}$)~\citep{glassbrenner1964thermal,inyushkin2004isotope} and varies with spot radius ($127\pm 9$ at $w_r \approx \SI{1.86}{\micro\metre}$ vs.\ $134\pm\SI{10}{\watt\per\metre\per\kelvin}$ at $w_r \approx \SI{10.44}{\micro\metre}$). Windowing the spectrum widens this spread: in the small-spot high-frequency window, where $\delta_p$ in Si shrinks to $\sim\SI{0.7}{\micro\metre}$ at \SI{50}{\mega\hertz}, SH-STM drops to $116\pm\SI{8}{\watt\per\metre\per\kelvin}$. Across all 24 local evaluations, $k_{\mathrm{Si}}$ ranges from $116\pm 8$ to $137\pm\SI{20}{\watt\per\metre\per\kelvin}$, and in all twelve model--window pairs the larger spot yields a higher value by $5\text{--}\SI{10}{\watt\per\metre\per\kelvin}$. Uncertainties in \Cref{tab:nonlocal_windows} widen in the large-spot low-frequency window ($w_r \approx \SI{10.44}{\micro\metre}$, $f \le \SI{1}{\mega\hertz}$), where quasi-one-dimensional heat flow couples $k_{\mathrm{Si}}$ and $G_{\mathrm{Ti/Si}}$ into nearly the same phase combination.

This spot-size and frequency-window dependence in Si ($\mathrm{Kn} \gtrsim 0.1$) leads the PRISM agent to select $\mathrm{OTh\text{-}TTM(NF)}$. With the GK and modal parameters fixed at the tabulated 311\,K values (Table~\ref{tab:archive_si_nf_params})~\citep{beardo2020phonon}, $\mathrm{OTh\text{-}TTM(NF)}$ fits $\bNF$ in place of $G_{\mathrm{Ti/Si}}$ with equal degrees of freedom. Across every model-matched comparison (\Cref{tab:ausi}B and \Cref{tab:nonlocal_windows}), the recovered equivalent conductance $G_{\mathrm{Ti/Si}}$ ($116\text{--}\SI{136}{\mega\watt\per\metre\squared\per\kelvin}$ in the full and high-frequency $\mathrm{OTh\text{-}TTM(NF)}$ fits, from $\bNF = 11\pm 3$ to $12\pm 3$) substantially exceeds $G_{\mathrm{Au/Si}}$ ($73\text{--}\SI{87}{\mega\watt\per\metre\squared\per\kelvin}$). Phonon densities of states calculated via the MatterSim machine-learning interatomic potential on PhononBench (Figure~\ref{fig:si_phonon_dos})~\citep{han2025phononbench} provide a spectral comparison relevant to this conductance difference: whereas Au vibrational modes cut off sharply by $\sim 4.2\text{ THz}$ (with peaks near $1.8$ and $3.5\text{ THz}$), Ti vibrational modes extend to $\sim 5\text{ THz}$ (peaking near $4\text{ THz}$), providing broader acoustic spectral overlap with the low-frequency acoustic branches of the Si substrate ($\sim 0\text{--}6\text{ THz}$) and facilitating phonon transmission across the $\text{Ti/Si}$ boundary~\citep{swartz1989thermal,hopkins2013thermal,duda2013influence,jeong2016enhancement}. Activating the nonlocal GK formulation in $\mathrm{OTh\text{-}TTM(NF)}$ eliminates the spot-size and frequency-window drift in the substrate (\Cref{fig:nonlocal}D, \Cref{tab:nonlocal_windows}): by coupling bulk spatial nonlocality ($\ell$) with the nonequilibrium boundary jump as the lowest-order continuum projection of the linearized BTE~\citep{torres2018emergence,beardo2020phonon,sendra2021derivation}, PRISM-FDTR reconciles all six spot-size and spectral-window combinations into a configuration-independent intrinsic bulk silicon conductivity of $\SI{141}{\watt\per\metre\per\kelvin}$, matching single-crystal standards~\citep{glassbrenner1964thermal,inyushkin2004isotope}.

\subsection{Agent-guided operator learning for continuous-profile reconstruction}
\label{sec:deeponet}

\begin{figure}[!htbp]
  \centering
  \includegraphics[width=0.99\textwidth,height=0.52\textheight,keepaspectratio]{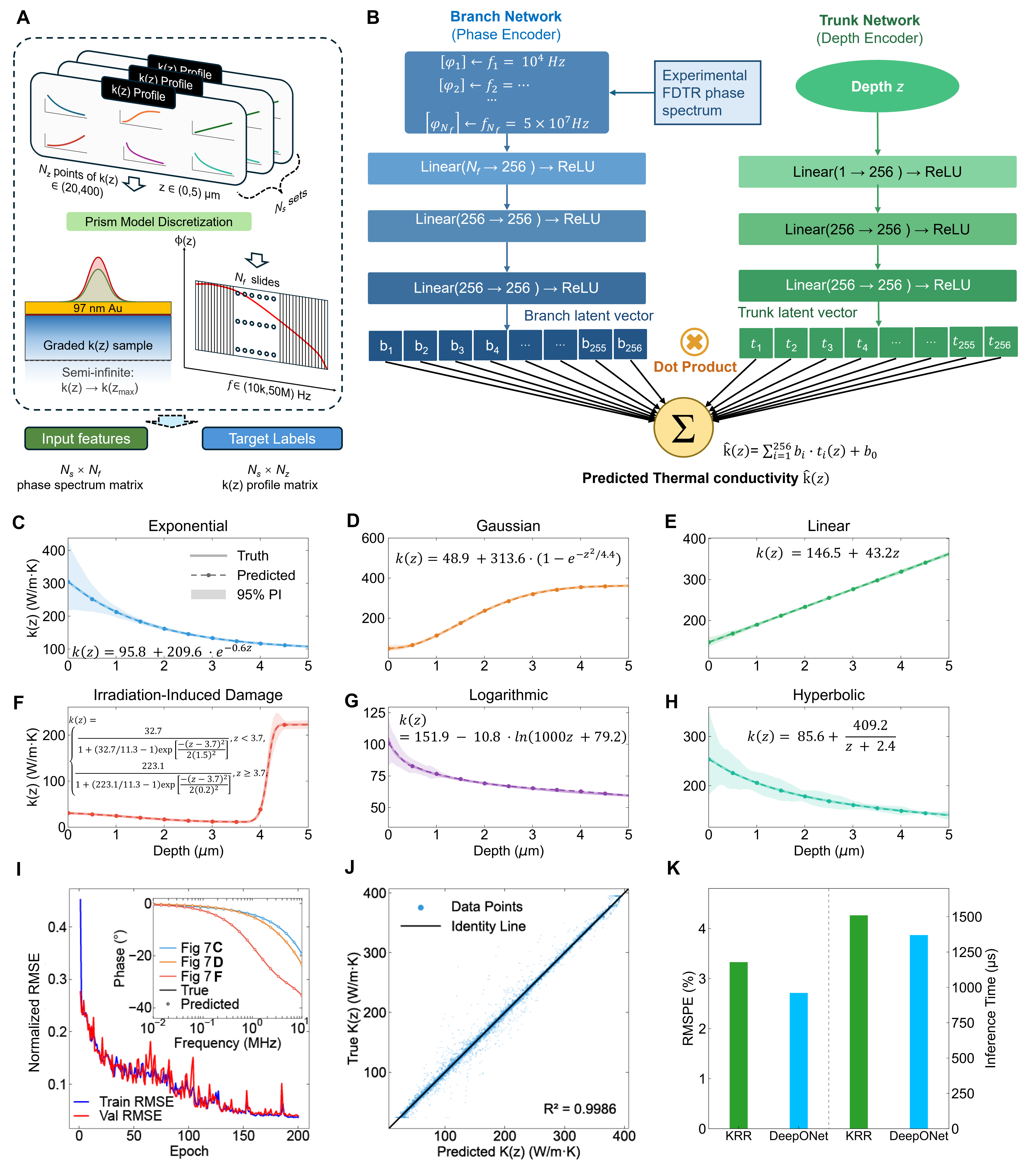}
  \caption{Continuous-profile branch of PRISM for $k(z)$ reconstruction. Three counts label the pipeline throughout: $N_s = 60{,}000$ is the number of paired phase-spectrum/profile samples in the synthetic dataset, $N_z = 500$ is the number of depth points at which every profile is sampled, and $N_f = 20$ is the number of modulation frequencies per spectrum. (A)~Synthetic training pipeline: $N_s$ profiles drawn from six analytic gradient families, each sampled at $N_z$ depth points over $z \in (0,5)\,\si{\micro\metre}$ with $k(z) \in [20,400]\,\si{\watt\per\metre\per\kelvin}$ ($[3,400]$ for the damage family), are discretized into 500 sublayers beneath a \SI{97}{\nano\metre} Au transducer and terminated by a semi-infinite base ($k(z) \to k(z_{\max})$), then evaluated with the graded OTh-STM solver, where the optically thick \SI{97}{\nano\metre} Au transducer ($h_{\mathrm{Au}} > 5\,d_{\mathrm{opt}}$) operates in the shallow-sensing limit so that the depth-weighted probe readout reduces analytically to the surface-temperature response (section~\ref{sec:volumetric_readout} of the supplemental information), at $N_f$ modulation frequencies spanning \SI{10}{\kilo\hertz}\text{--}\SI{50}{\mega\hertz}. The dataset is stored as an $N_s \times N_f$ phase-spectrum matrix (input features) and an $N_s \times N_z$ profile matrix (target labels), then split $7{:}2{:}1$ into 42,000 training, 12,000 validation, and 6,000 test pairs. (B)~DeepONet architecture: the branch network encodes the $N_f$-point phase spectrum---synthetic during training, an experimental FDTR phase spectrum at deployment---into latent vector $\mathbf{b}$; the trunk network maps query depth $z$ to basis functions $\mathbf{t}(z)$; their inner product reconstructs $\widehat{k}(z) = \sum_{i=1}^{256} b_i\, t_i(z) + b_0$. (C--H)~Representative profile reconstructions: ground-truth (solid) and predicted (dashed) profiles with nominal pointwise 95\% prediction intervals (shaded; section~\ref{sec:deeponet_kz} of the supplemental information) across (C)~exponential, (D)~Gaussian, (E)~linear, (F)~irradiation-induced damage profile with a steep recovery edge, (G)~logarithmic, and (H)~hyperbolic profiles. (I)~Training and validation normalized RMSE vs.\ epoch. Inset: physical re-simulation check for representative exponential, Gaussian, and damage-band profile pairs over the displayed \SI{10}{\kilo\hertz}\text{--}\SI{10}{\mega\hertz} range, achieving $\mathrm{RMSE} \le 0.024^\circ$. (J)~Parity plot of true vs.\ predicted $k(z)$ ($R^2 = 0.9986$). (K)~Comparison with kernel ridge regression (KRR): RMSPE is 3.3\% for KRR and 2.7\% for DeepONet; inference times are \SI{1510}{\micro\second} and \SI{1370}{\micro\second}, respectively, per batch of 100 spectra. RMSPE measures relative conductivity-profile error, whereas the normalized RMSE in (I) tracks training convergence (section~\ref{sec:krr} of the supplemental information).}
  \label{fig:deeponet}
\end{figure}

In functionally graded materials, ion-irradiated semiconductors, and thermally aged coatings, thermal conductivity varies continuously with depth, $k(z)$~\citep{pfeifer2022measuring,huynh2023depth,xiang2023machine,bao2025machine,zhang2026depth}. Reconstructing $k(z)$ through traditional transfer-matrix inversion requires discretizing the graded region into dozens of thin sublayers and fitting their conductivities simultaneously, creating a severely ill-posed inverse problem with more than 30 independently fitted layer conductivities ($N_p > 30$) prone to parameter cross-correlation, overfitting, and numerical instability~\citep{xiang2023machine}. Existing experimental studies of depth-damaged or graded specimens in continuous-wave FDTR typically bypass this ill-posedness by lumping the graded region into a single uniform layer with an effective conductivity $k_{\mathrm{eff}}$~\citep{treweek2024inversion,yan2024au,yan2025thermal}, whereas continuous $k(z)$ reconstructions on real graded samples have relied on pulsed TDTR or square-pulsed-source (SPS) hardware constrained to pre-assumed Gaussian or exponential functional forms~\citep{pfeifer2022measuring,huynh2023depth,bao2025machine,zhang2026depth}. Previous machine-learning studies have used synthetic FDTR phase libraries to benchmark continuous $k(z)$ inference with KRR~\citep{xiang2023machine} and dropout neural networks~\citep{ikeda2025}. For comparison with KRR on a shared synthetic dataset, PRISM constructs a deep operator network (DeepONet) for continuous-profile inversion~\citep{lu2021learning,wang2021learning} through the user-supervised extension route of Module~2 (\Cref{fig:routing_framework,fig:deeponet}; Figure~\ref{fig:si_deeponet_log}). The execution layer generates the training code, trains the neural operator, and checks the reconstructed profiles with the forward model (Figures~\ref{fig:infer-material} and~\ref{fig:compiled-state-transitions}). To generate the training library (\Cref{fig:deeponet}A), the agent discretizes $N_s = 60{,}000$ synthetic $k(z)$ profiles into $N=500$ sublayers, equal to the number $N_z$ of depth labels in this benchmark, over $0 < z < \SI{5}{\micro\metre}$ beneath a \SI{97}{\nano\metre} Au transducer and computes their $N_f = 20$-frequency phase spectra (\SI{10}{\kilo\hertz}\text{--}\SI{50}{\mega\hertz}) using the graded OTh-STM solver, where the optically thick \SI{97}{\nano\metre} Au film ($h_{\mathrm{Au}} > 5\,d_{\mathrm{opt}}$) operates in the shallow-sensing limit of the optical readout kernel (sections~\ref{sec:volumetric_readout} and~\ref{sec:deeponet_kz} of the supplemental information). At deployment, DeepONet (\Cref{fig:deeponet}B) factorizes the inverse operator $\mathcal{G}: \boldsymbol{\phi} \mapsto k(z)$ into a branch network ($N_f \to 256 \to 256 \to 256$, ReLU) that encodes the input phase spectrum $\boldsymbol{\phi} \in \mathbb{R}^{N_f}$ into latent coefficients $\mathbf{b} \in \mathbb{R}^{256}$ and a trunk network ($1 \to 256 \to 256 \to 256$, ReLU) that maps arbitrary query coordinates $z \in [0, \SI{5}{\micro\metre}]$ to continuous depth basis functions $\mathbf{t}(z) \in \mathbb{R}^{256}$, synthesizing the continuous profile via inner product, $\widehat{k}(z) = \mathbf{b}\cdot\mathbf{t}(z) + b_0$~\citep{lu2021learning}.

The synthetic library spans six functional $k(z)$ families over $0 < z < \SI{5}{\micro\metre}$ (\Cref{fig:deeponet}C--H; Table~\ref{tab:kz_profiles}): exponential, Gaussian, linear, irradiation-induced damage, logarithmic, and hyperbolic. Across all six families, the reconstructed profiles $\widehat{k}(z)$ closely track the ground-truth curves within the nominal pointwise 95\% prediction intervals. The irradiation-induced damage profile (\Cref{fig:deeponet}F) provides a synthetic test of operator expressivity, motivated by ion-implanted semiconductors where Stopping and Range of Ions in Matter (SRIM) recoil cascades generate a broad near-surface defect region culminating in a buried Bragg-peak damage valley followed by a sharp recovery to the undamaged crystal bulk~\citep{scott2020orders,yan2024au,yan2025thermal}. Parameterized as a piecewise asymmetric reciprocal-Gaussian valley (Equation~\eqref{eq:kz_damage}), accumulated displacement disorder depresses $k(z)$ from $\SI{30}{\watt\per\metre\per\kelvin}$ at the surface ($\sigma_L = \SI{1.5}{\micro\metre}$) to a minimum of $k_{\mathrm{min}} = \SI{11.3}{\watt\per\metre\per\kelvin}$ at the stopping depth $z_0 = \SI{3.7}{\micro\metre}$, immediately followed by a steep recovery edge ($\sigma_R = \SI{0.2}{\micro\metre}$) that restores the undamaged bulk conductivity of $\SI{223}{\watt\per\metre\per\kelvin}$ across a $10\text{--}90\%$ rise of only \SI{0.34}{\micro\metre} (a peak gradient of \SI{693}{\watt\per\metre\per\kelvin\per\micro\metre} and a valley-to-bulk contrast of $20$). From the 20-point phase spectrum, DeepONet reconstructs both the broad damaged shoulder and the abrupt recovery transition. Comparing the shaded 95\% prediction intervals across \Cref{fig:deeponet}C--H shows depth-dependent variation in the calibrated intervals: the intervals remain narrow across low-conductivity regions (such as the $11\text{--}\SI{30}{\watt\per\metre\per\kelvin}$ damage zone in \Cref{fig:deeponet}F and the near-surface region in \Cref{fig:deeponet}D), but widen in high-conductivity zones near the surface ($k \sim 200\text{--}\SI{310}{\watt\per\metre\per\kelvin}$ at $z < \SI{1}{\micro\metre}$ in \Cref{fig:deeponet}C,H) or at deep plateaus ($z > \SI{4}{\micro\metre}$ in \Cref{fig:deeponet}F). This contrast is consistent with the lower incremental thermal resistance of high-conductivity sublayers ($\mathrm{d}R_{\mathrm{th}} = \mathrm{d}z / k(z)$)---rendering high-$k(z)$ sublayers less thermally resistive while shallow sub-micrometer depths ($z \to 0$) are constrained only by the highest modulation frequencies approaching the $\SI{50}{\mega\hertz}$ bandwidth limit---and with the fact that the log-residual quantile intervals ($\widehat{k}(z)\exp[\pm q(z)]$; section~\ref{sec:deeponet_kz} of the supplemental information) scale multiplicatively with $\widehat{k}(z)$.

The $60{,}000$-pair library is partitioned $7{:}2{:}1$ into training ($42{,}000$), validation ($12{,}000$; \Cref{fig:deeponet}I), and test ($6{,}000$) subsets. The parity evaluation in \Cref{fig:deeponet}J reports a coefficient of determination $R^2 = 0.9986$. In the comparison with KRR~\citep{xiang2023machine} shown in \Cref{fig:deeponet}K, DeepONet achieves an RMSPE of 2.7\%, compared with 3.3\% for KRR: a reduction of 0.6 percentage points, or approximately 18\% relative to KRR. For a batch of 100 spectra, inference time decreases from \SI{1510}{\micro\second} to \SI{1370}{\micro\second}, a reduction of approximately 9.3\%. Following phase-to-profile inference~\citep{xiang2023machine,ikeda2025}, PRISM checks inferred profiles by forward re-simulation (\Cref{fig:deeponet}I inset): every predicted profile $\widehat{k}(z)$ is discretized and passed back through the 500-sublayer OTh-STM solver to verify spectral self-consistency. For the representative exponential, Gaussian, and damage-band profile pairs in the inset, applying the same forward operator to the reconstructed and ground-truth profiles gives phase differences of $\mathrm{RMSE} \le 0.024^\circ$ over the displayed \SI{10}{\kilo\hertz}\text{--}\SI{10}{\mega\hertz} range, well below the experimental noise floor ($\sigma_\phi = 0.1^\circ$). If a deployment reconstruction yields a re-simulation residual exceeding $\chi^2_{\mathrm{red}} = 3.0$, the agent automatically rejects the continuous profile and routes the dataset to discrete-layer inversion (Figure~\ref{fig:si_deeponet_log}).

\section{Discussion}
\label{sec:discussion}

The experimental benchmarks show that small phase-fit residuals can coexist with substantial errors in the extracted thermal properties. When a critical transport mechanism is omitted, non-linear regression absorbs the unmodeled phase lag into the remaining free parameters while still reproducing the experimental spectrum within noise tolerances. In thin-transducer $\text{Au/MoS}_2$, neglecting subsurface optical absorption and volumetric probe sensing depresses $k_{\mathrm{MoS}_2,z}$ by $40\%$ and $G$ by $29\%$ (\Cref{fig:optical}A,B)~\citep{yang2016analysis,guo2026light}; in $\text{Au/Si}$, treating the noble-metal transducer as a single-temperature medium lumps the near-interface electron--phonon equilibration resistance into the boundary, underestimating $G_{\mathrm{Au/Si}}$ by $24\text{--}28\%$ relative to two-temperature inversions (\Cref{tab:ausi}B)~\citep{majumdar2004role,regner2015interpretation}; and in $\text{Au/Si}$ and $\text{Au/Ti/Si}$, assuming local Fourier diffusion causes the apparent silicon conductivity to track the laser spot radius, beam offset, and modulation-frequency window, with the $\text{Au/Ti/Si}$ values spanning $116\text{--}\SI{137}{\watt\per\metre\per\kelvin}$ (\Cref{fig:ausi}D, \Cref{fig:nonlocal}D, \Cref{tab:nonlocal_windows})~\citep{minnich2011thermal,beardo2020phonon}. In practical device heterostructures such as $\text{Au/Ti/SiO}_2$ and $\text{Au/Ti/Si}$, these mechanisms act simultaneously, so correcting only one physical axis leaves residual biases along the others. PRISM combines the multiphysics model hierarchy with pre-fit SVD identifiability screening (\Cref{fig:ausi}C,F) and cross-configuration validation---assessing parameter identifiability and consistency across measurement configurations, as demonstrated by the convergence of $k_{\mathrm{Si}}$ to $\SI{141}{\watt\per\metre\per\kelvin}$ across all six spot-size and frequency-window combinations in $\text{Au/Ti/Si}$ (\Cref{fig:nonlocal}D).

PRISM combines transfer-matrix and admittance solvers with automated model selection, identifiability screening, and parameter estimation. Prior sensitivity-matrix reductions~\citep{tu2023universal,chen2024svd} identified measurable parameter groups within a static diffusive SH-STM model by rounding singular vectors to heuristic exponents; PRISM generalizes noise-whitened SVD screening across every branch of its optothermal hierarchy to prune unidentifiable parameters before fitting (\Cref{fig:workflow}). Building on rapid property inference using machine-learning surrogates~\citep{pang2021machine,xiang2023machine,mao2024deep} and agent-assisted analysis workflows such as Vibe-FDTR~\citep{yang2026vibefdtr} (\Cref{fig:arch}E), PRISM's execution layer selects and assembles the forward model (\Cref{fig:routing_framework}). By linking a structured domain knowledge graph of optothermal literature to modular C++/CUDA and MATLAB solver blocks, the agent evaluates dimensionless regime ratios and phase-deviation gates to assemble the appropriate transport branch, executes reduced-space global optimization via DTDE, and---when structured residuals or cross-condition drifts persist---proposes literature-backed model revisions or constructs and validates new solver branches under researcher supervision, as demonstrated by the DeepONet extension (\Cref{fig:arch}F; sections~\ref{sec:si_knowledge_base} and~\ref{sec:si_kg_revision} of the supplemental information).

The transfer-matrix formulation also provides a basis for adapting PRISM to other thermal-wave techniques. Time-domain thermoreflectance (TDTR) samples the same frequency-domain transfer-matrix response through a coherent sum over modulation sidebands~\citep{cahill2004analysis,mohan2025tdtr}, steady-state thermoreflectance (SSTR) probes its zero-frequency limit~\citep{braun2019steady}, the square-pulsed-source (SPS) method superposes its odd harmonics~\citep{song2025sps}, spatial-domain and modulated thermoreflectance map its lateral decay~\citep{fournier2020mtr,jiang2022spatial}, and electrothermal $3\omega$ measurements replace the optical pump with a narrow metal-line heater--thermometer~\citep{cahill1990thermal,borcatasciuc2001data}. Similarly, in photothermal radiometry and lock-in thermography of semitransparent or multilayer media, both pump absorption and detected infrared emission originate volumetrically beneath the surface~\citep{nordal1979photothermal,kusiak2013using,breitenstein2018lockin,chatterjee2024photothermal}. Since PRISM's OTh module already couples a transfer-matrix electromagnetic source with a depth-weighted readout kernel---a formulation equally applicable to interference-governed heating under extreme-ultraviolet irradiation~\citep{he2026interference}---infrared radiometry requires a depth-dependent emission kernel with the appropriate wavelength integration and boundary optics. These extensions reuse the layerwise TTM and non-Fourier models and the SVD analysis, with technique-specific source, readout, and noise models.

The current formulation can be extended in three directions. First, our non-Fourier silicon analyses employ the mesoscopic GK continuum equation with bulk relaxation parameters ($\tau, \ell$) fixed to the 311\,K first-principles values of Beardo \textit{et al.}~\citep{beardo2020phonon}. While this formulation reconciles multi-spot and multi-window coaxial measurements on bulk Si without adding free parameters (\Cref{fig:nonlocal}D), a difference remains between the coaxial and laterally offset $\text{Au/Si}$ configurations ($k_{\mathrm{Si}} = 147\pm 13$ vs.\ $132\pm\SI{8}{\watt\per\metre\per\kelvin}$ and equivalent conductance $G_{\mathrm{equiv}} = 87\pm 9$ vs.\ $73\pm\SI{8}{\mega\watt\per\metre\squared\per\kelvin}$; \Cref{tab:ausi}B). Because the substrate enters the multilayer recursion through its boundary admittance $Y_{\mathrm{sub}}^{\mathrm{NF}}$, the GK substrate model with an isotropic diffuse-slip boundary condition can be extended to include anisotropic tensor slip conditions, first-principles mode-resolved interfacial transmission~\citep{hua2017experimental,desanfeliciano2025metalsilicon}, or variance-reduced spectral BTE solvers~\citep{romano2021openbte,hu2024giftbte} when stronger ballistic anisotropy is probed. Second, whereas optical absorption in $\text{Au/Ti/Si}$ is confined to the metal bilayer (\Cref{fig:nonlocal}B) and ultrashort exciton lifetimes minimize photoexcited carrier transport in $\text{Au/MoS}_2$, bare or semitransparent narrow-bandgap semiconductors (such as $\text{InSb}$, $\text{Ge}$, or $\text{Bi}_2\text{O}_2\text{Se}$) exhibit pronounced ambipolar electron--hole diffusion and surface recombination that modulate optical reflectance alongside temperature~\citep{galovic2025carrier,park2025surfacerecomb}; embedding a coupled carrier-diffusion state vector into the transfer-matrix cascade will extend PRISM to these optoelectronic materials. Finally, the DeepONet benchmark uses a synthetic corpus generated by the graded OTh-STM solver. While this operator library accurately resolves smooth gradients and steep ion-irradiation damage valleys (\Cref{fig:deeponet}C--H), specimens containing sharp step discontinuities---such as buried voids or delaminations---fall outside smooth operator priors~\citep{kovachki2023neural}. PRISM's forward re-simulation check ($\chi^2_{\mathrm{red}} \le 3.0$) routes reconstructions that fail forward-residual or plausibility checks back to discrete-layer inversion, while coupling discontinuity-preserving physics-informed neural operators~\citep{wang2021learning,qiu2025novel} and TTM/NF-enabled training generators to experimental depth-profiling campaigns would extend the conductivity profiles and transport regimes represented in the training data.

\section{Materials and methods}
\label{sec:methods}

\subsection{Resource availability}
\subsubsection{Lead contact}
Further information and requests for resources should be directed to the lead contact, Qiye Zheng (\texttt{qiyezheng@ust.hk}).

\subsubsection{Materials availability}
This study did not generate unique biological or physical reagents.

\subsubsection{Data and code availability}
Primary measurement datasets, forward transport solvers, SVD screening scripts, and agentic orchestration routines are available from the corresponding author upon reasonable request.

\subsection{Method details}
\subsubsection{Frequency-domain thermoreflectance measurements}
A sinusoidally modulated pump laser beam ($\lambda^{\mathrm{pump}} = \SI{445}{\nano\metre}$) generates periodic thermal waves at the sample surface, while a probe laser beam ($\lambda^{\mathrm{probe}} = \SI{532}{\nano\metre}$) measures the resulting thermoreflectance response. Lock-in detection extracts the phase lag $\phi(f)$ relative to pump modulation across \SI{1}{\kilo\hertz}\text{--}\SI{50}{\mega\hertz}, with the analysed window set per sample by the data-quality gate. Both beams are focused by the same $20\times$ objective. For every FDTR measurement the beam radius is calibrated on a low-conductivity Au/SiO\textsubscript{2} reference sample by scanning the pump--probe offset and fitting the resulting cross-correlation profile; the calibration constrains the combined pump--probe widths along each axis. For our calibrated measurements, $w_x,w_y$ are the raw $1/e^2$ radii of the cross-correlation profile, so $w_a^2=w_{\mathrm{pump},a}^2+w_{\mathrm{probe},a}^2$ for $a=x,y$. These widths enter the combined Gaussian kernel of Equation~\eqref{eq:fourier_combined_kernel}. If equal pump/probe radii are used as equivalent solver inputs, each is $w_a/\sqrt{2}$ on its own axis; this calibration does not separately determine the two individual beam radii. The calibration was also checked independently by knife-edge scans. The Au/SiO\textsubscript{2}/Si, Au/Ti/SiO\textsubscript{2}/Si, and Au/Ti/Si benchmarks are measured at two coaxial spot sizes, with circular cross-correlation radii of \SI{1.86}{\micro\metre} and \SI{10.44}{\micro\metre}; the larger spot is obtained with the same objective by inserting a convex lens so that the beam underfills the back focal plane. The Au/Ti/SiO\textsubscript{2} control on bulk silica is measured at the small spot only. For the beam offset experiments, we used slightly elliptical beams ($w_x = \SI{2.19}{\micro\metre}$, $w_y = \SI{1.84}{\micro\metre}$), and the Au/Si analysis retains both radii in the two-dimensional Fourier quadrature. The Au/Si benchmark is measured in two configurations: coaxial ($\Delta = 0$) and offset, in which a steering mirror displaces the pump and probe centres by $\Delta \approx \SI{2.0}{\micro\metre}$; $\Delta$ denotes the centre-to-centre beam offset. The benchmark beam radii and offsets used throughout are collected in Table~\ref{tab:archive_si_beam_sizes}.   The pump and probe laser powers were calibrated with an optical power meter. The Au/Si and Au/Ti/Si samples were prepared on pieces of the same single-crystal Si wafer, and the FDTR system is the one used in our previous studies~\citep{lu2026process,lee2026disorder}. The ambient temperature was \SI{296.15}{\kelvin}. Steady-state laser heating raises the mean temperature in the probed region to approximately \SI{311}{\kelvin}. We therefore evaluate the silicon heat capacity and the non-Fourier transport coefficients at this representative operating temperature, using the 311\,K values reported by Beardo \textit{et al.}~\cite{beardo2020phonon} directly. The modulated thermal response is calculated about this steady-state temperature. The metal transducer films were deposited by magnetron sputtering, and their thicknesses were measured with a KLA-Tencor Alpha-Step D300 profilometer. The Au and Ti thermal conductivities were determined from van der Pauw electrical measurements of films co-deposited on a fused-silica reference substrate. Electrical resistivity $\rho$ was converted to thermal conductivity using $k=L_0T/\rho+k_{\mathrm{ph}}$, with the Sommerfeld Lorenz number $L_0=2.44\times10^{-8}\,\mathrm{V^2\,K^{-2}}$ and literature first-principles phonon conductivity $k_{\mathrm{ph}}$~\citep{wang2016first}. 

\subsubsection{Multilayer optical transfer-matrix absorption and volumetric sensing}
For layer $n$ with thickness $d_n$ and complex refractive index $\widetilde N_n=n_n+i\kappa_n$, the optical transfer-matrix method (TMM) propagates forward ($E_n^+$) and backward ($E_n^-$) electric-field amplitudes across interfaces and layers~\citep{yeh1988optical,pettersson1999modeling}. Evaluating the Poynting-flux divergence yields the depth-resolved pump absorption profile $Q_n(z)$ in $\mathrm{m^{-1}}$ per unit incident irradiance defined in \Cref{sec:inference_framework} (section~\ref{sec:optical_tmm} of the supplemental information, Equations~\eqref{eq:heat_source_general}--\eqref{eq:source_factorization}), with forward, backward, and interference coefficients:
\begin{equation}
Q_{n,\mathrm{fwd}}=\frac{\alpha_{\mathrm{opt},n}^{\mathrm{pump}}n_n}{n_{\mathrm{air}}}|E_n^+|^2,\qquad
Q_{n,\mathrm{bwd}}=\frac{\alpha_{\mathrm{opt},n}^{\mathrm{pump}}n_n}{n_{\mathrm{air}}}|E_n^- e^{-i\beta_n^{\mathrm{pump}}}|^2,\qquad
Q_{n,\mathrm{int}}=\frac{2\alpha_{\mathrm{opt},n}^{\mathrm{pump}}n_n}{n_{\mathrm{air}}}|(E_n^+)^* E_n^-|,
\label{eq:optical_source_meth}
\end{equation}
where $\alpha_{\mathrm{opt},n}^{\mathrm{pump}}=4\pi\kappa_n^{\mathrm{pump}}/\lambda^{\mathrm{pump}}$, $\beta_{\mathrm{opt},n}^{\mathrm{pump}}=4\pi n_n^{\mathrm{pump}}/\lambda^{\mathrm{pump}}$, and $\varphi_n=-\arg[(E_n^+)^* E_n^-]$~\citep{yang2016analysis,guo2026light}. Multiplying $Q_n(z)$ by the two-dimensional Fourier-transformed Gaussian pump envelope $\widetilde{I}_{\mathrm{pump}}(u,v)$ supplies the layerwise source $\widetilde{S}_n(u,v,z)$ (Equations~\eqref{eq:fourier_transform_pair} and~\eqref{eq:fourier_combined_kernel}). At the probe wavelength $\lambda^{\mathrm{probe}}$, the first-order Born reflectance perturbation integrates the oscillating temperature field against the complex optical sensitivity kernel across all absorbing layers (section~\ref{sec:volumetric_readout} of the supplemental information, Equations~\eqref{eq:probe_gf_def} and~\eqref{eq:couple_observable_volumetric}).

\subsubsection{Single- and two-temperature models}
For a nonmetallic layer $n$, harmonic time dependence and a two-dimensional in-plane spatial Fourier transform give:
\begin{equation}
i\omega C_n\hat T_n=k_{z,n}\frac{\partial^2\hat T_n}{\partial z^2}-4\pi^2(k_{x,n}u^2+k_{y,n}v^2)\hat T_n+\widetilde S_n(u,v,z),
\label{eq:stm_meth}
\end{equation}
with decay constant $\zeta_n=\sqrt{[i\omega C_n+4\pi^2(k_{x,n}u^2+k_{y,n}v^2)]/k_{z,n}}$ ($\operatorname{Re}\zeta_n>0$). For a transversely isotropic layer with $k_{x,n}=k_{y,n}$ and circular Gaussian beams, the thermal response depends on $\mu_\rho^2=4\pi^2(u^2+v^2)$ and the spatial integral reduces to its Hankel form. The layer state vectors $[\widehat T_n,\widehat q_n]^\mathsf{T}$, with cross-plane heat flux $\hat{q}_n = -k_{z,n}\partial\hat{T}_n/\partial z$, propagate top-down via $2\times2$ transfer matrices $\mathbf M_n^{\mathrm{STM}}$~\citep{cahill2004analysis,feldman1999algorithm}. The STM matrix is given in section~\ref{sec:tmm} of the supplemental information, with its transfer/admittance form in Equations~\eqref{eq:couple_layer_transfer}--\eqref{eq:couple_layer_admittance}.

In a metallic layer $n$ with electron--phonon nonequilibrium, applying the in-plane Fourier transform to Equations~\eqref{eq:ttm_el_main}--\eqref{eq:ttm_ph_main} yields the coupled $2\times 2$ homogeneous differential operator $\mathbf{A}_n$ with diagonal entries $\alpha_{a,n} = (i\omega C_{a,n} + g_{\mathrm{ep},n} + 4\pi^2(k_{a,x,n}u^2+k_{a,y,n}v^2))/k_{a,z,n}$ ($a \in \{\mathrm{el}, \mathrm{ph}\}$) and off-diagonal coupling terms $-g_{\mathrm{ep},n}/k_{a,z,n}$~\citep{regner2015interpretation}. Diagonalizing $\mathbf{A}_n$ yields two spatial decay eigenvalues $\lambda_{n,1}^2, \lambda_{n,2}^2$ with phonon-to-electron modal amplitude ratios $p_{n,m} = (k_{\mathrm{el},z,n}/g_{\mathrm{ep},n})(\alpha_{\mathrm{el},n} - \lambda_{n,m}^2)$. Rather than cascading exponentials that can overflow in thick layers, PRISM propagates the four-component state vector $\boldsymbol{\Theta}_n = [\widehat{T}_{\mathrm{el}}, \widehat{T}_{\mathrm{ph}}, \widehat{q}_{\mathrm{el}}, \widehat{q}_{\mathrm{ph}}]^\mathsf{T}$ using unconditionally stable self- and mutual-admittance matrices (section~\ref{sec:tmm} of the supplemental information, Equations~\eqref{eq:self_admittance}--\eqref{eq:offset_recursion}):
\begin{equation}
\mathbf{Y}_{s,n} = \mathbf{K}_n\mathbf{P}_n\boldsymbol{\Lambda}_n\,\coth(\boldsymbol{\Lambda}_n d_n)\,\mathbf{P}_n^{-1}, \qquad
\mathbf{Y}_{m,n} = \mathbf{K}_n\mathbf{P}_n\boldsymbol{\Lambda}_n\,\csch(\boldsymbol{\Lambda}_n d_n)\,\mathbf{P}_n^{-1},
\label{eq:ttm_matrix_meth}
\end{equation}
where $\mathbf{P}_n = \begin{pmatrix} 1 & 1 \\ p_{n,1} & p_{n,2} \end{pmatrix}$, $\boldsymbol{\Lambda}_n = \mathrm{diag}(\lambda_{n,1},\lambda_{n,2})$, and $\mathbf{K}_n = \mathrm{diag}(k_{\mathrm{el},z,n},k_{\mathrm{ph},z,n})$. Single-temperature nonmetallic layers are embedded into the phonon slots of this $4\times 4$ representation, and the full stack response is assembled via bottom-up Schur-complement admittance recursion across layers and interfaces (Equations~\eqref{eq:interface_bc_upper}--\eqref{eq:interface_operator_inverse} and~\eqref{eq:couple_interface_admittance}--\eqref{eq:couple_top_bc}).

\subsubsection{Nonlocal Guyer--Krumhansl transport formulation}
For a layer with non-diffusive phonon transport (e.g., the Si substrate), PRISM uses the isotropic GK formulation of Equation~\eqref{eq:gk_transport_main} when local branches leave configuration-dependent fitted properties~\citep{beardo2020phonon}. The non-Fourier parameters are taken directly from the 311\,K first-principles reference data of Beardo \textit{et al.}~\cite{beardo2020phonon}, with $\ell=\SI{185}{\nano\metre}$ and $\tau=\SI{42}{\pico\second}$; the full parameter set is listed in Table~\ref{tab:archive_si_nf_params}. Harmonic excitation and the two-dimensional Fourier transform of Equation~\eqref{eq:fourier_transform_pair} give compressive and shear decay rates through Helmholtz decomposition (section~\ref{sec:nf_helmholtz} of the supplemental information, Equations~\eqref{eq:lambda_p}--\eqref{eq:lambda_s}):
\begin{equation}
\lambda_p = \sqrt{4\pi^2(u^2+v^2) + \frac{i\omega(1+i\omega\tau)}{D + i\omega\ell^2(\alpha_{\mathrm{GK}}+1)}}, \qquad
\lambda_s = \sqrt{4\pi^2(u^2+v^2) + \frac{1+i\omega\tau}{\ell^2}},
\label{eq:decay_constants_meth}
\end{equation}
Here $C=C_{\mathrm{Si}}$ is volumetric heat capacity, $D=k_{\mathrm{Si}}/C_{\mathrm{Si}}$ is thermal diffusivity, $\omega=2\pi f$, $\tau$ is the flux-relaxation time and $\ell$ is the nonlocal length; decaying branches satisfy $\operatorname{Re}\lambda_{p,s}>0$. The tangential flux obeys $q_t=\mathcal C_{\mathrm{slip}}\ell\,\partial_nq_t$, with the interface-normal convention of Equation~\eqref{eq:bc_tangential_slip}. The subscript \emph{slip} denotes tangential heat-flux slip; $\mathcal C_{\mathrm{slip}}=(1+p)/(1-p)$ is dimensionless, $p$ is the phonon specular-reflection fraction, and this work uses $p=0$. The bulk impedance is
\begin{equation}
\begin{aligned}
Z_{\mathrm{bulk}}(u,v,\omega)&=
\frac{-i[4\pi^2(u^2+v^2)-\lambda_p^2]\lambda_s(1+\mathcal C_{\mathrm{slip}}\ell\lambda_s)}
{C\omega\,\mathcal D(u,v,\omega)},\\
\mathcal D(u,v,\omega)&=
4\pi^2(u^2+v^2)(1+\mathcal C_{\mathrm{slip}}\ell\lambda_p)
-\lambda_s\lambda_p(1+\mathcal C_{\mathrm{slip}}\ell\lambda_s).
\end{aligned}
\label{eq:gk_impedance_meth}
\end{equation}
Interface resistance and gradient corrections give $Z_{\mathrm{jump}}=R-Z_{\mathrm{corr},1}-Z_{\mathrm{corr},2}$ and $Y_{\mathrm{NF}}=(Z_{\mathrm{bulk}}+Z_{\mathrm{jump}})^{-1}$ (Equations~\eqref{eq:Z_corr1}--\eqref{eq:Y_NF_modeIII}). Eliminating the semi-infinite GK modes yields the bottom boundary admittance matrix for the substrate:
\begin{equation}
\widehat q_{\mathrm{el}}=0,\qquad
\widehat q_{\mathrm{ph}}=Y_{\mathrm{NF}}\widehat T_{\mathrm{ph}},\qquad
\mathbf Y_{b,N}=\begin{pmatrix}0&0\\0&Y_{\mathrm{NF}}\end{pmatrix}.
\label{eq:gk_ttm_load_meth}
\end{equation}
This boundary admittance matrix initializes the upward TTM Schur-complement recursion (section~\ref{sec:nf_ttm_embedding} of the supplemental information, Algorithm~\ref{alg:gk_embedding}). The ordinary substrate-interface conductance is bypassed because its resistance is already included in $Z_{\mathrm{jump}}$.

\subsubsection{Agentic execution layer}
The execution layer is implemented as a stateful LangGraph execution graph~\citep{langgraph2024}. The PRISM solvers and analysis tools are exposed to the language model, DeepSeek V4 Flash, as Model Context Protocol (MCP) tools, and numerical outputs are logged together with the tool inputs. The knowledge graph consulted for routing and recovery is a wiki of typed, cited pages that the agent compiles from the optothermal transport literature and the solver code \siref{sections~\ref{sec:si_knowledge_base} and \ref{sec:si_kg_revision} of the supplemental information}.

\subsubsection{Noise-scaled sensitivity, SVD screening, and global optimization}
Logarithmic sensitivities $S_{p_j}(f_i)=\partial\phi(f_i)/\partial\ln p_j$ are evaluated by symmetric central differences with step $\delta=\ln1.01$. Decomposing the noise-whitened Jacobian $\mathbf{S}_{\mathrm{norm}}=(\delta/\sigma_\phi)\mathbf{S}=\mathbf{U}\boldsymbol{\Sigma}\mathbf{V}^\mathsf{T}$ (with baseline noise floor $\sigma_\phi=0.1^\circ$ or calibrated $\sigma_\phi(f)$)~\citep{tu2023universal,chen2024svd} yields the retained singular-mode subspace $\mathcal{I}_{\mathrm{SVD}}=\{m:\sigma_m/\sigma_1>\tau_{\mathrm{SNR}}\}$ ($\tau_{\mathrm{SNR}}=0.1$, $r_{\mathrm{SVD}}=|\mathcal{I}_{\mathrm{SVD}}|$). Conditional on independently fixed priors, each candidate parameter $p_j$ is scored by its retained-subspace projection $\eta_j$ and sensitivity energy $E_j$ (Algorithm~\ref{alg:svd} in the supplemental information):
\begin{equation}
\eta_j = \sum_{m\in\mathcal{I}_{\mathrm{SVD}}} |V_{jm}|^2 \ge \frac{1}{P}, \qquad
E_j = \sum_{m\in\mathcal{I}_{\mathrm{SVD}}} \left(\frac{\sigma_m}{\sigma_1}\right)^2 |V_{jm}|^2, \qquad
\mathcal{L}_{\mathrm{ident}} = \mathcal{L}_{\mathrm{SVD}} \cap \mathcal{L}_{\mathrm{sensitivity}},
\label{eq:identifiable_list_main}
\end{equation}
where $\mathcal{L}_{\mathrm{SVD}}$ orders candidates passing $\eta_j \ge 1/P$ by descending $E_j$, and $\mathcal{L}_{\mathrm{sensitivity}}=\{p_j:\bar{S}_{p_j,\mathrm{rms}}\ge S_{\min}\}$ ($S_{\min}=0.2^\circ$, $\bar{S}_{p_j,\mathrm{rms}}=\sqrt{N_f^{-1}\sum_i|S_{p_j}(f_i)|^2}$) enforces minimum band-averaged phase sensitivity.
For routine inference, requested fitting targets are admitted by membership (Algorithm~\ref{alg:target_adjust} in the supplemental information) in this list. The bulk-silica co-fit is a diagnostic model comparison described in section~\ref{sec:optical_result}. Parameters outside the list retain their specified prior values. $N_{\mathrm{ident}}$ counts eligible list entries, while $r_{\mathrm{SVD}}$ counts retained sensitivity directions; the confirmed request specifies the subset fitted together. Table~\ref{tab:fit_ledger} records released targets and fixed inputs. We benchmarked seven derivative-free global optimizers---CMA-ES, DTDE~\citep{zhang2023dtde}, jSO, EA4Eig, L-SHADE~\citep{brest2017single,bujok2022ea4eig,tanabe2014lshade}, HPSO~\citep{xiao2025hybridpso}, and canonical DE---under a shared 600-evaluation budget and terminal L-BFGS-B refinement protocol in section~\ref{sec:optimizer_benchmark} of the supplemental information (Figure~\ref{fig:si_optimizer}). While all seven methods converge to an identical terminal phase RMSE ($\approx 0.208^\circ$) and CMA-ES requires the fewest function evaluations by adapting its covariance matrix, DTDE achieves the shortest evolutionary wall-clock time ($\approx\SI{127}{\second}$) by dynamically contracting the search domain without inner-loop matrix factorizations; because analytical FDTR forward evaluations are computationally fast, wall-clock runtime governs throughput and DTDE is selected as the default optimizer.

\subsubsection{DeepONet operator learning architecture}
The continuous-profile inversion employs a dual-MLP DeepONet architecture~\citep{lu2021learning,wang2021learning}. The branch network ($N_f \rightarrow 256 \rightarrow 256 \rightarrow 256$, ReLU) encodes the $N_f$-frequency phase spectrum---$N_f = 20$ points spanning \SI{10}{\kilo\hertz}\text{--}\SI{50}{\mega\hertz} for phase-to-profile training and inference---into latent vector $\mathbf{b} \in \mathbb{R}^{256}$. The trunk network ($1 \rightarrow 256 \rightarrow 256 \rightarrow 256$, ReLU) maps query depth $z \in [0, \SI{5}{\micro\metre}]$ to depth basis vector $\mathbf{t}(z) \in \mathbb{R}^{256}$. Thermal conductivity is reconstructed as $\widehat{k}(z) = \mathbf{b} \cdot \mathbf{t}(z) + b_0$~\citep{lu2021learning}. The synthetic corpus of $N_s = 60{,}000$ graded OTh-STM spectra, whose target profiles are sampled at $N_z = 500$ depth points, is partitioned $7{:}2{:}1$ into training (42,000), validation (12,000), and test (6,000) subsets, so that training uses the 42,000 training pairs, with Adam at initial learning rate $10^{-3}$ and weight decay $10^{-5}$, a decaying learning-rate schedule, batch size 128, 200 epochs, and mean-squared profile error.

\section{Author contributions}
Y.Q.: conceptualization, methodology, software, model training, investigation, data curation, formal analysis, visualization, and writing -- original draft. Z.Z. and Y.J.H.: investigation and writing -- review \& editing. Z.Q. and Y.L.: resources (FDTR experimental data). C.L.: funding acquisition and writing -- review \& editing. Q.Z.: conceptualization, supervision, funding acquisition, and writing -- review \& editing.

\section{Declaration of generative AI and AI-assisted technologies in the writing process}
During the preparation of this work, the authors used large-language-model assistants to polish the language of the manuscript and to help debug code. After using these tools, the authors reviewed and edited the content as needed and take full responsibility for the content of the publication. The large language model DeepSeek V4 Flash is a component of the PRISM framework itself and its use is described in \hyperref[sec:methods]{materials and methods}.

\section{Declaration of interests}
The authors declare no competing interests.

\section{Acknowledgments}
{\RaggedRight\rightskip=0pt plus 1fil\relax
This work was supported by the National Key R\&D Program of China (Grant No. 2024YFB4405700), the Research Grants Council of Hong Kong (Project No. 16214626), and the National Natural Science Foundation of China (Grant No. 52676057). It was also funded by the Guangdong Basic and Applied Basic Research Foundation (Grant Nos. 2025A1515012898 and 2026A1515011290), the Opening Project of the State Key Laboratory of Advanced Glass Materials (Grant No. 2025KF27), and the Frontier Technology Research for Joint Institutes with Industry Scheme sponsored by the Center on Smart Sensors and Environmental Technologies at HKUST. Furthermore, the authors gratefully acknowledge the HKUST Fok Ying Tung Research Institute and the National Supercomputing Center in Guangzhou (Nansha Sub-center) for providing the high-performance computational resources used in this study. C. Liu acknowledges support from the National Natural Science Foundation of China (Grant No. 52676072), the Basic Research Program of Jiangsu (BK20250035), and the National Key R\&D Program of China (2024YFF0508900).
\par}

\section*{Supplemental information}
Additional derivations, parameter tables, validation figures, and execution details are provided in the \hyperlink{supplemental-information}{supplemental information}.

\begin{singlespace}
\AtNextBibliography{\setlength{\RaggedRightParindent}{0pt}\RaggedRight\rightskip=0pt plus 1fil\relax}
\printbibliography
\end{singlespace}

\clearpage

\SInumbering
\hypertarget{supplemental-information}{}
\begin{center}
  \rule{\textwidth}{1.5pt}\\[14pt]
  {\LARGE\bfseries\color{sectioncolor} Supplemental Information}\\[10pt]
  {\Large\bfseries PRISM: Unifying multiphysics optothermal transport and identifiability-guided inference for nanoscale thermal metrology}\\[12pt]
  {\normalsize Yicong~QIU\textsuperscript{1}, Ziyang~ZHONG\textsuperscript{1,2,3}, Ziyan~QIAN\textsuperscript{1}, Yang~LU\textsuperscript{1}, Yong~Jian~HONG\textsuperscript{1}, Chenhan~LIU\textsuperscript{4}, and Qiye~ZHENG\textsuperscript{1,5,*}}\\[6pt]
  {\footnotesize
  \textsuperscript{1}Department of Mechanical and Aerospace Engineering, The Hong Kong University of Science and Technology, Clear Water Bay, Hong Kong SAR, China 999077\par
  \textsuperscript{2}Department of Computer Science and Engineering, The Hong Kong University of Science and Technology, Clear Water Bay, Hong Kong SAR, China 999077\par
  \textsuperscript{3}Department of Electronic and Computer Engineering, The Hong Kong University of Science and Technology, Clear Water Bay, Hong Kong SAR, China 999077\par
  \textsuperscript{4}School of Energy and Mechanical Engineering, Nanjing Normal University, Nanjing 210023, China\par
  \textsuperscript{5}Lead contact\par
  \textsuperscript{*}Correspondence: [qiyezheng@ust.hk]
}
  \\[12pt]
  \rule{\textwidth}{1.5pt}
\end{center}

\clearpage
\tableofcontents
\clearpage
\addtocontents{toc}{\protect\setcounter{tocdepth}{3}}

\section{Nomenclature and Symbol Conventions}
\label{sec:nomenclature}

Tables~\ref{tab:ifm_terms}--\ref{tab:module_symbols} define the terminology, beam widths, material parameters, matrices, and module symbols used in the main text and supplemental information.

\paragraph{PRISM terminology conventions.}

PRISM (Physics-Resolved Inference and Screening Meta-framework) comprises the forward-model hierarchy, screening tools, and agentic execution layer.

\begin{table}[H]
\centering
\caption{PRISM terminology conventions.}
\label{tab:ifm_terms}
\small
\begin{tabularx}{\textwidth}{@{} l >{\raggedright\arraybackslash}X @{}}
\toprule
\textbf{Term} & \textbf{Definition} \\
\midrule
PRISM & Physics-Resolved Inference and Screening Meta-framework; the unified, physically ordered FDTR framework used in this work. \\
PRISM branch hierarchy & The discrete chain SH-STM, VH-STM, OTh-STM, OTh-TTM, and OTh-TTM(NF), together with the continuous-profile route when activated. \\
PRISM screening logic & The data-diagnosis, sensitivity, identifiability, and target-reduction logic used before fitting. \\
PRISM execution layer & The agentic execution layer that parses tasks, routes branches, orchestrates solver calls, tracks state, and applies verification and recovery. \\
PRISM solver and diagnostic toolset & The set of tools exposed to the router:
\texttt{diagnose\_data}, \texttt{estimate\_sensitivity}, \texttt{analyze\_identifiability},
\newline \texttt{compute\_phase}, \texttt{fit\_parameters}, \texttt{predict\_kz}, and \texttt{infer\_material}. \\
\bottomrule
\end{tabularx}
\end{table}

\subsection{Physical and Material Parameters}

FDTR uses Gaussian pump and probe beams in coaxial or offset configurations. Individual beam radii are measured at the $1/e^2$ intensity contour; directional cross-correlation widths are defined in Table~\ref{tab:beam_params}.

\begin{table}[H]
\centering
\caption{Laser beam parameter symbols.}
\label{tab:beam_params}
\small
\begin{tabularx}{\textwidth}{@{} l l l >{\raggedright\arraybackslash}X @{}}
\toprule
\textbf{Symbol} & \textbf{Unit} & \textbf{Name} & \textbf{Definition} \\
\midrule
$w_{\mathrm{pump}}$ & $\mu\mathrm{m}$ & Pump beam radius & $1/e^2$ intensity radius of pump laser \\
$w_{\mathrm{probe}}$ & $\mu\mathrm{m}$ & Probe beam radius & $1/e^2$ intensity radius of probe laser \\
$w_x,w_y$ & $\mu\mathrm{m}$ & Directional cross-correlation radii & Raw $1/e^2$ calibration widths along $x,y$, with $w_a^2=w_{\mathrm{pump},a}^2+w_{\mathrm{probe},a}^2$, $a=x,y$; both axes remain independent in Equation~\eqref{eq:fourier_combined_kernel}. \\
$w,w_r$ & $\mu\mathrm{m}$ & Circular correlation radius (tag) & Common raw calibration width for our circular spots, $w_x=w_y=w_r$; individual pump/probe radii are written explicitly when needed. \\
\midrule
\multicolumn{4}{l}{\textit{Note:}} \\
\multicolumn{4}{p{12cm}}{For a Gaussian field with intensity $I\propto|E|^2$, the $1/e$ field-amplitude radius equals the $1/e^2$ intensity radius. A $1/e$ intensity radius is smaller by $\sqrt{2}$ than the $1/e^2$ intensity radius. Individual and cross-correlation radii are related by the squared-width sum above.} \\
\bottomrule
\end{tabularx}
\end{table}

\paragraph{Thermal conductivity.}

Scalar thermal conductivity is denoted by lowercase $k$ throughout this document. A depth-dependent profile is $k(z)$, its prediction is $\widehat{k}(z)$, and its sublayer value is $k_j=k(z_j)$. In scalar component equations, $z$ is normal to the interfaces and $x,y$ are the two in-plane directions; $k_z,k_x,k_y$ denote the corresponding conductivity components. Additional subscripts identify the material, layer, or electron/phonon channel.

\begin{table}[H]
\centering
\caption{Thermal conductivity symbols.}
\label{tab:thermal_conductivity}
\small
\begin{tabularx}{\textwidth}{@{} l l l >{\raggedright\arraybackslash}X @{}}
\toprule
\textbf{Symbol} & \textbf{Unit} & \textbf{Name} & \textbf{Definition} \\
\midrule
$k$ & $\mathrm{W\,m^{-1}\,K^{-1}}$ & Thermal conductivity (generic) & Fourier thermal conductivity \\
$k_z$ & $\mathrm{W\,m^{-1}\,K^{-1}}$ & Cross-plane conductivity & Conductivity along the interface-normal $z$ direction \\
$k_x$ & $\mathrm{W\,m^{-1}\,K^{-1}}$ & In-plane $x$ conductivity & Conductivity along the in-plane $x$ direction \\
$k_y$ & $\mathrm{W\,m^{-1}\,K^{-1}}$ & In-plane $y$ conductivity & Conductivity along the in-plane $y$ direction \\
$k_{\mathrm{Si}}$ & $\mathrm{W\,m^{-1}\,K^{-1}}$ & Silicon thermal conductivity & Isotropic bulk Si thermal conductivity \\
$k_{\mathrm{Si},r}$ & $\mathrm{W\,m^{-1}\,K^{-1}}$ & Si in-plane thermal conductivity & Anisotropic Si in-plane conductivity, used only where the offset geometry is analysed anisotropically \\
$k_{\mathrm{Si},z}$ & $\mathrm{W\,m^{-1}\,K^{-1}}$ & Si cross-plane thermal conductivity & Anisotropic Si cross-plane conductivity, used only where the offset geometry is analysed anisotropically \\
$k_{\mathrm{SiO_2}}$ & $\mathrm{W\,m^{-1}\,K^{-1}}$ & Silica thermal conductivity & Isotropic thermal conductivity of SiO$_2$ \\
$k_{\mathrm{Au}}$ & $\mathrm{W\,m^{-1}\,K^{-1}}$ & Gold thermal conductivity & Isotropic thermal conductivity of Au \\
$k_{\mathrm{Ti}}$ & $\mathrm{W\,m^{-1}\,K^{-1}}$ & Titanium thermal conductivity & Isotropic thermal conductivity of Ti \\
$k_{\mathrm{Au,el}}$ & $\mathrm{W\,m^{-1}\,K^{-1}}$ & Au electron thermal conductivity & Electron channel in TTM \\
$k_{\mathrm{Au,ph}}$ & $\mathrm{W\,m^{-1}\,K^{-1}}$ & Au phonon thermal conductivity & Phonon channel in TTM \\
$k_{\mathrm{MoS_2},z}$ & $\mathrm{W\,m^{-1}\,K^{-1}}$ & MoS$_2$ cross-plane conductivity & Anisotropic; perpendicular to layers \\
$k_{\mathrm{MoS_2},r}$ & $\mathrm{W\,m^{-1}\,K^{-1}}$ & MoS$_2$ in-plane conductivity & Anisotropic; parallel to layers \\
$a_{\mathrm{ani}}$ & --- & Transverse anisotropy factor & $a_{\mathrm{ani}}=k_x/k_z=k_y/k_z$ only when $k_x=k_y$ \\
\midrule
\multicolumn{4}{l}{\textit{Note:}} \\
\multicolumn{4}{p{12cm}}{In isotropic cases, $k_x=k_y=k_z\equiv k$. Scalar component equations retain $x,y,z$ explicitly whenever directional conductivities differ. The reported material-specific symbol $k_{\mathrm{Si},r}$ or $k_{\mathrm{MoS_2},r}$ denotes the common in-plane value only in a transversely isotropic case, where the corresponding $x$ and $y$ components are equal. Thermal-wave spatial frequencies are denoted by $u,v$ or $\mu_\rho$. The $k$ in the microscopic phonon phase-space measure $\mathrm{d}^3k$ denotes a crystal wavevector, and discrete $k$ in summations, singular-value labels, or optimization algorithms is an integer index; these are not thermal conductivities.} \\
\bottomrule
\end{tabularx}
\end{table}

\paragraph{Interface thermal conductance.}

The interface thermal conductance $G$ represents the heat flux per unit area per unit temperature difference across an interface.

\begin{table}[H]
\centering
\caption{Interface thermal conductance symbols.}
\label{tab:conductance}
\small
\begin{tabularx}{\textwidth}{@{} l l l >{\raggedright\arraybackslash}X @{}}
\toprule
\textbf{Symbol} & \textbf{Unit} & \textbf{Name} & \textbf{Definition} \\
\midrule
$G$ & $\mathrm{MW\,m^{-2}\,K^{-1}}$ & Total interface conductance & Total heat flux per temperature drop across interface \\
$G_{\mathrm{ph}}$ & $\mathrm{MW\,m^{-2}\,K^{-1}}$ & Phonon interface conductance & Phonon-mediated contribution to $G$ \\
$G_{\mathrm{el}}$ & $\mathrm{MW\,m^{-2}\,K^{-1}}$ & Electron interface conductance & Electron-mediated contribution to $G$ (metallic interfaces) \\
$G_{\mathrm{Au/Si}}$ & $\mathrm{MW\,m^{-2}\,K^{-1}}$ & Au/Si interface conductance & Au-to-Si interface conductance \\
$G_{\mathrm{Au/MoS_2}}$ & $\mathrm{MW\,m^{-2}\,K^{-1}}$ & Au/MoS$_2$ interface conductance & Au-to-MoS$_2$ interface conductance \\
$G_{\mathrm{Ti/SiO_2}}$ & $\mathrm{MW\,m^{-2}\,K^{-1}}$ & SiO$_2$/Ti interface conductance & SiO$_2$-to-Ti interface conductance \\
\midrule
\multicolumn{4}{l}{\textit{Note:}} \\
\multicolumn{4}{p{12cm}}{In the TTM formalism, $G_{ee}$, $G_{ep}$, $G_{pe}$, $G_{pp}$ denote the electron--electron, electron--phonon, phonon--electron, and phonon--phonon components of the interface conductance matrix $\mathbf{G}$. For the diagonal two-channel metallic-interface inputs used here, $G_{\mathrm{total}}=G_{\mathrm{ph}}+G_{\mathrm{el}}$.} \\
\bottomrule
\end{tabularx}
\end{table}

\paragraph{Volumetric heat capacity.}

\begin{table}[H]
\centering
\caption{Volumetric heat capacity symbols.}
\label{tab:heat_capacity}
\small
\begin{tabularx}{\textwidth}{@{} l l l >{\raggedright\arraybackslash}X @{}}
\toprule
\textbf{Symbol} & \textbf{Unit} & \textbf{Name} & \textbf{Definition} \\
\midrule
$C$ & $\mathrm{J\,m^{-3}\,K^{-1}}$ & Volumetric heat capacity (generic) & Energy storage per unit volume \\
$C_{\mathrm{Si}}$ & $\mathrm{J\,m^{-3}\,K^{-1}}$ & Si volumetric heat capacity & Bulk Si heat capacity \\
$C_{\mathrm{Au}}$ & $\mathrm{J\,m^{-3}\,K^{-1}}$ & Au volumetric heat capacity & Bulk Au heat capacity \\
$C_{\mathrm{el}}$ & $\mathrm{J\,m^{-3}\,K^{-1}}$ & Electron heat capacity & Electron channel in TTM (metal) \\
$C_{\mathrm{ph}}$ & $\mathrm{J\,m^{-3}\,K^{-1}}$ & Phonon heat capacity & Phonon channel in TTM (metal) \\
\bottomrule
\end{tabularx}
\end{table}

\paragraph{Electron--phonon coupling and transport parameters.}

\begin{table}[H]
\centering
\caption{Electron--phonon coupling and transport parameters.}
\label{tab:transport_params}
\small
\begin{tabularx}{\textwidth}{@{} l l l >{\raggedright\arraybackslash}X @{}}
\toprule
\textbf{Symbol} & \textbf{Unit} & \textbf{Name} & \textbf{Definition} \\
\midrule
$g_{\mathrm{ep}}$ & $\mathrm{W\,m^{-3}\,K^{-1}}$ & Electron--phonon coupling coefficient & Energy transfer rate per volume per temperature \\
$L_{\mathrm{ep}}$ & $\mathrm{nm}$ & Electron--phonon coupling length & $L_{\mathrm{ep}} = \sqrt{k_{\mathrm{ph}}/g_{\mathrm{ep}}}$ \\
\midrule
\multicolumn{4}{l}{\textit{Non-Fourier (NF) hydrodynamic parameters for Si:}} \\
$\tau$ & $\mathrm{ps}$ & Relaxation time & Heat-flux memory time; controls onset of wave-like behavior \\
$\ell$ & $\mathrm{nm}$ & Nonlocal length & Spatial nonlocality characteristic length \\
$\alpha_{\mathrm{GK}}$ & --- & Nonlocal coefficient & Relative weight of compressive vs. total nonlocal corrections in the GK-type phonon-hydrodynamic substrate closure \\
$b_{\mathrm{NF}}$ & --- & Contact-quality factor & Scales interface coefficients for imperfect contact \\
$\mathcal{C}_{\mathrm{slip}}$ & --- & Tangential-slip coefficient & $\mathcal{C}_{\mathrm{slip}} = (1+p)/(1-p)$; diffuse limit $\mathcal{C}_{\mathrm{slip}}=1$ \\
$\gamma_0$ & $\mathrm{MW\,m^{-2}\,K^{-1}}$ & Substrate modal transfer coefficient & $\gamma_0 = \frac{1}{4}v_g C$ from Si-side distribution \\
$\gamma_{0\Gamma}$ & $\mathrm{MW\,m^{-2}\,K^{-1}}$ & Transducer modal transfer coefficient & $\gamma_{0\Gamma} = \frac{1}{4}v_{g\Gamma}C_\Gamma$ from the transducer-side distribution \\
$\beta_{\mathrm{NF}}$ & $\mathrm{nm}$ & Nonlocal divergence coefficient & From divergence term $\nabla\cdot\mathbf{q}$ in the nonequilibrium interface jump condition \\
$\chi_{xx}$ & $\mathrm{nm}$ & Cross-plane gradient coefficient & Interface-normal component in the local GK coordinates of section~\ref{sec:non_fourier_si}, where $x$ is the normal direction \\
$\chi_{zz}$ & $\mathrm{nm}$ & In-plane gradient coefficient & Tangential component in the same local GK coordinates \\
\bottomrule
\end{tabularx}
\end{table}

\subsection{Mathematical and Module Symbols}

\begin{table}[H]
\centering
\caption{Matrix and operator symbols.}
\label{tab:matrices}
\small
\begin{tabularx}{\textwidth}{@{} l l >{\raggedright\arraybackslash}X @{}}
\toprule
\textbf{Symbol} & \textbf{Dimension} & \textbf{Name} \\
\midrule
$\mathbf{M}_n$ & $2\times2$ or $4\times4$ & Top-to-bottom layer matrix; the upward map is its inverse \\
$\mathbf{G}_j$ & $2\times2$ & Physical interface conductance matrix for interface $j$ \\
$\widetilde{\mathbf{G}}_j$ & $4\times4$ & Interface transfer operator for interface $j$ \\
$\mathbf{Y}_{s,n}$ & $2\times2$ & Self-admittance matrix of layer $n$ \\
$\mathbf{Y}_{m,n}$ & $2\times2$ & Mutual-admittance matrix of layer $n$ \\
$\mathbf{Y}_{b,n}$ & $2\times2$ & Equivalent admittance loading the bottom of layer $n$ \\
$\mathbf{Y}_{t,n}$ & $2\times2$ & Equivalent admittance seen at the top of layer $n$ \\
$\mathbf J\equiv\mathbf S$ & $N_f\times P$ & Logarithmic phase-sensitivity Jacobian \\
$\hat{\mathbf J}\equiv\mathbf S_{\mathrm{norm}}$ & $N_f\times P$ & Dimensionless normalized Jacobian \\
$\mu_{\rho}$ & $\mathrm{m^{-1}}$ & Angular in-plane magnitude, $\mu_\rho^2=4\pi^2(u^2+v^2)$; $u,v$ remain independent Fourier variables \\
$Q^\ast$ & --- & Data-quality score \\
$\mathbf{U}, \boldsymbol{\Sigma}, \mathbf{V}^\top$ & --- & SVD components of $\hat{\mathbf{J}}$ \\
$\sigma_k$ & --- & $k$-th singular value \\
$N_{\mathrm{ident}}$ & --- & Number of entries in the ordered intersection list $\mathcal{L}_{\mathrm{ident}}$ \\
$r_{\mathrm{SVD}}$ & --- & Number of modes above the relative SVD threshold \\
$\mathcal{L}_{\mathrm{ident}}$ & --- & Ordered list passing SVD and absolute-sensitivity screening \\
\midrule
\multicolumn{3}{l}{\textit{TTM state vector:}} \\
$\boldsymbol{\Theta}_n=[\hat T_{e,n},\hat T_{p,n},\hat q_{e,n},\hat q_{p,n}]^T$ & $4\times1$ & Thermal state vector of layer $n$ (electron/phonon temperatures and fluxes) \\
\bottomrule
\end{tabularx}
\end{table}

\paragraph{Module-specific symbols.}

\begin{table}[H]
\centering
\caption{Optical, TTM, interface, and NF module symbols.}
\label{tab:module_symbols}
\small
\begin{tabularx}{\textwidth}{@{} l l >{\raggedright\arraybackslash}X @{}}
\toprule
\textbf{Symbol} & \textbf{Module} & \textbf{Definition} \\
\midrule
$Q_n(z)$ & Optical source & Depth deposition per unit incident irradiance, $\mathrm{m^{-1}}$ \\
$S_n(x,y,z)$ & Optical source & Physical volumetric source $I_{\mathrm{pump}}Q_n$, $\mathrm{W\,m^{-3}}$ \\
$\tilde S_n(u,v,z)$ & Optical source & Two-dimensional Fourier source $\tilde I_{\mathrm{pump}}Q_n$, $\mathrm{W\,m^{-1}}$ \\
$Q_{n,\mathrm{fwd}},Q_{n,\mathrm{bwd}},Q_{n,\mathrm{int}}$ & Optical source & Per-irradiance depth coefficients, $\mathrm{m^{-1}}$ \\
$\Phi_{n,a}(u,v)$ & Optical/thermal coupling & Source coefficient $\tilde I_{\mathrm{pump}}Q_{n,a}$ for $a=\mathrm{fwd,bwd,int}$, $\mathrm{W\,m^{-1}}$ \\
$\lambda_{n,1},\lambda_{n,2}$ & TTM & Two thermal decay constants of the metallic TTM layer $n$ \\
$p_{n,m}$ & TTM & Phonon-to-electron temperature ratio of TTM mode $m$ in layer $n$ \\
$\mathbf{q}^{\mathrm{off}}_{\mathrm{top},n},\mathbf{q}^{\mathrm{off}}_{\mathrm{bot},n}$ & TTM/admittance & Source-offset vectors at the top and bottom faces of layer $n$ \\
$Y_{\mathrm{sub}}^{\mathrm{STM}}$ & Substrate boundary module & Diffusive STM admittance of the semi-infinite substrate \\
$Y_{\mathrm{sub}}^{\mathrm{NF}}$ & Substrate boundary module & Phonon-hydrodynamic non-Fourier admittance that replaces $Y_{\mathrm{sub}}^{\mathrm{STM}}$ in the benchmark-conditioned high-thermal-conductivity substrate branch \\
$\alpha_{\mathrm{opt},n}^{(a)},\beta_{\mathrm{opt},n}^{(a)},\lambda^{(a)}$ & Optical source/readout & Absorption coefficient, standing-wave wavenumber and wavelength for $a=\mathrm{pump}$ or $\mathrm{probe}$ \\
$\alpha_{\mathrm{GK}},\beta_{\mathrm{NF}}$ & NF/hydrodynamic & GK-type nonlocal coefficient and nonequilibrium divergence coefficient used only in the substrate closure \\
\bottomrule
\end{tabularx}
\end{table}

\clearpage

\section{PRISM Branch Hierarchy for Metallic and Semiconductor FDTR Stacks}
\label{sec:tmm}

This section details the governing equations, matrix formulations, and layer-cascading algorithms that comprise the forward transport hierarchy of PRISM. For planar multilayer heterostructures, PRISM systematically constructs the forward solver along three coupled physical axes: optical electrodynamics ($\mathcal{E}$), carrier nonequilibrium ($\mathcal{X}$), and substrate phonon conduction ($\mathcal{T}$):
\begin{equation}
\mathrm{SH}\text{-}\mathrm{STM} \;\longrightarrow\; \{\mathrm{VH}\text{-}\mathrm{STM},\;\mathrm{OTh}\text{-}\mathrm{STM}\} \;\longrightarrow\; \mathrm{OTh}\text{-}\mathrm{TTM} \;\longrightarrow\; \mathrm{OTh}\text{-}\mathrm{TTM(NF)}.
\label{eq:si_branch_chain}
\end{equation}
The hierarchy progresses logically from the classical diffusive baseline to fully coupled non-Fourier formulations:
\begin{itemize}
    \item \textbf{Diffusive backbone ($\mathcal{X} = \mathrm{STM}$):} Non-metallic layers are solved via the $2\times 2$ single-temperature transfer-matrix formulation, enforcing local Fourier diffusion under surface or volumetric heating.
    \item \textbf{Electrodynamic source extension ($\mathcal{E} = \mathrm{OTh}$):} Optical transfer matrices replace idealized surface-flux assumptions with depth-resolved volumetric absorption profiles $Q(z)$ and depth-weighted probe reflectance kernels $\widetilde H(u,v,\omega)$, resolving standing-wave interference and subsurface heating.
    \item \textbf{Carrier nonequilibrium ($\mathcal{X} = \mathrm{TTM}$):} Metallic films are expanded into a $4\times 4$ two-temperature matrix differential system that tracks independent electron ($T_{\mathrm{el}}$) and phonon ($T_{\mathrm{ph}}$) fields, coupled to a four-element boundary conductance matrix $\mathbf{G}_j$.
    \item \textbf{Mesoscopic nonlocal formulation ($\mathcal{T} = \mathrm{NF}$):} In crystalline semiconductor substrates where long-mean-free-path phonons induce quasi-ballistic suppression, the semi-infinite Fourier boundary admittance is replaced by the nonlocal Guyer--Krumhansl (GK) continuum. This boundary formulation is derived via Helmholtz decomposition of the heat flux into compressive and shear modes, yielding closed-form expressions for the frequency- and wavevector-dependent bulk impedance $Z_{\mathrm{bulk}}$, nonequilibrium boundary jump $Z_{\mathrm{jump}}$, and effective substrate admittance $Y_{\mathrm{sub}}^{\mathrm{NF}}$.
\end{itemize}
Diffusive dielectric media (e.g., amorphous $\mathrm{SiO}_2$) remain on the local Fourier branch ($\mathcal{T} = \mathrm{Fourier}$), whereas the nonlocal closure ($\mathcal{T} = \mathrm{NF}$) is activated only when persistent spot-size or frequency dependencies signal the breakdown of local diffusion. Full symbol definitions and units are tabulated in section~\ref{sec:nomenclature}.

\subsection{Diffusive Transfer-Matrix Backbone: STM and TTM}

The diffusive backbone combines single-temperature (STM) and two-temperature (TTM) layer models. Dielectric and nonmetallic layers use STM with directional thermal conductivities. Metallic layers can also use STM in the local-equilibrium limit; TTM resolves separate electron and phonon temperatures when their nonequilibrium must be retained. Both descriptions enter the same multilayer transfer-matrix framework.

\subsubsection{Single-Temperature Model (STM) for Nonmetal Layers}
\label{sec:stm_model}

The in-plane transform is two-dimensional, with
\begin{equation}
\begin{aligned}
    \tilde f(u,v,z)&=\iint_{\mathbb R^2}f(x,y,z)e^{-2\pi i(ux+vy)}\,\mathrm{d}x\,\mathrm{d}y,\\
    f(x,y,z)&=\iint_{\mathbb R^2}\tilde f(u,v,z)e^{2\pi i(ux+vy)}\,\mathrm{d}u\,\mathrm{d}v.
\end{aligned}
    \label{eq:fourier_transform_pair}
\end{equation}
Here $u,v$ are cyclic spatial frequencies, with angular wavenumbers $\mu_x=2\pi u$ and $\mu_y=2\pi v$. Thermal fields and layer/interface operators are evaluated at each $(u,v,\omega)$. For in-plane-isotropic material kernels, $\mu_\rho^2=4\pi^2(u^2+v^2)$ is a convenient invariant, while independent beam widths and offsets enter the two-dimensional synthesis. The conductivity and beam axes are aligned; a rotated in-plane tensor retains its mixed derivative term. The same spatial transform is used for local and nonlocal constitutive models.

In nonmetallic layer $n$, a single lattice temperature $\hat{T}_n$ suffices to describe thermal transport. The anisotropic Fourier law relates the cross-plane heat flux to the temperature gradient through the cross-plane thermal conductivity $k_{z,n}$:
\begin{equation}
    \hat{q}_n = -k_{z,n}\frac{\partial \hat{T}_n}{\partial z}.
    \label{eq:stm_fourier}
\end{equation}
In the frequency--wavevector domain, energy conservation takes the form
\begin{equation}
    i\omega C_n\,\hat{T}_n = k_{z,n}\frac{\partial^2\hat{T}_n}{\partial z^2} - 4\pi^2(k_{x,n}u^2+k_{y,n}v^2)\,\hat{T}_n + \tilde{S}_n(u,v,z),
    \label{eq:stm_energy}
\end{equation}
where $C_n$ is the volumetric heat capacity and $k_{x,n},k_{y,n},k_{z,n}$ are the directional conductivities. In the transversely isotropic limit $k_{x,n}=k_{y,n}$, the in-plane loss term becomes $k_{x,n}\mu_\rho^2$ and $a_{\mathrm{ani},n}=k_{x,n}/k_{z,n}=k_{y,n}/k_{z,n}$. Seeking $\hat T_n\propto e^{\pm\zeta_nz}$ gives
\begin{equation}
    \zeta_n^2 = \frac{i\omega C_n + 4\pi^2(k_{x,n}u^2+k_{y,n}v^2)}{k_{z,n}},
    \label{eq:stm_gamma}
\end{equation}
The decaying branch satisfies $\operatorname{Re}\zeta_n>0$. Equation~\eqref{eq:stm_gamma} retains the cross-plane conductivity in the denominator and the two directional in-plane conductivities in the numerator. For $k_{x,n}=k_{y,n}=k_{z,n}\equiv k_n$, it reduces to the isotropic scalar model.

Using the state vector $[\hat{T}_n,\hat{q}_n]^\mathsf{T}$ with $\hat{q}_n=-k_{z,n}\partial\hat{T}_n/\partial z$, the $2\times 2$ transfer matrix across a layer of thickness $h_n$ is
\begin{equation}
    \mathbf{M}_n^{\mathrm{STM}} =
    \begin{pmatrix}
        \cosh(\zeta_n h_n) & -\dfrac{\sinh(\zeta_n h_n)}{k_{z,n}\zeta_n} \\[6pt]
        -k_{z,n}\zeta_n\sinh(\zeta_n h_n) & \cosh(\zeta_n h_n)
    \end{pmatrix}.
    \label{eq:stm_matrix}
\end{equation}
This matrix maps the layer-top state at $z=0$ to the layer-bottom state at $z=h_n$. With a volumetric source, the forward relation is $\boldsymbol\Theta_n(h_n)=\mathbf M_n^{\mathrm{STM}}\boldsymbol\Theta_n(0)+\mathbf b_n$. A multilayer stack is assembled by ordered multiplication of layer and interface operators.

\subsubsection{Two-Temperature Model (TTM) for Metallic Films}
\label{sec:ttm_model}

When layer $n$ is metallic, the single-temperature description is extended to a two-temperature model that resolves the electron and phonon subsystems separately. In real space, the layer equations are
\begin{align}
    C_{\mathrm{el},n}\frac{\partial T_{\mathrm{el},n}}{\partial t} &= k_{\mathrm{el},z,n}\frac{\partial^2 T_{\mathrm{el},n}}{\partial z^2} + k_{\mathrm{el},x,n}\partial_x^2T_{\mathrm{el},n}+k_{\mathrm{el},y,n}\partial_y^2T_{\mathrm{el},n} - g_{\mathrm{ep},n}\bigl(T_{\mathrm{el},n}-T_{\mathrm{ph},n}\bigr) + S_n(x,y,z),\label{eq:ttm_realspace_e} \\
    C_{\mathrm{ph},n}\frac{\partial T_{\mathrm{ph},n}}{\partial t} &= k_{\mathrm{ph},z,n}\frac{\partial^2 T_{\mathrm{ph},n}}{\partial z^2} + k_{\mathrm{ph},x,n}\partial_x^2T_{\mathrm{ph},n}+k_{\mathrm{ph},y,n}\partial_y^2T_{\mathrm{ph},n} + g_{\mathrm{ep},n}\bigl(T_{\mathrm{el},n}-T_{\mathrm{ph},n}\bigr).\label{eq:ttm_realspace_p}
\end{align}
After harmonic time dependence and the two-dimensional Fourier transform of Equation~\eqref{eq:fourier_transform_pair}, the layer equations become
\begin{align}
    i\omega C_{\mathrm{el},n}\hat{T}_{\mathrm{el},n} &= k_{\mathrm{el},z,n}\frac{\partial^2 \hat{T}_{\mathrm{el},n}}{\partial z^2} - 4\pi^2(k_{\mathrm{el},x,n}u^2+k_{\mathrm{el},y,n}v^2)\hat{T}_{\mathrm{el},n} - g_{\mathrm{ep},n}\bigl(\hat{T}_{\mathrm{el},n}-\hat{T}_{\mathrm{ph},n}\bigr) + \tilde{S}_n(u,v,z),\label{eq:ttm_transform_e} \\
    i\omega C_{\mathrm{ph},n}\hat{T}_{\mathrm{ph},n} &= k_{\mathrm{ph},z,n}\frac{\partial^2 \hat{T}_{\mathrm{ph},n}}{\partial z^2} - 4\pi^2(k_{\mathrm{ph},x,n}u^2+k_{\mathrm{ph},y,n}v^2)\hat{T}_{\mathrm{ph},n} + g_{\mathrm{ep},n}\bigl(\hat{T}_{\mathrm{el},n}-\hat{T}_{\mathrm{ph},n}\bigr).\label{eq:ttm_transform_p}
\end{align}
where $k_{\mathrm{el/ph},z,n}$ and $k_{\mathrm{el/ph},x,n},k_{\mathrm{el/ph},y,n}$ are cross-plane and directional in-plane thermal conductivities, $C_{\mathrm{el/ph},n}$ are volumetric heat capacities, $g_{\mathrm{ep},n}$ is the electron--phonon coupling coefficient, and the volumetric source couples exclusively to the electron system.

The homogeneous part of the transformed TTM equations is
\begin{equation}
    \frac{d^2}{dz^2} \begin{pmatrix} \hat{T}_{\mathrm{el},n} \\ \hat{T}_{\mathrm{ph},n} \end{pmatrix} = \mathbf{A}_n \begin{pmatrix} \hat{T}_{\mathrm{el},n} \\ \hat{T}_{\mathrm{ph},n} \end{pmatrix}, \qquad
    \mathbf{A}_n = \begin{pmatrix}
        \alpha_{\mathrm{el},n} & -g_{\mathrm{ep},n}/k_{\mathrm{el},z,n} \\
        -g_{\mathrm{ep},n}/k_{\mathrm{ph},z,n} & \alpha_{\mathrm{ph},n}
    \end{pmatrix}
    \label{eq:ttm_matrix_form}
\end{equation}
with
\[
\alpha_{a,n} = \frac{i\omega C_{a,n}+g_{\mathrm{ep},n}+4\pi^2(k_{a,x,n}u^2+k_{a,y,n}v^2)}{k_{a,z,n}}, \qquad a\in\{\mathrm{el},\mathrm{ph}\}.
\]

\subsubsection{Eigenmodes, Layer Matrices, and Interface/Admittance Recursion}
\label{sec:interface_coupling}

The coupled TTM system is diagonalized through the eigenvalues $\lambda_{n,1}^2$ and $\lambda_{n,2}^2$ of $\mathbf{A}_n$,
\begin{equation}
    \lambda_{n,m}^2 = \frac{\mathrm{Tr}(\mathbf{A}_n) + (-1)^{m+1}\sqrt{\mathrm{Tr}(\mathbf{A}_n)^2 - 4\det(\mathbf{A}_n)}}{2}, \qquad m\in\{1,2\},
    \label{eq:ttm_lambda}
\end{equation}
and the corresponding phonon-to-electron temperature ratios
\begin{equation}
    p_{n,m} = \frac{\hat{T}_{\mathrm{ph},n}^{(m)}}{\hat{T}_{\mathrm{el},n}^{(m)}} = \frac{k_{\mathrm{el},z,n}}{g_{\mathrm{ep},n}}\bigl(\alpha_{\mathrm{el},n}-\lambda_{n,m}^2\bigr), \qquad m\in\{1,2\}.
    \label{eq:ttm_modal_ratio}
\end{equation}
Using these two decaying modes, the layer-resolved analytic temperature profiles are
\begin{align}
    \hat{T}_{\mathrm{el},n}(z) &= \sum_{m=1}^{2}\left[A_{n,m}e^{-\lambda_{n,m}z}+B_{n,m}e^{-\lambda_{n,m}(h_n-z)}\right] + \hat{T}_{\mathrm{el},n}^{(p)}(z),\label{eq:ttm_te_profile} \\
    \hat{T}_{\mathrm{ph},n}(z) &= \sum_{m=1}^{2} p_{n,m}\left[A_{n,m}e^{-\lambda_{n,m}z}+B_{n,m}e^{-\lambda_{n,m}(h_n-z)}\right] + \hat{T}_{\mathrm{ph},n}^{(p)}(z),\label{eq:ttm_tp_profile}
\end{align}
where $h_n$ is the layer thickness, $A_{n,m}$ and $B_{n,m}$ are modal amplitudes fixed by its boundary conditions, and $\hat T_{\mathrm{el/ph},n}^{(p)}$ are source-generated particular solutions. Define the full state $\boldsymbol\Theta_n$ and its homogeneous part $\boldsymbol\Theta_n^{(h)}=\boldsymbol\Theta_n-\boldsymbol\Theta_{p,n}$. The two propagation directions are
\begin{equation}
\begin{aligned}
\boldsymbol\Theta_n(z)&=[\hat T_{\mathrm{el},n},\,\hat T_{\mathrm{ph},n},\,\hat q_{\mathrm{el},n},\,\hat q_{\mathrm{ph},n}]^\mathsf T,\\
\mathbf M_n^{\mathrm{TTM},\uparrow}&=\boldsymbol\chi_n\mathbf E_n\boldsymbol\chi_n^{-1},\\
\mathbf M_n^{\mathrm{TTM},\downarrow}
&=(\mathbf M_n^{\mathrm{TTM},\uparrow})^{-1}
=\boldsymbol\chi_n\mathbf E_n^{-1}\boldsymbol\chi_n^{-1},\\
\boldsymbol\Theta_n^{(h)}(h_n)&=\mathbf M_n^{\mathrm{TTM},\downarrow}\boldsymbol\Theta_n^{(h)}(0).
\end{aligned}
\label{eq:ttm_layer_matrix}
\end{equation}
with
\begin{equation}
\begin{aligned}
    \boldsymbol{\chi}_n&=\begin{pmatrix}
        1 & 1 & 1 & 1 \\
        p_{n,1} & p_{n,2} & p_{n,1} & p_{n,2} \\
        k_{\mathrm{el},z,n}\lambda_{n,1} & k_{\mathrm{el},z,n}\lambda_{n,2} & -k_{\mathrm{el},z,n}\lambda_{n,1} & -k_{\mathrm{el},z,n}\lambda_{n,2} \\
        k_{\mathrm{ph},z,n}\lambda_{n,1}p_{n,1} & k_{\mathrm{ph},z,n}\lambda_{n,2}p_{n,2} & -k_{\mathrm{ph},z,n}\lambda_{n,1}p_{n,1} & -k_{\mathrm{ph},z,n}\lambda_{n,2}p_{n,2}
    \end{pmatrix},
    \\
    \mathbf{E}_n&=\mathrm{diag}\!\left(
        e^{\lambda_{n,1}h_n},
        e^{\lambda_{n,2}h_n},
        e^{-\lambda_{n,1}h_n},
        e^{-\lambda_{n,2}h_n}
    \right).
\end{aligned}
\label{eq:ttm_chi_matrix}
\end{equation}

\paragraph{Interface assembly and admittance recursion.}

For interface $j$ between the upper layer and the lower layer, the physical interfacial conductance operator is
\begin{equation}
    \mathbf{G}_j=\begin{pmatrix} G_{\mathrm{ee},j} & G_{\mathrm{ep},j} \\ G_{\mathrm{pe},j} & G_{\mathrm{pp},j} \end{pmatrix}.
    \label{eq:interface_conductance_matrix}
\end{equation}
For ordinary diffusive interfaces, the nonzero entries of $\mathbf G_j$ are empirical channel conductances. At the Ti/Si boundary of the Au/Ti/Si example, Mode~I uses $G_{\mathrm{pp},j}$. Modes~II and III instead use the non-Fourier substrate admittance of section~\ref{sec:non_fourier_si}, which includes the resistance $R$ of Equation~\eqref{eq:R_kapitza}; the ordinary phonon-interface operator is bypassed.
Define the upper- and lower-channel sums
\begin{equation}
    \mathbf{G}_{u,j}=\mathrm{diag}(G_{\mathrm{ee},j}+G_{\mathrm{ep},j},\,G_{\mathrm{pe},j}+G_{\mathrm{pp},j}), \qquad
    \mathbf{G}_{\ell,j}=\mathrm{diag}(G_{\mathrm{ee},j}+G_{\mathrm{pe},j},\,G_{\mathrm{ep},j}+G_{\mathrm{pp},j}).
    \label{eq:interface_channel_sums}
\end{equation}
Take $z$ and the interfacial heat flux as positive downward, and write $\mathbf{T}_{u,j}$ and $\mathbf{T}_{\ell,j}$ for the temperatures immediately above and below interface $j$. Each entry $G_{ab,j}$ connects upper channel $a$ to lower channel $b$. The boundary conditions are
\begin{align}
    \mathbf{q}_{u,j} &= \mathbf{G}_{u,j}\mathbf{T}_{u,j}-\mathbf{G}_j\mathbf{T}_{\ell,j},\label{eq:interface_bc_upper}\\
    \mathbf{q}_{\ell,j} &= \mathbf{G}_j^\mathsf{T}\mathbf{T}_{u,j}-\mathbf{G}_{\ell,j}\mathbf{T}_{\ell,j}.\label{eq:interface_bc_lower}
\end{align}
These equations conserve total heat flux, $\mathbf{1}^\mathsf{T}\mathbf{q}_{u,j}=\mathbf{1}^\mathsf{T}\mathbf{q}_{\ell,j}$, while allowing conversion between electron and phonon channels. In the scalar STM limit they reduce to $\hat q_u=\hat q_\ell=G_j(\hat T_u-\hat T_\ell)$: finite $G_j$ permits a temperature jump $\hat q/G_j$.

For the diffusive metal/Si boundary used here, the electron channel is closed and the phonon conditions are
\begin{equation}
    \hat q_{\mathrm{el}}=0,\qquad
    \hat q_{\mathrm{ph}}=G_{\mathrm{pp}}(\hat T_{\mathrm{ph},u}-\hat T_{\mathrm{Si}})
    =Y_{\mathrm{sub}}^{\mathrm{STM}}\hat T_{\mathrm{Si}},\qquad
    Y_{\mathrm{eff}}=\frac{G_{\mathrm{pp}}Y_{\mathrm{sub}}^{\mathrm{STM}}}{G_{\mathrm{pp}}+Y_{\mathrm{sub}}^{\mathrm{STM}}}.
    \label{eq:interface_scalar_load}
\end{equation}
Thus $\hat q_{\mathrm{ph}}=Y_{\mathrm{eff}}\hat T_{\mathrm{ph},u}$ is the load applied to the metal-side boundary. As $G_{\mathrm{pp}}\to\infty$, the temperatures become continuous and $Y_{\mathrm{eff}}\to Y_{\mathrm{sub}}^{\mathrm{STM}}$; as $G_{\mathrm{pp}}\to0$, the boundary becomes insulating. For OTh-TTM(NF), $Y_{\mathrm{sub}}^{\mathrm{NF}}=(Z_{\mathrm{bulk}}+Z_{\mathrm{jump}})^{-1}$ already includes the Ti/Si interfacial jump and is applied directly: $G_{\mathrm{pp}}$ is bypassed, so the same resistance is not counted twice. All interface and admittance operations are performed separately at each in-plane Fourier mode before spatial integration.

For nonsingular $\mathbf{G}_j$, solving Equations~\eqref{eq:interface_bc_upper}--\eqref{eq:interface_bc_lower} gives the $4\times4$ interface transfer operator that maps the state immediately below the interface to the state immediately above the interface:
\begin{equation}
    \widetilde{\mathbf{G}}_j=
    \begin{pmatrix}
    (\mathbf{G}_j^\mathsf{T})^{-1}\mathbf{G}_{\ell,j} & (\mathbf{G}_j^\mathsf{T})^{-1} \\
    \mathbf{G}_{u,j}(\mathbf{G}_j^\mathsf{T})^{-1}\mathbf{G}_{\ell,j} - \mathbf{G}_j & \mathbf{G}_{u,j}(\mathbf{G}_j^\mathsf{T})^{-1}
    \end{pmatrix},
    \label{eq:interface_operator_forward}
\end{equation}
and the inverse operator used in the top-down recursion is
\begin{equation}
    \widetilde{\mathbf{G}}_j^{-1}=
    \begin{pmatrix}
        \mathbf{G}_j^{-1}\mathbf{G}_{u,j} & -\mathbf{G}_j^{-1} \\
        \mathbf{G}_j^\mathsf{T}-\mathbf{G}_{\ell,j}\mathbf{G}_j^{-1}\mathbf{G}_{u,j} & \mathbf{G}_{\ell,j}\mathbf{G}_j^{-1}
    \end{pmatrix}.
    \label{eq:interface_operator_inverse}
\end{equation}
The inverse forms above require $\det\mathbf{G}_j\ne0$. Closed channels, including the metal/Si electron channel, are imposed directly through the boundary constraints of Equation~\eqref{eq:interface_scalar_load}, rather than by inverting a singular conductance matrix.

For a metallic TTM layer $n$, define the modal matrix
\[
\mathbf{P}_n=\begin{pmatrix}1 & 1 \\ p_{n,1} & p_{n,2}\end{pmatrix}, \qquad
\boldsymbol{\Lambda}_n=\mathrm{diag}(\lambda_{n,1},\lambda_{n,2}), \qquad
\mathbf{K}_n=\mathrm{diag}(k_{\mathrm{el},z,n},k_{\mathrm{ph},z,n}).
\]
The self- and mutual-admittance matrices appearing in the stable recursion are
\begin{align}
    \mathbf{Y}_{s,n} &= \mathbf{K}_n\mathbf{P}_n\boldsymbol{\Lambda}_n\,\coth(\boldsymbol{\Lambda}_n h_n)\,\mathbf{P}_n^{-1},\label{eq:self_admittance} \\
    \mathbf{Y}_{m,n} &= \mathbf{K}_n\mathbf{P}_n\boldsymbol{\Lambda}_n\,\csch(\boldsymbol{\Lambda}_n h_n)\,\mathbf{P}_n^{-1},\label{eq:mutual_admittance}
\end{align}
where the hyperbolic functions act elementwise on the diagonal matrix $\boldsymbol{\Lambda}_n h_n$. Let $\mathbf{Y}_{b,n}$ and $\mathbf{q}_{b,n}^{\mathrm{off}}$ denote the equivalent admittance and source-offset vector presented to the bottom of layer $n$ by the layers below. The bottom-up Schur-complement recursion is then
\begin{align}
    \Delta\mathbf{Y}_n &= \mathbf{Y}_{s,n}+\mathbf{Y}_{b,n},\label{eq:delta_admittance} \\
    \mathbf{Y}_{t,n} &= \mathbf{Y}_{s,n}-\mathbf{Y}_{m,n}\Delta\mathbf{Y}_n^{-1}\mathbf{Y}_{m,n},\label{eq:top_admittance}
\end{align}
and the matching source-offset recursion is
\begin{equation}
    \mathbf{q}_{t,n}^{\mathrm{off}} = \mathbf{q}_{\mathrm{top},n}^{\mathrm{off}} + \mathbf{Y}_{m,n}\Delta\mathbf{Y}_n^{-1}\left(\mathbf{q}_{b,n}^{\mathrm{off}}-\mathbf{q}_{\mathrm{bot},n}^{\mathrm{off}}\right).
    \label{eq:offset_recursion}
\end{equation}
When a nonmetal (STM) layer is adjacent to a metal (TTM) layer, its single temperature and heat flux occupy the phonon slots of the mixed-stack representation. The unused electron flux is closed; the auxiliary electron-temperature slot does not impose a physical temperature condition on the nonmetal. This preserves a single cascade structure for optical-STM, optical-TTM, and mixed stacks.

\subsection{Non-Fourier Substrate Boundary Formulation for Si: OTh-TTM(NF)}
\label{sec:non_fourier_si}

The OTh-TTM(NF) branch replaces the diffusive substrate boundary admittance while retaining the metallic STM/TTM layers and optical modules. It is activated here when the full diffusive hierarchy leaves a persistently low, beam-size-dependent fitted Si conductivity. The derivation follows the FDTR boundary formulation of Beardo \textit{et al.}~\cite{beardo2020phonon}, with related transport interpretations in Refs.~\cite{beardo2026theory,kovacs2024nonfourier,turprats2025anharmonic,xiang2022time}. Applicability at room temperature depends on modulation frequency, beam geometry and interface conditions~\cite{liu2025boundary,turprats2025anharmonic}. The Au/Ti/Si example applies the nonequilibrium closure at Ti/Si. Its characteristic roots, bulk impedance and total admittance are evaluated together with their Fourier limits.

\subsubsection{Governing Equations and Modal Decomposition}
\label{sec:nf_helmholtz}
\label{sec:nf_characteristic_roots}

For the silicon substrate considered here, the diffusive baseline is Fourier's law,
\begin{equation}
    \mathbf{q} = -k_{\mathrm{Si}}\nabla T,
    \label{eq:si_fourier_baseline}
\end{equation}
where $k_{\mathrm{Si}}$ is the bulk thermal conductivity of Si. The substrate energy conservation equation is
\begin{equation}
    C_{\mathrm{Si}}\frac{\partial T}{\partial t} + \nabla\cdot\mathbf{q} = 0,
    \label{eq:si_energy_conservation}
\end{equation}
with $C_{\mathrm{Si}}$ the volumetric heat capacity. Together, Equations~\eqref{eq:si_fourier_baseline}--\eqref{eq:si_energy_conservation} yield the classical diffusion equation $\partial T/\partial t = D_{\mathrm{Si}}\nabla^2 T$ with thermal diffusivity $D_{\mathrm{Si}} = k_{\mathrm{Si}}/C_{\mathrm{Si}}$.

In the transformed STM limit, the semi-infinite diffusive substrate enters through the boundary admittance
\begin{equation}
    Y_{\mathrm{sub}}^{\mathrm{STM}}(u,v,\omega)=k_{\mathrm{Si}}q_F, \qquad q_F=\sqrt{4\pi^2(u^2+v^2)+\frac{i\omega C_{\mathrm{Si}}}{k_{\mathrm{Si}}}}.
    \label{eq:diffusive_substrate_admittance}
\end{equation}

In the non-Fourier extension, the substrate constitutive relation is promoted to the Guyer--Krumhansl-type relation~\cite{guyer1966solution} used in the kinetic-collective phonon hydrodynamic model of Beardo \textit{et al.}~[Equation~(1) of Ref.~\cite{beardo2020phonon}]:
\begin{equation}
    \mathbf{q} + \tau \dot{\mathbf{q}} - \ell^2\left[\nabla^2\mathbf{q}+\alpha_{\mathrm{GK}}\nabla(\nabla\cdot\mathbf{q})\right] = -k_{\mathrm{Si}}\nabla T,
    \label{eq:si_non_fourier_constitutive}
\end{equation}
where $\tau$ is the phonon relaxation time, $\ell$ is the nonlocal length, and $\alpha_{\mathrm{GK}}$ is the GK-type hydrodynamic nonlocal coefficient, taken as $\alpha_{\mathrm{GK}} = 1/3$ following Ref.~\cite{beardo2020phonon} (Table~\ref{tab:archive_si_nf_params}). Relative to Fourier conduction, Equation~\eqref{eq:si_non_fourier_constitutive} introduces two types of corrections: (i) $\tau\dot{\mathbf{q}}$---the heat-flux response to a temperature gradient has a temporal lag, reflecting thermal memory effects; (ii) $\ell^2[\nabla^2\mathbf{q}+\alpha_{\mathrm{GK}}\nabla(\nabla\cdot\mathbf{q})]$---the heat flux is also affected by its spatial curvature and divergence, reflecting nonlocal transport.

The characteristic scales are $\omega\tau$ for temporal memory and $\ell/w$ for spatial nonlocality, where $w$ denotes the relevant beam radius. Values such as $\omega\tau\gtrsim0.1$ or $\ell\gtrsim w/10$ flag regimes in which the diffusive approximation should be assessed against the measured spectra; the need for the non-Fourier branch is determined by the configuration-dependent validation above.

\paragraph{Helmholtz decomposition and mode decoupling.}

Differentiating Equation~\eqref{eq:si_non_fourier_constitutive} in time and using the energy-conservation law Equation~\eqref{eq:si_energy_conservation}, in the form $\nabla\dot{T}=-\nabla(\nabla\cdot\mathbf{q})/C_{\mathrm{Si}}$, to eliminate $\nabla\dot T$ from the resulting $-k_{\mathrm{Si}}\nabla\dot T$ term, eliminates $T$ entirely. The result is a closed PDE for the heat flux, matching Equation~(D3) of Ref.~\cite{beardo2020phonon}:
\begin{equation}
    \tau\ddot{\mathbf{q}} + \dot{\mathbf{q}} = D_{\mathrm{Si}}\nabla(\nabla\cdot\mathbf{q}) + \ell^2\left[\nabla^2\dot{\mathbf{q}} + \alpha_{\mathrm{GK}}\nabla(\nabla\cdot\dot{\mathbf{q}})\right].
    \label{eq:nf_heat_flux_pde}
\end{equation}
Because $\mathbf{q}$ is a vector field, Equation~\eqref{eq:nf_heat_flux_pde} couples longitudinal and transverse components in general. The Helmholtz decomposition
\begin{equation}
    \mathbf{q} = \nabla\Phi + \nabla\times\!\left[\nabla\times(\Psi\,\hat{\mathbf{n}})\right]
    \label{eq:helmholtz_decomposition}
\end{equation}
separates the longitudinal potential $\Phi$ from the transverse potential $\Psi$, with $\hat{\mathbf n}$ the substrate-normal direction. Taking the divergence and curl of Equation~\eqref{eq:nf_heat_flux_pde} gives their respective modal equations.

\emph{Divergence: the compressive branch.} The curl contribution is divergence-free, so $\nabla\cdot\mathbf q=\nabla^2\Phi$. Taking the divergence of Equation~\eqref{eq:nf_heat_flux_pde} gives $\nabla^2[\tau\ddot\Phi+\dot\Phi-D_{\mathrm{Si}}\nabla^2\Phi-\ell^2(\alpha_{\mathrm{GK}}+1)\nabla^2\dot\Phi]=0$. The decaying compressive branch used here satisfies
\begin{equation}
    \tau\ddot{\Phi} + \dot{\Phi} = D_{\mathrm{Si}}\nabla^2\Phi + \ell^2(\alpha_{\mathrm{GK}}+1)\nabla^2\dot{\Phi},
    \label{eq:phi_equation}
\end{equation}
matching Equation~(D4a) of Ref.~\cite{beardo2020phonon}.

\emph{Curl: the shear branch.} Since $\nabla\times\nabla\Phi=0$, taking $\nabla\times(\cdot)$ of Equation~\eqref{eq:nf_heat_flux_pde} eliminates the $D_{\mathrm{Si}}\nabla(\nabla\cdot\mathbf{q})$ and $\alpha_{\mathrm{GK}}$ terms entirely (both are gradients), leaving an equation for the vorticity $\boldsymbol{\Omega}\equiv\nabla\times\mathbf{q}$:
\begin{equation}
    \tau\ddot{\boldsymbol{\Omega}}+\dot{\boldsymbol{\Omega}} = \ell^2\nabla^2\dot{\boldsymbol{\Omega}}.
    \label{eq:vorticity_eq}
\end{equation}
Writing $\mathbf{B}\equiv\nabla\times(\Psi\hat{\mathbf{n}})$ so that $\mathbf{q}=\nabla\Phi+\nabla\times\mathbf{B}$, the vorticity is $\boldsymbol{\Omega}=\nabla\times\nabla\times\mathbf{B}=\nabla(\nabla\cdot\mathbf{B})-\nabla^2\mathbf{B}$; since $\nabla\cdot\mathbf{B}=0$ automatically (any curl is divergence-free), $\boldsymbol{\Omega}=-\nabla^2\mathbf{B}=-\nabla^2[\nabla\times(\Psi\hat{\mathbf{n}})]=-\nabla\times[(\nabla^2\Psi)\hat{\mathbf{n}}]$, using that $\nabla^2$ commutes with $\nabla\times$ and, for the constant vector $\hat{\mathbf{n}}$, that $\nabla^2(\Psi\hat{\mathbf{n}})=(\nabla^2\Psi)\hat{\mathbf{n}}$. Substituting into Equation~\eqref{eq:vorticity_eq} and using that $\partial_t$ and $\nabla^2$ commute with the outer curl gives
\begin{equation}
    \nabla\times\Bigl\{\nabla^2\bigl[\tau\dot{\Psi}+\Psi-\ell^2\nabla^2\Psi\bigr]\,\hat{\mathbf{n}}\Bigr\} = 0.
\end{equation}
For the nontrivial in-plane two-dimensional Fourier modes of interest ($u^2+v^2\neq0$), a vector of the form (scalar)$\times\hat{\mathbf{n}}$ has vanishing curl only if that scalar is itself identically zero, so the bracketed quantity's Laplacian must vanish:
\begin{equation}
    \nabla^2\!\left(\tau\dot{\Psi} + \Psi - \ell^2\nabla^2\Psi\right) = 0,
    \label{eq:psi_equation}
\end{equation}
matching the further-reduced streamfunction equation (D5) of Ref.~\cite{beardo2020phonon} [equivalently Equation~(S12)/(S13) of Ref.~\cite{xiang2022time}].
The compressive equation has the frequency-dependent coefficient $D_{\mathrm{Si}}+i\omega\ell^2(\alpha_{\mathrm{GK}}+1)$, whereas the transverse equation contains $\tau$ and $\ell$ without a Fourier-diffusion term. These branches correspond to the compressive and shear modes of the Kelvin--Voigt analogy discussed in Appendix~D of Ref.~\cite{beardo2020phonon}.

\paragraph{Characteristic decay rates.}

For the semi-infinite isotropic Si substrate, this subsection retains the reference's local normal coordinate $x$ and tangential coordinates $y,z$; they correspond to the depth and the two beam-plane directions, respectively. The spatial frequencies $u,v$ are conjugate to these tangential coordinates, so $\Phi=\Phi_0e^{-\lambda_px}e^{2\pi i(uy+vz)}$ and $\Psi=\Psi_0e^{-\lambda_sx}e^{2\pi i(uy+vz)}$ under $e^{i\omega t}$. Thus $\nabla^2\Phi=[\lambda_p^2-4\pi^2(u^2+v^2)]\Phi$, with the analogous expression for $\Psi$. The constitutive coefficients remain isotropic, while the two Fourier coordinates retain the elliptical or offset beam geometry. The characteristic equations are
\begin{align}
    \lambda_p^2 &= 4\pi^2(u^2+v^2) + \frac{i\omega(1+i\omega\tau)}{D_{\mathrm{Si}} + i\omega\ell^2(\alpha_{\mathrm{GK}}+1)},\label{eq:lambda_p} \\
    \lambda_s^2 &= 4\pi^2(u^2+v^2) + \frac{1+i\omega\tau}{\ell^2}.\label{eq:lambda_s}
\end{align}
Equations~\eqref{eq:lambda_p}--\eqref{eq:lambda_s} follow the compressive and shear construction of Equations~(D8a)--(D8b) in Ref.~\cite{beardo2020phonon}. Under the $e^{i\omega t}$ convention, the decaying branches are selected by
\begin{equation}
    \lambda_{p,s} = \sqrt{\lambda_{p,s}^2}\quad \text{with}\quad \operatorname{Re}(\lambda_{p,s}) > 0.
    \label{eq:branch_selection}
\end{equation}
The substrate temperature field is recovered from the compressive potential through the energy conservation equation~\eqref{eq:si_energy_conservation} [cf.\ Equation~(D10) of Ref.~\cite{beardo2020phonon}]:
\begin{equation}
    \widetilde{T}_{\mathrm{Si}}(x) = \frac{-i}{C_{\mathrm{Si}}\omega}\left(4\pi^2(u^2+v^2) - \lambda_p^2\right)\Phi(x).
    \label{eq:T_from_Phi}
\end{equation}

As a consistency check on Equations~\eqref{eq:lambda_p}--\eqref{eq:lambda_s}, the diffusive limit $\tau\to 0$, $\ell\to 0$ recovers the classical Fourier decay behavior directly at the level of the characteristic roots. The compressive root reduces to the standard thermal wavevector,
\begin{equation}
    \lambda_p \;\xrightarrow{\tau,\ell\to 0}\; q_F \equiv \sqrt{4\pi^2(u^2+v^2) + \frac{i\omega C_{\mathrm{Si}}}{k_{\mathrm{Si}}}},
    \label{eq:fourier_limit_lambda_p}
\end{equation}
matching Equation~\eqref{eq:diffusive_substrate_admittance}, while the shear root diverges,
\begin{equation}
    \lambda_s \;\xrightarrow{\ell\to 0}\; \infty,
    \label{eq:fourier_limit_lambda_s}
\end{equation}
so the transverse contribution decays within a vanishingly thin region and is absent from the Fourier substrate response.

\subsubsection{Boundary Conditions and Effective Substrate Admittance}
\label{sec:nf_boundary_potentials}
\label{sec:nf_Z_bulk}
\label{sec:nf_jump_decomposition}
\label{sec:nf_admittance_modes}

The semi-infinite Si substrate is closed by two boundary conditions at the Ti/Si interface ($x=0^+$), matching Equations~(A5) and (A7) of Ref.~\cite{beardo2020phonon}. First, normal-flux continuity requires that the normal heat flux on the Si side equals the phonon heat flux from the Ti layer:
\begin{equation}
    q_{\mathrm{Si}}\cdot\hat{\mathbf{n}}\big|_{0^+} = q_\Gamma,
    \label{eq:bc_normal_flux}
\end{equation}
where $q_\Gamma \equiv (q_{p,\mathrm{Ti}}\cdot\hat{\mathbf{n}})|_{0^-}$ is the phonon-channel heat flux from the Ti side; Equation~\eqref{eq:bc_normal_flux} is simply energy conservation carried across the interface and introduces no free parameter. Second, the tangential heat flux satisfies a velocity-slip condition analogous to the Maxwell slip boundary condition of rarefied-gas kinetic theory, controlled by the specular reflection fraction $p$ of phonons scattering off the interface:
\begin{equation}
    q_t = \mathcal{C}_{\mathrm{slip}}\ell\,\partial_n q_t,\qquad \mathcal{C}_{\mathrm{slip}} = \frac{1+p}{1-p}.
    \label{eq:bc_tangential_slip}
\end{equation}
This work adopts fully diffusive phonon reflection, $p=0$, giving $\mathcal C_{\mathrm{slip}}=1$. Specularity is fixed rather than fitted.
To apply these boundary conditions, $q_{\mathrm{Si}}\cdot\hat{\mathbf{n}}$ (normal) and $q_t$ (tangential, with components along $2\pi(u,v)$) must first be written explicitly in terms of $\Phi_0,\Psi_0$. Using $\mathbf{q}=\nabla\Phi+\nabla\times\mathbf{B}$ with $\mathbf{B}=\nabla\times(\Psi\hat{\mathbf{n}})$ (section~\ref{sec:nf_helmholtz}) and the identity $\nabla\times\mathbf{B}=\nabla(\partial_x\Psi)-\hat{\mathbf{n}}\nabla^2\Psi$ [from $\nabla\times\nabla\times(\Psi\hat{\mathbf{n}})=\nabla(\nabla\cdot(\Psi\hat{\mathbf{n}}))-\nabla^2(\Psi\hat{\mathbf{n}})$, with $\hat{\mathbf{n}}\cdot\nabla\Psi=\partial_x\Psi$ for $\hat{\mathbf{n}}=\hat{\mathbf{x}}$], the normal component is $q_x=\partial_x\Phi+\partial_x^2\Psi-\nabla^2\Psi$. Since $\nabla^2\Psi=\partial_x^2\Psi-4\pi^2(u^2+v^2)\Psi$ for the in-plane harmonic ansatz, the $\partial_x^2\Psi$ terms cancel, leaving
\begin{equation}
    q_x = \partial_x\Phi + 4\pi^2(u^2+v^2)\Psi = -\lambda_p\Phi + 4\pi^2(u^2+v^2)\Psi,
    \label{eq:qx_from_potentials}
\end{equation}
The two tangential components are $q_y=2\pi iu(\Phi-\lambda_s\Psi)$ and $q_z=2\pi iv(\Phi-\lambda_s\Psi)$, with normal derivatives $2\pi iu(-\lambda_p\Phi+\lambda_s^2\Psi)$ and $2\pi iv(-\lambda_p\Phi+\lambda_s^2\Psi)$, respectively. The same slip condition is applied componentwise.

Substituting either nonzero tangential component into Equation~\eqref{eq:bc_tangential_slip} and cancelling its common Fourier factor gives $\Phi-\lambda_s\Psi=\mathcal C_{\mathrm{slip}}\ell(-\lambda_p\Phi+\lambda_s^2\Psi)$, hence
\begin{equation}
    \Phi_0(1+\mathcal{C}_{\mathrm{slip}}\ell\lambda_p) = \Psi_0\lambda_s(1+\mathcal{C}_{\mathrm{slip}}\ell\lambda_s).
    \label{eq:slip_ratio}
\end{equation}
Substituting Equation~\eqref{eq:qx_from_potentials} into the normal-flux condition Equation~\eqref{eq:bc_normal_flux} gives the second equation, $-\lambda_p\Phi_0+4\pi^2(u^2+v^2)\Psi_0=q_\Gamma$. Eliminating $\Phi_0$ using Equation~\eqref{eq:slip_ratio} yields
\begin{equation}
\begin{aligned}
\Psi_0\left[4\pi^2(u^2+v^2)-\frac{\lambda_p\lambda_s(1+\mathcal C_{\mathrm{slip}}\ell\lambda_s)}{1+\mathcal C_{\mathrm{slip}}\ell\lambda_p}\right]&=q_\Gamma,\\
\Longrightarrow\quad\Psi_0\frac{\mathcal D(u,v,\omega)}{1+\mathcal C_{\mathrm{slip}}\ell\lambda_p}&=q_\Gamma.
\end{aligned}
\end{equation}
and the bracketed numerator is precisely the denominator $\mathcal{D}(u,v,\omega)$ defined below, giving $\Psi_0$ directly; back-substituting into Equation~\eqref{eq:slip_ratio} then gives $\Phi_0$. This two-equation solve reproduces, matching Equations~(D7a)--(D7b) of Ref.~\cite{beardo2020phonon} [equivalently Equations~(S17)--(S18) of Ref.~\cite{xiang2022time}],
\begin{align}
    \Phi_0 &= \frac{\lambda_s(1+\mathcal{C}_{\mathrm{slip}}\ell\lambda_s)}{\mathcal{D}(u,v,\omega)}\,q_\Gamma,
    \label{eq:Phi0} \\
    \Psi_0 &= \frac{1+\mathcal{C}_{\mathrm{slip}}\ell\lambda_p}{\mathcal{D}(u,v,\omega)}\,q_\Gamma,
    \label{eq:Psi0}
\end{align}
with the common denominator
\begin{equation}
    \mathcal{D}(u,v,\omega) = 4\pi^2(u^2+v^2)(1+\mathcal{C}_{\mathrm{slip}}\ell\lambda_p) - \lambda_s\lambda_p(1+\mathcal{C}_{\mathrm{slip}}\ell\lambda_s).
    \label{eq:D_denom}
\end{equation}
Note that both $\Phi_0$ and $\Psi_0$ are linear in $q_\Gamma$, which is essential for deriving a scalar impedance closure.

\paragraph{Bulk impedance \texorpdfstring{$Z_{\mathrm{bulk}}$}{Z\_bulk}.}

Evaluating the substrate temperature at the interface ($x=0$) using Equation~\eqref{eq:T_from_Phi} and substituting the potential amplitude from Equation~\eqref{eq:Phi0} gives the interfacial admittance relation, of the same form as the boundary-value construction of Ref.~\cite{beardo2020phonon}:
\begin{equation}
    \widetilde{T}_{\mathrm{Si}}(0) = Z_{\mathrm{bulk}}(u,v,\omega)\,q_\Gamma,
    \label{eq:Z_bulk_definition}
\end{equation}
where the bulk impedance is
\begin{equation}
    Z_{\mathrm{bulk}}(u,v,\omega) = \frac{-i(4\pi^2(u^2+v^2) - \lambda_p^2)\,\lambda_s\,(1+\mathcal{C}_{\mathrm{slip}}\ell\lambda_s)}{C_{\mathrm{Si}}\,\omega\,\mathcal{D}(u,v,\omega)}.
    \label{eq:Z_bulk}
\end{equation}
Physically, $Z_{\mathrm{bulk}}$ is the frequency- and wavevector-dependent thermal impedance of the semi-infinite hydrodynamic substrate. It encodes both the compressive diffusion mode (through $\lambda_p$) and the purely nonlocal shear mode (through $\lambda_s$), weighted by the common denominator that couples them through the boundary conditions.

As a consistency check, divide the numerator and denominator of Equation~\eqref{eq:Z_bulk} by $\lambda_s(1+\mathcal{C}_{\mathrm{slip}}\ell\lambda_s)$:
\begin{equation*}
    Z_{\mathrm{bulk}} = \frac{-i(4\pi^2(u^2+v^2)-\lambda_p^2)}{C_{\mathrm{Si}}\,\omega\left[\dfrac{4\pi^2(u^2+v^2)(1+\mathcal{C}_{\mathrm{slip}}\ell\lambda_p)}{\lambda_s(1+\mathcal{C}_{\mathrm{slip}}\ell\lambda_s)} - \lambda_p\right]}.
\end{equation*}
In the diffusive limit $\tau,\ell\to 0$, the shear root diverges [Equation~\eqref{eq:fourier_limit_lambda_s}] while $\lambda_p\to q_F$ [Equation~\eqref{eq:fourier_limit_lambda_p}] and $\mathcal{C}_{\mathrm{slip}}\ell\lambda_p\to 0$ both stay finite, so the bracketed fraction is bounded in the numerator while its denominator diverges; the fraction therefore vanishes and the bracket reduces to $-\lambda_p\to -q_F$. Combined with the exact identity $4\pi^2(u^2+v^2)-\lambda_p^2\to4\pi^2(u^2+v^2)-q_F^2=-i\omega C_{\mathrm{Si}}/k_{\mathrm{Si}}$ [from the definition of $q_F$, Equation~\eqref{eq:diffusive_substrate_admittance}], Equation~\eqref{eq:Z_bulk} collapses to
\begin{equation}
    Z_{\mathrm{bulk}} \;\xrightarrow{\tau,\ell\to 0}\; \frac{1}{k_{\mathrm{Si}}\,q_F},
    \label{eq:fourier_limit_Z_bulk}
\end{equation}
the standard Fourier substrate impedance of Equation~\eqref{eq:diffusive_substrate_admittance}, confirming that $Z_{\mathrm{bulk}}$ reduces correctly to the diffusive result once the hydrodynamic corrections are switched off.

\paragraph{Nonequilibrium interface temperature jump.}

At the Ti/Si boundary, the interface condition contains a Kapitza resistance and nonequilibrium gradient corrections. The closure follows the maximum-entropy expansion in Appendix~B of Ref.~\cite{beardo2020phonon}, also used in section~S2 of the supplemental information accompanying Ref.~\cite{xiang2022time}. On the Si side, the distribution is expanded in the heat flux and its divergence and gradient, with relaxation time $\tau$ and volumetric heat capacity $C\equiv C_{\mathrm{Si}}$,
\begin{equation}
    f = f^{\mathrm{eq}} + \frac{3}{C v_g^2}\,(\mathbf{v}_g\cdot\mathbf{q})\,\frac{\partial f^{\mathrm{eq}}}{\partial T}
    + \frac{\tau}{C}(\nabla\cdot\mathbf{q})\,\frac{\partial f^{\mathrm{eq}}}{\partial T}
    - \frac{3\tau}{C v_g^2}\,(\mathbf{v}_g\mathbf{v}_g:\nabla\mathbf{q})\,\frac{\partial f^{\mathrm{eq}}}{\partial T},
    \label{eq:nf_f_expansion}
\end{equation}
[Equation~(B1) of Ref.~\cite{beardo2020phonon}], while the Ti side is treated diffusively via its own first-order (Chapman--Enskog-type) nonequilibrium distribution $f_\Gamma$ [Equation~(B4)]. Assuming diffusive (not specular) reflection of phonons incident on the interface, the incident hemisphere on the Si side matches the actual local distribution $f$, while the diffusely reflected outgoing hemisphere on each side is populated using a shared interface contact temperature $T_C$ [Equations~(B3a)--(B3b)] that is itself eliminated once the energy-balance condition is imposed; enforcing energy-flux balance across the interface, $q_x=\int f\,\hbar\omega_{\mathbf k}\, v_{gx}\,\mathrm{d}^3k$ (with $\omega_{\mathbf k}$ the phonon angular frequency) evaluated over the appropriate hemispheres [Equation~(B2)], and linearizing the result in $T_{\mathrm{Si}}-T_\Gamma$ and $\mathbf{q}$ yields the interface closure in real space,
\begin{equation}
    T_{\mathrm{Si}} - T_\Gamma = -\frac{1}{2}b_{\mathrm{NF}}\left(\gamma_0^{-1}+\gamma_{0\Gamma}^{-1}\right)\mathbf{q}\cdot\hat{\mathbf{n}}
    + b_{\mathrm{NF}}\gamma_0^{-1}\bigl(\beta_{\mathrm{NF}}\,\nabla\cdot\mathbf{q} - \nabla\mathbf{q}:\chi\bigr),
    \label{eq:nf_jump}
\end{equation}
where $\gamma_0=\tfrac14v_gC$ and $\gamma_{0\Gamma}=\tfrac14v_{g\Gamma}C_\Gamma$ are substrate- and transducer-side modal transfer coefficients [Equations~(B7a)--(B7b) of Ref.~\cite{beardo2020phonon}]. The divergence coefficient $\beta_{\mathrm{NF}}$ and gradient tensor $\chi$ follow Equations~(B7d)--(B7e) of that reference. The dimensionless contact factor $b_{\mathrm{NF}}$ scales the ideal-contact resistance: $R_{\min}$ is obtained at $b_{\mathrm{NF}}=1$ and $R=b_{\mathrm{NF}}R_{\min}$, following the imperfect-contact interpretation in Section~V of Ref.~\cite{beardo2020phonon}. It is fitted for each stack and geometry together with the released thermal-property targets. For Au/Ti/Si, $\Gamma$ denotes the Ti-phonon side in direct contact with Si, so $T_\Gamma=T_{\mathrm{ph},\mathrm{Ti}}|_{\mathrm{Ti/Si}}$; for Au/Si, $T_\Gamma=T_{\mathrm{ph},\mathrm{Au}}|_{\mathrm{Au/Si}}$.

\paragraph{Kinetic origin of the modal transfer coefficients.} For an isotropic phonon population with characteristic group speed $v_g$ and volumetric energy density $u(T)$, the one-sided flux is obtained by integrating $v_g\cos\theta\,u(T)/(4\pi)$ over the emitting hemisphere:
\begin{equation}
    q^{\mathrm{one\text{-}sided}}(T) = \frac{v_g}{4\pi}\int_0^{2\pi}\!\!\mathrm{d}\varphi\int_0^{\pi/2}\!\!\cos\theta\sin\theta\,\mathrm{d}\theta\;u(T) = \frac{v_g}{4\pi}\cdot 2\pi\cdot\frac{1}{2}\,u(T) = \frac{1}{4}v_g\,u(T),
    \label{eq:hemisphere_flux}
\end{equation}
using $\int_0^{\pi/2}\cos\theta\sin\theta\,\mathrm{d}\theta=1/2$. Linearizing in a small temperature perturbation $\Delta T$ about the reference state, $\Delta q^{\mathrm{one\text{-}sided}} = \tfrac{1}{4}v_g(\mathrm{d}u/\mathrm{d}T)\Delta T = \tfrac{1}{4}v_gC\,\Delta T \equiv \gamma_0\Delta T$, since $C\equiv \mathrm{d}u/\mathrm{d}T$ is precisely the volumetric heat capacity; the transducer-side coefficient $\gamma_{0\Gamma}=\tfrac14 v_{g\Gamma}C_\Gamma$ follows identically with the transducer-side group velocity and heat capacity.

The interface resistance combines the inverse modal transfer coefficients as $(b_{\mathrm{NF}}/2)(\gamma_0^{-1}+\gamma_{0\Gamma}^{-1})$. The hemispherical-flux calculation above gives the origin of $\gamma_0,\gamma_{0\Gamma}$; the factor $1/2$ and gradient coefficients $\beta_{\mathrm{NF}},\chi$ follow the joint linearization of both distributions in the interface energy balance [Appendix~B, Equations~(B4)--(B8) of Ref.~\cite{beardo2020phonon}].

The jump condition decomposes explicitly into three physically distinct contributions. First, define the Kapitza-like resistance
\begin{equation}
    R = \frac{b_{\mathrm{NF}}}{2}\left(\gamma_0^{-1} + \gamma_{0\Gamma}^{-1}\right),
    \label{eq:R_kapitza}
\end{equation}
which is the standard scalar interface resistance that would survive even in a purely diffusive model. This is the same physical Ti/Si interface resistance represented, in the ordinary diffusive branch of this manuscript, by the phonon-channel entry $G_{\mathrm{pp},j}$ of the interface conductance matrix $\mathbf{G}_j$ [Equation~\eqref{eq:interface_conductance_matrix}]; the two are alternative closures for the same physical resistance rather than independent quantities, and Step~2 below (see also the ``Mode~I'' discussion) makes explicit which of the two is active in a given branch. In the two-dimensional Fourier/frequency domain, the two nonequilibrium correction impedances are
\begin{align}
    Z_{\mathrm{corr},1} &= b_{\mathrm{NF}}\gamma_0^{-1}(\chi_{xx}-\chi_{zz})\,4\pi^2(u^2+v^2)\,\lambda_s\,\frac{1+\mathcal{C}_{\mathrm{slip}}\ell\lambda_p}{\mathcal{D}},
    \label{eq:Z_corr1} \\
    Z_{\mathrm{corr},2} &= b_{\mathrm{NF}}\gamma_0^{-1}\left[(\beta_{\mathrm{NF}}-\chi_{xx})\lambda_p^2 + (\chi_{zz}-\beta_{\mathrm{NF}})4\pi^2(u^2+v^2)\right]\lambda_s\,\frac{1+\mathcal{C}_{\mathrm{slip}}\ell\lambda_s}{\mathcal{D}},
    \label{eq:Z_corr2}
\end{align}
where $Z_{\mathrm{corr},1}$ arises from the anisotropy of the $\chi$ tensor (the difference between the cross-plane $\chi_{xx}$ and in-plane $\chi_{zz}$ components, with $x$ the interface-normal direction), and $Z_{\mathrm{corr},2}$ captures the combined nonlocal stress correction from both $\beta_{\mathrm{NF}}$ and $\chi$; both are the transformed-domain counterparts of the second-order gradient terms in Equation~\eqref{eq:nf_jump}, obtained explicitly as follows, following the same reduction as Equation~(D11) of Ref.~\cite{beardo2020phonon}. Since $\chi$ is diagonal with equal in-plane components $\chi_{yy}=\chi_{zz}$ (by the axial symmetry of the problem about the interface normal) and a distinct normal component $\chi_{xx}$, the contraction $\nabla\mathbf{q}:\chi=\chi_{xx}\partial_x q_x+\chi_{zz}(\partial_y q_y+\partial_z q_z)$ can be rewritten using $\partial_yq_y+\partial_zq_z=\nabla\cdot\mathbf{q}-\partial_xq_x$ as
\begin{equation}
    \nabla\mathbf{q}:\chi = (\chi_{xx}-\chi_{zz})\,\partial_xq_x + \chi_{zz}\,\nabla\cdot\mathbf{q}.
    \label{eq:chi_contraction}
\end{equation}
Using $q_x=\partial_x\Phi+4\pi^2(u^2+v^2)\Psi$ [Equation~\eqref{eq:qx_from_potentials}] together with the ansatz derivatives $\partial_x^2\Phi=\lambda_p^2\Phi$, $\partial_x\Psi=-\lambda_s\Psi$, and $\nabla\cdot\mathbf{q}=\nabla^2\Phi=(\lambda_p^2-4\pi^2(u^2+v^2))\Phi$ (section~\ref{sec:nf_helmholtz}), the two building blocks evaluate to $\partial_xq_x=\lambda_p^2\Phi-4\pi^2(u^2+v^2)\lambda_s\Psi$ and $\nabla\cdot\mathbf{q}=(\lambda_p^2-4\pi^2(u^2+v^2))\Phi$. Substituting both into Equation~\eqref{eq:chi_contraction} and then into $\beta_{\mathrm{NF}}\nabla\cdot\mathbf{q}-\nabla\mathbf{q}:\chi$, and collecting terms by $\Phi$ and $\Psi$, gives
\begin{equation}
    \beta_{\mathrm{NF}}\nabla\cdot\mathbf{q}-\nabla\mathbf{q}:\chi \;=\; \bigl[(\beta_{\mathrm{NF}}-\chi_{xx})\lambda_p^2+(\chi_{zz}-\beta_{\mathrm{NF}})4\pi^2(u^2+v^2)\bigr]\Phi \;+\; (\chi_{xx}-\chi_{zz})4\pi^2(u^2+v^2)\lambda_s\,\Psi.
    \label{eq:jump_gradient_evaluated}
\end{equation}
Writing the jump condition Equation~\eqref{eq:nf_jump} at $x=0$ as $T_{\mathrm{Si}}-T_\Gamma=-Z_{\mathrm{jump}}\,q_\Gamma$ (the sign consistent with heat flowing from the warmer to the cooler side) and using $\mathbf{q}\cdot\hat{\mathbf{n}}|_0=q_\Gamma$ [Equation~\eqref{eq:bc_normal_flux}] for the zeroth-order term, Equation~\eqref{eq:jump_gradient_evaluated} supplies the second-order contribution; substituting the boundary amplitudes $\Phi_0,\Psi_0$ [Equations~\eqref{eq:Phi0}--\eqref{eq:Psi0}], both proportional to $q_\Gamma$, and dividing out the common factor of $q_\Gamma$ reproduces Equations~\eqref{eq:Z_corr1}--\eqref{eq:Z_corr2} above term by term: the $\Psi$ coefficient in Equation~\eqref{eq:jump_gradient_evaluated} gives $Z_{\mathrm{corr},1}$ and the $\Phi$ coefficient gives $Z_{\mathrm{corr},2}$, in each case multiplied by $b_{\mathrm{NF}}\gamma_0^{-1}$ and by the respective potential's amplitude prefactor from Equations~\eqref{eq:Phi0}--\eqref{eq:Psi0}. The total jump impedance is then
\begin{equation}
    Z_{\mathrm{jump}}(u,v,\omega) = R - Z_{\mathrm{corr},1} - Z_{\mathrm{corr},2}.
    \label{eq:Z_jump}
\end{equation}
Equation~\eqref{eq:nf_jump} makes clear that the interface response is not captured by a single scalar conductance once the substrate is treated hydrodynamically: the correction terms $Z_{\mathrm{corr},1}$ and $Z_{\mathrm{corr},2}$ introduce frequency- and wavevector-dependent modifications to the classical Kapitza picture.

\paragraph{Total admittance and three boundary modeling levels.}

Combining the bulk impedance (Equation~\eqref{eq:Z_bulk}) with the interface impedance, the substrate admittance is evaluated at one of three modeling levels, ordered by increasing physical completeness. \textbf{Mode~I (Fourier reference)} is the diffusive OTh-TTM baseline already used elsewhere in this Section: the Ti/Si interface is closed by the ordinary phonon-channel entry $G_{\mathrm{pp},j}$ of the interface conductance matrix $\mathbf{G}_j$ [Equation~\eqref{eq:interface_conductance_matrix}], and the Si substrate is represented by the diffusive admittance of Equation~\eqref{eq:diffusive_substrate_admittance}; no hydrodynamic quantity from this subsection enters. Mode~I is the classical reference against which the non-Fourier boundary formulations below are validated in the diffusive limit.

\paragraph{Mode~II (bulk + Kapitza only).} The hydrodynamic bulk impedance is combined with a simplified Kapitza jump that retains only the $R$ term, discarding the $\beta_{\mathrm{NF}}$ and $\chi$ corrections:
\begin{equation}
    Y_{\mathrm{NF}}^{(\mathrm{II})} = \frac{1}{Z_{\mathrm{bulk}} + R}.
    \label{eq:Y_NF_modeII}
\end{equation}
Equation~\eqref{eq:Y_NF_modeII} is the $\beta_{\mathrm{NF}}=0,\,\chi=0$ special case of the complete admittance Equation~\eqref{eq:Y_NF_modeIII}, matching the simplified jump condition of Appendix~D.2 of Ref.~\cite{beardo2020phonon}; it serves as a diagnostic branch for isolating the bulk nonlocal contribution.

\paragraph{Mode~III (full interface).} The complete boundary formulation retains both the bulk impedance and all interface nonequilibrium terms:
\begin{equation}
    Y_{\mathrm{NF}}^{(\mathrm{III})} = \frac{1}{Z_{\mathrm{bulk}} + Z_{\mathrm{jump}}}.
    \label{eq:Y_NF_modeIII}
\end{equation}
Setting $\beta_{\mathrm{NF}}=0$ and $\chi=0$ in Equations~\eqref{eq:Z_corr1}--\eqref{eq:Z_corr2} gives $Z_{\mathrm{corr},1}=Z_{\mathrm{corr},2}=0$, so that $Z_{\mathrm{jump}}\to R$ [Equation~\eqref{eq:Z_jump}] and Equation~\eqref{eq:Y_NF_modeIII} reduces exactly to Equation~\eqref{eq:Y_NF_modeII}; Mode~II is therefore the bulk-only approximation of Mode~III rather than a separately derived model. The substrate replacement used in the benchmark-conditioned high-thermal-conductivity substrate branch is
\begin{equation}
    Y_{\mathrm{sub}}^{\mathrm{STM}}(u,v,\omega)\;\mapsto\;Y_{\mathrm{sub}}^{\mathrm{NF}}(u,v,\omega)\equiv Y_{\mathrm{NF}}^{(\mathrm{III})}(u,v,\omega).
    \label{eq:nf_replacement_map}
\end{equation}
This replacement leaves the optical module, the upper metallic TTM layers, and the interface/admittance recursion unchanged; it replaces only the semi-infinite substrate boundary admittance. The empirical $G_{\mathrm{pp},\mathrm{Ti/Si}}$ path used by Mode~I is bypassed whenever the non-Fourier formulation is enabled, in either Mode~II or Mode~III, because the Kapitza-like resistance $R$ is already embedded in $Y_{\mathrm{NF}}$ and Mode~III additionally retains the $\beta_{\mathrm{NF}}$- and $\chi$-dependent jump corrections. Retaining a separate $G_{\mathrm{pp},\mathrm{Ti/Si}}$ term in either non-Fourier branch would therefore double-count the Ti/Si interface resistance: $G_{\mathrm{pp},\mathrm{Ti/Si}}$ (Mode~I) and $R$ (Modes~II--III) are two mutually exclusive representations of the same physical Ti/Si phonon interface resistance, never summed.

As a final consistency check, the diffusive limit $\tau,\ell,\beta_{\mathrm{NF}},\chi\to 0$ applied to Equation~\eqref{eq:Y_NF_modeIII} recovers the classical diffusive admittance with a Kapitza interface resistance in series,
\begin{equation}
    Y_{\mathrm{NF}} \;\xrightarrow{\tau,\ell,\beta_{\mathrm{NF}},\chi\to 0}\; \frac{k_{\mathrm{Si}}\,q_F}{1 + k_{\mathrm{Si}}\,q_F\,R},
    \label{eq:fourier_limit_Y_NF}
\end{equation}
Here $Z_{\mathrm{bulk}}\to1/(k_{\mathrm{Si}}q_F)$ and $Z_{\mathrm{jump}}\to R$. This has the same form as the diffusive metal/Si boundary admittance of Equation~\eqref{eq:interface_scalar_load} when $G_{\mathrm{pp}}=1/R$.

\subsubsection{FDTR Observable, Parameter Groups, and Implementation}
\label{sec:nf_fdtr_observable}
\label{sec:nf_parameter_tables}
\label{sec:nf_ttm_embedding}

\paragraph{FDTR boundary coupling and observable.}

The non-Fourier substrate admittance determines the temperature field in the metal layers. For surface-sensing branches, the probe-weighted response is
\begin{equation}
    \Delta T(\omega)=\iint_{\mathbb R^2}\widetilde{T}_{\Gamma,\mathrm{surf}}(u,v,\omega)\,W_{\mathrm{pr}}(u,v)\,e^{2\pi i(u\Delta_x+v\Delta_y)}\,\mathrm{d}u\,\mathrm{d}v,
    \label{eq:nf_surface_response}
\end{equation}
Here $u,v$ follow Equation~\eqref{eq:fourier_transform_pair}; $W_{\mathrm{pr}}$ is the normalized probe weight of Equation~\eqref{eq:couple_observable}, and the surface response includes the pump envelope. The phase is $\phi(\omega)=\arg\Delta T(\omega)$. In OTh branches, including OTh-TTM(NF), the depth-weighted response $\widetilde H$ replaces $\widetilde T_{\Gamma,\mathrm{surf}}$; its integrated observable is $H_{\mathrm{OTh}}$ with $\phi=\arg H_{\mathrm{OTh}}$, as in Equation~\eqref{eq:couple_observable_volumetric}. The NF formulation enters through the substrate admittance and interface condition. Circular beams with an in-plane-isotropic kernel admit the radial Hankel reduction.

\paragraph{Non-Fourier parameter groups.}

The non-Fourier parameters are organized into three physically distinct groups to avoid mixing different levels of the model.

\paragraph{(i) Si substrate bulk parameters.} These govern the hydrodynamic transport within the semi-infinite Si substrate and are listed in Table~\ref{tab:nf_bulk_params}:

\begin{table}[ht]
\centering
\caption{Si bulk transport parameters for the non-Fourier formulation.}
\label{tab:nf_bulk_params}
\small
\begin{tabular*}{\textwidth}{@{\extracolsep{\fill}} l l l p{5cm}}
\toprule
\textbf{Symbol} & \textbf{Unit} & \textbf{Name} & \textbf{Physical meaning} \\
\midrule
$k_{\mathrm{Si}}$ & $\mathrm{W\,m^{-1}\,K^{-1}}$ & Thermal conductivity & Bulk phonon thermal conductivity of Si \\
$C_{\mathrm{Si}}$ & $\mathrm{J\,m^{-3}\,K^{-1}}$ & Volumetric heat capacity & Energy storage per unit volume \\
$\tau$ & s & Relaxation time & Heat-flux memory time; controls onset of wave-like behavior \\
$\ell$ & m & Nonlocal length & Characteristic length for spatial nonlocality in heat flux \\
$\alpha_{\mathrm{GK}}$ & --- (dimensionless) & Nonlocal coefficient & Relative weight of compressive vs.\ total nonlocal corrections \\
\bottomrule
\end{tabular*}
\end{table}

\paragraph{(ii) Ti/Si interface parameters.} These parameters define the nonequilibrium jump condition across the metal/semiconductor boundary and are listed in Table~\ref{tab:nf_interface_params}:

\begin{table}[ht]
\centering
\caption{Ti/Si interface nonequilibrium parameters.}
\label{tab:nf_interface_params}
\small
\begin{tabular*}{\textwidth}{@{\extracolsep{\fill}} l l l >{\raggedright\arraybackslash}p{5cm}}
\toprule
\textbf{Symbol} & \textbf{Unit} & \textbf{Name} & \textbf{Physical meaning} \\
\midrule
$\gamma_0$ & $\mathrm{MW\,m^{-2}\,K^{-1}}$ & Substrate modal transfer coeff. & $\gamma_0 = \tfrac{1}{4}v_g C$ from Si-side distribution \\
$\gamma_{0\Gamma}$ & $\mathrm{MW\,m^{-2}\,K^{-1}}$ & Transducer modal transfer coeff. & $\gamma_{0\Gamma} = \tfrac{1}{4}v_{g\Gamma}C_\Gamma$ from the transducer-side distribution \\
$\beta_{\mathrm{NF}}$ & m & Nonlocal divergence coeff. & From divergence term $\nabla\cdot\mathbf{q}$ in jump condition \\
$\chi_{xx}$ & m & Cross-plane gradient coeff. & Cross-plane (interface-normal, $x$) component of $\chi$ tensor \\
$\chi_{zz}$ & m & In-plane gradient coeff. & In-plane component of $\chi$ tensor \\
$b_{\mathrm{NF}}$ & --- (dimensionless) & Contact-quality factor & Scales interface coefficients for imperfect contact \\
\bottomrule
\end{tabular*}
\end{table}

\paragraph{(iii) Temperature-dependent parameter values.} The non-Fourier parameters are strongly temperature-dependent. Table~\ref{tab:nf_temp_PRB} lists the values from Beardo \textit{et al.}~\cite{beardo2020phonon} at four representative temperatures, with the collective-transport interpretation anchored in the de Tomas literature~\cite{detomas2014collective,detomas2014groupiv}. The bulk Si parameters ($C_{\mathrm{Si}}$, $k_{\mathrm{Si}}$, $\tau$, $\ell$, $\gamma_0$) are properties of the Si substrate alone and transfer directly. The transducer-side interface values ($\gamma_{0\Gamma}$ and, through it, $1/R$) were originally reported by Ref.~\cite{beardo2020phonon} for its own Au-with-Cr-adhesion-layer transducer on Si; the Au/Si and Au/Ti/Si benchmarks use the same literature-reference $\gamma_{0\Gamma}$, with $b_{\mathrm{NF}}$ scaling the resulting reference interface resistance. The coefficients $\beta_{\mathrm{NF}}$ and $\chi$ retain their Si-side origin:

\begin{table}[htbp]
\centering
\caption{Temperature-dependent non-Fourier parameters for Si (Table~I of Beardo \textit{et al.}~\cite{beardo2020phonon}). In the notation of Equation~\eqref{eq:R_kapitza}, the tabulated reference $1/R$ values imply $b_{\mathrm{NF}}\approx2.7$; the sample-specific contact factors fitted in this work are reported separately.}
\label{tab:nf_temp_PRB}
\small
\begin{tabular*}{\textwidth}{@{\extracolsep{\fill}} l c c c c}
\toprule
 & 81~K & 154~K & 311~K & 416~K \\
\midrule
$C_{\mathrm{Si}}$ ($\mathrm{kJ\,m^{-3}K^{-1}}$) & 466 & 1061 & 1692 & 1840 \\
$k_{\mathrm{Si}}$ ($\mathrm{W\,m^{-1}K^{-1}}$) & 1260 & 315 & 150 & 109 \\
$\tau$ (ps) & 1002 & 148 & 42 & 27 \\
$\ell$ (nm) & 3127 & 1047 & 185 & 142 \\
$\gamma_0$ ($\mathrm{MW\,m^{-2}K^{-1}}$) & 359 & 732 & 1068 & 1145 \\
$\gamma_{0\Gamma}$ ($\mathrm{MW\,m^{-2}K^{-1}}$) & 975 & 1801 & 2237 & 2313 \\
$\beta_{\mathrm{NF}}$ (nm) & $-745$ & $-87$ & $-21$ & $-14$ \\
$\chi_{xx}$ (nm) & $-1066$ & $-129$ & $-31$ & $-21$ \\
$\chi_{zz}$ (nm) & $-586$ & $-66$ & $-16$ & $-11$ \\
$1/R$ ($\mathrm{MW\,m^{-2}K^{-1}}$) & 194 & 384 & 534 & 565 \\
\bottomrule
\end{tabular*}
\end{table}

At 81~K, $\ell\approx3.1\,\mu\mathrm m$ and $\tau\approx1\,\mathrm{ns}$; at 311~K, they decrease to approximately $185\,\mathrm{nm}$ and $42\,\mathrm{ps}$. The correction depends on temperature together with modulation frequency, spot size and interface conditions. The benchmarks use the 311\,K reference parameters to assess that configuration dependence against the diffusive branches.

\paragraph{Parameter evaluation.} The Ti/Si and Au/Si benchmarks use diffuse phonon reflection ($p=0$, $\mathcal C_{\mathrm{slip}}=1$) with $\alpha_{\mathrm{GK}}=1/3$ and fix the bulk Si and reference modal parameters directly to the 311\,K first-principles values of Beardo \textit{et al.} in Table~\ref{tab:nf_temp_PRB} (Table~\ref{tab:archive_si_nf_params})~\cite{beardo2020phonon,guo2018generalized}. Because the nonlocal substrate admittance $Y_{\mathrm{NF}}$ already incorporates the interfacial resistance $R$, the ordinary $G_{\mathrm{pp},\mathrm{Ti/Si}}$ operator is bypassed.

\paragraph{Embedding in the TTM recursion: four-step algorithm.}

For the benchmarked high-thermal-conductivity substrate realization used here (Au/Ti/Si), the metallic overlayer is still propagated with the TTM or STM matrices of section~\ref{sec:tmm}, but the bottom boundary condition is replaced by the non-Fourier substrate admittance. Because the non-Fourier admittance depends on both frequency and wavevector, it must be evaluated and injected at each quadrature point before the TTM cascade begins. The integration therefore proceeds in four sequential steps:
\paragraph{Step~1: Compute non-Fourier substrate admittance.}
For each $(u,v,\omega)$ quadrature point, the characteristic roots $\lambda_p$ and $\lambda_s$ are evaluated from Equations~\eqref{eq:lambda_p}--\eqref{eq:lambda_s}, the denominator $\mathcal{D}$ is constructed from Equation~\eqref{eq:D_denom}, and the substrate admittance $Y_{\mathrm{sub}} = Y_{\mathrm{NF}}(u,v,\omega)$ is computed from Equation~\eqref{eq:Y_NF_modeII} or \eqref{eq:Y_NF_modeIII} depending on the selected mode.

\paragraph{Step~2: Bypass the Mode-I interface conductance.}
For either NF branch, omit the ordinary Ti/Si phonon-interface operator and use the substrate admittance that already includes $R$. Upper interfaces retain their ordinary conductance operators.

\paragraph{Step~3: Bottom-up TTM cascade.}
Using $Y_{\mathrm{sub}}$ as the $2\times 2$ bottom boundary---with phonon-channel entry $Y_{b,22} = Y_{\mathrm{sub}}$ and electron-channel entry $Y_{b,11} = 0$, reflecting the closed-electron condition $q_{e,\mathrm{Ti}}\cdot\hat{\mathbf{n}}=0$---the $4\times 4$ TTM transfer matrices are propagated upward through each metal layer.

\paragraph{Step~4: Apply top-surface boundary condition.}
At the top surface, SH heating enters as a prescribed flux in the electron channel of a TTM metal or the single STM channel. For VH/OTh heating, the source is carried by layer offsets and the free top boundary has $\mathbf q_{\mathrm{surf}}=\mathbf0$. The temperature field provides the surface or depth-weighted response according to the selected readout branch.
This four-step procedure preserves the modularity of the existing TTM recursion: the non-Fourier physics is entirely encapsulated in the substrate admittance $Y_{\mathrm{sub}}$, so that no modification to the metal-layer propagation framework is required. The same algorithm applies irrespective of whether Mode~II or Mode~III is selected for the interface admittance.

\begin{algorithm}[H]
\caption{GK Substrate Admittance Coupled to the Optothermal TTM Stack}
\label{alg:gk_embedding}
\begin{algorithmic}
\REQUIRE Frequencies, independent $u,v$ quadrature, stack and beam inputs, GK bulk and interface priors
\ENSURE Probe-weighted complex response and phase
\FOR{each $(u,v,\omega)$}
    \STATE Evaluate decaying $\lambda_p,\lambda_s$ from Equations~\eqref{eq:lambda_p}--\eqref{eq:lambda_s}
    \STATE Form $\mathcal D$, $Z_{\mathrm{bulk}}$, $R$ and $Z_{\mathrm{corr},1},Z_{\mathrm{corr},2}$
    \STATE $Z_{\mathrm{jump}}\leftarrow R-Z_{\mathrm{corr},1}-Z_{\mathrm{corr},2}$
    \STATE $Y_{\mathrm{NF}}\leftarrow(Z_{\mathrm{bulk}}+Z_{\mathrm{jump}})^{-1}$
    \STATE Bypass the ordinary substrate-interface conductance; load $\mathbf Y_{b,N}\leftarrow\operatorname{diag}(0,Y_{\mathrm{NF}})$
    \STATE Cascade the upper STM/TTM and interface blocks by the existing admittance recursion
    \STATE Solve the source-driven top boundary and reconstruct the depth-weighted response for OTh
\ENDFOR
\STATE Integrate with the probe kernel and offset factor in Equation~\eqref{eq:couple_observable_volumetric}
\STATE Return $\phi(\omega)=\arg\{H_{\mathrm{OTh}}(\omega)\}$
\end{algorithmic}
\end{algorithm}

\subsection{Forward Solution and Full Module Coupling}
\label{sec:module_coupling}

Once the source and thermal branch are selected, layer and interface operators are cascaded with the semi-infinite substrate boundary admittance. SH heating is supplied by top-surface flux; VH/OTh heating enters through layer source vectors with a zero-flux free top boundary. TTM sources couple to electrons, whereas STM sources couple to the single thermal field. The selected probe readout yields a complex response $H(\omega_m)$ and phase
\begin{equation}
    \phi(\omega_m)=\arg\!\left(H(\omega_m)\right),
\end{equation}
implemented numerically through \texttt{angle(H)} followed by phase unwrapping and branch normalization; under the $e^{+i\omega t}$ convention adopted here a diffusive thermal lag yields $\arg H<0$, so $\phi$ is already negative and no sign inversion is applied.

\paragraph{Module boundaries and full forward coupling.}

The six equations below trace the complete signal path from the optical module to the measured FDTR observable, making explicit every hand-off between subsystems. The coupling chain is
\[
\resizebox{0.98\textwidth}{!}{$\displaystyle \text{Optical TMM} \;\to\; \{Q_n,\tilde{S}_n\} \;\to\; \{\mathbf M_n,\mathbf b_n\}\leftrightarrow\{\mathbf Y_{s,n},\mathbf Y_{m,n},\mathbf q_{\mathrm{top},n}^{\mathrm{off}},\mathbf q_{\mathrm{bot},n}^{\mathrm{off}}\} \;\to\; \widetilde{\mathbf{G}}_j \;\to\; Y_{\mathrm{sub}} \;\to\; \widetilde{T}_{\Gamma,\mathrm{surf}} \;\to\; \Delta T(\omega),\,\phi(\omega)$}
\]

\subsubsection*{Optical module output}
The optical transfer-matrix module outputs the two-dimensional Fourier-domain volumetric source for each layer:
\begin{equation}
    \tilde{S}_n(u,v,z) = \tilde{I}_{\mathrm{pump}}(u,v)\,Q_n(z), \qquad n=1,\dots,N,
    \label{eq:couple_optical_output}
\end{equation}
where $Q_n(z)$ is the depth-only deposition profile per unit incident irradiance from the optical TMM (Equation~\eqref{eq:heat_source_decomposition_si}) and $\tilde{I}_{\mathrm{pump}}(u,v)$ is the two-dimensional Fourier-transformed Gaussian pump envelope (Equation~\eqref{eq:optical_gaussian_fourier}). This factorization separates the lateral beam profile from the depth-resolved absorption, allowing the thermal module to operate on $Q_n(z)$ alone.

\subsubsection*{Layer module output --- transfer-matrix / admittance equivalence}
Each layer $n$ is represented either by the affine transfer relation
\begin{equation}
    \boldsymbol{\Theta}_n(h_n) = \mathbf{M}_n\,\boldsymbol{\Theta}_n(0) + \mathbf{b}_n,
    \label{eq:couple_layer_transfer}
\end{equation}
or equivalently by the self- and mutual-admittance matrices
\begin{equation}
    \mathbf{Y}_{s,n} = \mathbf{K}_n\mathbf{P}_n\boldsymbol{\Lambda}_n\coth(\boldsymbol{\Lambda}_n h_n)\mathbf{P}_n^{-1}, \qquad
    \mathbf{Y}_{m,n} = \mathbf{K}_n\mathbf{P}_n\boldsymbol{\Lambda}_n\csch(\boldsymbol{\Lambda}_n h_n)\mathbf{P}_n^{-1},
    \label{eq:couple_layer_admittance}
\end{equation}
together with the source-offset vectors $\mathbf{q}_{\mathrm{top},n}^{\mathrm{off}}$ and $\mathbf{q}_{\mathrm{bot},n}^{\mathrm{off}}$ generated by the particular solutions of $\tilde{S}_n$ (Equations~\eqref{eq:self_admittance}--\eqref{eq:offset_recursion}). The homogeneous matrix $\mathbf{M}_n$ and source vector $\mathbf{b}_n$ are defined in section~\ref{sec:affine_transfer}. The admittance representation is algebraically equivalent but numerically preferred, because it avoids exponentiation of large arguments that arise for thick or high-conductance layers.

\subsubsection*{Interface module}
At interface $j$, the $4\times4$ interface transfer operator $\widetilde{\mathbf{G}}_j$ (Equation~\eqref{eq:interface_operator_forward}) maps the state vector from below to above,
\begin{equation}
    \boldsymbol{\Theta}_n(z_j^-) = \widetilde{\mathbf{G}}_j\,\boldsymbol{\Theta}_{n+1}(z_j^+),
    \label{eq:couple_interface}
\end{equation}
transforming the boundary admittance matrix accordingly:
\begin{equation}
    \mathbf{Y}_{\mathrm{above}} =
    (\mathbf{G}_C+\mathbf{G}_D\mathbf{Y}_{\mathrm{below}})
    (\mathbf{G}_A+\mathbf{G}_B\mathbf{Y}_{\mathrm{below}})^{-1}.
    \label{eq:couple_interface_admittance}
\end{equation}
Here $\mathbf{G}_A,\mathbf{G}_B,\mathbf{G}_C,\mathbf{G}_D$ are the $2\times2$ blocks of $\widetilde{\mathbf{G}}_j$. The interface operator therefore simultaneously enforces thermal boundary-resistance conditions and transforms the admittance representation so that the Schur-complement recursion can continue upward through the stack.

\subsubsection*{Substrate boundary admittance coupling}
The semi-infinite substrate provides the bottom boundary admittance. For a diffusive metal/substrate boundary, elimination of the substrate-side temperature gives $Y_{\mathrm{load}}=Y_{\mathrm{eff}}$ of Equation~\eqref{eq:interface_scalar_load}; for the non-Fourier Ti/Si branch, $Y_{\mathrm{load}}=Y_{\mathrm{sub}}^{\mathrm{NF}}$, which already includes the interface jump. The boundary admittance and source-offset vector applied to the bottom finite metal layer are
\begin{equation}
    \mathbf Y_{b,N}=\begin{pmatrix}0&0\\0&Y_{\mathrm{load}}\end{pmatrix},\qquad \mathbf q_{b,N}^{\mathrm{off}}=\begin{pmatrix}0\\q_{\mathrm{load}}^{\mathrm{off}}\end{pmatrix},
    \label{eq:couple_substrate}
\end{equation}
The zero electron-flux slot enforces the closed-electron boundary. For a diffusive substrate with volumetric heating, $q_{\mathrm{sub}}=Y_{\mathrm{sub}}T_{\mathrm{sub}}+q_{\mathrm{sub}}^{\mathrm{off}}$, where $Y_{\mathrm{sub}}$ is the homogeneous substrate admittance and $q_{\mathrm{sub}}^{\mathrm{off}}=q_p(0)-Y_{\mathrm{sub}}T_p(0)$ follows from its particular solution. Elimination of the substrate temperature across the scalar interface conductance $G$ gives $q_{\mathrm{load}}^{\mathrm{off}}=Gq_{\mathrm{sub}}^{\mathrm{off}}/(G+Y_{\mathrm{sub}})$. For a source-free substrate, $q_{\mathrm{load}}^{\mathrm{off}}=0$. The terminal interface is included once in $Y_{\mathrm{load}}$ and its matching source offset; the NF substrate admittance already contains its interfacial jump and receives no second conductance operator. A bottom STM layer uses the corresponding scalar admittance and source offset.
\subsubsection*{Top boundary condition and surface temperature}
After the bottom-up Schur-complement recursion (Equations~\eqref{eq:delta_admittance}--\eqref{eq:offset_recursion}) propagates through all layers and interfaces, the top-surface boundary condition is applied. For volumetric VH/OTh heating, power is encoded in the source-offset vectors and $\mathbf q_{\mathrm{surf}}=\mathbf0$. For SH heating, $\mathbf q_{\mathrm{surf}}$ instead supplies the transformed pump flux to the electron channel of a TTM metal or to the single STM field, without a duplicate volumetric source. With downward-positive flux, the boundary solve is
\begin{equation}
    \mathbf{Y}_{t,1}\,\mathbf{T}_0 = \mathbf q_{\mathrm{surf}}-\mathbf{q}_{t,1}^{\mathrm{off}}, \qquad \widetilde{T}_{\Gamma,\mathrm{surf}}(u,v,\omega) = \mathbf{e}_{\mathrm{probe}}^\mathsf{T}\,\mathbf{T}_0,
    \label{eq:couple_top_bc}
\end{equation}
where $\mathbf{T}_0=[\hat{T}_{\mathrm{el},1}(0),\,\hat{T}_{\mathrm{ph},1}(0)]^\mathsf{T}$ is the two-channel surface temperature vector, $\mathbf{Y}_{t,1}$ and $\mathbf{q}_{t,1}^{\mathrm{off}}$ are the final recursion outputs, and $\mathbf{e}_{\mathrm{probe}}$ selects the thermoreflectance-sensed channel (phonon for standard FDTR, electron for metals with dominant electronic thermoreflectance). Inverting $\mathbf{Y}_{t,1}$ at each $(u,v,\omega)$ quadrature point yields the surface response. For the surface-sensing SH-STM and VH-STM branches this response enters the two-dimensional Fourier integration directly; for the OTh branches it seeds the depth-resolved back-substitution of section~\ref{sec:volumetric_readout}.

\subsubsection*{FDTR observable}
The normalized probe-weighted frequency response is obtained from the full two-dimensional Fourier synthesis,
\begin{equation}
\begin{aligned}
    \Delta T(\omega)&=\iint_{\mathbb R^2}\widetilde T_{\Gamma,\mathrm{surf}}(u,v,\omega)\,W_{\mathrm{pr}}(u,v)\,
       e^{2\pi i(u\Delta_x+v\Delta_y)}\,\mathrm{d}u\,\mathrm{d}v,\\
    W_{\mathrm{pr}}(u,v)&=\exp\!\left[-\frac{\pi^2}{2}\left(w_{\mathrm{probe},x}^2u^2+w_{\mathrm{probe},y}^2v^2\right)\right],\qquad
    \phi(\omega)=\arg\{\Delta T(\omega)\}.
\end{aligned}
    \label{eq:couple_observable}
\end{equation}
The probe intensity is normalized to unit area, and $\Delta_x,\Delta_y$ give its centre relative to the pump. Responses $\widetilde T_{\Gamma,\mathrm{surf}}$ and $\widetilde H$ include the pump envelope, so their integrals use the probe weight alone. A response per unit pump Fourier amplitude uses the combined kernel $W$ of Equation~\eqref{eq:fourier_combined_kernel}. SH-STM/VH-STM use the surface response; OTh uses $\widetilde H$. The elliptical Au/Si geometry retains both spatial directions.

\subsection{Depth-Weighted Thermoreflectance Readout for the OTh Branches}
\label{sec:volumetric_readout}

The observable defined by Equations~\eqref{eq:couple_top_bc} and~\eqref{eq:couple_observable} reads out the surface temperature alone and is the readout used by the SH-STM and VH-STM branches. The OTh-STM and OTh-TTM branches instead resolve the finite optical penetration of the probe: the probe field forms a partial standing wave inside the stack, the temperature oscillation modulates the local permittivity at every depth, and the reflectance perturbation is a depth integral of the temperature field weighted by an optical sensitivity kernel. This subsection derives that kernel within the same optical transfer-matrix formalism as the pump module (section~\ref{sec:optical_tmm}), following the opto-thermal treatment of Ref.~\cite{guo2026light}, and defines the sensitivity coefficients $C_{f,j}^{(\pm)}$ and depth exponents $g_{f,j}^{(\pm)}$ that specify the volumetric detection mode of the OTh branches. All optical quantities below are evaluated at the probe wavelength $\lambda^{\mathrm{probe}}$ and carry the superscript $\mathrm{probe}$; in particular $\beta_{\mathrm{vac}}^{\mathrm{probe}}=2\pi/\lambda^{\mathrm{probe}}$, $\tilde{N}_j^{\mathrm{probe}}=n_j^{\mathrm{probe}}+i\kappa_j^{\mathrm{probe}}$, and $\widetilde{K}_j\equiv\beta_{\mathrm{vac}}^{\mathrm{probe}}\tilde{N}_j^{\mathrm{probe}}$ is the complex probe wavenumber in layer $j$.

\subsubsection*{Thermoreflectance perturbation of the stack reflectance}
Linearising the relative permittivity $\epsilon_{r,j}=(\tilde{N}_j^{\mathrm{probe}})^2$ with respect to temperature gives the depth-resolved permittivity perturbation
\begin{equation}
    \delta\hat{\epsilon}_{r,j}(z;u,v,\omega)
    = \frac{\mathrm{d}\epsilon_{r,j}}{\mathrm{d}T}\,\hat{T}_j(z;u,v,\omega),
    \qquad
    \frac{\mathrm{d}\epsilon_{r,j}}{\mathrm{d}T}
    = 2\tilde{N}_j^{\mathrm{probe}}\,\frac{\mathrm{d}\tilde{N}_j}{\mathrm{d}T},
    \label{eq:probe_eps_linearization}
\end{equation}
where $\mathrm{d}\tilde{N}_j/\mathrm{d}T=\mathrm{d}n_j/\mathrm{d}T+i\,\mathrm{d}\kappa_j/\mathrm{d}T$ is the complex thermo-optic coefficient at $\lambda^{\mathrm{probe}}$ and $\hat{T}_j(z)$ is the two-dimensional Fourier-domain temperature of the thermoreflectance-active channel selected by $\mathbf{e}_{\mathrm{probe}}$ (the phonon channel for standard FDTR; the STM branches carry a single channel). In the experimental OTh branches, the depth-weighted probe observable includes all absorbing layers through their layer-wise optical sensitivity kernels. In the graded DeepONet library ($h_{\mathrm{Au}}=\SI{97}{\nano\metre}>5\,d_{\mathrm{opt}}$), this depth-weighted readout operates in the shallow-sensing limit and reduces to the probe-averaged surface temperature (section~\ref{sec:deeponet_training_data}). A slab of thickness $\mathrm{d}z$ carrying the excess permittivity $\delta\hat{\epsilon}_{r,j}$, embedded in the host medium of layer $j$, back-scatters the probe with the thin-film (Airy) limit amplitude
\begin{equation}
    \delta r_{\mathrm{loc}}
    = \frac{i\,(\beta_{\mathrm{vac}}^{\mathrm{probe}})^2}{2\widetilde{K}_j}\,
      \delta\hat{\epsilon}_{r,j}(z)\,\mathrm{d}z ,
    \label{eq:probe_thin_slab}
\end{equation}
where $(\beta_{\mathrm{vac}}^{\mathrm{probe}})^2=\omega_{\mathrm{probe}}^2/c^2$ with $\omega_{\mathrm{probe}}$ the probe optical angular frequency. To be observed in reflection, this locally scattered amplitude must complete a round trip between the free surface and depth $z$. Let $\mathbf{M}_{\mathrm{to},j}$ denote the partial transfer matrix built from the interface and propagation matrices, Equations~\eqref{eq:interface_matrix} and~\eqref{eq:propagation_matrix} evaluated at $\lambda^{\mathrm{probe}}$, running from air down to just inside the top face of layer $j$. Two-port algebra gives the downward transmission $t_{\downarrow}=1/(\mathbf{M}_{\mathrm{to},j})_{11}$ and the upward transmission $t_{\uparrow}=\det\mathbf{M}_{\mathrm{to},j}/(\mathbf{M}_{\mathrm{to},j})_{11}$, so the round-trip amplitude transmittance of the overlying stack is
\begin{equation}
    \mathcal{T}_j
    = t_{\downarrow}\,t_{\uparrow}
    = \frac{\det\mathbf{M}_{\mathrm{to},j}}{\bigl[(\mathbf{M}_{\mathrm{to},j})_{11}\bigr]^{2}} .
    \label{eq:probe_roundtrip}
\end{equation}
Because $\det\mathbf{D}_n=\tilde{N}_{n+1}/\tilde{N}_n$ (Stokes relations) and $\det\mathbf{P}_n=1$, the determinant telescopes to $\det\mathbf{M}_{\mathrm{to},j}=\tilde{N}_j^{\mathrm{probe}}/\tilde{N}_{\mathrm{air}}$, which provides a convenient numerical check. Propagation from the top of layer $j$ to depth $z$ and back contributes the round-trip factor $e^{2i\widetilde{K}_j z}$, whose modulus $e^{-\alpha_{\mathrm{opt},j}^{\mathrm{probe}}z}$ encodes the double-pass probe attenuation. Summing over layers and integrating over depth, the first-order reflectance-amplitude perturbation is
\begin{equation}
    \delta\hat{r}(u,v,\omega)
    = \sum_{j}\mathcal{T}_j\,
      \frac{i\,(\beta_{\mathrm{vac}}^{\mathrm{probe}})^2}{2\widetilde{K}_j}\,
      \frac{\mathrm{d}\epsilon_{r,j}}{\mathrm{d}T}
      \int_{0}^{h_j} e^{2i\widetilde{K}_j z}\,\hat{T}_j(z;u,v,\omega)\,\mathrm{d}z ,
    \label{eq:probe_dr_kernel}
\end{equation}
{\RaggedRight\rightskip=0pt plus 1fil\relax
with $z$ measured downward from the top face of each layer. Equation~\eqref{eq:probe_dr_kernel} is the co-propagating, single-pass restriction of the exact first-order Born perturbation $\delta\hat{r}=(i\beta_{\mathrm{vac}}^{\mathrm{probe}}/2)\int\delta\hat{\epsilon}_r\,[E(z)/E_0]^2\,\mathrm{d}z$: the two coincide whenever no probe amplitude returns from below the sensed depth. Expanding the full field square introduces terms linear and quadratic in the returning amplitude, as shown below. These terms vanish in the opaque limit; semitransparent stacks require the full-field kernel of Equations~\eqref{eq:probe_born_general}--\eqref{eq:probe_kappa_general}.
\par}

\subsubsection*{Lock-in projection and the $(C_f,g_f)$ readout coefficients}
The photodetector measures the reflectance perturbation $\delta R=\delta|r|^2=r_s^{*}\,\delta r+r_s\,\delta r^{*}$, where $r_s$ is the unperturbed probe reflection amplitude of the full stack, Equation~\eqref{eq:reflection_amplitude} evaluated at $\lambda^{\mathrm{probe}}$. In the time domain the temperature field is real, so the conjugation acts on the optical factors only, while the complex two-dimensional Fourier-domain temperature amplitude $\hat{T}_j(z;u,v,\omega)$ is carried through unconjugated; the depth kernel applied to $\hat{T}_j$ is therefore the real function $w_j(z)=2\,\mathrm{Re}\{C_{f,j}^{(+)}e^{g_{f,j}^{(+)}z}\}$, which splits into two mutually conjugate exponential terms. The resulting depth-weighted FDTR observable is
\begin{equation}
    \widetilde{H}(u,v,\omega)
    = \sum_{j}\sum_{s=\pm} C_{f,j}^{(s)}
      \int_{0}^{h_j} e^{\,g_{f,j}^{(s)} z}\,\hat{T}_j(z;u,v,\omega)\,\mathrm{d}z ,
    \label{eq:volumetric_observable}
\end{equation}
with the layer-wise sensitivity coefficients and depth exponents
\begin{align}
    C_{f,j}^{(+)} &= r_s^{*}\,\mathcal{T}_j\,
      \frac{i\,(\beta_{\mathrm{vac}}^{\mathrm{probe}})^2}{2\widetilde{K}_j}\,
      2\tilde{N}_j^{\mathrm{probe}}\frac{\mathrm{d}\tilde{N}_j}{\mathrm{d}T},
    \qquad
    C_{f,j}^{(-)} = \bigl(C_{f,j}^{(+)}\bigr)^{*},
    \label{eq:probe_Cf_def}\\
    g_{f,j}^{(+)} &= 2i\widetilde{K}_j
    = -\alpha_{\mathrm{opt},j}^{\mathrm{probe}} + i\beta_{\mathrm{opt},j}^{\mathrm{probe}},
    \qquad
    g_{f,j}^{(-)} = \bigl(g_{f,j}^{(+)}\bigr)^{*},
    \label{eq:probe_gf_def}
\end{align}
where $\alpha_{\mathrm{opt},j}^{\mathrm{probe}}=4\pi\kappa_j^{\mathrm{probe}}/\lambda^{\mathrm{probe}}$ and $\beta_{\mathrm{opt},j}^{\mathrm{probe}}=4\pi n_j^{\mathrm{probe}}/\lambda^{\mathrm{probe}}$ are the absorption coefficient and standing-wave wavenumber of section~\ref{sec:optical_tmm} evaluated at the probe wavelength. The coefficients $C_{f,j}^{(\pm)}$ carry units of $\mathrm{m^{-1}\,K^{-1}}$ and $g_{f,j}^{(\pm)}$ of $\mathrm{m^{-1}}$, so the real-space reflectance modulation is dimensionless; its Fourier transform $\widetilde H$ carries the area factor of Equation~\eqref{eq:fourier_transform_pair}. Both optical coefficients are independent of $(u,v)$ (normal-incidence paraxial optics, consistent with the pump module) and of the modulation frequency, so all frequency dependence of $\widetilde{H}$ enters through the thermal field.

\subsubsection*{Closed-form depth integral}
Inside each thermal layer the two-dimensional Fourier-domain temperature of the sensed channel is a finite sum of exponentials: collecting the homogeneous modes of the layer admittance relation, Equation~\eqref{eq:couple_layer_admittance}, referenced to the top face, together with the particular solutions generated by the optical source exponentials,
\begin{equation}
    \hat{T}_j(z) = \sum_{k} D_{j,k}\, e^{\vartheta_{j,k} z},
    \qquad
    \vartheta_{j,k}\in\{-\lambda_{j,1},-\lambda_{j,2},+\lambda_{j,1},+\lambda_{j,2}\}
    \cup\{\beta_{\mathrm{src}}\ \text{exponents of }\tilde{S}_j\},
    \label{eq:probe_modal_expansion}
\end{equation}
where the modal amplitudes $D_{j,k}$ are recovered by a top-down back-substitution of the admittance recursion: the surface solve of Equation~\eqref{eq:couple_top_bc} fixes the top-face temperatures, and the stored recursion operators of Equations~\eqref{eq:couple_layer_transfer}--\eqref{eq:couple_substrate} then yield the interface temperatures of every layer without re-solving the cascade. Substituting Equation~\eqref{eq:probe_modal_expansion} into Equation~\eqref{eq:volumetric_observable}, each depth integral is elementary and the observable takes the closed form
\begin{equation}
    \widetilde{H}(u,v,\omega)
    = \sum_{j}\sum_{s=\pm}\sum_{k}
      C_{f,j}^{(s)}\,D_{j,k}\,
      \frac{e^{\left(g_{f,j}^{(s)}+\vartheta_{j,k}\right)h_j}-1}
           {g_{f,j}^{(s)}+\vartheta_{j,k}} ,
    \label{eq:probe_H_closed_form}
\end{equation}
with the removable singularity at $g_{f,j}^{(s)}+\vartheta_{j,k}\to0$ evaluated as $h_j$. For the OTh-STM and OTh-TTM branches, the depth-weighted response replaces the surface response in Equation~\eqref{eq:couple_observable}:
\begin{equation}
    H_{\mathrm{OTh}}(\omega)=\iint_{\mathbb R^2}\widetilde H(u,v,\omega)\,W_{\mathrm{pr}}(u,v)\,
    e^{2\pi i(u\Delta_x+v\Delta_y)}\,\mathrm{d}u\,\mathrm{d}v,\qquad
    \phi(\omega)=\arg\{H_{\mathrm{OTh}}(\omega)\}.
    \label{eq:couple_observable_volumetric}
\end{equation}
Here $H_{\mathrm{OTh}}$ is the dimensionless probe-averaged reflectance modulation. Only the readout is generalized; the source construction, layer admittances, interface operators, substrate closure, and top-surface solve are applied mode by mode as before. The SH-STM and VH-STM branches retain the surface readout.

\subsubsection*{Shallow-sensing limit and consistency with the surface readout}
When the probe absorption in the transducer is strong, $\alpha_{\mathrm{opt},1}^{\mathrm{probe}}h_1\gg1$ and $\alpha_{\mathrm{opt},1}^{\mathrm{probe}}\gg|\vartheta_{1,k}|$ for all thermal modes, each depth integral in Equation~\eqref{eq:probe_H_closed_form} reduces to $-1/g_{f,1}^{(s)}$, the modal sum collapses to the surface value $\sum_k D_{1,k}=\hat{T}_1(0)$, and the contributions of deeper layers are suppressed through $|\mathcal{T}_j|\lesssim e^{-\alpha_{\mathrm{opt},1}^{\mathrm{probe}}h_1}$. The observable then degenerates to
\begin{equation}
    \widetilde{H}(u,v,\omega)
    \;\longrightarrow\;
    C_R\,\hat{T}_1(0;u,v,\omega),
    \qquad
    C_R = -2\,\mathrm{Re}\!\left\{\frac{C_{f,1}^{(+)}}{g_{f,1}^{(+)}}\right\}\in\mathbb{R},
    \label{eq:probe_surface_limit}
\end{equation}
{\RaggedRight
Here $C_R$ is the real, frequency-independent thermoreflectance coefficient. For a bare air/medium interface, $C_R=\mathrm dR/\mathrm dT=2\operatorname{Re}\{r_s^*(\mathrm dr_s/\mathrm d\tilde N)\,\mathrm d\tilde N/\mathrm dT\}$. A positive $C_R$ leaves the phase unchanged; a negative $C_R$ adds a constant $\pi$ offset. Calibrating this optical sign and using a common phase branch gives the surface-temperature readout in the shallow-sensing limit.
\par}

\subsubsection*{General multi-pass kernel for semitransparent stacks}
Equations~\eqref{eq:probe_dr_kernel}--\eqref{eq:probe_gf_def} retain only the co-propagating round trip and are therefore exact when no probe amplitude returns from beneath the sensed depth. That restriction is lifted by evaluating the first-order perturbation with the full internal probe field rather than with its downward component alone. Writing the field inside layer $j$ in terms of the amplitudes at its top face,
\begin{equation}
    E_j(z) = E_j^{+}\,e^{\,i\widetilde{K}_j z} + E_j^{-}\,e^{-i\widetilde{K}_j z},
    \label{eq:probe_internal_field}
\end{equation}
where $E_j^{\pm}$ are obtained from the same partial transfer matrices used for the pump source, Equation~\eqref{eq:field_solve}, evaluated at $\lambda^{\mathrm{probe}}$, the exact first-order (Born) reflectance perturbation reads
\begin{equation}
    \delta\hat{r}(u,v,\omega)
    = \frac{i\,\beta_{\mathrm{vac}}^{\mathrm{probe}}}{2}
      \sum_j \frac{\mathrm{d}\epsilon_{r,j}}{\mathrm{d}T}
      \int_{0}^{h_j}\!\left[\frac{E_j(z)}{E_0}\right]^{2}
      \hat{T}_j(z;u,v,\omega)\,\mathrm{d}z .
    \label{eq:probe_born_general}
\end{equation}
Squaring Equation~\eqref{eq:probe_internal_field} produces \emph{three} depth exponents instead of one,
\begin{equation}
    \left[E_j(z)\right]^{2}
    = \bigl(E_j^{+}\bigr)^{2} e^{+2i\widetilde{K}_j z}
    + \bigl(E_j^{-}\bigr)^{2} e^{-2i\widetilde{K}_j z}
    + 2E_j^{+}E_j^{-},
    \label{eq:probe_field_squared}
\end{equation}
namely the co-propagating round trip, its counter-propagating counterpart, and a depth-independent cross term generated by the standing wave. Applying the same lock-in projection as in Equation~\eqref{eq:volumetric_observable} pairs each with its conjugate, so the general readout has six $(C_f,g_f)$ terms per layer,
\begin{align}
    C_{f,j}^{(1,2)} &= \Bigl\{r_s^{*}\,\varkappa_j \bigl(E_j^{+}\bigr)^{2},\;
                              r_s\,\varkappa_j^{*}\bigl(E_j^{+}\bigr)^{*2}\Bigr\},
    & g_{f,j}^{(1,2)} &= \Bigl\{+2i\widetilde{K}_j,\; -2i\widetilde{K}_j^{*}\Bigr\},
    \label{eq:probe_Cf_general_a}\\
    C_{f,j}^{(3,4)} &= \Bigl\{r_s^{*}\,\varkappa_j \bigl(E_j^{-}\bigr)^{2},\;
                              r_s\,\varkappa_j^{*}\bigl(E_j^{-}\bigr)^{*2}\Bigr\},
    & g_{f,j}^{(3,4)} &= \Bigl\{-2i\widetilde{K}_j,\; +2i\widetilde{K}_j^{*}\Bigr\},
    \label{eq:probe_Cf_general_b}\\
    C_{f,j}^{(5,6)} &= \Bigl\{r_s^{*}\,\varkappa_j\, 2E_j^{+}E_j^{-},\;
                              r_s\,\varkappa_j^{*}\,2\bigl(E_j^{+}E_j^{-}\bigr)^{*}\Bigr\},
    & g_{f,j}^{(5,6)} &= \{0,\;0\},
    \label{eq:probe_Cf_general_c}
\end{align}
with the common factor
\begin{equation}
    \varkappa_j
    = \frac{i\,\beta_{\mathrm{vac}}^{\mathrm{probe}}}{2}\,
      \frac{\mathrm{d}\epsilon_{r,j}}{\mathrm{d}T}
    = i\,\beta_{\mathrm{vac}}^{\mathrm{probe}}\,\tilde{N}_j^{\mathrm{probe}}
      \frac{\mathrm{d}\tilde{N}_j}{\mathrm{d}T}.
    \label{eq:probe_kappa_general}
\end{equation}
The general readout uses the depth integral of Equation~\eqref{eq:probe_H_closed_form}, with six $(C_f,g_f)$ terms per layer. For $g_{f,j}^{(5,6)}=0$ and $\vartheta_{j,k}=0$, the integral is $h_j$. Field amplitudes $E_j^\pm$ are evaluated from the probe optical constants.

Two limits fix the relationship between this general form and Equation~\eqref{eq:probe_dr_kernel}. First, for a bare air/medium interface Equation~\eqref{eq:probe_born_general} reproduces the exact Fresnel derivative: with $E^{+}=2/(1+\tilde{N})$, $E^{-}=0$ one obtains $\partial r/\partial T = -2(\mathrm{d}\tilde{N}/\mathrm{d}T)/(1+\tilde{N})^{2}$, identical to differentiating $r=(1-\tilde{N})/(1+\tilde{N})$, which fixes the normalization of Equation~\eqref{eq:probe_kappa_general}. Second, when the transducer is optically thick at the probe wavelength the returning amplitude is suppressed as $|E_j^{-}/E_j^{+}|\sim e^{-\alpha^{\mathrm{probe}}_{\mathrm{opt},j}h_j}$: terms $(3,4)$ and $(5,6)$ vanish and $\varkappa_j (E_j^{+})^{2}$ reduces to the round-trip form $\mathcal{T}_j\,i(\beta^{\mathrm{probe}}_{\mathrm{vac}})^{2}\,\mathrm{d}\epsilon_{r,j}/\mathrm{d}T\,/(2\widetilde{K}_j)$ of Equation~\eqref{eq:probe_Cf_def}. The general branch is therefore the one to use whenever the transducer transmits appreciably, and the single-pass branch is retained as its opaque limit.

\subsection{Summary of the PRISM Branch Hierarchy and Regime Escalation Rules}

The public PRISM branch chain used for the benchmark results in this paper is therefore
\begin{equation}
\begin{aligned}
\text{Optical TMM} &\rightarrow \{\text{SH-STM, VH-STM}\} \rightarrow \text{OTh-STM} \\
&\rightarrow \text{OTh-TTM} \rightarrow \text{OTh-TTM(NF) (high-conductivity substrate branch)}.
\end{aligned}
\label{eq:public_hierarchy}
\end{equation}
For standard diffusive non-Si stacks, the public PRISM route terminates at the optical TMM + STM/TTM thermal backbone or, when metallic nonequilibrium and volumetric heating must both be retained, at OTh-TTM. For the benchmark-conditioned high-thermal-conductivity substrate case discussed here, OTh-TTM(NF) (Two-Temperature Model with Non-Fourier substrate extension and Optical Transfer-Matrix Heating) is the deepest public PRISM branch and is exposed only for the benchmark family in which the diffusive PRISM branches demonstrably fail.

\clearpage
\section{Optical Transfer-Matrix Method (TMM) for Volumetric Heat Source}
\label{sec:optical_tmm}

For $a\in\{\mathrm{pump},\mathrm{probe}\}$, the optical TMM is evaluated separately at $\lambda^{(a)}$ using the corresponding optical constants. The phase $\beta_n^{(a)}$ is dimensionless, whereas $\beta_{\mathrm{opt},n}^{(a)}$ below is a standing-wave wavenumber. The volumetric heat source uses the pump solution; the probe solution supplies the readout kernel of section~\ref{sec:volumetric_readout}.

The optical transfer-matrix formulation resolves depth-dependent absorption, multiple reflections and interference in metallic and dielectric multilayers~\cite{guo2026light}. These effects alter heat deposition in ultrathin or semitransparent stacks relative to the surface-heating approximation~\cite{yang2016analysis}.

\subsection{Physical Model and Transfer-Matrix Construction}
\label{sec:optical_model}

The lateral pump envelope retains its two directional $1/e^2$ intensity radii,
\begin{equation}
I_{\mathrm{pump}}(x,y)=I_{\mathrm{pump},0}\exp\!\left(-\frac{2x^2}{w_{\mathrm{pump},x}^2}-\frac{2y^2}{w_{\mathrm{pump},y}^2}\right),
\label{eq:optical_gaussian_real}
\end{equation}
with the pump centre at the origin. Under Equation~\eqref{eq:fourier_transform_pair},
\begin{equation}
\tilde I_{\mathrm{pump}}(u,v)=\tilde I_{\mathrm{pump},0}\exp\!\left[-\frac{\pi^2}{2}\left(w_{\mathrm{pump},x}^2u^2+w_{\mathrm{pump},y}^2v^2\right)\right],
\label{eq:optical_gaussian_fourier}
\end{equation}
where $\tilde I_{\mathrm{pump},0}=\pi w_{\mathrm{pump},x}w_{\mathrm{pump},y}I_{\mathrm{pump},0}/2$ is the zero-wavevector amplitude. Multiplication by the normalized probe transform produces the combined beam kernel
\begin{equation}
W(u,v)=\exp\!\left[-\frac{\pi^2}{2}\sum_{a=x,y}(w_{\mathrm{pump},a}^2+w_{\mathrm{probe},a}^2)s_a^2\right]
=\exp\!\left[-\frac{\pi^2}{2}(w_x^2u^2+w_y^2v^2)\right],
\label{eq:fourier_combined_kernel}
\end{equation}
where $s_x=u$, $s_y=v$, and $w_a^2=w_{\mathrm{pump},a}^2+w_{\mathrm{probe},a}^2$ defines the raw $1/e^2$ cross-correlation radius $w_a$ on each axis. An equal-pump/probe implementation must therefore use $w_a/\sqrt{2}$ for each individual beam on that axis. The $x$ and $y$ widths remain independent. For in-plane-isotropic layer kernels and circular pump/probe beams, angular integration reduces the two-dimensional integral to the familiar radial Hankel form (with $J_0(\mu_\rho\Delta)$ for an offset distance $\Delta$). An elliptical beam or an in-plane-anisotropic layer retains the two-dimensional integral.

\paragraph{Compact response kernel in the screening workflow.}
The scalar $G(u,v,\omega)$ in Figure~\ref{fig:workflow} is the complex response per unit Fourier amplitude of the lateral pump intensity. In terms of the responses defined in section~\ref{sec:volumetric_readout},
\begin{equation}
G(u,v,\omega)=
\begin{cases}
\widetilde T_{\Gamma,\mathrm{surf}}(u,v,\omega)/\widetilde I_{\mathrm{pump}}(u,v), & \mathrm{SH/VH},\\
\widetilde H(u,v,\omega)/\widetilde I_{\mathrm{pump}}(u,v), & \mathrm{OTh}.
\end{cases}
\label{eq:workflow_response_kernel}
\end{equation}
The quotient denotes a linear-response coefficient: $G$ retains the selected depth-dependent heat source and readout, while the lateral pump and probe Gaussian envelopes are supplied together by Equation~\eqref{eq:fourier_combined_kernel}. Here $u,v$ are cyclic spatial frequencies conjugate to $x,y$, $\omega=2\pi f$ is angular modulation frequency, and $\Delta_x,\Delta_y$ are the probe-centre offsets relative to the pump. The phase integral is
\begin{equation}
\phi(\omega)=\arg\!\left\{
\iint_{\mathbb R^2}G(u,v,\omega)W(u,v)
e^{2\pi i(u\Delta_x+v\Delta_y)}\,\mathrm du\,\mathrm dv
\right\}.
\end{equation}
With the pump modulation as phase reference, $\widetilde I_{\mathrm{pump},0}>0$ is a real normalization factor. The original observable equals this factor times the normalized integral, so the phase is unchanged. This figure-specific response function is distinct from the interface thermal conductance $G$.

Consider a plane wave normally incident on a multilayer thin-film structure consisting of $N$ layers with complex refractive indices $\tilde{N}_n = n_n + i\kappa_n$ and thicknesses $d_n$, where $n_n$ is the real refractive index and $\kappa_n$ is the extinction coefficient. The electromagnetic field in layer $n$ consists of forward-propagating and backward-propagating components:

\begin{equation}
E_n(z) = E_n^+ e^{i \beta_{\mathrm{vac}}^{(a)} \tilde{N}_n z} + E_n^- e^{-i \beta_{\mathrm{vac}}^{(a)} \tilde{N}_n z}
\label{eq:electric_field_si}
\end{equation}

where $\beta_{\mathrm{vac}}^{(a)} = 2\pi/\lambda^{(a)}$ is the vacuum optical wavenumber, $E_n^+$ is the forward wave amplitude, and $E_n^-$ is the backward wave amplitude. The complex refractive index convention $\tilde{N}_n = n_n + i\kappa_n$ with $\kappa_n > 0$ ensures that forward-propagating waves decay exponentially as $\exp(-\beta_{\mathrm{vac}}^{(a)} \kappa_n z)$.

\paragraph{Transfer-matrix construction.}
\hypertarget{par:transfer_matrix_construction}{}

\paragraph{Interface matrix.}

At the interface between layer $n$ and layer $n+1$, boundary conditions require continuity of tangential electric and magnetic fields. This gives rise to Fresnel reflection and transmission coefficients:

\begin{align}
 r_n &= \frac{\tilde{N}_n - \tilde{N}_{n+1}}{\tilde{N}_n + \tilde{N}_{n+1}} \label{eq:fresnel_r} \\
 t_n &= \frac{2\tilde{N}_n}{\tilde{N}_n + \tilde{N}_{n+1}} \label{eq:fresnel_t}
\end{align}

The interface matrix $\mathbf{D}_n$ relates the field amplitudes on either side of the interface:

\begin{equation}
\begin{pmatrix} E_n^+ \\ E_n^- \end{pmatrix} = \mathbf{D}_n \begin{pmatrix} E_{n+1}^+ \\ E_{n+1}^- \end{pmatrix}, \quad
\mathbf{D}_n = \frac{1}{t_n} \begin{pmatrix} 1 & r_n \\ r_n & 1 \end{pmatrix}
\label{eq:interface_matrix}
\end{equation}

\paragraph{Propagation matrix.}

Within layer $n$ of thickness $d_n$, the optical phase accumulated is $\beta_n^{(a)} = \beta_{\mathrm{vac}}^{(a)} \tilde{N}_{n} d_n$. The propagation matrix is

\begin{equation}
\mathbf{P}_n = \begin{pmatrix} e^{-i\beta_n^{(a)}} & 0 \\ 0 & e^{+i\beta_n^{(a)}} \end{pmatrix}
\label{eq:propagation_matrix}
\end{equation}

The choice of sign convention ensures consistency with the Airy formula for thin-film reflectance and properly accounts for absorption-induced amplitude decay in the forward direction.

\paragraph{Total transfer matrix.}

The total optical matrix maps the substrate state to air. For $N_{\mathrm{film}}$ finite films between air and the semi-infinite substrate, it is

\begin{equation}
\mathbf M_{\mathrm{total}}=\mathbf D_0\mathbf P_1\mathbf D_1\cdots\mathbf P_{N_{\mathrm{film}}}\mathbf D_{N_{\mathrm{film}}}
\label{eq:total_matrix}
\end{equation}

The reflectance $R$ and transmittance $T$ are extracted from the total matrix elements using boundary conditions: $E_{\text{air}}^+ = 1$ (normalized incident field) and $E_{\text{substrate}}^- = 0$ (no back-reflection from semi-infinite substrate):

\begin{align}
t &= \frac{1}{M_{11}} \label{eq:transmission_amplitude} \\
r &= \frac{M_{21}}{M_{11}} \label{eq:reflection_amplitude} \\
R &= |r|^2 \label{eq:reflectance} \\
T &= \frac{n_{\text{substrate}}}{n_{\text{air}}} |t|^2 \label{eq:transmittance}
\end{align}

\subsection{Electric-Field Distribution and Volumetric Heat-Source Derivation}
\label{sec:field_distribution}

The partial matrices are defined by the state relations below: $\mathbf M_{\mathrm{to},n}$ maps the top-of-layer-$n$ optical state to air, whereas $\mathbf M_{\mathrm{from},n}$ maps it to the substrate.

The field amplitudes $E_n^+$ and $E_n^-$ at the top of layer $n$ are obtained by solving:

\begin{equation}
\begin{pmatrix} 1 \\ r \end{pmatrix} = \mathbf{M}_{\text{to},n} \begin{pmatrix} E_n^+ \\ E_n^- \end{pmatrix}
\label{eq:field_solve}
\end{equation}

subject to the constraint from the backward matrix:

\begin{equation}
\begin{pmatrix} E_n^+ \\ E_n^- \end{pmatrix} = \mathbf{M}_{\text{from},n}^{-1} \begin{pmatrix} t \\ 0 \end{pmatrix}
\label{eq:field_backward}
\end{equation}

These state relations determine the forward and backward amplitudes used in the layer absorption profile.

\paragraph{Volumetric heat-source derivation.}
\hypertarget{par:volumetric_heat_source}{}

The optical fields are normalized to incident amplitude, $E_{\mathrm{air}}^+=1$. Let $E_{\mathrm{inc}}$ be the physical incident electric-field amplitude and $I_{\mathrm{inc}}=n_{\mathrm{air}}|E_{\mathrm{inc}}|^2/(2\mu_0c)$ its irradiance, with $c$ the speed of light and $\mu_0$ the vacuum permeability. The deposition profile $Q_n$ is absorbed volumetric power per unit incident irradiance:
\begin{equation}
Q_n(z)=\frac{-\nabla\cdot\langle\mathbf S\rangle}{I_{\mathrm{inc}}}
=\frac{\omega_{\mathrm{pump}}}{2I_{\mathrm{inc}}}\operatorname{Im}\epsilon_n\,|E_{\mathrm{inc}}E_n(z)|^2,
\label{eq:poynting_divergence}
\end{equation}
where $\omega_{\mathrm{pump}}=2\pi c/\lambda^{\mathrm{pump}}$ is the optical angular frequency and $\epsilon_n=\epsilon_0(\tilde N_n^{\mathrm{pump}})^2$ is absolute permittivity. Thus $Q_n$ has units $\mathrm{m^{-1}}$; multiplying by local incident irradiance gives $\mathrm{W\,m^{-3}}$.

Using the complex refractive index gives
\begin{equation}
Q_n(z)=\frac{\alpha_{\mathrm{opt},n}^{\mathrm{pump}}n_n}{n_{\mathrm{air}}}
\left[
 |E_n^+|^2e^{-\alpha_{\mathrm{opt},n}^{\mathrm{pump}}z}
 +|E_n^-|^2e^{+\alpha_{\mathrm{opt},n}^{\mathrm{pump}}z}
 +2\operatorname{Re}\{(E_n^+)^*E_n^-e^{-i\beta_{\mathrm{opt},n}^{\mathrm{pump}}z}\}
\right].
\label{eq:heat_source_general}
\end{equation}
Here $\alpha_{\mathrm{opt},n}^{\mathrm{pump}}=4\pi\kappa_n/\lambda^{\mathrm{pump}}$ and $\beta_{\mathrm{opt},n}^{\mathrm{pump}}=4\pi n_n/\lambda^{\mathrm{pump}}$ are absorption and standing-wave coefficients, with $n_n,\kappa_n$ evaluated at the pump wavelength. The equivalent decomposition is
\begin{equation}
Q_n(z)=Q_{n,\mathrm{fwd}}e^{-\alpha_{\mathrm{opt},n}^{\mathrm{pump}}z}
+Q_{n,\mathrm{bwd}}e^{-\alpha_{\mathrm{opt},n}^{\mathrm{pump}}(d_n-z)}
+Q_{n,\mathrm{int}}\cos(\beta_{\mathrm{opt},n}^{\mathrm{pump}}z+\varphi_n),
\label{eq:heat_source_decomposition_si}
\end{equation}
with
\begin{align}
Q_{n,\mathrm{fwd}}&=\frac{\alpha_{\mathrm{opt},n}^{\mathrm{pump}}n_n}{n_{\mathrm{air}}}|E_n^+|^2,\label{eq:Q_fwd}\\
Q_{n,\mathrm{bwd}}&=\frac{\alpha_{\mathrm{opt},n}^{\mathrm{pump}}n_n}{n_{\mathrm{air}}}|E_n^-e^{-i\beta_n^{\mathrm{pump}}}|^2,\label{eq:Q_bwd}\\
Q_{n,\mathrm{int}}&=\frac{2\alpha_{\mathrm{opt},n}^{\mathrm{pump}}n_n}{n_{\mathrm{air}}}|(E_n^+)^*E_n^-|,\label{eq:Q_int}\\
\varphi_n&=-\arg[(E_n^+)^*E_n^-].\label{eq:phase_int}
\end{align}
Both optical field amplitudes $E_n^+$ and $E_n^-$ are defined at the layer top face. The deposition coefficient $Q_{n,\mathrm{fwd}}$ is referenced to that face, while $Q_{n,\mathrm{bwd}}$ uses the bottom-face field amplitude $E_n^-e^{-i\beta_n^{\mathrm{pump}}}$. Integrating the per-irradiance profile gives the dimensionless absorptions in Equations~\eqref{eq:A_fwd}--\eqref{eq:A_int}. The physical and Fourier-domain thermal sources are
\begin{equation}
S_n(x,y,z)=I_{\mathrm{pump}}(x,y)Q_n(z),\qquad
\tilde S_n(u,v,z)=\tilde I_{\mathrm{pump}}(u,v)Q_n(z),
\label{eq:source_factorization}
\end{equation}
which provide the optical-to-thermal interface.

\subsection{Consistency Checks and Coordinate Convention}
\label{sec:energy_conservation}
\label{sec:coordinate_transform}

The dimensionless absorption rate in layer $n$ is obtained by integrating the normalized heat source over the layer thickness:

\begin{align}
A_{n,\text{fwd}} &= \frac{n_n}{n_{\text{air}}} |E_n^+|^2 \left( 1 - e^{-\alpha_{\mathrm{opt},n}^{\mathrm{pump}} d_n} \right) \label{eq:A_fwd} \\
A_{n,\text{bwd}} &= \frac{n_n}{n_{\text{air}}} |E_n^- e^{-i\beta_n^{\mathrm{pump}}}|^2 \left( 1 - e^{-\alpha_{\mathrm{opt},n}^{\mathrm{pump}} d_n} \right) \label{eq:A_bwd} \\
A_{n,\text{int}} &= \frac{2\alpha_{\mathrm{opt},n}^{\mathrm{pump}} n_n}{n_{\text{air}}} \text{Re}\left[ (E_n^+)^* E_n^- \, \frac{1 - e^{-i\beta_{\mathrm{opt},n}^{\mathrm{pump}} d_n}}{i\,\beta_{\mathrm{opt},n}^{\mathrm{pump}}} \right] \label{eq:A_int}
\end{align}

where $\beta_n^{\mathrm{pump}} = \beta_{\mathrm{vac}}^{\mathrm{pump}} \tilde{N}_n d_n$ is the complex optical phase. The total absorption in layer $n$ is:

\begin{equation}
A_n = A_{n,\text{fwd}} + A_{n,\text{bwd}} + A_{n,\text{int}}
\label{eq:A_total}
\end{equation}

Energy conservation for the thin-film structure requires:

\begin{equation}
R + \sum_{n=1}^{N_{\text{film}}} A_n + T = 1
\label{eq:energy_conservation}
\end{equation}

Here the sum is over finite film layers, and $T$ is the power entering the semi-infinite substrate. For an absorbing substrate, including its absorbed power gives $R+\sum A_n=1$. Finite-depth integration gives this equality when remaining transmitted power is negligible, for example $\alpha_{\mathrm{substrate}}d_{\mathrm{substrate}}\gg1$.

\paragraph{Coordinate transformation for thermal modeling.}

The optical and thermal coordinates start at each layer top face and increase downward, $z_{\mathrm{opt}}=z_{\mathrm{th}}$. The cross-plane flux is $q=-k_z\,\mathrm dT/\mathrm dz$, positive downward. Absorption profiles use this coordinate directly, and the affine thermal relation maps top to bottom: $\boldsymbol\Theta_{\mathrm{bot}}=\mathbf M_{\mathrm{layer}}\boldsymbol\Theta_{\mathrm{top}}+\mathbf b$.

\subsection{Affine Transfer Matrix with Volumetric Sources}
\label{sec:affine_transfer}

The presence of volumetric heat sources modifies the homogeneous transfer matrix relation to an affine transformation. For a single layer with heat source $S(z)$ deposited into the electron system, the thermal state vector at the bottom is related to the top by:

\begin{equation}
\boldsymbol{\Theta}_{\text{bot}} = \mathbf{M}_{\text{layer}} \boldsymbol{\Theta}_{\text{top}} + \mathbf{b}
\label{eq:affine_transfer}
\end{equation}

where $\boldsymbol{\Theta} = [T_{\mathrm{el}}, T_{\mathrm{ph}}, q_{\mathrm{el}}, q_{\mathrm{ph}}]^T$ for TTM layers, $\mathbf M_{\mathrm{layer}}\equiv\mathbf M_n^{\mathrm{TTM},\downarrow}$ is the homogeneous top-to-bottom transfer matrix of Equation~\eqref{eq:ttm_layer_matrix}, and $\mathbf{b}$ is the source vector computed via particular solutions. In this source construction, $k_{\mathrm{el/ph}}\equiv k_{\mathrm{el/ph},z}$ denotes the cross-plane component; each thermal field and source amplitude is evaluated at the independent Fourier pair $(u,v)$.

In Figure~\ref{fig:workflow}, $\boldsymbol\Psi\equiv\boldsymbol\Theta$ and $\boldsymbol\Psi_p\equiv\boldsymbol\Theta_p$ are compact diagram labels for the thermal state and its source-generated particular solution. The suffixes $\mathrm{in}$ and $\mathrm{out}$ identify the top and bottom of the propagated layer, and $\mathbf M\equiv\mathbf M_{\mathrm{layer}}$ maps the input homogeneous state to the output. The STM state has temperature and normal heat-flux components; the TTM state has the four components stated above. In $\boldsymbol\Psi_p$, the subscript $p$ means particular solution. Substituting Equation~\eqref{eq:source_vector} into Equation~\eqref{eq:affine_transfer} gives the diagram's expression
\begin{equation}
\boldsymbol\Psi_{\mathrm{out}}
=\mathbf M(\boldsymbol\Psi_{\mathrm{in}}-\boldsymbol\Psi_{p,\mathrm{in}})
+\boldsymbol\Psi_{p,\mathrm{out}}.
\end{equation}
The scalar $\Psi$ in the GK modal decomposition is a shear potential.

\paragraph{Source vector computation via particular solutions.}

For an exponential heat source $S(z) = S_0 e^{\beta_{\text{src}} z}$ injected into the electron system, the TTM equations become:

\begin{equation}
\frac{\mathrm d^2}{\mathrm dz^2}
\begin{pmatrix}T_{\mathrm{el}}\\T_{\mathrm{ph}}\end{pmatrix}
=\mathbf A\begin{pmatrix}T_{\mathrm{el}}\\T_{\mathrm{ph}}\end{pmatrix}
-\begin{pmatrix}S_0e^{\beta_{\mathrm{src}}z}/k_{\mathrm{el}}\\0\end{pmatrix}.
\label{eq:inhomogeneous_ttm}
\end{equation}

where the TTM system matrix is:
\begin{equation}
\begin{aligned}
\mathbf{A} &= \begin{pmatrix}
\alpha_{\mathrm{el}} & -g_{\mathrm{ep}}/k_{\mathrm{el}} \\
-g_{\mathrm{ep}}/k_{\mathrm{ph}} & \alpha_{\mathrm{ph}}
\end{pmatrix},\\[4pt]
\alpha_{\mathrm{el/ph}} &= \frac{i\omega C_{\mathrm{el/ph}} + g_{\mathrm{ep}} + 4\pi^2\left(k_{\mathrm{el/ph},x}u^2+k_{\mathrm{el/ph},y}v^2\right)}{k_{\mathrm{el/ph}}}.
\end{aligned}
\end{equation}

Assuming a particular solution of the form $\mathbf{T}_p(z) = \mathbf{v} e^{\beta_{\text{src}} z}$, substitution yields:

\begin{equation}
\left( \beta_{\text{src}}^2 \mathbf{I} - \mathbf{A} \right) \mathbf{v} = -\begin{pmatrix} S_0 / k_{\mathrm{el}} \\ 0 \end{pmatrix}
\label{eq:particular_algebra}
\end{equation}

The corresponding particular-solution amplitude is

\begin{equation}
\mathbf{v} = -\left( \beta_{\text{src}}^2 \mathbf{I} - \mathbf{A} \right)^{-1} \begin{pmatrix} S_0 / k_{\mathrm{el}} \\ 0 \end{pmatrix}
\label{eq:particular_solution}
\end{equation}

The particular heat fluxes follow from $q=-k\,\mathrm dT/\mathrm dz$:
\begin{equation}
q_{\mathrm{el}}^{(p)}(z)=-k_{\mathrm{el}}\beta_{\mathrm{src}}v_1e^{\beta_{\mathrm{src}}z},\qquad
q_{\mathrm{ph}}^{(p)}(z)=-k_{\mathrm{ph}}\beta_{\mathrm{src}}v_2e^{\beta_{\mathrm{src}}z}.
\label{eq:particular_flux}
\end{equation}

The complete particular solution state vector is:
\begin{equation}
\boldsymbol{\Theta}_p(z) = \begin{pmatrix} v_1 \\ v_2 \\ -k_{\mathrm{el}} \beta_{\text{src}} v_1 \\ -k_{\mathrm{ph}} \beta_{\text{src}} v_2 \end{pmatrix} e^{\beta_{\text{src}} z}
\label{eq:particular_state}
\end{equation}

The source vector is then computed as:
\begin{equation}
\mathbf{b} = \boldsymbol{\Theta}_p(h) - \mathbf{M}_{\text{layer}} \boldsymbol{\Theta}_p(0)
\label{eq:source_vector}
\end{equation}

\paragraph{Processing multiple heat-source components.}

At a fixed $(u,v)$, write $\Phi_{n,a}=\tilde I_{\mathrm{pump}}(u,v)Q_{n,a}$ for $a\in\{\mathrm{fwd},\mathrm{bwd},\mathrm{int}\}$. The forward component has $\beta_{\mathrm{src}}=-\alpha_{\mathrm{opt},n}^{\mathrm{pump}}$ and $S_0=\Phi_{n,\mathrm{fwd}}$. The bottom-referenced backward component has $\beta_{\mathrm{src}}=+\alpha_{\mathrm{opt},n}^{\mathrm{pump}}$ and $S_0=\Phi_{n,\mathrm{bwd}}e^{-\alpha_{\mathrm{opt},n}^{\mathrm{pump}}d_n}$. The interference component is decomposed into two spatial exponentials below.

For the interference term, the cosine is decomposed using Euler's formula:
\begin{equation}
\cos(\beta_{\mathrm{opt},n}^{\mathrm{pump}} z + \varphi_n) = \frac{1}{2}\left[ e^{i\varphi_n} e^{i\beta_{\mathrm{opt},n}^{\mathrm{pump}} z} + e^{-i\varphi_n} e^{-i\beta_{\mathrm{opt},n}^{\mathrm{pump}} z} \right]
\end{equation}

The two spatial exponents are $\beta_{\mathrm{src}}=\pm i\beta_{\mathrm{opt},n}^{\mathrm{pump}}$, with source amplitudes $(\Phi_{n,\mathrm{int}}/2)e^{\pm i\varphi_n}$. Solve the particular response for each exponent at the same modulation frequency and add the complex source vectors:
\begin{equation}
\mathbf b_{\mathrm{total}}
=\mathbf b_{\mathrm{fwd}}+\mathbf b_{\mathrm{bwd}}
+\mathbf b_{\mathrm{int},+}+\mathbf b_{\mathrm{int},-}.
\label{eq:source_total}
\end{equation}
The cosine source is real in space, but its frequency-domain thermal response is complex; both quadratures are retained in the linear assembly.

\paragraph{Multi-layer source accumulation.}

For the complete multi-layer structure with Top-Down propagation (from layer 1 at the top to layer $N$ at the substrate), source vectors are accumulated through cascaded matrix multiplication. Processing layers from the top down ($n = 1, 2, \ldots, N$):

\begin{equation}
\mathbf{b}_{\text{accum}} \leftarrow \mathbf{M}_n \cdot \mathbf{b}_{\text{accum}} + \mathbf{b}_n
\label{eq:source_accumulation}
\end{equation}

Between adjacent layers, the interface transfer matrix is inserted as a zero-source affine block and acts on both the accumulated homogeneous state and source vector. The total matrix and source therefore include the interleaved layer and interface blocks. After this complete cascade, the total affine relation is:
\begin{equation}
\boldsymbol{\Theta}_{\text{substrate}} = \mathbf{M}_{\text{total}} \boldsymbol{\Theta}_{\text{surface}} + \mathbf{b}_{\text{total}}
\label{eq:total_affine}
\end{equation}

\subsection{Algorithm Pseudocode}
\label{sec:optical_algorithms}

\begin{algorithm}[H]
\caption{Optical TMM and Electric Field Computation}
\label{alg:optical_tmm}
\begin{algorithmic}
\REQUIRE Complex refractive indices $\{\tilde{N}_n\}_{n=0}^{N}$, Layer thicknesses $\{d_n\}_{n=1}^{N}$, Wavelength $\lambda^{(a)}$
\ENSURE Reflectance $R$, Transmittance $T$, Field amplitudes $\{E_n^+, E_n^-\}_{n=1}^{N}$
\STATE $\beta_{\mathrm{vac}}^{(a)} \leftarrow 2\pi/\lambda^{(a)}$
\STATE $\mathbf{M}_{\text{total}} \leftarrow \mathbf{I}_{2 \times 2}$
\FOR{$j = N-1$ \textbf{downto} $0$}
    \STATE Compute Fresnel coefficients $r_j$, $t_j$ using Equation~\eqref{eq:fresnel_r}, \eqref{eq:fresnel_t}
    \STATE Construct interface matrix $\mathbf{D}_j$ using Equation~\eqref{eq:interface_matrix}
    \IF{$j < N-1$}
        \STATE $n \leftarrow j+1$
        \STATE $\beta_n^{(a)} \leftarrow \beta_{\mathrm{vac}}^{(a)} \tilde{N}_{n} d_n$
        \STATE Construct propagation matrix $\mathbf{P}_n$ using Equation~\eqref{eq:propagation_matrix}
    \ELSE
        \STATE $\mathbf{P}_{\mathrm{id}} \leftarrow \mathbf{I}_{2 \times 2}$
    \ENDIF
    \IF{$j < N-1$}
        \STATE $\mathbf{M}_{\text{total}} \leftarrow \mathbf{D}_j \mathbf{P}_n \mathbf{M}_{\text{total}}$
    \ELSE
        \STATE $\mathbf{M}_{\text{total}} \leftarrow \mathbf{D}_j \mathbf{P}_{\mathrm{id}} \mathbf{M}_{\text{total}}$
    \ENDIF
\ENDFOR
\STATE Extract $r$, $t$ using Equation~\eqref{eq:transmission_amplitude}, \eqref{eq:reflection_amplitude}
\STATE Compute $R$, $T$ using Equation~\eqref{eq:reflectance}, \eqref{eq:transmittance}
\FOR{$n = 1$ \textbf{to} $N$}
    \STATE Construct forward partial matrix $\mathbf{M}_{\text{to},n}$
    \STATE Construct backward partial matrix $\mathbf{M}_{\text{from},n}$
    \STATE Solve for $E_n^+$, $E_n^-$ using Equation~\eqref{eq:field_solve}, \eqref{eq:field_backward}
\ENDFOR
\RETURN $R$, $T$, $\{E_n^+, E_n^-\}_{n=1}^{N}$
\end{algorithmic}
\end{algorithm}

\begin{algorithm}[H]
\caption{Volumetric Heat Source Computation}
\label{alg:heat_source}
\begin{algorithmic}
\REQUIRE Layer index $n$, Field amplitudes $E_n^+$, $E_n^-$, Optical constants $n_n$, $\kappa_n$, Wavelength $\lambda^{\mathrm{pump}}$
\ENSURE Depth coefficients $Q_{n,\mathrm{fwd}}$, $Q_{n,\mathrm{bwd}}$, $Q_{n,\mathrm{int}}$, phase $\varphi_n$, and attenuation/standing-wave coefficients $\alpha_{\mathrm{opt},n}^{\mathrm{pump}}$, $\beta_{\mathrm{opt},n}^{\mathrm{pump}}$
\STATE $\alpha_{\mathrm{opt},n}^{\mathrm{pump}} \leftarrow 4\pi \kappa_n / \lambda^{\mathrm{pump}}$
\STATE $\beta_{\mathrm{opt},n}^{\mathrm{pump}} \leftarrow 4\pi n_n / \lambda^{\mathrm{pump}}$
\STATE $c \leftarrow 3 \times 10^8$ m/s, $\mu_0 \leftarrow 4\pi \times 10^{-7}$ H/m
\STATE Compute $Q_{n,\mathrm{fwd}}$ using Equation~\eqref{eq:Q_fwd}
\STATE Compute $Q_{n,\mathrm{bwd}}$ using Equation~\eqref{eq:Q_bwd}
\STATE Compute $Q_{n,\mathrm{int}}$ using Equation~\eqref{eq:Q_int}
\STATE Compute $\varphi_n$ using Equation~\eqref{eq:phase_int}
\STATE Compute absorption rates $A_{n,\text{fwd}}$, $A_{n,\text{bwd}}$, $A_{n,\text{int}}$ using Equation~\eqref{eq:A_fwd}, \eqref{eq:A_bwd}, \eqref{eq:A_int}
\STATE Verify $A_n = A_{n,\text{fwd}} + A_{n,\text{bwd}} + A_{n,\text{int}} \geq 0$
\RETURN $Q_{n,\mathrm{fwd}}$, $Q_{n,\mathrm{bwd}}$, $Q_{n,\mathrm{int}}$, $\varphi_n$, $\alpha_{\mathrm{opt},n}^{\mathrm{pump}}$, $\beta_{\mathrm{opt},n}^{\mathrm{pump}}$
\end{algorithmic}
\end{algorithm}

\begin{algorithm}[H]
\caption{Affine Transfer Matrix Source Vector Computation}
\label{alg:source_vector}
\begin{algorithmic}
\REQUIRE Heat source parameters $S_0$, $\beta$, Layer thickness $h$, TTM parameters $k_{\mathrm{el}}$, $k_{\mathrm{ph}}$, $C_{\mathrm{el}}$, $C_{\mathrm{ph}}$, $g_{\mathrm{ep}}$
\REQUIRE System matrix $\mathbf{A}$, Homogeneous transfer matrix $\mathbf{M}_{\text{layer}}$
\ENSURE Source vector $\mathbf{b}$
\STATE Solve Equation~\eqref{eq:particular_solution} for $\mathbf v$ with a conditioned linear-system solver
\STATE $\boldsymbol{\Theta}_p(0) \leftarrow [\mathbf{v}^T, -k_{\mathrm{el}} \beta v_1, -k_{\mathrm{ph}} \beta v_2]^T$
\STATE $\boldsymbol{\Theta}_p(h) \leftarrow e^{\beta h} \boldsymbol{\Theta}_p(0)$
\STATE Compute $\mathbf{b}$ using Equation~\eqref{eq:source_vector}
\RETURN $\mathbf{b}$
\end{algorithmic}
\end{algorithm}

\begin{algorithm}[H]
\caption{Particular Solution for TTM with Exponential Source}
\label{alg:particular_solution_ttm}
\begin{algorithmic}
\REQUIRE Source exponent $\beta$, Source amplitude $S_0$, Layer thickness $h$, TTM parameters
\ENSURE Particular solution $\boldsymbol{\Theta}_p(0)$, $\boldsymbol{\Theta}_p(h)$
\STATE Construct system matrix $\mathbf{A}$ from TTM parameters (section~\ref{sec:tmm})
\STATE $\alpha_{\mathrm{el}} \leftarrow (i\omega C_{\mathrm{el}} + g_{\mathrm{ep}} + 4\pi^2(k_{\mathrm{el},x}u^2+k_{\mathrm{el},y}v^2)) / k_{\mathrm{el}}$
\STATE $\alpha_{\mathrm{ph}} \leftarrow (i\omega C_{\mathrm{ph}} + g_{\mathrm{ep}} + 4\pi^2(k_{\mathrm{ph},x}u^2+k_{\mathrm{ph},y}v^2)) / k_{\mathrm{ph}}$
\STATE $\mathbf{A} \leftarrow \begin{pmatrix} \alpha_{\mathrm{el}} & -g_{\mathrm{ep}}/k_{\mathrm{el}} \\ -g_{\mathrm{ep}}/k_{\mathrm{ph}} & \alpha_{\mathrm{ph}} \end{pmatrix}$
\STATE Compute $\mathbf{v} \leftarrow -(\beta^2 \mathbf{I} - \mathbf{A})^{-1} [S_0/k_{\mathrm{el}}, 0]^T$
\STATE $v_1 \leftarrow \mathbf{v}(1)$, $v_2 \leftarrow \mathbf{v}(2)$ \COMMENT{Electron and phonon temperatures}
\STATE $\boldsymbol{\Theta}_p(0) \leftarrow [v_1, v_2, -k_{\mathrm{el}} \beta v_1, -k_{\mathrm{ph}} \beta v_2]^T$
\STATE $\boldsymbol{\Theta}_p(h) \leftarrow e^{\beta h} \boldsymbol{\Theta}_p(0)$
\RETURN $\boldsymbol{\Theta}_p(0)$, $\boldsymbol{\Theta}_p(h)$
\end{algorithmic}
\end{algorithm}

The optical TMM and volumetric-source construction above feed the thermal recursion directly: these sources are passed into the STM/TTM cascade and, for benchmark-conditioned high-thermal-conductivity Si substrates, into the OTh-TTM(NF) substrate closure.

\clearpage
\section{Sensitivity and Identifiability Analysis}
\label{sec:sensitivity_svd}

This section defines the data-quality gate, logarithmic sensitivity matrix, and SVD and absolute-sensitivity screens used to construct the fitting-target list.

\subsection{Data Diagnosis and Quality Gate}
\label{sec:quality_gate}

Before any Jacobian-based identifiability analysis is attempted, the PRISM screening logic evaluates whether the measured spectrum is suitable for inversion. Given the raw FDTR arrays $\{f_i,\phi_i\}_{i=1}^{N_f}$, it first estimates the signal-to-noise ratio as
\begin{equation}
    \mathrm{SNR} = 20 \log_{10}\!\left(\frac{\sigma_{\mathrm{sig}}}{\sigma_{\mathrm{noise}}}\right),
\end{equation}
where $\sigma_{\mathrm{sig}}$ denotes the characteristic inlier signal variation across the retained spectrum and $\sigma_{\mathrm{noise}}$ denotes the estimated phase-noise level. The diagram's $\sigma_s$ and $\sigma_n$ abbreviate these two quantities, respectively; SNR is expressed in decibels. The diagnostic noise estimate is recorded separately from the fixed SVD normalization floor $\sigma_\phi=0.1^\circ$. For outlier screening, a preliminary smooth phase trend $\phi_{\mathrm{fit}}(f)$ defines
\begin{equation}
 r_i=\phi_i-\phi_{\mathrm{fit}}(f_i),\qquad |r_i|>3\sigma_r\quad\text{flags an outlier},
 \label{eq:quality_residuals}
\end{equation}
where $r_i$ is the diagnostic residual and $\sigma_r$ is its estimated standard deviation, both in degrees. The figure's $\phi_{\mathrm{fit}}$ suppresses the frequency argument $f_i$. After robust outlier detection, the data-quality score is defined as
\begin{equation}
    Q^\ast = 1 - \frac{N_{\mathrm{outliers}}}{N_{\mathrm{total}}},
\end{equation}
where $N_{\mathrm{outliers}}$ is the number of rejected spectral points and $N_{\mathrm{total}}$ is the original sample count. Spectra below the configured $Q^\ast$ or SNR threshold undergo outlier pruning, bandwidth restriction, or data reacquisition. Spectra that pass the quality gate proceed to sensitivity construction, as shown in the left panel of Figure~\ref{fig:workflow}.

\subsection{Sensitivity Estimation}

After a spectrum passes the quality gate, the workflow constructs the logarithmic sensitivity Jacobian whose entries are
\begin{equation}
    J_{ij} = \frac{\partial \phi(f_i)}{\partial \ln p_j},
\end{equation}
for candidate parameters $\{p_j\}_{j=1}^{N_p}$ and measured frequencies $\{f_i\}_{i=1}^{N_f}$. The logarithmic sensitivity of phase $\phi$ to a single parameter $p$ is therefore
\begin{equation}
    S_p = \frac{\partial \phi}{\partial \ln p} = p \frac{\partial \phi}{\partial p}
    \label{eq:sensitivity_def}
\end{equation}
so that $J_{ij}=S_{p_j}(f_i)=S_{ij}$. The diagram and this section use the equivalent notations $\mathbf S\equiv\mathbf J$ and $P\equiv N_p$. For sensitivity construction, $N_f$ counts the accepted frequency points.

The sensitivity is computed numerically using central finite differences with symmetric logarithmic perturbation:
\begin{equation}
    S_p \approx \frac{\phi(p \cdot e^{\delta}) - \phi(p \cdot e^{-\delta})}{2\delta}
\end{equation}
where $\delta = \ln(1.01) \approx 0.00995$. This choice ensures symmetric perturbation in log-space: $e^{+\delta} \approx 1.01$ and $e^{-\delta} \approx 0.9901$.

The resulting sensitivity has units of degrees. A 1\% change in parameter $p$ (i.e., $\Delta \ln p \approx 0.01$) produces a phase change of approximately $S_p \times 0.01$ degrees.

The RMS magnitude $\bar{S}_{p,\mathrm{rms}}=\sqrt{N_f^{-1}\sum_i|S_p(f_i)|^2}$ determines membership in the absolute-sensitivity list $\mathcal L_{\mathrm{sensitivity}}$. Section~\ref{sec:dual_criterion} defines its ordered intersection with the SVD list, and Algorithm~\ref{alg:target_adjust} filters the requested targets by membership.

\subsection{Identifiability Metrics and Screening}
\label{sec:fisher_ranking}
\label{sec:dual_criterion}

SVD characterizes locally constrained parameter combinations~\cite{chen2024svd}. The retained-mode count is $r_{\mathrm{SVD}}$, whereas $N_{\mathrm{ident}}$ counts entries in the final ordered parameter list. Each declared candidate is assessed through its projection onto the retained right-singular-vector subspace. Independently known beam, thickness and material inputs can appear in diagnostic charts without being released as requested fit targets. The Jacobian is normalized as follows:
\begin{equation}
    \mathbf{J}_{\text{norm}} = \mathbf{J} \cdot \frac{\delta}{\sigma_\phi}
\end{equation}
where $\delta=\ln(1.01)$ and $\sigma_\phi=0.1^\circ$ is the baseline phase-noise floor used for normalization. This uniform scaling leaves the relative singular-value ratios unchanged; the absolute phase-sensitivity screen is applied separately. Here $N_p\equiv P$ is the number of declared diagnostic candidates, including independently fixed inputs shown in the charts. The normalized Jacobian $\mathbf J_{\mathrm{norm}}\in\mathbb R^{N_f\times N_p}$ is decomposed as $\mathbf J_{\mathrm{norm}}=\mathbf U\boldsymbol\Sigma\mathbf V^T$. The notations $\mathbf S_{\mathrm{norm}}$, $\mathbf J_{\mathrm{norm}}$ and $\hat{\mathbf J}$ refer to this same dimensionless matrix. The columns of $\mathbf U$ and $\mathbf V$ are frequency-space and parameter-space singular vectors, respectively, and $\boldsymbol\Sigma$ contains the ordered singular values $\sigma_m$. The dimensionless $\tau$ in Figure~\ref{fig:workflow} denotes $\tau_{\mathrm{SNR}}$; the GK flux-relaxation time is defined separately in section~\ref{sec:non_fourier_si}.

\begin{algorithm}[H]
\caption{SVD Screening and Ordered Parameter-List Construction}
\label{alg:svd}
\begin{algorithmic}
\REQUIRE Declared model and fixed inputs; candidate set of size $P$, Jacobian $\mathbf J$, $\delta=\ln1.01$, $\sigma_\phi=0.1^\circ$, $\tau_{\mathrm{SNR}}=0.1$
\ENSURE $\mathcal L_{\mathrm{SVD}}$, $r_{\mathrm{SVD}}$, singular spectrum and coupling diagnostics
\IF{$P=0$}
    \RETURN Empty list and $r_{\mathrm{SVD}}=0$
\ENDIF
\STATE $\mathbf J_{\mathrm{norm}}\leftarrow(\delta/\sigma_\phi)\mathbf J$
\STATE $\mathbf U,\boldsymbol\Sigma,\mathbf V^{\mathsf T}\leftarrow\operatorname{SVD}(\mathbf J_{\mathrm{norm}})$
\IF{$\sigma_1=0$}
    \RETURN Empty list and $r_{\mathrm{SVD}}=0$
\ENDIF
\STATE $\mathcal R\leftarrow\{m:\sigma_m/\sigma_1>\tau_{\mathrm{SNR}}\}$; $r_{\mathrm{SVD}}\leftarrow|\mathcal R|$
\FOR{each candidate $p_j$}
    \STATE $\eta_j\leftarrow\sum_{m\in\mathcal R}|V_{jm}|^2$
    \STATE $E_j\leftarrow\sum_{m\in\mathcal R}(\sigma_m/\sigma_1)^2|V_{jm}|^2$
\ENDFOR
\STATE Form $\mathcal L_{\mathrm{SVD}}$ from $\eta_j\ge1/P$, in descending $E_j$ order
\STATE Resolve ties by the declared candidate order; record weak-mode loadings as diagnostics
\RETURN $\mathcal L_{\mathrm{SVD}}$, $r_{\mathrm{SVD}}$, $\boldsymbol\Sigma$, coupling diagnostics
\end{algorithmic}
\end{algorithm}

The relative criterion $\sigma_m/\sigma_1>\tau_{\mathrm{SNR}}$ selects singular modes. Algorithm~\ref{alg:svd} screens individual parameters by their retained-subspace projections, using the declared projection cutoff. Its intersection with the absolute-sensitivity list determines the final list conditional on the model and fixed inputs.

\paragraph{Workflow target reduction after SVD.}
For a routine request $\mathcal Q_0\equiv\mathcal Q_{\mathrm{requested}}$ in Figure~\ref{fig:workflow}, the execution layer returns
\begin{equation}
\mathcal{Q}_{\mathrm{fit}}=\bigl(p\in\mathcal{L}_{\mathrm{ident}}:\ p\in\mathcal{Q}_0\bigr),
\qquad \mathcal{Q}_{\mathrm{unfit}}=\mathcal{Q}_0\setminus\mathcal{Q}_{\mathrm{fit}}.
\label{eq:request_membership}
\end{equation}
The membership check applies to every request size. Requested parameters outside $\mathcal L_{\mathrm{ident}}$ form $\mathcal Q_{\mathrm{unfit}}$ and require independently specified fixed inputs. Any proposed composite reparameterization is assessed as a separate task through the same screens.

Table~\ref{tab:fit_ledger} summarizes, for every benchmark, the targets released to the fit and the parameters held at their input values.

\begin{table}[H]
\centering
\caption{Fit-target ledger of the benchmarks. For each stack the table lists the targets released to the fit, the parameters held at their input values (all remaining layer, optical, and beam parameters are also held; Tables~\ref{tab:archive_si_thermal_props}--\ref{tab:archive_si_nf_params}), and the figures and tables that document the decision.}
\label{tab:fit_ledger}
\footnotesize
\begin{tabularx}{\textwidth}{@{} >{\raggedright\arraybackslash}p{3.3cm} >{\raggedright\arraybackslash}X >{\raggedright\arraybackslash}X >{\raggedright\arraybackslash}p{3.1cm} @{}}
\toprule
\textbf{Benchmark} & \textbf{Released to the fit} & \textbf{Held at input values} & \textbf{Evidence} \\
\midrule
Au/SiO$_2$/Si (\SI{112}{\nano\metre} Au; $w_r = 1.86$ and \SI{10.44}{\micro\metre}) & $k_{\mathrm{SiO}_2}$ & $k_{\mathrm{Si}}$; $G_{\mathrm{Au/SiO}_2}$ (prior \SI{50}{\mega\watt\per\metre\squared\per\kelvin}) & Figure~\ref{fig:ausi}A--C; Table~\ref{tab:silica}A \\
Au/Ti/SiO$_2$ (bulk silica) & Diagnostic co-fit: $k_{\mathrm{SiO}_2}$; $G_{\mathrm{Ti/SiO}_2}$ & Au and Ti layer properties; $G_{\mathrm{Au/Ti}}$ & Table~\ref{tab:silica}A; workflow in Figure~\ref{fig:si_chat_ui} \\
Au/Ti/SiO$_2$/Si (buried-oxide target-reduction benchmark) & $k_{\mathrm{SiO}_2}$ & $k_{\mathrm{Si}}$, $G_{\mathrm{Ti/SiO}_2}$, $G_{\mathrm{SiO}_2/\mathrm{Si}}$, $G_{\mathrm{Au/Ti}}$ & Figures~\ref{fig:si_microscopic_1}--\ref{fig:si_microscopic_3} \\
Au/Si, offset and coaxial (\SI{97}{\nano\metre} Au) & Local branches: $k_{\mathrm{Si}}$, $G_{\mathrm{Au/Si}}$; OTh-TTM(NF): $k_{\mathrm{Si}}$, $\bNF$ & Nonlocal GK parameters (Table~\ref{tab:archive_si_nf_params}) & Figure~\ref{fig:ausi}D--F; Figure~\ref{fig:si_spatial_overlap}A--C; Table~\ref{tab:silica}B \\
Au/Ti/Si (\SI{62}{\nano\metre} Au / \SI{40}{\nano\metre} Ti; $w_r = 1.86$ and \SI{10.44}{\micro\metre}) & Local branches: $k_{\mathrm{Si}}$, $G_{\mathrm{Ti/Si}}$; OTh-TTM(NF): $k_{\mathrm{Si}}$, $\bNF$ & $G_{\mathrm{Au/Ti}}$; nonlocal GK parameters (Table~\ref{tab:archive_si_nf_params}) & Figure~\ref{fig:nonlocal}; Figure~\ref{fig:si_spatial_overlap}D--F; Table~\ref{tab:nonlocal_windows} \\
Au/MoS$_2$ (\SI{43}{\nano\metre} Au; spectrum of Ref.~\cite{guo2026light}) & $k_{\mathrm{MoS}_2,z}$, $k_{\mathrm{MoS}_2,r}$, $G$ & Au optical constants $n_{\mathrm{Au}}$, $\kappa_{\mathrm{Au}}$; $h_{\mathrm{Au}}$ & Figure~\ref{fig:optical}B,D; Table~\ref{tab:archive_aumos2} \\
\bottomrule
\end{tabularx}
\end{table}

\paragraph{SVD-based parameter screening and ranking.}
For retained modes $\mathcal R=\{m:\sigma_m/\sigma_1>\tau_{\mathrm{SNR}}\}$, define
\begin{equation}
\eta_j=\sum_{m\in\mathcal R}|V_{jm}|^2,\qquad
E_j=\sum_{m\in\mathcal R}\left(\frac{\sigma_m}{\sigma_1}\right)^2|V_{jm}|^2.
\label{eq:svd_parameter_score}
\end{equation}
The projection screen retains $\eta_j\ge\eta_{\min}$, with the declared baseline $\eta_{\min}=1/P$ for $P$ candidates. These parameters form $\mathcal L_{\mathrm{SVD}}$ in descending retained sensitivity energy $E_j$; ties preserve the declared candidate order. Filtering this list by absolute-sensitivity membership preserves that order to obtain $(p_{(1)},\ldots,p_{(N_{\mathrm{ident}})})$. $N_{\mathrm{ident}}$ is the final list length, and $r_{\mathrm{SVD}}$ is the retained-mode count.

\paragraph{Reading conventions for the SVD bubble chart.}
Figure~\ref{fig:workflow} defines the bubble-chart reading conventions. Numerical diagnostic colors use raw-Jacobian singular values $\sigma_m^{\mathrm{raw}}$, related to the baseline-normalized values by $\sigma_m=(\delta/\sigma_\phi)\sigma_m^{\mathrm{raw}}$; this constant rescaling leaves $\mathbf V$ and relative singular values unchanged. Requested members are released to fitting under the fixed inputs.

\paragraph{Dual-criterion screen: SVD threshold and sensitivity verification.}

Both screens apply to the declared candidate set $\mathcal C=\{p_1,\ldots,p_P\}$. The SVD pass list is the projection-screened list from Algorithm~\ref{alg:svd}. The absolute pass list is
\begin{equation}
\mathcal{L}_{\mathrm{sensitivity}}
=\{p\in\mathcal{C}:\bar{S}_{p,\mathrm{rms}}\ge S_{\min}\},
\qquad
\bar{S}_{p,\mathrm{rms}}=\sqrt{\frac{1}{N_f}\sum_i|S_p(f_i)|^2}.
\label{eq:absolute_sensitivity_list}
\end{equation}
The configuration described in the main text uses $S_{\min}=0.2^\circ$ on the RMS magnitude of the raw logarithmic sensitivity matrix $\mathbf S=\mathbf J$. SVD uses the dimensionless normalized matrix with $\sigma_\phi=0.1^\circ$. Candidates and thresholds are declared before optimization.
\begin{equation}
\mathcal{L}_{\mathrm{ident}}
=\bigl(p\in\mathcal{L}_{\mathrm{SVD}}:\ p\in\mathcal{L}_{\mathrm{sensitivity}}\bigr),
\qquad N_{\mathrm{ident}}=|\mathcal{L}_{\mathrm{ident}}|.
\label{eq:ordered_intersection_si}
\end{equation}
Filtering preserves the SVD-list order and completes the conditional eligibility decision. Sensitivity curves show the response to each coefficient, the SVD chart shows coupling, and the final intersection gives the parameters eligible for fitting under the declared model and fixed priors. The retained-mode count measures constrained parameter combinations; the list length counts eligible entries, and the requested subset determines the fitted covariance and propagated uncertainties.

\paragraph{Illustrative example: $k_{\mathrm{Si}}$ in OTh-TTM analysis.}
For the Au(62\,nm)/\allowbreak Ti(40\,nm)/\allowbreak SiO$_2$(2\,$\mu$m)/\allowbreak Si benchmark, the 20-coefficient OTh-TTM diagnostic Jacobian gives a dominant $k_{\mathrm{Si}}$ loading $|V_{\mathrm{Si},6}|=0.814$ in mode $M_6$. The unnormalized singular values $\sigma_6^{\mathrm{raw}}\approx2.77$ and $\sigma_1^{\mathrm{raw}}\approx100.6$ give $\sigma_6^{\mathrm{raw}}/\sigma_1^{\mathrm{raw}}\approx0.0275<0.1$, placing this dominant loading in a weak mode. The reported RMS sensitivity is approximately $0.55^\circ$ (Figure~\ref{fig:si_microscopic_1}), above the $0.2^\circ$ absolute-sensitivity threshold.

In the final screening record, $k_{\mathrm{Si}}$ is absent from $\mathcal L_{\mathrm{SVD}}$ and is held at the independently supplied value $140$\,W\,m$^{-1}$\,K$^{-1}$ in Table~\ref{tab:fit_ledger}.

\clearpage
\section{DTDE-Centered Benchmark for Reduced-Space Global Optimization}
\label{sec:optimizer_benchmark}

The reduced-space benchmark of Figure~\ref{fig:si_optimizer} compares seven global optimizers under a common FDTR fitting protocol. DTDE is the default because it has the shortest evolutionary wall-clock time at comparable accuracy; CMA-ES reaches the target RMSE with the fewest function evaluations.

\subsection{Unified Benchmark Protocol}

The optimizer benchmark compares EA4Eig, jSO, DTDE, L-SHADE, CMA-ES, HPSO, and canonical DE on exactly the same reduced-space Au/Si FDTR inverse problem. Each objective evaluation uses the same GPU-accelerated multilayer forward model and the same overlap-geometry phase spectrum over \SI{10}{\kilo\hertz}--\SI{50}{\mega\hertz}, with the Au/Si elliptical spot radii $w_x = \SI{2.19}{\micro\metre}$ and $w_y = \SI{1.84}{\micro\metre}$, and reduced-space bounds $k \in [50,300]~\si{\watt\per\metre\per\kelvin}$ and $G \in [10,200]~\si{\mega\watt\per\square\metre\per\kelvin}$.

The fitting objective is the root-mean-square phase error
\begin{equation}
    J(\mathbf{x}) = \sqrt{\frac{1}{N_f} \sum_{k=1}^{N_f} \left[\phi_{\text{model}}(f_k, \mathbf{x}) - \phi_{\text{exp}}(f_k)\right]^2},
    \label{eq:optimizer_rmse_si}
\end{equation}
where $\mathbf{x}$ collects the reduced-space fit parameters and $N_f$ counts accepted frequencies. The diagram's $\phi_{\mathrm{exp}}$ and $\phi_{\mathrm{model}}$ are the measured and predicted phases; their root-mean-square difference is the phase RMSE in degrees. In its Results box, $p^*$ is a best-fit parameter value and $\epsilon_n$ is the propagated standard uncertainty of the $n$-th reported parameter, with matching units. The displayed $\pm$ example reports one standard deviation under the fixed-input propagation convention of Table~\ref{tab:silica}; confidence intervals state their coverage separately. All stochastic optimizers use $NFE_{\max}=600$ and five seeds (42--46). Figure~\ref{fig:si_optimizer}A and the runtime statistics cover the evolutionary search. Each optimizer then receives the same terminal \texttt{L-BFGS-B} refinement. Population-batch counting gives exported evolutionary counts of typically 600--603.

\subsection{DTDE: Literature Family and Benchmark Implementation}

DTDE belongs to the domain-transform differential evolution family introduced by Zhang \textit{et al.}~\citep{zhang2023dtde}. In that family, a transform is applied to smooth the effective fitness landscape and to help the population escape premature convergence or stagnation on complicated multimodal objective surfaces.

The FDTR benchmark uses the DTDE-aligned implementation specified in Algorithm~\ref{alg:dtde}. All seven optimizers share the interface
\begin{equation}
    \mathcal{A}:\left(J,\mathbf{lb},\mathbf{ub},NFE_{\max}\right)
    \mapsto
    \left(\mathbf{x}^{\star}_{\mathrm{evo}},J^{\star}_{\mathrm{evo}},NFE_{\mathrm{evo}},t_{\mathrm{evo}}\right),
    \label{eq:optimizer_interface_si}
\end{equation}
where $J$ is the phase-RMSE objective of Equation~\eqref{eq:optimizer_rmse_si}, $\mathbf{lb}$ and $\mathbf{ub}$ are the bounds, and $NFE_{\max}=600$ is the evaluation budget. The outputs are the best evolutionary-stage point $\mathbf x^\star_{\mathrm{evo}}$, its objective value $J^\star_{\mathrm{evo}}$, the evaluation count $NFE_{\mathrm{evo}}$, and runtime $t_{\mathrm{evo}}$. Runs use the full budget, followed by the common terminal refinement described above.

For the differential-evolution family (EA4Eig, jSO, DTDE, L-SHADE, and DE), the optimizer first constructs a mutant vector $\mathbf{v}_i$ and then assembles the trial vector coordinatewise as
\begin{equation}
    u_{ij}=
    \begin{cases}
        v_{ij}, & \xi_{ij}<CR_i \ \text{or}\ j=j_{\mathrm{rand}},\\
        x_{ij}, & \text{otherwise},
    \end{cases}
    \label{eq:de_trial_assembly_si}
\end{equation}
where $\xi_{ij}\sim\mathcal U(0,1)$ and $j_{\mathrm{rand}}$ selects a coordinate inherited from the mutant. The mutation factor $F_i$ and crossover probability $CR_i$ control the DE update. Greedy replacement accepts $\mathbf u_i$ when $J(\mathbf u_i)<J(\mathbf x_i)$. In the optimizer algorithms, $P$ denotes the population, $N_P$ its size, $n=\dim\mathbf x$ the fitting dimension, $A$ the archive, and $M_F,M_{CR}$ the success-history memories; $r_i$ indexes a selected memory entry.

\begin{algorithm}[H]
\caption{DTDE Kernel under the Unified FDTR Benchmark Protocol}
\label{alg:dtde}
\begin{algorithmic}
\STATE \textbf{Input:} $J$, $\mathbf{lb}$, $\mathbf{ub}$, $NFE_{\max}=600$
\STATE Initialize $P=\{\mathbf{x}_i\}_{i=1}^{N_P}$ uniformly in $[\mathbf{lb},\mathbf{ub}]$ with $N_P=30$
\STATE Evaluate $f_i=J(\mathbf{x}_i)$, set archive $A=\emptyset$, and initialize $M_F=M_{CR}=0.5$
\WHILE{$NFE < NFE_{\max}$}
    \IF{generation index is a multiple of $5$}
        \STATE Compute weights $\omega_i \propto f_{\max}-f_i$ and normalize $\sum_i \omega_i = 1$
        \STATE Compute weighted center $\mathbf{c}=\sum_i \omega_i \mathbf{x}_i$
        \STATE Apply the domain transform $\mathbf{x}_i \leftarrow \mathrm{clip}\!\left(\mathbf{c}+0.95(\mathbf{x}_i-\mathbf{c}),\mathbf{lb},\mathbf{ub}\right)$
        \STATE Re-evaluate $f_i=J(\mathbf{x}_i)$
    \ENDIF
    \FOR{each target vector $\mathbf{x}_i$}
        \STATE Sample $F_i$ from a truncated Cauchy law centered at $M_F[r_i]$ and $CR_i$ from a clipped normal law centered at $M_{CR}[r_i]$
        \STATE Choose $\mathbf{x}_{p\mathrm{best}}$ from the best $\max(2,\lfloor 0.15N_P \rfloor)$ members
        \STATE Choose $\mathbf{x}_{r_1}$ from the population/archive union and $\mathbf{x}_{r_2}$ from the same union with distinct indices
        \STATE Form the mutant
        \[
            \mathbf{v}_i=\mathbf{x}_i+F_i(\mathbf{x}_{p\mathrm{best}}-\mathbf{x}_i)+F_i(\mathbf{x}_{r_1}-\mathbf{x}_{r_2})
        \]
        \STATE Build $\mathbf{u}_i$ using Equation~\eqref{eq:de_trial_assembly_si} and evaluate $J(\mathbf{u}_i)$
        \IF{$J(\mathbf{u}_i)<J(\mathbf{x}_i)$}
            \STATE Accept $\mathbf{u}_i$, push the replaced parent into $A$, and record successful $F_i,CR_i$
        \ENDIF
    \ENDFOR
    \STATE Update $M_F$ and $M_{CR}$ by improvement-weighted success history
    \STATE Reduce $N_P$ linearly from $30$ toward $4$ as the NFE budget is consumed
\ENDWHILE
\RETURN $\mathbf{x}^{\star}_{\mathrm{evo}}=\arg\min_{\mathbf{x}\in P} J(\mathbf{x})$
\end{algorithmic}
\end{algorithm}

\subsection{Comparator Algorithms in the Same Benchmark}

\paragraph{EA4Eig.}
EA4Eig is represented here by an EA4Eig-style adaptive differential evolution implementation motivated by eigen-crossover methods that embed evolutionary updates in the covariance eigenbasis~\citep{bujok2022ea4eig, brest2017single}. In this benchmark, it shares the adaptive-memory, archive, and linear population-reduction backbone of the jSO-like comparators, but it occasionally performs trial assembly after rotating the parent and mutant into the eigenbasis of the running covariance estimate.

\begin{algorithm}[H]
\caption{EA4Eig Kernel under the Unified FDTR Benchmark Protocol}
\label{alg:ea4eig}
\begin{algorithmic}
\STATE \textbf{Input:} $J$, $\mathbf{lb}$, $\mathbf{ub}$, $NFE_{\max}=600$
\STATE Initialize population $P$, archive $A$, memories $M_F,M_{CR}$, covariance $C=I$, and mean $\bar{\mathbf{x}}$
\WHILE{$NFE < NFE_{\max}$}
    \STATE Update the covariance estimate
    \[
        C \leftarrow 0.9\,C + 0.1\frac{(P-\bar{\mathbf{x}})^{\top}(P-\bar{\mathbf{x}})}{N_P},
        \qquad
        C = Q\Lambda Q^{\top}
    \]
    \FOR{each target vector $\mathbf{x}_i$}
        \STATE Sample $F_i,CR_i$ from the success-history memories
        \STATE Choose $\mathbf{x}_{p\mathrm{best}}$ from the best $\max(2,\lfloor 0.1N_P \rfloor)$ members
        \STATE Choose $\mathbf{x}_{r_1}$ from the population/archive union and $\mathbf{x}_{r_2}$ from the same union with distinct indices
        \STATE Form
        \[
            \mathbf{v}_i=\mathbf{x}_i+F_i(\mathbf{x}_{p\mathrm{best}}-\mathbf{x}_i)+F_i(\mathbf{x}_{r_1}-\mathbf{x}_{r_2})
        \]
        \IF{with probability $0.4$}
            \STATE Rotate to eigenspace: $\tilde{\mathbf{x}}_i=Q^{\top}\mathbf{x}_i$, $\tilde{\mathbf{v}}_i=Q^{\top}\mathbf{v}_i$
            \STATE Assemble $\tilde{\mathbf{u}}_i$ coordinatewise from $\tilde{\mathbf{x}}_i$ and $\tilde{\mathbf{v}}_i$, then map back with $\mathbf{u}_i=Q\tilde{\mathbf{u}}_i$
        \ELSE
            \STATE Assemble $\mathbf{u}_i$ directly from $\mathbf{x}_i$ and $\mathbf{v}_i$ using Equation~\eqref{eq:de_trial_assembly_si}
        \ENDIF
        \STATE Apply greedy replacement, update $A$, and record successful $F_i,CR_i$
    \ENDFOR
    \STATE Update $M_F,M_{CR}$ and reduce $N_P$ linearly from $30$ to $4$
\ENDWHILE
\RETURN $\mathbf{x}^{\star}_{\mathrm{evo}}$
\end{algorithmic}
\end{algorithm}

\paragraph{jSO.}
jSO is a success-history-based differential-evolution variant~\citep{brest2017single}. The benchmark implementation uses an early-stage weighting schedule for $F$ and $CR$ that changes as the evaluation budget is consumed.

\begin{algorithm}[H]
\caption{jSO Kernel under the Unified FDTR Benchmark Protocol}
\label{alg:jso_si}
\begin{algorithmic}
\STATE \textbf{Input:} $J$, $\mathbf{lb}$, $\mathbf{ub}$, $NFE_{\max}=600$
\STATE Initialize population $P$, archive $A$, and success-history memories $M_F,M_{CR}$
\WHILE{$NFE < NFE_{\max}$}
    \STATE Set the elite fraction
    \[
        p = 0.25 - (0.25-0.05)\frac{NFE}{NFE_{\max}}
    \]
    \FOR{each target vector $\mathbf{x}_i$}
        \STATE Raise the local memory centers during early search with
        \[
            F_{\mathrm{mean}} \leftarrow
            \begin{cases}
                \max(M_F[r_i],0.7), & NFE < 0.25NFE_{\max},\\
                \max(M_F[r_i],0.8), & 0.25NFE_{\max} \le NFE < 0.5NFE_{\max},\\
                \max(M_F[r_i],0.9), & 0.5NFE_{\max} \le NFE < 0.75NFE_{\max},\\
                M_F[r_i], & \text{otherwise},
            \end{cases}
        \]
        \[
            CR_{\mathrm{mean}} \leftarrow
            \begin{cases}
                \max(M_{CR}[r_i],0.7), & NFE < 0.25NFE_{\max},\\
                \max(M_{CR}[r_i],0.6), & 0.25NFE_{\max} \le NFE < 0.5NFE_{\max},\\
                M_{CR}[r_i], & \text{otherwise}
            \end{cases}
        \]
        \STATE Sample $F_i$ from a positive Cauchy perturbation around $F_{\mathrm{mean}}$ and $CR_i$ from a clipped normal perturbation around $CR_{\mathrm{mean}}$
        \STATE Choose $\mathbf{x}_{p\mathrm{best}}$ from the best $\max(2,\lfloor pN_P \rfloor)$ members
        \STATE Choose $\mathbf{x}_{r_1}$ from the population/archive union and $\mathbf{x}_{r_2}$ from the same union with distinct indices
        \STATE Scale the attraction term with
        \[
            F_i^{(w)} =
            \begin{cases}
                0.7F_i, & NFE < 0.2NFE_{\max},\\
                0.8F_i, & 0.2NFE_{\max} \le NFE < 0.4NFE_{\max},\\
                F_i, & \text{otherwise}
            \end{cases}
        \]
        \STATE Form
        \[
            \mathbf{v}_i=\mathbf{x}_i+F_i^{(w)}(\mathbf{x}_{p\mathrm{best}}-\mathbf{x}_i)+F_i(\mathbf{x}_{r_1}-\mathbf{x}_{r_2})
        \]
        \STATE Build $\mathbf{u}_i$ with Equation~\eqref{eq:de_trial_assembly_si}, apply greedy replacement, update $A$, and record successful $F_i,CR_i$
    \ENDFOR
    \STATE Update $M_F,M_{CR}$ by improvement-weighted success history and reduce $N_P$ linearly from $30$ to $4$
\ENDWHILE
\RETURN $\mathbf{x}^{\star}_{\mathrm{evo}}$
\end{algorithmic}
\end{algorithm}

\paragraph{L-SHADE.}
L-SHADE is the canonical success-history adaptive differential evolution algorithm with linear population size reduction~\citep{tanabe2014lshade}. The benchmark implementation retains the standard ingredients of that family without the DTDE transform step or the jSO-specific early weighted schedule.

\begin{algorithm}[H]
\caption{L-SHADE Kernel under the Unified FDTR Benchmark Protocol}
\label{alg:lshade_si}
\begin{algorithmic}
\STATE \textbf{Input:} $J$, $\mathbf{lb}$, $\mathbf{ub}$, $NFE_{\max}=600$
\STATE Initialize population $P$, archive $A$, archive capacity $|A|_{\max}=N_P$, and memories $M_F,M_{CR}$
\WHILE{$NFE < NFE_{\max}$}
    \FOR{each target vector $\mathbf{x}_i$}
        \STATE Sample $F_i$ from a positive Cauchy perturbation around $M_F[r_i]$ and $CR_i$ from a clipped normal perturbation around $M_{CR}[r_i]$
        \STATE Choose $\mathbf{x}_{p\mathrm{best}}$ from the best $\max(2,\lfloor 0.1N_P \rfloor)$ members
        \STATE Choose $\mathbf{x}_{r_1}$ from the population/archive union and $\mathbf{x}_{r_2}$ from the same union with distinct indices
        \STATE Form
        \[
            \mathbf{v}_i=\mathbf{x}_i+F_i(\mathbf{x}_{p\mathrm{best}}-\mathbf{x}_i)+F_i(\mathbf{x}_{r_1}-\mathbf{x}_{r_2})
        \]
        \STATE Build $\mathbf{u}_i$ with Equation~\eqref{eq:de_trial_assembly_si}
        \STATE Apply greedy replacement, update the archive with the displaced parent, and record successful $F_i,CR_i$
    \ENDFOR
    \STATE Update $M_F,M_{CR}$ by improvement-weighted success history
    \STATE Reduce $N_P$ linearly from $30$ to $4$ and truncate the archive to the new $|A|_{\max}=N_P$
\ENDWHILE
\RETURN $\mathbf{x}^{\star}_{\mathrm{evo}}$
\end{algorithmic}
\end{algorithm}

\paragraph{CMA-ES.}
CMA-ES adapts a multivariate Gaussian search distribution through covariance updates, evolution paths, and step-size control~\citep{hansen2006cma}. Algorithm~\ref{alg:cmaes_si} uses the common FDTR objective and bounds. Here $n=\dim\mathbf x$, $\lambda$ is the number of sampled candidates, $\mu$ is the number selected for recombination, and $w_k$ are normalized recombination weights. The notation $\mathbf x_{k:\lambda}$ denotes the $k$-th ranked candidate. In $C=B\,\mathrm{diag}(D^2)B^\top$, $B$ contains covariance eigenvectors and $D$ contains square roots of the corresponding eigenvalues. The step-size coefficients $c_\sigma,d_\sigma$ control cumulation and damping, and $\chi_n=\mathbb E\|\mathbf z\|$ for $\mathbf z\sim\mathcal N(0,I_n)$.

\begin{algorithm}[H]
\caption{CMA-ES Kernel under the Unified FDTR Benchmark Protocol}
\label{alg:cmaes_si}
\begin{algorithmic}
\STATE \textbf{Input:} $J$, $\mathbf{lb}$, $\mathbf{ub}$, $NFE_{\max}=600$
\STATE Initialize mean $\mathbf{m}_0=(\mathbf{lb}+\mathbf{ub})/2$, covariance $C_0=I$, step size $\sigma_0=0.3\,\mathrm{mean}(\mathbf{ub}-\mathbf{lb})$, evolution paths $\mathbf{p}_c=\mathbf{p}_{\sigma}=0$, and recombination weights $w_1,\ldots,w_{\mu}$
\WHILE{$NFE < NFE_{\max}$}
    \STATE Factorize $C = B\,\mathrm{diag}(D^2)\,B^{\top}$
    \STATE Sample $\mathbf{z}_k \sim \mathcal{N}(0,I)$ and generate candidates
    \[
        \mathbf{y}_k = B\,\mathrm{diag}(D)\,\mathbf{z}_k,
        \qquad
        \mathbf{x}_k = \mathrm{clip}\!\left(\mathbf{m}+\sigma\mathbf{y}_k,\mathbf{lb},\mathbf{ub}\right)
    \]
    \STATE Evaluate $J(\mathbf{x}_k)$, sort the elite set, and update the mean
    \[
        \mathbf{m} \leftarrow \sum_{k=1}^{\mu} w_k \mathbf{x}_{k:\lambda},
        \qquad
        \mathbf{y}_w \leftarrow \sum_{k=1}^{\mu} w_k \mathbf{y}_{k:\lambda}
    \]
    \STATE Update the step-size path $\mathbf p_\sigma$ and covariance path $\mathbf p_c$
    \STATE Update $C$ using the previous covariance, the covariance-path outer product, and the weighted outer products of the selected steps
    \STATE Update the global step size
    \[
        \sigma \leftarrow \sigma \exp\!\left[\frac{c_{\sigma}}{d_{\sigma}}\left(\frac{\|\mathbf{p}_{\sigma}\|}{\chi_n}-1\right)\right]
    \]
\ENDWHILE
\RETURN Best sampled point $\mathbf{x}^{\star}_{\mathrm{evo}}$
\end{algorithmic}
\end{algorithm}

\paragraph{HPSO.}
The HPSO-style comparator uses particle-swarm velocity and position updates under the common FDTR objective and bounds~\citep{kennedy1995pso,xiao2025hybridpso}. Algorithm~\ref{alg:hpso_si} specifies the benchmark implementation.

\begin{algorithm}[H]
\caption{HPSO Kernel under the Unified FDTR Benchmark Protocol}
\label{alg:hpso_si}
\begin{algorithmic}
\STATE \textbf{Input:} $J$, $\mathbf{lb}$, $\mathbf{ub}$, $NFE_{\max}=600$
\STATE Initialize particle positions $\mathbf{x}_i$, velocities $\mathbf{v}_i$, personal bests $\mathbf{p}_i=\mathbf{x}_i$, and global best $\mathbf{g}$
\STATE Fix inertia and attraction constants $(w,c_1,c_2)=(0.7,1.5,1.5)$
\WHILE{$NFE < NFE_{\max}$}
    \FOR{each particle $i$}
        \STATE Draw random vectors $\mathbf{r}_{1,i},\mathbf{r}_{2,i}\sim\mathcal{U}(0,1)^n$
        \STATE Update velocity and position
        \[
            \mathbf{v}_i \leftarrow w\mathbf{v}_i + c_1\,\mathbf{r}_{1,i}\odot(\mathbf{p}_i-\mathbf{x}_i) + c_2\,\mathbf{r}_{2,i}\odot(\mathbf{g}-\mathbf{x}_i)
        \]
        \[
            \mathbf{x}_i \leftarrow \mathrm{clip}\!\left(\mathbf{x}_i+\mathbf{v}_i,\mathbf{lb},\mathbf{ub}\right)
        \]
        \STATE Evaluate $J(\mathbf{x}_i)$ and update $\mathbf{p}_i$ if the particle improved
    \ENDFOR
    \STATE Update the global best $\mathbf{g}$ from the current personal-best set
\ENDWHILE
\RETURN $\mathbf{x}^{\star}_{\mathrm{evo}}=\mathbf{g}$
\end{algorithmic}
\end{algorithm}

\paragraph{DE.}
Canonical DE provides a fixed-parameter differential-evolution baseline under the common FDTR objective and bounds~\citep{storn1997differential}. Algorithm~\ref{alg:de_si} uses classical mutation, binomial crossover, and greedy replacement.

\begin{algorithm}[H]
\caption{Classical DE Kernel under the Unified FDTR Benchmark Protocol}
\label{alg:de_si}
\begin{algorithmic}
\STATE \textbf{Input:} $J$, $\mathbf{lb}$, $\mathbf{ub}$, $NFE_{\max}=600$
\STATE Initialize a population $P=\{\mathbf{x}_i\}_{i=1}^{20}$ uniformly in $[\mathbf{lb},\mathbf{ub}]$
\STATE Fix $F=0.8$ and $CR=0.9$
\WHILE{$NFE < NFE_{\max}$}
    \FOR{each target vector $\mathbf{x}_i$}
        \STATE Choose three distinct population members $\mathbf{x}_a,\mathbf{x}_b,\mathbf{x}_c$ with $a,b,c\neq i$
        \STATE Form the mutant
        \[
            \mathbf{v}_i=\mathbf{x}_a+0.8(\mathbf{x}_b-\mathbf{x}_c)
        \]
        \STATE Build $\mathbf{u}_i$ from $\mathbf{x}_i$ and $\mathbf{v}_i$ using Equation~\eqref{eq:de_trial_assembly_si} with $CR_i=0.9$
        \STATE Accept $\mathbf{u}_i$ only if $J(\mathbf{u}_i)<J(\mathbf{x}_i)$
    \ENDFOR
\ENDWHILE
\RETURN $\mathbf{x}^{\star}_{\mathrm{evo}}=\arg\min_{\mathbf{x}\in P} J(\mathbf{x})$
\end{algorithmic}
\end{algorithm}

\subsection{Interpretation of Figure~\ref{fig:si_optimizer} and Default Choice}

Figure~\ref{fig:si_optimizer}A shows phase RMSE against the number of function evaluations, each corresponding to one FDTR forward solve. CMA-ES reaches $\mathrm{RMSE}\le0.21^\circ$ first, at approximately NFE 140.

Figure~\ref{fig:si_optimizer}B shows the mean evolutionary wall-clock time (left axis, solid bars) and terminal phase RMSE (right axis, hatched bars). DTDE has the shortest mean runtime, about \SI{127.20}{\second}, with mean terminal RMSE $0.2083^\circ$, matching CMA-ES ($0.2083^\circ$). DTDE is therefore selected as the default by runtime at comparable terminal accuracy. CMA-ES is preferred when the cost of each forward evaluation makes evaluation count the governing metric.

\clearpage
\section{Continuous-Profile Branch of PRISM: DeepONet for Gradient Thermal Conductivity k(z) Prediction}
\label{sec:deeponet_kz}

The DeepONet branch reconstructs depth-dependent thermal conductivity $k(z)$ from 20-point FDTR phase vectors for continuous profiles that require more than a low-order discrete-layer parameterization. The benchmark of Figure~\ref{fig:deeponet} uses $N_s=60{,}000$ synthetic phase--profile pairs: $10{,}000$ from each of six analytic families. Each pair contains $N_f=20$ phase features and $N_z=500$ conductivity labels. The $7{:}2{:}1$ split gives 42,000 training, 12,000 validation, and 6,000 held-out test pairs. The following subsections define data generation, architecture, training, and deployment.

\subsection{Problem Setup and Training-Data Generation}
\label{sec:deeponet_training_data}

Given FDTR phase measurements $\boldsymbol{\phi} = [\phi(f_1), \ldots, \phi(f_{20})]$, predict the thermal conductivity profile sampled over 500 depth points on $0 < z < 5~\mu\mathrm{m}$:
\begin{equation}
    k(z): [0, z_{\max}] \rightarrow \mathbb{R}^+
\end{equation}

This is an operator learning problem:
\begin{equation}
    \mathcal{G}: \boldsymbol{\phi} \mapsto k(z)
\end{equation}

\paragraph{Forward model for training-data generation.}
\hypertarget{par:deeponet_forward}{}

Synthetic phase spectra are generated with the graded OTh-STM path: optical-TMM volumetric heating drives a single thermal field in the Au transducer and the graded Si-based sample. The sample is represented by sublayers carrying local conductivities $k_j$ and continuous temperature and heat flux between adjacent cells. Because the \SI{97}{\nano\metre} Au transducer is optically thick ($h_{\mathrm{Au}} > 5\,d_{\mathrm{opt}}$), the depth-weighted OTh probe readout operates in the shallow-sensing limit of section~\ref{sec:volumetric_readout} and reduces analytically to the probe-averaged surface-temperature observable defined below.

\paragraph{Sublayer discretization.}
Each sampled profile $k(z)$, with $k\in[20,400]$~W\,m$^{-1}$K$^{-1}$ throughout and a lower bound of $3$~W\,m$^{-1}$K$^{-1}$ for the damage-band family of Equation~\eqref{eq:kz_damage}, is discretized into $N=500$ uniform sublayers of thickness $\Delta z=z_{\max}/N$ over $0<z<z_{\max}=5\;\mu\text{m}$. Cell $j=0,\ldots,N-1$ carries $k_j=k(z_j)$ at $z_j=(j+\tfrac12)\Delta z$. All cells have volumetric heat capacity $C=1.632\times10^6$~J\,m$^{-3}$K$^{-1}$, the 296.15~K silicon value in Table~\ref{tab:archive_si_thermal_props}. Continuous temperature and heat flux impose $G_{j,j+1}\to\infty$, so adjacent cells connect directly. For $z>z_{\max}$, a semi-infinite continuation has conductivity $k(z_{\max})$ and the same heat capacity.

\paragraph{Per-sublayer transfer matrix.}
For this subsection define the two-component state $\mathbf{v}=[\widetilde{T},F]^\mathsf{T}$, where $F=k\,\partial\widetilde{T}/\partial z=-\hat q$ and $z$ increases downward. This local choice differs from the heat-flux state $[\hat T,\hat q]^\mathsf{T}$ used in Equation~\eqref{eq:stm_matrix} only by the sign of the second component. The top-to-bottom transfer matrix for sublayer $j$ is
\begin{equation}
    \mathbf{M}_j =
    \begin{pmatrix}
        \cosh(q_j\,\Delta z) & \dfrac{\sinh(q_j\,\Delta z)}{k_j\,q_j} \\[6pt]
        k_j\,q_j\,\sinh(q_j\,\Delta z) & \cosh(q_j\,\Delta z)
    \end{pmatrix},
    \label{eq:sublayer_matrix}
\end{equation}
where the local thermal decay constant is
\begin{equation}
    q_j = \sqrt{\mu_\rho^2 + \frac{i\,\omega\,C}{k_j}}.
    \label{eq:sublayer_q}
\end{equation}
Equation~\eqref{eq:sublayer_matrix} is the similarity transform of Equation~\eqref{eq:stm_matrix} under $F=-\hat q$. When $k_j$ is uniform, the cascade therefore reduces exactly to the homogeneous STM layer of total thickness $z_{\max}$.

\paragraph{Cascaded sample matrix.}
Because each matrix maps the top face of a cell to its bottom face and $j=0$ denotes the topmost cell, the ordered sample cascade is
\begin{equation}
    \mathbf{M}_{\text{sample}}
    = \mathbf{M}_{N-1}\mathbf{M}_{N-2}\cdots\mathbf{M}_{0},
    \label{eq:sample_cascade}
\end{equation}

\paragraph{Au transducer and interface.}
Above the graded sample sits a $h_{\mathrm{Au}}=97$-nm Au transducer treated within the same two-component STM state, with $k_{\mathrm{Au}}=150$~W\,m$^{-1}$K$^{-1}$ and $C_{\mathrm{Au}}=2.48\times10^{6}$~J\,m$^{-3}$K$^{-1}$ from Table~\ref{tab:archive_si_thermal_props}. Its homogeneous matrix $\mathbf{M}_{\mathrm{Au}}$ has the form of Equation~\eqref{eq:sublayer_matrix}. For the unit-incident-irradiance response, the source excludes the lateral Gaussian: $S_0$ is a per-irradiance coefficient of $Q_{\mathrm{Au}}$ in $\mathrm{m^{-1}}$. For one pump-source exponential component (index $s$, with $\alpha_s^{\mathrm{pump}}$ its exponent) $S(z)=S_0e^{-\alpha_s^{\mathrm{pump}} z}$, the particular temperature is $T_p(z)=v_p e^{-\alpha_s^{\mathrm{pump}} z}$, where
\begin{equation}
    q_{\mathrm{Au}}=\sqrt{4\pi^2(u^2+v^2)+\frac{i\omega C_{\mathrm{Au}}}{k_{\mathrm{Au}}}},
    \qquad
    v_p=\frac{S_0}{k_{\mathrm{Au}}(q_{\mathrm{Au}}^2-(\alpha_s^{\mathrm{pump}})^2)}.
    \label{eq:gradient_au_particular}
\end{equation}
The corresponding top-to-bottom affine relation and source vector are
\begin{align}
    \mathbf{v}(h_{\mathrm{Au}})
    &=\mathbf{M}_{\mathrm{Au}}\mathbf{v}(0)+\mathbf{b}_{\mathrm{Au}},
    \label{eq:gradient_au_affine}\\
    \mathbf{b}_{\mathrm{Au}}
    &=v_p\begin{pmatrix}
    e^{-\alpha_s^{\mathrm{pump}} h_{\mathrm{Au}}}-\cosh(q_{\mathrm{Au}}h_{\mathrm{Au}})
    +(\alpha_s^{\mathrm{pump}}/q_{\mathrm{Au}})\sinh(q_{\mathrm{Au}}h_{\mathrm{Au}})\\[3pt]
    k_{\mathrm{Au}}\!\left[\alpha_s^{\mathrm{pump}}\cosh(q_{\mathrm{Au}}h_{\mathrm{Au}})
    -q_{\mathrm{Au}}\sinh(q_{\mathrm{Au}}h_{\mathrm{Au}})
    -\alpha_s^{\mathrm{pump}} e^{-\alpha_s^{\mathrm{pump}} h_{\mathrm{Au}}}\right]
    \end{pmatrix}.
    \label{eq:gradient_au_source}
\end{align}
The optical TMM expresses $Q_{\mathrm{Au}}(z)$ as a finite sum of forward, backward, and interference exponential components; Equation~\eqref{eq:gradient_au_source} is applied to each component and the resulting vectors are summed. For the dominant forward component, $\alpha_s^{\mathrm{pump}}=\alpha_{\mathrm{opt},\mathrm{Au}}^{\mathrm{pump}}=4\pi\kappa/\lambda^{\mathrm{pump}}$ with $n=1.4002$, $\kappa=1.9238$, and $\lambda^{\mathrm{pump}}=445$~nm~\cite{Johnson1972Optical}, giving $(\alpha_s^{\mathrm{pump}})^{-1}\approx18.4$~nm. Because $\operatorname{Im}(q_{\mathrm{Au}}^2)=\omega C_{\mathrm{Au}}/k_{\mathrm{Au}}>0$ and $\alpha_s^{\mathrm{pump}}$ is real, the denominator in Equation~\eqref{eq:gradient_au_particular} does not vanish for $\omega>0$.

The Au/sample interface has $G_{\mathrm{Au/sample}}=50$~MW\,m$^{-2}$K$^{-1}$ (Au/Si entry of Table~\ref{tab:archive_si_interface_G}). In the present $F=-\hat q$ convention its top-to-bottom matrix is
\begin{equation}
    \mathbf{M}_{G}=\begin{pmatrix}1&G_{\mathrm{Au/sample}}^{-1}\\0&1\end{pmatrix}.
    \label{eq:gradient_interface}
\end{equation}
With the volumetric source already contained in $\mathbf{b}_{\mathrm{Au}}$, the free top surface obeys $F(0)=0$. Writing $\widetilde{T}_0=\widetilde T(0)$ for the surface temperature response per unit incident irradiance, with units $\mathrm{K\,m^2\,W^{-1}}$, the complete affine cascade is therefore
\begin{equation}
    \mathbf{v}(z_{\max})
    =\mathbf{M}_{\mathrm{sample}}\mathbf{M}_{G}
    \left[\mathbf{M}_{\mathrm{Au}}\begin{pmatrix}\widetilde{T}_0\\0\end{pmatrix}
    +\mathbf{b}_{\mathrm{Au}}\right]
    \equiv \mathbf{A}\widetilde{T}_0+\mathbf{B}.
    \label{eq:gradient_cascade}
\end{equation}

Below $z_{\max}$, the graded medium continues as a semi-infinite material with $k_s=k(z_{\max})$ and the same heat capacity. The decaying boundary condition at $z_{\max}$ imposes
\begin{equation}
    F(z_{\max})=-Y_s\widetilde T(z_{\max}),\qquad
    Y_s=k_s q_s,\qquad
    q_s=\sqrt{\mu_\rho^2+\frac{i\omega C}{k_s}}.
    \label{eq:gradient_substrate_closure}
\end{equation}
If $\mathbf{A}=(A_T,A_F)^\mathsf{T}$ and $\mathbf{B}=(B_T,B_F)^\mathsf{T}$, the surface temperature kernel of each in-plane Fourier mode is
\begin{equation}
    \widetilde{T}_0(\mu_\rho,\omega)
    =-\frac{B_F+Y_s B_T}{A_F+Y_s A_T}.
    \label{eq:gradient_surface_mode}
\end{equation}

\paragraph{Phase extraction.}
For the coaxial synthetic benchmark, spatial synthesis and normalized probe averaging retain $w_x=\SI{2.19}{\micro\metre}$ and $w_y=\SI{1.84}{\micro\metre}$ separately. The graded material kernel is isotropic in plane, so $\widetilde{T}_0$ may be parameterized by $\mu_\rho(u,v)=2\pi\sqrt{u^2+v^2}$; the elliptical Gaussian weight is nevertheless two-dimensional:
\begin{equation}
\begin{aligned}
    \overline T(\omega_m)&=P_{\mathrm{pump}}\iint_{\mathbb R^2}\widetilde{T}_0(\mu_\rho(u,v),\omega_m)
       \exp[-\tfrac{\pi^2}{2}(w_x^2u^2+w_y^2v^2)]\,\mathrm{d}u\,\mathrm{d}v,\\
    \phi(\omega_m)&=\arg\{\overline T(\omega_m)\}.
\end{aligned}
    \label{eq:gradient_phase}
\end{equation}
Here $P_{\mathrm{pump}}>0$ is the incident pump-power amplitude. The modal surface response $\widetilde{T}_0$ excludes this power and the lateral Gaussian, which enter the synthesis once. The two widths follow Equation~\eqref{eq:fourier_combined_kernel}; circular Hankel synthesis requires $w_x=w_y$.

Equation~\eqref{eq:gradient_phase} is evaluated at 20 logarithmically spaced frequencies from $10\;\text{kHz}$ to $50\;\text{MHz}$ to produce the branch input $\boldsymbol{\phi}=[\phi(f_1),\ldots,\phi(f_{20})]$.

\paragraph{Summary of the gradient-forward distinction.}
The graded OTh-STM cascade produces the input phases from the two-dimensional Fourier-integrated surface response of Equation~\eqref{eq:gradient_phase}.

\paragraph{Phase-consistency check.}
For a fixed predicted profile, define
\[
\chi^2_{\mathrm{red}}
=\frac{1}{N_f}\sum_{i=1}^{N_f}
\left[\frac{\phi_{\mathrm{pred}}(f_i)-\phi_{\mathrm{in}}(f_i)}
{\sigma_\phi}\right]^2,
\qquad \sigma_\phi=0.1^\circ,
\]
where $\phi_{\mathrm{in}}$ is the input phase spectrum and $\phi_{\mathrm{pred}}$ is its forward re-simulation. Scores above $3.0$ trigger return to discrete-layer inversion.

\subsection{DeepONet Architecture}

The DeepONet consists of two sub-networks:

\paragraph{Branch network (phase encoder).}
\begin{equation}
    \mathbf{b} = f_{\text{branch}}(\boldsymbol{\phi}) = \sigma(W_3 \sigma(W_2 \sigma(W_1 \boldsymbol{\phi} + b_1) + b_2) + b_3) \in \mathbb{R}^p
\end{equation}

where $\sigma$ is the ReLU activation and $p=256$ is the latent dimension. The matrices $W_\ell$ and vectors $b_\ell$, $\ell=1,2,3$, are the trainable layer weights and biases. The branch widths are $20\rightarrow256\rightarrow256\rightarrow256$.

\paragraph{Trunk network (depth encoder).}
\begin{equation}
    \mathbf{t}(z) = f_{\text{trunk}}(z) = \sigma(W_3' \sigma(W_2' \sigma(W_1' z + b_1') + b_2') + b_3') \in \mathbb{R}^p
\end{equation}

The primed weights and biases belong to the trunk network, whose widths are $1\rightarrow256\rightarrow256\rightarrow256$. Its output is a depth-dependent latent basis at query coordinate $z$.

\paragraph{Output reconstruction.}
\begin{equation}
    \hat{k}(z) = \sum_{i=1}^{p} b_i\,t_i(z) + b_0 = \mathbf{b}^T\mathbf{t}(z) + b_0
    \label{eq:deeponet_output}
\end{equation}
where $b_0$ is the trainable scalar output bias. The normalized RMSE in Figure~\ref{fig:deeponet}I is $\mathrm{RMSE}/s_k$, where $s_k$ is the standard deviation of conductivity values in the training set; the same scale is used for training and validation.

\begin{algorithm}[H]
\caption{DeepONet Training for k(z) Prediction}
\label{alg:deeponet_train}
\begin{algorithmic}
\REQUIRE Training data $\{(\boldsymbol{\phi}_n,k_n(z))\}_{n=1}^{N_{\mathrm{train}}}$, depth grid $\{z_j\}_{j=1}^{N_z}$
\ENSURE Trained model parameters $\Theta$
\STATE Initialize branch net $f_{\text{branch}}$, trunk net $f_{\text{trunk}}$
\STATE $\Theta \leftarrow \text{RandomInit}()$
\FOR{epoch $= 1$ to $E$}
    \FOR{each batch $\mathcal{B}$}
        \STATE $\mathcal{L} \leftarrow 0$
        \FOR{each $(\boldsymbol{\phi}_n, k_n) \in \mathcal{B}$}
            \STATE $\mathbf{b}_n \leftarrow f_{\text{branch}}(\boldsymbol{\phi}_n)$
            \FOR{each depth $z_j$}
                \STATE $\mathbf{t}_j \leftarrow f_{\text{trunk}}(z_j)$
                \STATE $\hat{k}_{n,j}\leftarrow\mathbf b_n^T\mathbf t_j+b_0$
                \STATE $\mathcal{L} \leftarrow \mathcal{L} + (\hat{k}_{n,j} - k_n(z_j))^2$
            \ENDFOR
        \ENDFOR
        \STATE $\mathcal{L} \leftarrow \mathcal{L} / (|\mathcal{B}| \cdot N_z)$
        \STATE Update $\Theta$ with Adam using $\nabla_\Theta\mathcal L$ and learning rate $\alpha_{\mathrm{lr}}$
    \ENDFOR
\ENDFOR
\RETURN $\Theta$
\end{algorithmic}
\end{algorithm}

\subsection{k(z) Profile Classes}

The training library contains six analytic profile families spanning increasing and decreasing gradients: five globally smooth single-expression gradients and one piecewise damage-band shape \cite{xiang2023machine, huynh2023depth}. The labels in Table~\ref{tab:kz_profiles} describe the motivating profile shapes. Coefficients of the five smooth families are sampled subject to $20\le k(z)\le400$~W\,m$^{-1}$K$^{-1}$ throughout $0<z<z_{\max}$; the damage-band family stays within the same upper bound but is sampled to a lower floor, $k_{\min}\ge3$~W\,m$^{-1}$K$^{-1}$, so that a strongly suppressed buried layer is representable. Specimens containing true internal interfaces, delaminations, or finite conductivity jumps remain outside this training distribution and are not interpreted through DeepONet alone; failure of the forward-residual or plausibility gate returns the case to a discrete PRISM branch with explicit interfaces.

\begin{table}[H]
\centering
\caption{k(z) Profile Types for Training Data Generation}
\label{tab:kz_profiles}
\begin{tabularx}{\textwidth}{@{} l l >{\raggedright\arraybackslash}X @{}}
\toprule
\textbf{Profile} & \textbf{Formula} & \textbf{Motivating profile shape} \\
\midrule
Exponential & $k(z) = A + B e^{-Cz}$ & Growth-related monotonic gradient \\
Gaussian & $k(z) = k_s+(k_b-k_s)\left[1-\exp\left(-z^2/(2\sigma^2)\right)\right]$ & Near-surface damage-recovery shape \\
Linear & $k(z) = A + Bz$ & Compositional gradient \\
Damage band & Equation~\eqref{eq:kz_damage} & Buried irradiation-damage layer \\
Logarithmic & $k(z) = A + B\ln(z + \epsilon)$ & Grain size gradient \\
Hyperbolic & $k(z) = A + B/(z + C)$ & Interface scattering \\
\bottomrule
\end{tabularx}
\end{table}

In Table~\ref{tab:kz_profiles}, $A,B,C$ are family-specific profile coefficients; the formulas use the numerical depth measured in \si{\micro\metre}, so the logarithm acts on a dimensionless numerical argument. In the Gaussian family, $k_s,k_b$ are the surface and bulk conductivity levels and $\sigma$ is the depth-width parameter. The logarithmic family uses a positive depth offset $\epsilon$. These coefficients are defined separately for each profile family.

The damage-band family is the piecewise member of the library:
\begin{equation}
k(z)=
\begin{cases}
\dfrac{k_L}{1+\left(k_L/k_{\min}-1\right)\exp\!\left[-\dfrac{(z-z_0)^2}{2\sigma_L^2}\right]}, & z<z_0,\\[12pt]
\dfrac{k_B}{1+\left(k_B/k_{\min}-1\right)\exp\!\left[-\dfrac{(z-z_0)^2}{2\sigma_R^2}\right]}, & z\ge z_0,
\end{cases}
\label{eq:kz_damage}
\end{equation}

\noindent with $z$, $z_0$, $\sigma_L$ and $\sigma_R$ in \si{\micro\metre} and $k_{\min}$, $k_L$, $k_B$ in \si{\watt\per\metre\per\kelvin}. Here $k_{\min}$ is the conductivity at the band centre, $k_L$ the near-surface shoulder plateau, $k_B$ the undamaged bulk level, $z_0$ the band depth, and $\sigma_L,\sigma_R$ the flank widths. The branches match in value and first derivative at $z=z_0$, while their different amplitudes and widths permit unequal curvatures. Asymmetric widths permit a broad defect-bearing shoulder and a sharper recovery toward the bulk. The six parameters are drawn by Latin-hypercube sampling within $k_{\min}\in[3,15]$, $k_L/k_{\min}\in[1.2,3]$, $k_B\in[150,300]$~W\,m$^{-1}$K$^{-1}$, $z_0\in[0.8,4.0]$, $\sigma_L\in[0.15,1.5]$ and $\sigma_R\in[0.06,0.2]$~\si{\micro\metre}. The instance in Figure~\ref{fig:deeponet}F has $k_{\min}=11.3$, $k_L=32.7$ and $k_B=223.1$~W\,m$^{-1}$K$^{-1}$ with $z_0=3.7$, $\sigma_L=1.5$ and $\sigma_R=0.2$~\si{\micro\metre}.

\paragraph{Prediction intervals.} The shaded bands in Figure~\ref{fig:deeponet}C--H are nominal pointwise 95\% prediction intervals calibrated separately for the known synthetic profile families. The $60{,}000$-pair library uses the same $7{:}2{:}1$ split throughout: 42,000 training, 12,000 validation, and 6,000 test pairs. Validation profiles supply the family-wise calibration scores, and test profiles are used for coverage assessment. At each depth, the calibration score is $s(z)=\lvert\ln k(z)-\ln\widehat k(z)\rvert$. The interval is $\widehat k(z)\exp[\pm q(z)]$, where $n$ is the number of calibration profiles for that family and $q(z)$ is the $\lceil0.95(n+1)\rceil$-th smallest score at that depth. The reported pointwise coverage on held-out profiles is 94.4--95.3\%. The intervals widen toward the surface, where thin near-surface regions are resolved by the highest modulation frequencies.

\clearpage
\section{Kernel Ridge Regression as a Comparator Within the Continuous-Profile PRISM Route}
\label{sec:krr}

Kernel ridge regression (KRR) provides a non-neural baseline for the same phase-to-profile task and is used for the accuracy and timing comparison with DeepONet. Both models use the same 42,000 training, 12,000 validation, and 6,000 test pairs from the $60{,}000$-pair library. KRR produces deterministic point predictions.\paragraph{Evaluation metrics and timing.} RMSPE is the root-mean-square of the relative conductivity errors $(\widehat{k}-k)/k$ over the evaluated profile values, expressed as a percentage. Figure~\ref{fig:deeponet}K reports 3.3\% for KRR and 2.7\% for DeepONet. The corresponding inference times are \SI{1510}{\micro\second} and \SI{1370}{\micro\second} per batch of 100 spectra. The normalized RMSE in Figure~\ref{fig:deeponet}I instead scales conductivity RMSE by the training-set standard deviation, as defined in section~\ref{sec:deeponet_kz}; Figure~\ref{fig:deeponet}J reports the parity coefficient of determination $R^2=0.9986$. The inset phase re-simulation RMSE is a separate spectral-consistency metric in degrees.

\subsection{Kernel, Solution, and Prediction}

The Laplacian kernel is defined as:
\begin{equation}
    K(\mathbf{x}_i, \mathbf{x}_j) = \exp(-\beta \|\mathbf{x}_i - \mathbf{x}_j\|_1)
\end{equation}

where $\|\cdot\|_1$ is the L1 (Manhattan) distance and $\beta$ is the inverse distance scale of the kernel.

\paragraph{KRR solution.}

Given training data $\{(\boldsymbol{\phi}_n, \mathbf{k}_n)\}_{n=1}^N$, where $\mathbf{k}_n=[k_n(z_j)]_{j=1}^{N_z}$ is the sampled conductivity profile:
\begin{equation}
    \boldsymbol{\alpha} = (\mathbf{K} + \lambda \mathbf{I})^{-1} \mathbf{Y}
\end{equation}

where $\mathbf K_{ij}=K(\boldsymbol\phi_i,\boldsymbol\phi_j)$ is the $N\times N$ kernel matrix, $\mathbf I$ is the identity, $\lambda$ is the ridge regularization parameter, and $\mathbf Y\in\mathbb R^{N\times N_z}$ contains sampled conductivity profiles as rows. The coefficient matrix $\boldsymbol\alpha\in\mathbb R^{N\times N_z}$ supplies one prediction coefficient per training sample and depth.

\paragraph{Prediction.}

For new phase data $\boldsymbol{\phi}^*$:
\begin{equation}
    \hat{k}(z_j) = \sum_{n=1}^N K(\boldsymbol{\phi}^*, \boldsymbol{\phi}_n)\,\alpha_{n,j},\qquad j=1,\ldots,N_z
\end{equation}

\clearpage
\section{Agentic Execution Layer and Branch Routing of PRISM}
\label{sec:agent_system_overview}

This section specifies task parsing, branch routing, solver dispatch, state management, verification, recovery, and auditability in the PRISM execution layer. It connects the optical, thermal, screening, optimization, and continuous-profile modules defined in sections~\ref{sec:tmm}, \ref{sec:non_fourier_si}, \ref{sec:sensitivity_svd}, \ref{sec:optimizer_benchmark}, and \ref{sec:deeponet_kz}.

\subsection{System Overview and Design Philosophy}

The execution layer has four components: (1) a \textbf{task-parsing layer} converts requests into a stack schema; (2) a \textbf{branch-routing layer} selects PRISM modules, retrieval actions, and solver-preparation tasks; (3) a \textbf{deterministic-solver layer} invokes numerical tools; and (4) a \textbf{state, verification, and recovery layer} retains intermediate results, checks, and analysis history. The router assembles a bounded path from optical source construction, layerwise STM or TTM propagation, interface rules, substrate closure, optimizer choice, and validation hooks.

The complete information flow through the execution layer is
\begin{equation}
\text{Task Description} \xrightarrow{\text{Parse}} \text{Schema} \xrightarrow{\text{Route}} \text{Tool Calls / Solver Path} \xrightarrow{\text{Execute}} \text{Results} \xrightarrow{\text{Audit}} \text{Report}.
\end{equation}
The design principles are \textit{the PRISM branch hierarchy}, which constrains the physical modules available to routing; \textit{deterministic numerical tools}, invoked through structured calls; \textit{stateful context}, retaining data quality, Jacobian, SVD, verification, and fit history; \textit{bounded solver preparation}, governed by explicit verification and recovery rules; and \textit{auditable recovery}, recording failures and fallback actions. Layerwise composition can retain STM in one layer, use metallic TTM in another, and replace a substrate closure within the validated module library.

{\RaggedRight
\paragraph{Implementation and knowledge base.} The execution graph is implemented in LangGraph~\cite{langgraph2024}, and the PRISM solvers and analysis tools are exposed to the language model, DeepSeek V4 Flash (model \texttt{deepseek-v4-flash}, accessed through the DeepSeek API with its default settings), as Model Context Protocol (MCP) tools; the tool names in Figure~\ref{fig:si_chat_ui} (\texttt{mcp\_\_fdtr\_\_\ldots}) are these MCP endpoints. The wiki-based knowledge graph consulted by the routing and recovery policies is described in section~\ref{sec:si_knowledge_base}, and its use in model revision in section~\ref{sec:si_kg_revision}. The layer operates in two modes. In routine operation it routes among and assembles existing, validated branches within the bounds described below; a new branch is constructed only in a user-supervised extension mode (Figure~\ref{fig:si_deeponet_log}) and enters routine use after its validation gates are passed.
\par}

\subsection{Wiki-Based Knowledge Graph}
\label{sec:si_knowledge_base}

The routing and recovery knowledge graph is stored in an Obsidian vault of Markdown files (Figure~\ref{fig:si_kg_structure}). The agent compiles sources into structured pages linked to one another and to their source records, preserving cited knowledge between analyses. Researchers can read the pages and graph, and the agent can parse the same content.

\paragraph{Three layers.} The \emph{source layer} holds the literature (articles with their BibTeX records and DOIs), the PRISM solver code, and the records of past analyses (measured spectra, fit logs, and audit trails). The agent reads it but never edits it, so every statement in the wiki can be checked against an unchanged original. The \emph{wiki layer} holds the pages that the agent writes and maintains. Each page carries YAML front matter that declares its type, its typed links to other pages, and its sources, followed by a body in prose and equations. Six page types are used: a \emph{mechanism} page describes a physical process; a \emph{model-form} page gives the constitutive law or boundary condition that represents it; a \emph{regime} page states when the model form is required, preferably in terms of dimensionless ratios that can be computed from the stack schema, such as $d_{\mathrm{trans}}/d_{\mathrm{opt}}$, $d_{\mathrm{trans}}/L_{\mathrm{ep}}$, and $\lambda_{\mathrm{ph}}/\min(w_r,\delta_p)$; a \emph{parameter} page collects material data together with the conditions under which they were measured; a \emph{code-module} page describes the solver function that implements a model form and its interface; and a \emph{case} page records one analysis. The typed links connect them: a mechanism is modeled by a model form; a model form is valid in a regime, requires parameters, and is implemented by a code module; and a case records the mechanism it was diagnosed as and the solver path it used. The \emph{schema layer} is written by the researchers. It defines the page types, the link types, and the front-matter fields; it sets the citation rule that every factual statement on a page cites a source in the source layer; it specifies the procedures for the operations below; and it maintains an index of all pages and a chronological log of every change.

\paragraph{Operations.} The agent maintains and uses the wiki through four operations defined in the schema. \emph{Ingest}: when a source is added, the agent writes a summary page for it, creates or updates the pages it bears on, adds the corresponding links, and records the change in the index and the log; a value that conflicts with an existing page is not overwritten but recorded next to it with its own source and flagged. \emph{Query}: the descriptors of a misfit (section~\ref{sec:si_kg_revision}) are matched against the regime pages, by numerical comparison where the validity condition is a computable ratio; the agent then follows the typed links, over a bounded number of steps, to the model forms, their parameters and code modules, and the past cases in which the same regime was diagnosed, and returns this subgraph together with its citations. \emph{Write-back}: every analysis is filed as a case page that records the diagnosis, the proposed revision, the re-fit, and the researcher's decision, so that a later diagnosis can retrieve it as a precedent. \emph{Lint}: a periodic check reports statements without a source, contradictory values across pages, orphan pages, links to missing pages, and code-module pages whose implementation has changed since they were last verified.

\paragraph{Provenance and maintenance.} Each knowledge-graph page links to its source, and typed links and computable validity conditions support repeatable queries. The researchers define the schema, the agent maintains the pages, and numerical solvers produce the fitted values. New literature and case records extend the graph used by the model-revision loop of Figure~\ref{fig:arch}F.

\subsection{KG-Guided Model Revision and Human Review}
\label{sec:si_kg_revision}

\paragraph{Model-revision loop.} The loop of Figure~\ref{fig:arch}F is entered from the recovery step of Algorithm~\ref{alg:error_recovery}, when a converged fit leaves a structured residual or a fitted property that changes between measurement configurations. (1)~\emph{KG-guided misfit diagnosis} expresses the failure as descriptors: which quantity is inconsistent, how it varies with the measurement variables (spot radius, frequency window, beam offset, and transducer thickness), and the values of the gating ratios for the stack. These descriptors form the query to the knowledge graph. (2)~\emph{Model-form screening} ranks the model forms in the returned subgraph by how well their regime pages match the descriptors and by the precedents in earlier case pages, and discards those that do not apply to the layer concerned, for instance a two-temperature form for a non-metallic layer. (3)~\emph{Descriptor-based term synthesis} turns the leading model form into a concrete change of the solver path: the block that is replaced (optical source, carrier description, conduction closure, or interface condition), the constitutive term that enters, the parameters of that term that are taken from parameter pages, and the fitted parameter that it replaces, so that the number of free parameters stays the same. If the term is already implemented in the PRISM block library, the agent assembles it directly; if it is not, the agent implements it in the user-supervised extension mode from its model-form page and checks the implementation against the conservation and limiting-case tests before use, as for the DeepONet branch (Figure~\ref{fig:si_deeponet_log}). (4)~\emph{Re-fit and cross-case validation} refits the revised branch across the available configurations to assess fitted-property consistency and target identifiability.

\paragraph{Worked example.} For Au/Ti/Si (section~\ref{sec:nonlocal_result} of the main text), diagnosis expressed the failure of the diffusive hierarchy as follows: up to OTh-TTM, the fitted $k_{\mathrm{Si}}$ increases with spot radius in all twelve model--window pairs and lies below the single-crystal reference band, and the Knudsen number at the small spot and at megahertz frequencies exceeds the quasi-ballistic threshold. The matching regime page is that of quasi-ballistic transport in crystalline Si; its linked model form is the nonlocal Guyer--Krumhansl closure with the interface jump condition of Ref.~\cite{beardo2020phonon}, and the linked parameter pages supply $\tau$, $\ell$, $\alpha_{\mathrm{GK}}$, $\gamma_0$, and $\gamma_{0\Gamma}$ at the measurement temperature. Term synthesis replaced only the substrate admittance by $Y_{\mathrm{NF}}$ [Equation~\eqref{eq:nf_replacement_map}], kept the optical source, the metallic TTM layers, and the upper-interface recursion, and replaced the fitted $G_{\mathrm{Ti/Si}}$ by $\bNF$. The re-fit returned $k_{\mathrm{Si}}$ between 140.54 and \SI{141.37}{\watt\per\metre\per\kelvin} at both spot sizes and in all three windows. The same loop, triggered by the disagreement between the offset and coaxial geometries, escalated the Au/Si case.

\paragraph{Human review.} The agent returns the revised solver path, its source pages and literature, the identifiability evidence, and the re-fit across configurations. The researcher reviews this proposal and records the adoption decision as a case page. The analyses reported here were reviewed by the authors.

\subsection{Task Parsing and Branch Selection}

The branch-routing layer bridges user language and the deterministic PRISM stack through the following prompt-build / tool-call / synthesis loop. In this context, routing includes selecting a PRISM branch, constructing the corresponding solver path, and deciding whether auxiliary retrieval or verification steps are required before execution.
\begin{algorithm}[H]
\caption{PRISM Branch-Routing Pipeline}
\label{alg:llm_router}
\begin{algorithmic}
\REQUIRE User message $m$, context $\mathcal{H}$, tool registry $\mathcal{K}$
\ENSURE Response $r$, updated context $\mathcal{H}'$
\STATE $\mathcal{O} \leftarrow \text{BuildSystemPrompt}(\mathcal{K}, \mathcal{H})$
\STATE $\text{history} \leftarrow \text{GetConversationHistory}(\mathcal{H})$
\STATE $\text{prompt} \leftarrow [\mathcal{O}, \text{history}, m]$
\STATE $\text{response} \leftarrow \text{LLM.Generate}(\text{prompt})$
\STATE $\text{calls} \leftarrow \text{ParseToolCalls}(\text{response})$
\IF{$\text{calls} \neq \emptyset$}
    \FOR{each $\text{call} \in \text{calls}$}
        \STATE $\text{tool} \leftarrow \mathcal{K}.\text{Get}(\text{call.name})$
        \STATE $\text{result} \leftarrow \text{tool.Execute}(\text{call.args})$
        \STATE $\mathcal{H} \leftarrow \text{UpdateContext}(\mathcal{H}, \text{call.name}, \text{result})$
    \ENDFOR
    \STATE $r \leftarrow \text{LLM.Synthesize}(\text{results}, \mathcal{H})$
\ELSE
    \STATE $r \leftarrow \text{response}$
\ENDIF
\RETURN $r$, $\mathcal{H}$
\end{algorithmic}
\end{algorithm}

The system prompt comprises the role description, tool schemas, current analysis-state summary, and output-format constraints. The router selects compatible optical-source, layerwise STM/TTM, interface, and substrate blocks. For OTh-TTM(NF), substrate-closure revision retains the optical source and upper-layer recursion.

\subsection{State Management and Execution Graph}

The execution graph is implemented as a LangGraph state machine. The persisted state records both user-facing conversation context and solver-facing numerical context, field by field in Table~\ref{tab:fdtr_state}:
\begin{table}[ht]
\centering
\caption{Fields of the persistent \texttt{FDTRState} typed dictionary.}
\label{tab:fdtr_state}
\begin{tabularx}{\textwidth}{@{}l>{\raggedright\arraybackslash}X@{}}
\toprule
\textbf{Field} & \textbf{Description} \\
\midrule
\texttt{messages} & Conversation history accumulated by LangGraph reducers. \\[2pt]
\texttt{current\_phase} & Workflow phase: \texttt{IDLE}, \texttt{DATA\_LOADING}, \texttt{MODEL\_SELECTION}, \texttt{SENSITIVITY},
\newline \texttt{IDENTIFIABILITY}, \texttt{FITTING}, \texttt{OPTIMIZATION}, \texttt{COMPLETED}, or \texttt{ERROR}. \\[2pt]
\texttt{layers} & Layer-property dictionaries forwarded to the public FDTR forward solvers. \\[2pt]
\texttt{interfaces} & Interface conductance values or equivalent boundary parameters. \\[2pt]
\texttt{frequencies} & Experimental modulation frequencies. \\[2pt]
\texttt{phase\_exp} & Experimental phase spectrum. \\[2pt]
\texttt{model\_type} & Selected public solver branch, e.g.\ \texttt{"auto"}, \texttt{"stm"}, or \texttt{"ttm"}, together with the corresponding execution path. \\[2pt]
\texttt{solver\_path} & Ordered record of the physics blocks activated for the current run, e.g.\ optical source, layerwise thermal formalism, interface policy, substrate closure, optimizer, and verification hooks. \\[2pt]
\texttt{branch\_history} & Audit trail of branch-retention, branch-rejection, and bounded escalation decisions accumulated across the workflow. \\[2pt]
\texttt{fit\_targets} & Candidate fit parameters after quality and identifiability screening. \\[2pt]
\texttt{analysis\_results} & Dictionary carrying diagnosis, sensitivity, SVD, verification, and fitting outputs. \\[2pt]
\texttt{validation\_artifacts} & Residual diagnostics, plausibility checks, Fourier-limit tests, and other branch-conditioned acceptance evidence. \\[2pt]
\texttt{interface\_policy} & Effective interface treatment used by the selected path, including any branch-conditioned bypasses needed to avoid double counting. \\[2pt]
\texttt{error} & Error payload propagated to recovery handlers. \\
\bottomrule
\end{tabularx}
\end{table}

The LangGraph controller alternates between reasoning and acting through a ReAct-style loop: a reasoning node invokes the language model, and the graph either terminates directly when no tool calls are requested or dispatches to a tools node, whose results are fed back into the reasoning node for the next iteration.

Within this loop, explicit task nodes perform the analysis steps summarized in Table~\ref{tab:workflow_nodes}.

\begin{table}[ht]
\centering
\caption{Execution-graph nodes and their functions. The \texttt{optimization} and \texttt{error\_handler} nodes activate for refinement or recovery; Figure~\ref{fig:complete-workflow} shows the workflow without these optional steps.}
\label{tab:workflow_nodes}
\begin{tabularx}{\textwidth}{@{}l>{\raggedright\arraybackslash}X@{}}
\toprule
\textbf{Node} & \textbf{Function} \\
\midrule
\texttt{data\_loading} & Validates input data structure, checks layer/interface consistency, and initializes the state. \\[2pt]
\texttt{model\_selection} & Selects the public branch, assembles the corresponding solver path from admissible physics blocks, and schedules any required retrieval or verification hooks. \\[2pt]
\texttt{sensitivity\_analysis} & Calls the finite-difference Jacobian engine documented in section~\ref{sec:sensitivity_svd}. \\[2pt]
\texttt{identifiability\_check} & Performs SVD-based target screening using the normalized Jacobian. \\[2pt]
\texttt{fitting} & Executes reduced-space inverse analysis with the configured reduced-space optimizer (default: DTDE) together with the documented DTDE/L-BFGS-B fallback from section~\ref{sec:optimizer_benchmark}. \\[2pt]
\texttt{optimization} & Optional post-fit refinement, bounded retry, or verification-triggered re-run. \\[2pt]
\texttt{error\_handler} & Classifies failures and applies the configured recovery strategy. \\
\bottomrule
\end{tabularx}
\end{table}

\begin{figure}[p]
\centering
\begin{adjustbox}{max width=\linewidth,max totalheight=0.82\textheight,center}
\begin{tikzpicture}[
    node distance=0.9cm,
    exec/.style={rectangle, rounded corners=4pt, draw=langgraph, fill=langgraph!12,
                 line width=1pt, minimum width=3.0cm, minimum height=0.72cm,
                 font=\small, align=center},
    route/.style={diamond, draw=mcpyellow!80!black, fill=mcpyellow!22,
                  line width=1pt, aspect=2.2, inner xsep=0.8em,
                  font=\small, align=center},
    terminal/.style={rectangle, rounded corners=6pt, draw=datagreen, fill=datagreen!18,
                     line width=1pt, minimum width=2.0cm, minimum height=0.72cm,
                     font=\small\bfseries, align=center},
    fail/.style={rectangle, rounded corners=4pt, draw=skillblue, fill=skillblue!14,
                 line width=1pt, minimum width=3.0cm, minimum height=0.72cm,
                 font=\small, align=center},
    flow/.style={-Stealth, thick}
]
    \node[terminal, draw=userpurple, fill=userpurple!18] (start) {START};
    \node[exec, below=of start] (loading) {\texttt{data\_loading}};
    \node[route, below=1.1cm of loading] (routeLoad) {route after\\loading};
    \node[fail, right=2.8cm of routeLoad] (error) {\texttt{error\_handler}};
    \node[exec, below=1.1cm of routeLoad] (model) {\texttt{model\_selection}};
    \node[exec, below=of model] (sens) {\texttt{sensitivity\_analysis}};
    \node[exec, below=of sens] (ident) {\texttt{identifiability\_check}};
    \node[exec, below=of ident] (fitting) {\texttt{fitting}};
    \node[route, below=1.1cm of fitting] (routeAnalysis) {route after\\analysis};
    \node[exec, left=2.8cm of routeAnalysis] (opt) {\texttt{optimization}};
    \node[terminal, below=1.2cm of routeAnalysis] (end) {END};

    \draw[flow, userpurple] (start) -- (loading);
    \draw[flow] (loading) -- (routeLoad);
    \draw[flow, skillblue] (routeLoad) -- node[midway, above, font=\scriptsize] {error} (error);
    \draw[flow, skillblue] (error) |- (end.east);
    \draw[flow] (routeLoad) -- node[midway, right, font=\scriptsize] {continue} (model);
    \draw[flow] (model) -- (sens);
    \draw[flow] (sens) -- (ident);
    \draw[flow] (ident) -- (fitting);
    \draw[flow] (fitting) -- (routeAnalysis);
    \draw[flow] (routeAnalysis) -- node[midway, above, font=\scriptsize] {refine} (opt);
    \draw[flow] (opt) |- (end.west);
    \draw[flow, datagreen] (routeAnalysis) -- node[midway, right, font=\scriptsize] {finish} (end);
\end{tikzpicture}
\end{adjustbox}
\caption{Compiled state transitions for the execution graph. The discrete analysis path can terminate directly after fitting or continue through bounded post-fit refinement, while error-triggered recovery is routed through a separate handler.}
\label{fig:compiled-state-transitions}
\end{figure}
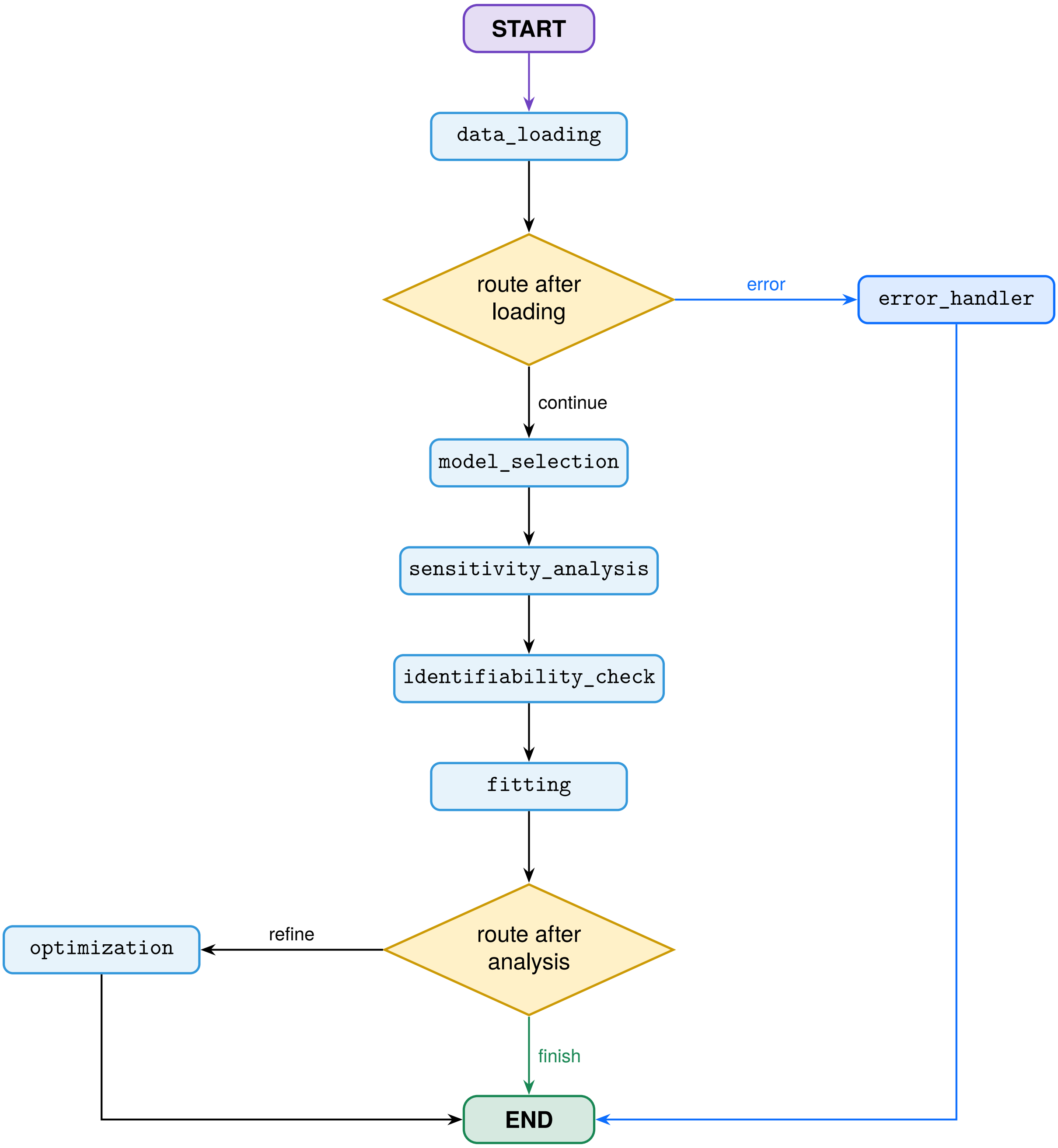

The compiled state transitions are shown schematically in Figure~\ref{fig:compiled-state-transitions}.
The execution nodes dispatch tools according to the recorded workflow state and execute branch revisions or auxiliary consistency checks.

\subsection{Dynamic Planning and Error Recovery}

The planner filters requested targets by membership in the final ordered intersection list:
\begin{algorithm}[H]
\caption{Target Admission from the Ordered Intersection List}
\label{alg:target_adjust}
\begin{algorithmic}
\REQUIRE Requested targets $\mathcal{Q}_0$, $\mathcal{L}_{\mathrm{SVD}}$, $\mathcal{L}_{\mathrm{sensitivity}}$
\ENSURE Final list, $N_{\mathrm{ident}}$, admitted targets, rejected requests
\STATE $\mathcal{L}_{\mathrm{ident}}\leftarrow\operatorname{Filter}(\mathcal{L}_{\mathrm{SVD}},p\in\mathcal{L}_{\mathrm{sensitivity}})$
\COMMENT{Preserve SVD-list order}
\STATE $N_{\mathrm{ident}}\leftarrow|\mathcal{L}_{\mathrm{ident}}|$
\STATE $\mathcal{Q}_{\mathrm{fit}}\leftarrow\operatorname{Filter}(\mathcal{L}_{\mathrm{ident}},p\in\mathcal{Q}_0)$
\STATE $\mathcal{Q}_{\mathrm{unfit}}\leftarrow\mathcal{Q}_0\setminus\mathcal{Q}_{\mathrm{fit}}$
\STATE Report rejected requests as not identifiable under current conditions
\RETURN $\mathcal{L}_{\mathrm{ident}}$, $N_{\mathrm{ident}}$, $\mathcal{Q}_{\mathrm{fit}}$, $\mathcal{Q}_{\mathrm{unfit}}$
\end{algorithmic}
\end{algorithm}

Recovery actions depend on the failure: optimizer switching (DTDE $\leftrightarrow$ L-BFGS-B) for convergence failures, bounded range expansion for boundary violations, renewed identifiability screening and confirmation of fixed inputs for singular-matrix failures, and parameter reset with physical-consistency checks for NaN/Inf failures. A structured residual can instead trigger replacement of the affected transport closure or interface block after branch-local verification.

\section{PRISM Solver Toolset and Agent Routing Policies}
\label{sec:agent_tool_surface}

This section specifies the input/output contracts and activation rules of the public PRISM tools used by the branch router.

\subsection{Capability Contracts and Execution-Graph Integration}

Each capability is an execution-graph node that reads \texttt{FDTRState}, validates its arguments, runs the numerical routine, and writes the outputs back to state for subsequent routing.

Table~\ref{tab:skill_summary} summarizes the core PRISM solver and diagnostic toolset used by the router, and Table~\ref{tab:mcp_io} states, for each capability, its required inputs and auditable outputs.

\begin{table}[H]
\centering
\caption{Core PRISM solver and diagnostic toolset and activation rules.}
\label{tab:skill_summary}
\small
\begin{tabularx}{\linewidth}{l >{\raggedright\arraybackslash}X >{\raggedright\arraybackslash}X}
\toprule
\textbf{Capability} & \textbf{Purpose} & \textbf{Activation Policy} \\
\midrule
\textbf{diagnose\_data} & Data-quality diagnosis and spectrum triage & Gates the workflow before expensive solver dispatch; can reject, prune, or restrict the fitting window. \\
\textbf{estimate\_sensitivity} & Jacobian generation and sensitivity diagnostics & Computes sensitivity magnitudes and frequency-window diagnostics. \\
\textbf{analyze\_identifiability} & SVD-based target reduction & Determines the resolvable parameter subspace and documents why weak targets are removed. \\
\textbf{compute\_phase} & Deterministic forward evaluation & Delegates only to the public PRISM branch chain: optical TMM / STM / TTM / OTh-TTM / OTh-TTM(NF), with the executed path represented internally as a compatible assembly of source, layer, interface, and substrate blocks. \\
\textbf{fit\_parameters} & Reduced-space inverse analysis & Optimizes only the retained target set within bounded priors and the selected solver branch or block assembly. \\
\textbf{predict\_kz} & Continuous-profile branch execution & Activated only when the discrete-layer hypothesis is rejected and the output remains in-distribution. \\
\textbf{infer\_material} & Branch recommendation & Suggests discrete, anisotropic, or continuous-profile routing; low-confidence cases trigger user confirmation. \\
\textbf{extension tools} & Supervised construction of a new branch (e.g.\ \texttt{generate\_training\_library}, \texttt{assemble\_deeponet}, \texttt{validate\_extension}, \texttt{attach\_deployment\_gates}) & Available only in the user-supervised extension mode; a new branch enters routine use only after its validation and deployment gates are passed (Figure~\ref{fig:si_deeponet_log}). \\
\bottomrule
\end{tabularx}
\end{table}

\begin{table}[H]
\centering
\caption{Capability contracts exposed through the public PRISM surface.}
\label{tab:mcp_io}
\small
\begin{tabularx}{\linewidth}{l >{\raggedright\arraybackslash}X >{\raggedright\arraybackslash}X}
\toprule
\textbf{Tool} & \textbf{Input Specification} & \textbf{Auditable Outputs} \\
\midrule
\textbf{diagnose\_data} &
Consumes the measured spectrum and evaluates whether the dataset is fit-worthy. &
SNR estimate, retained fitting window, outlier map, quality score, and warnings. \\
\midrule
\textbf{estimate\_sensitivity} &
Builds the Jacobian for named parameters under the current branch assumptions. &
Sensitivity curves, magnitude diagnostics, and branch-conditioned window recommendations. \\
\midrule
\textbf{analyze\_identifiability} &
Applies normalized-SVD screening to the current target set. &
Identifiable parameter count, singular spectrum, coupling modes, and retained targets. \\
\midrule
\textbf{compute\_phase} &
Executes the deterministic branch-specific forward model selected by the router. &
Phase response, selected branch metadata, executed solver path, and any closure or interface overrides applied during assembly. \\
\midrule
\textbf{fit\_parameters} &
Runs bounded optimization on the reduced target set under the selected branch. &
Best-fit parameters, RMSE, uncertainty estimates, convergence state, fit history, and the accepted block assembly used for the final inverse solve. \\
\midrule
\textbf{predict\_kz} &
Invokes the continuous-profile branch when the discrete-layer hypothesis is rejected. &
Depth profile, uncertainty bounds, profile class, and residual-based acceptance diagnostics. \\
\midrule
\textbf{infer\_material} &
Maps diagnostic features into a routing recommendation without redefining the underlying solver physics. &
Suggested branch, confidence score, and rationale for escalation or fallback. \\
\bottomrule
\end{tabularx}
\end{table}

\subsection{Diagnostic and Model-Selection Entry Points}

\paragraph{Diagnostic entry point: \texttt{diagnose\_data}.}

\begin{figure}[htbp]
\centering
\begin{tikzpicture}[scale=0.90, transform shape,
    node distance=1cm,
    box/.style={rectangle, rounded corners=5pt, draw=skillblue, fill=skillblue!10,
                line width=1pt, minimum width=2.8cm, minimum height=0.8cm,
                font=\small, align=center},
    process/.style={rectangle, rounded corners=3pt, draw=physicsgreen, fill=physicsgreen!10,
                    line width=1pt, minimum width=3.5cm, minimum height=0.7cm,
                    font=\small, align=center},
    io/.style={trapezium, trapezium left angle=70, trapezium right angle=110,
               draw=mcpyellow!80!black, fill=mcpyellow!20, minimum width=2cm,
               minimum height=0.6cm, font=\small, align=center},
    arrow/.style={-Stealth, thick}
]
    \node[io] (input) {Input: Frequency $f$, Phase $\phi$};
    \node[process, below=0.8cm of input] (snr) {\textbf{SNR Estimation}\\Diff-2 Method};
    \node[process, below=0.6cm of snr] (coverage) {\textbf{Frequency Verification}\\Range Assessment};
    \node[process, below=0.6cm of coverage] (outlier) {\textbf{Outlier Detection}\\Polynomial Fit Residuals};
    \node[process, below=0.6cm of outlier] (shape) {\textbf{Feature Extraction}\\Slope \& Curvature};
    \node[process, below=0.6cm of shape] (quality) {\textbf{Quality Scoring}\\Retained-point fraction};
    \node[io, below=0.8cm of quality] (output) {Output: Diagnosis Report};

    \draw[arrow] (input) -- (snr);
    \draw[arrow] (snr) -- (coverage);
    \draw[arrow] (coverage) -- (outlier);
    \draw[arrow] (outlier) -- (shape);
    \draw[arrow] (shape) -- (quality);
    \draw[arrow] (quality) -- (output);

    \node[box, right=3cm of quality, minimum height=3.5cm, minimum width=4.5cm] (result) {
        \textbf{Diagnosis Metric Vector $\mathbf{X}$:}\\[3pt]
        $SNR$: Signal-to-Noise Ratio (dB)\\
        $R_{freq}$: Frequency-range status\\
        $I_{out}$: Outlier indices\\
        $\mathbf{Z}$: Shape-feature vector\\
        $Q^\ast$: Quality score $\in [0,1]$\\
        $\text{Rec}$: Analysis recommendation
    };

    \draw[ArrDash, Ortho] (quality.east) -- (result.west);
\end{tikzpicture}
\caption{Data-quality diagnostic tool used as the initial PRISM routing gate.}
\label{fig:diagnose-data}
\end{figure}
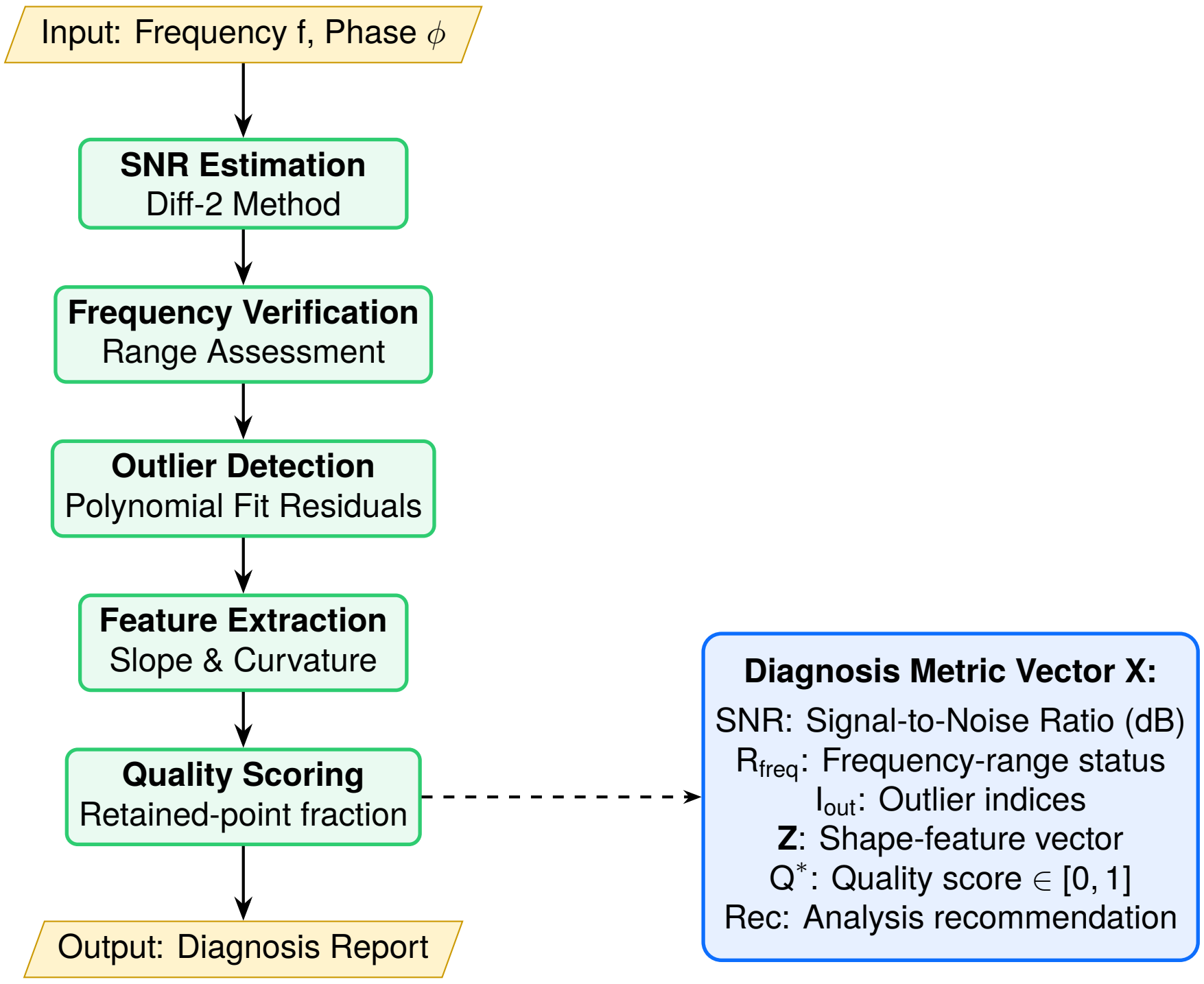

The diagnostic tool evaluates spectrum noise, frequency coverage, and outliers before solver dispatch. Figure~\ref{fig:diagnose-data} shows the diagnostic sequence and its outputs, including the shape-feature vector $\mathbf{Z}$, quality score $Q^\ast$, and analysis recommendation.

Using the quality criteria of section~\ref{sec:sensitivity_svd}, this tool recommends proceeding, pruning outliers, restricting the fitting window, or acquiring an improved dataset.

\paragraph{Model-selection entry point: \texttt{infer\_material}.}

\begin{figure}[htbp]
\centering
\begin{tikzpicture}[
    node distance=0.8cm,
    box/.style={rectangle, rounded corners=5pt, draw=skillblue, fill=skillblue!10,
                line width=1pt, minimum width=2.5cm, minimum height=0.6cm,
                font=\small, align=center},
    signature/.style={rectangle, rounded corners=3pt, draw=physicsgreen, fill=physicsgreen!10,
                      line width=1pt, minimum width=3.5cm, minimum height=0.8cm,
                      font=\footnotesize, align=left},
    arrow/.style={-Stealth, thick},
    jct/.style={circle, fill=black, inner sep=1.5pt}
]
    \node[box, fill=mcpyellow!20] (input) {Input: Feature Vector $\mathbf{x}$};
    \node[signature, below left=1.2cm and 1.5cm of input] (uniform) {
        \textbf{Uniform Material}\\
        Uniform-like spectral trend\\
        Low curvature
    };
    \node[signature, below=1.2cm of input] (gradient) {
        \textbf{Gradient Material}\\
        Depth-dependent spectral trend\\
        Pronounced curvature
    };
    \node[signature, below right=1.2cm and 1.5cm of input] (aniso) {
        \textbf{Anisotropic}\\
        Directional response\\
        Geometry-dependent sensitivity
    };

    \coordinate (fork_y) at ($(input.south)!0.5!(gradient.north)$);
    \node[jct] (fork) at (input.south |- fork_y) {};
    \draw[thick] (input.south) -- (fork);
    \draw[arrow] (fork) -| (uniform.north);
    \draw[arrow] (fork) -- (gradient.north);
    \draw[arrow] (fork) -| (aniso.north);

    \node[box, below=1.0cm of gradient] (score) {Compare Branch Signatures};
    \coordinate (merge_y) at ($(gradient.south)!0.5!(score.north)$);
    \node[jct] (merge) at (score.north |- merge_y) {};
    \draw[thick] (uniform.south) |- (merge);
    \draw[thick] (gradient.south) -- (merge);
    \draw[thick] (aniso.south) |- (merge);
    \draw[arrow] (merge) -- (score);

    \node[box, below=0.6cm of score] (norm) {Rank Route Candidates};
    \node[box, fill=mcpyellow!20, below=0.6cm of norm] (output) {Suggested Analysis Branch};
    \draw[arrow] (score) -- (norm);
    \draw[arrow] (norm) -- (output);
\end{tikzpicture}
\caption{Qualitative material-routing schematic. Spectral-shape and directional-response cues suggest a discrete-layer, anisotropic, or continuous-profile analysis branch; physical regime checks and forward-model comparisons determine the branch used for fitting.}
\label{fig:infer-material}
\end{figure}
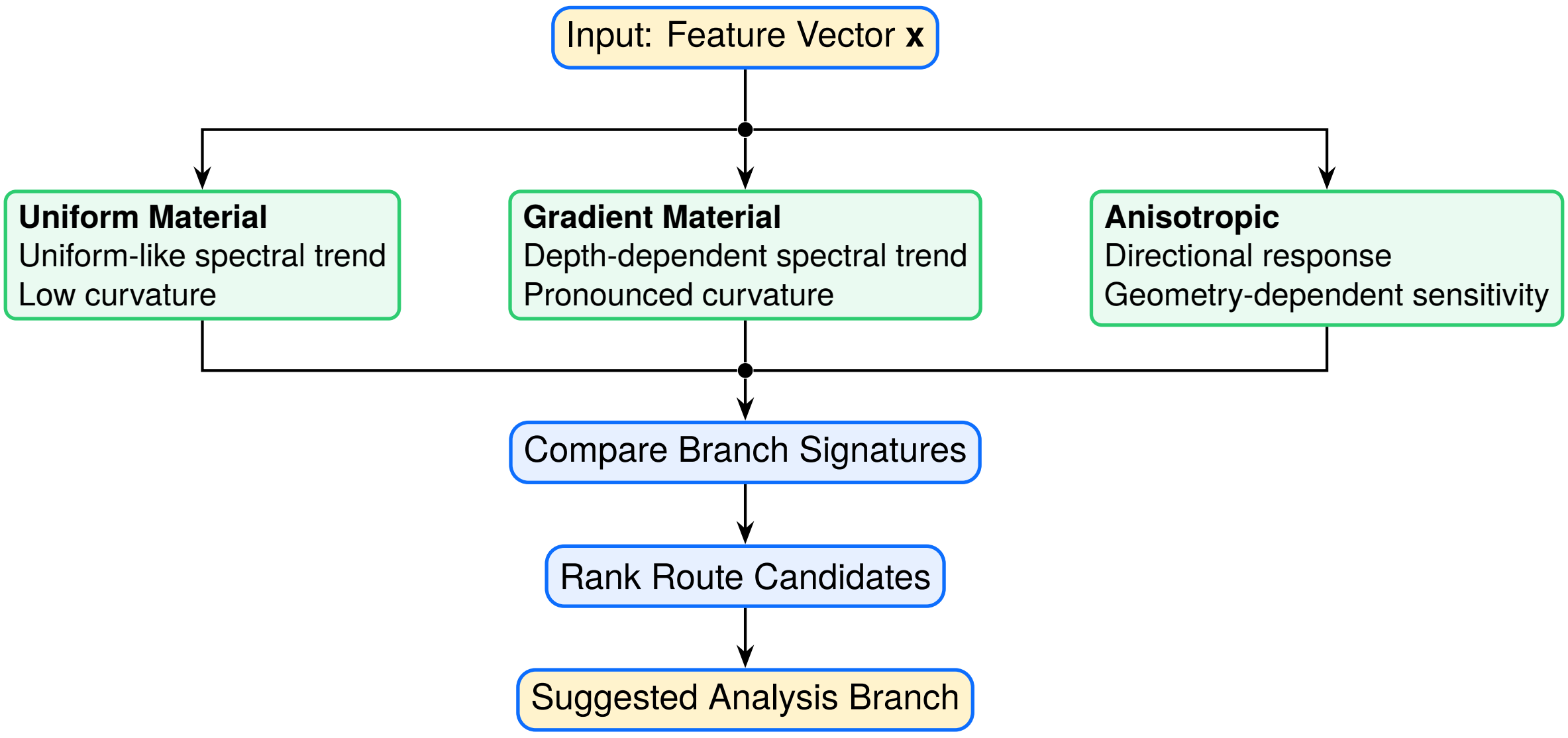

The material-inference tool converts diagnostic features into a branch-routing recommendation. Here $\mathbf{x}$ collects spectral-shape and directional-response cues. Figure~\ref{fig:infer-material} gives a qualitative routing schematic; branch admission follows the physical regime and forward-phase checks described in section~\ref{sec:inference_framework} of the main text.

The tool consumes the diagnostic cues shown in the schematic and returns a routing recommendation. Uniform materials stay on the public PRISM route for discrete-branch fitting; gradient materials can trigger the DeepONet/KRR profile branch from sections~\ref{sec:deeponet_kz} and \ref{sec:krr}; anisotropic stacks keep the discrete-layer route but alter the target set and interpretation of sensitivity.

\subsection{Delegated Tools and Adaptation Rules}

\paragraph{Delegated solver tools.}

The solver tools expose the numerical procedures specified in the preceding sections:
\begin{itemize}
\item \textbf{\texttt{estimate\_sensitivity}} calls the finite-difference Jacobian engine in section~\ref{sec:sensitivity_svd} and returns the Jacobian and sensitivity diagnostics.
\item {\setlength{\RaggedRightParindent}{0pt}\RaggedRight
\textbf{\texttt{analyze\_identifiability}} constructs the ordered SVD pass list and its absolute-sensitivity intersection (section~\ref{sec:sensitivity_svd}). It returns $\mathcal{L}_{\mathrm{ident}}$, $N_{\mathrm{ident}}=|\mathcal{L}_{\mathrm{ident}}|$, admitted and rejected requests, and the separate $r_{\mathrm{SVD}}$ and coupling diagnostics. The router selects requested members of this ordered list.
\par}
\item \textbf{\texttt{compute\_phase}} delegates to the public PRISM branch chain documented in sections~\ref{sec:tmm}, \ref{sec:non_fourier_si}, and \ref{sec:optical_tmm}.
  \item \textbf{\texttt{fit\_parameters}} wraps the reduced-space optimizer from section~\ref{sec:optimizer_benchmark}; the execution layer sees only bounds, algorithm choice, fit targets, and fit outputs, with DTDE as the default public configuration.
  \item \textbf{\texttt{predict\_kz}} exposes the continuous-profile inference branch defined in sections~\ref{sec:deeponet_kz} and \ref{sec:krr}; the branch selector treats it as a callable alternative when the discrete-layer hypothesis is rejected.
\end{itemize}

The execution layer can assemble driver scripts, update the interface settings of a validated branch, prepare dependencies, and generate checks for energy conservation, heat-flux continuity, boundary conditions, module coupling, and parameter plausibility. New physics branches require the user-supervised extension and validation process of Figure~\ref{fig:si_deeponet_log}.

\paragraph{Execution adaptation rules.}

The routing layer converts intermediate tool outputs into explicit workflow adjustments:
\begin{algorithm}[H]
\caption{Schematic PRISM Data-Quality Adaptation}
\label{alg:quality_adapt}
\begin{algorithmic}
\REQUIRE SNR, quality\_score
\ENSURE Adjusted fitting configuration
\IF{SNR $< 20$ dB OR quality\_score $< 0.5$}
    \STATE Reconfirm fixed priors and the accepted spectral window
    \STATE Recompute the final identifiable list for the retained dataset
    \STATE Request user confirmation before fitting
\ELSIF{SNR $\in [20, 30]$ dB AND quality\_score $\in [0.5, 0.7]$}
    \STATE Retain confirmed priors and the accepted spectral window
    \STATE Select requested members of the final identifiable list
\ELSE
    \STATE Use the standard configuration
\ENDIF
\end{algorithmic}
\end{algorithm}

\begin{algorithm}[H]
\caption{PRISM Routing from the Final Identifiable List}
\label{alg:svd_adapt}
\begin{algorithmic}
\REQUIRE $\mathcal{L}_{\mathrm{ident}}$, $N_{\mathrm{ident}}=|\mathcal{L}_{\mathrm{ident}}|$, requested targets $\mathcal{Q}_0$
\ENSURE Final targets $\mathcal{Q}_{\mathrm{final}}$, rejected requests
\STATE $\mathcal{Q}_{\mathrm{final}}\leftarrow\operatorname{Filter}(\mathcal{L}_{\mathrm{ident}},p\in\mathcal{Q}_0)$
\STATE Rejected requests $\leftarrow\mathcal{Q}_0\setminus\mathcal{Q}_{\mathrm{final}}$
\STATE Report rejected targets and screening reasons
\STATE Present the fit-target and fixed-input ledger for confirmation
\RETURN $\mathcal{Q}_{\mathrm{final}}$, rejected requests
\end{algorithmic}
\end{algorithm}

\begin{algorithm}[H]
\caption{Schematic recovery actions within the PRISM execution layer. Substrate-closure escalation was invoked for the Au/Si and Au/Ti/Si benchmarks.}
\label{alg:error_recovery}
\begin{algorithmic}
\REQUIRE Error message $e$, current configuration $\mathcal{H}$
\ENSURE Recovery action
\IF{``convergence'' $\in e$}
    \STATE Switch algorithm: DTDE $\leftrightarrow$ L-BFGS-B
    \STATE Retry with a different initial guess
\ELSIF{``boundary'' $\in e$}
    \STATE Expand bounds by a factor of 2
    \STATE Retry fitting
\ELSIF{``singular matrix'' $\in e$}
    \STATE Re-evaluate the final identifiable list for the current configuration
    \STATE Confirm revised targets or independently fixed inputs before retrying
\ELSIF{branch-conditioned validation shows persistent beam-size or window-banded residual inconsistency}
    \STATE Preserve the validated optical-source and metallic-layer blocks
    \STATE Replace only the inadequate closure or interface block within the admissible PRISM hierarchy
    \STATE If the residual inconsistency tracks the heating length scale and the diffusive hierarchy is exhausted, escalate the substrate closure from OTh-TTM to OTh-TTM(NF) (instantiated in this work by the Au/Si and Au/Ti/Si benchmark conditions)
    \STATE Record the escalation reason and the revised solver path in the audit trail
\ELSIF{``NaN/Inf'' $\in e$}
    \STATE Check for unphysical parameter combinations
    \STATE Reset to the default initial guess
\ELSE
    \STATE Surface the error and request manual intervention
\ENDIF
\end{algorithmic}
\end{algorithm}

Branch routing follows the evidence-to-action mapping:
\begin{equation}
\begin{aligned}
\text{Schema / Evidence} &\rightarrow \text{Workflow Action} \\
\begin{array}[t]{@{}l@{}}\text{``uniform'' stack with}\\ \text{adequate diffusive fit}\end{array} &\rightarrow
\begin{array}[t]{@{}l@{}}\text{Retain reduced-space fitting}\\ \text{on the public discrete hierarchy}\end{array} \\
\begin{array}[t]{@{}l@{}}\text{``gradient'' stack or}\\ \text{rejected discrete hypothesis}\end{array} &\rightarrow
\begin{array}[t]{@{}l@{}}\text{Invoke } \textbf{predict\_kz}\\ \text{with the continuous-profile prior}\end{array} \\
\begin{array}[t]{@{}l@{}}\text{``anisotropic''}\\ \text{interpretation required}\end{array} &\rightarrow
\begin{array}[t]{@{}l@{}}\text{Enable anisotropic targets}\\ \text{within the discrete-layer route}\end{array} \\
\begin{array}[t]{@{}l@{}}\text{validated diffusive inconsistency}\\ \text{under an eligible stack}\end{array} &\rightarrow
\begin{array}[t]{@{}l@{}}\text{Reassemble the solver path by upgrading}\\ \text{only the failing block}\end{array}
\end{aligned}
\end{equation}
with a confidence threshold of 0.7 for automatic schema-based branch selection; lower-confidence recommendations require user confirmation. Subsequent validation can trigger substrate-closure revision, as in the Au/Ti/Si case, while retaining the optical source, metallic TTM layers, and upper-interface recursion.

In an end-to-end run, the router calls \textbf{diagnose\_data}, \textbf{estimate\_sensitivity}, and \textbf{analyze\_identifiability}; assembles the selected solver path; invokes \textbf{fit\_parameters} or \textbf{predict\_kz}; and reports the fit or profile with residual and plausibility diagnostics.

\paragraph{Per-case execution traces.}
The following case summaries describe target screening and model-revision decisions. The Au/MoS\textsubscript{2} benchmark compares the four forward branches using the same published phase spectrum. On \emph{clean Au/Si}, the router admitted the full spectral window after \textbf{diagnose\_data} and retained $k_{\mathrm{Si}}$ and $G_{\mathrm{Au/Si}}$ as fit targets under \textbf{analyze\_identifiability}, in the isotropic substrate parameterization of Table~\ref{tab:ausi}; \textbf{fit\_parameters} was dispatched along the local hierarchy up to OTh-TTM, the validation layer then flagged the residual disagreement between the offset and coaxial geometries, and a bounded recovery step escalated the substrate closure to OTh-TTM(NF), where the interface degree of freedom becomes $\bNF$ in place of $G_{\mathrm{Au/Si}}$. On \emph{buried-silica Au/Ti/SiO\textsubscript{2}/Si}, the router received the broader target list $\{k_{\mathrm{SiO_2}}, G_{\mathrm{Ti/SiO_2}}, G_{\mathrm{SiO_2/Si}}, k_{\mathrm{Si}}\}$; the identifiability gate autonomously rejected $G_{\mathrm{Ti/SiO_2}}$, $G_{\mathrm{SiO_2/Si}}$, and $k_{\mathrm{Si}}$ because they were absent from the final intersection of the SVD and absolute-sensitivity pass lists, \textbf{fit\_parameters} was then dispatched on the reduced $\{k_{\mathrm{SiO_2}}\}$ target, and the validation-layer plausibility check accepted without recovery. The separate bulk-silica comparison in Figure~\ref{fig:si_chat_ui} illustrates the user-confirmed two-parameter fit of $k_{\mathrm{SiO_2}}$ and $G_{\mathrm{Ti/SiO_2}}$ reported in \Cref{tab:silica}A. On \emph{Au/Ti/Si}, the diffusive branches initially converged with an anomalously low fitted $k_{\mathrm{Si}}$; the validation layer flagged window-banded and spot-size residual inconsistencies, which triggered a bounded recovery step that escalated the substrate closure to OTh-TTM(NF); the subsequent fit was accepted, and Figure~\ref{fig:nonlocal}D of the main text records the five branches visited along that escalation path. 

\section{End-to-End PRISM Workflow}
\label{sec:agent_deployment}

The execution layer accepts a task description and measurement data, selects a solver path, records intermediate numerical outputs and diagnostics, and returns a report.

\subsection{Complete System Workflow}

\begin{figure}[htbp]
\centering
\begin{adjustbox}{max width=\linewidth,center}
\begin{tikzpicture}[
    node distance=1.2cm,
    user/.style={ellipse, draw=userpurple, fill=userpurple!20, minimum width=2cm,
                 minimum height=0.8cm, font=\small\bfseries},
    action/.style={rectangle, rounded corners=5pt, draw=langgraph, fill=langgraph!10,
                   line width=1pt, minimum width=3cm, minimum height=0.7cm,
                   font=\small, align=center},
    tool/.style={rectangle, rounded corners=5pt, draw=skillblue, fill=skillblue!15,
                 line width=1pt, minimum width=2.5cm, minimum height=0.7cm,
                 font=\small, align=center},
    result/.style={rectangle, rounded corners=5pt, draw=datagreen, fill=datagreen!15,
                   line width=1pt, minimum width=3.5cm, minimum height=0.8cm,
                   font=\small, align=center},
    arrow/.style={-Stealth, thick},
    phase/.style={font=\bfseries\small, text=gray}
]
    \node[user] (user) at (0, 0) {User};
    \node[action, right=2.5cm of user] (upload) {Provide Task\\and Data};
    \node[tool, right=2.5cm of upload] (diagnose) {Quality Gate};

    \node[phase, above=0.2cm of upload] {Phase 1};
    \node[phase, above=0.2cm of diagnose] {Phase 2};

    \node[action, below=1.8cm of diagnose] (choose) {Schema Parsing\\\& Branch Routing};
    \node[phase, above=0.2cm of choose] {Phase 3};

    \node[tool, below left=1.8cm and 2cm of choose] (sens) {Sensitivity \&\\Identifiability};
    \node[tool, below right=1.8cm and 2cm of choose] (infer) {Prior Retrieval\\\& Profile Check};

    \node[tool, below=1.5cm of sens] (fit) {Parameter\\Fitting};
    \node[tool, below=1.5cm of infer] (predict) {Profile\\Prediction};

    \node[result] (result) at ($(fit)!0.5!(predict) + (0, -1.8)$) {Final Report:\\Parameters or $k(z)$};

    \draw[arrow] (user) -- (upload);
    \draw[arrow] (upload) -- (diagnose);
    \draw[arrow] (diagnose) -- (choose);

    \coordinate (busW) at ($(choose.south)+(0,-0.45)$);
    \draw[Ortho] (choose.south) -- (busW);
    \draw[Arr, Ortho] (busW) -| node[pos=0.25, above, fill=white, inner sep=1.5pt, font=\small] {Discrete branch} (sens.north);
    \draw[Arr, Ortho] (busW) -| node[pos=0.25, above, fill=white, inner sep=1.5pt, font=\small] {Continuous branch?} (infer.north);

    \draw[arrow] (sens) -- (fit);
    \draw[arrow] (infer) -- (predict) node[midway, right, font=\small] {Yes};
    \draw[arrow] (fit.south) |- (result.west);
    \draw[arrow] (predict.south) |- (result.east);
    \draw[arrow] (result.south) -- ++(0, -0.6) coordinate (bottomLine)
        -- node[pos=0.5, above, fill=white, inner sep=1.5pt, font=\small] {Display Results} (bottomLine -| user.south)
        -- (user.south);
\end{tikzpicture}
\end{adjustbox}
\caption{End-to-end PRISM workflow. A user task and dataset pass through quality diagnosis and schema-based branch routing. The discrete route proceeds through sensitivity and identifiability screening to parameter fitting; the continuous route uses a profile prior to predict $k(z)$. Both routes return the numerical outputs and diagnostics in a report.}
\label{fig:complete-workflow}
\end{figure}
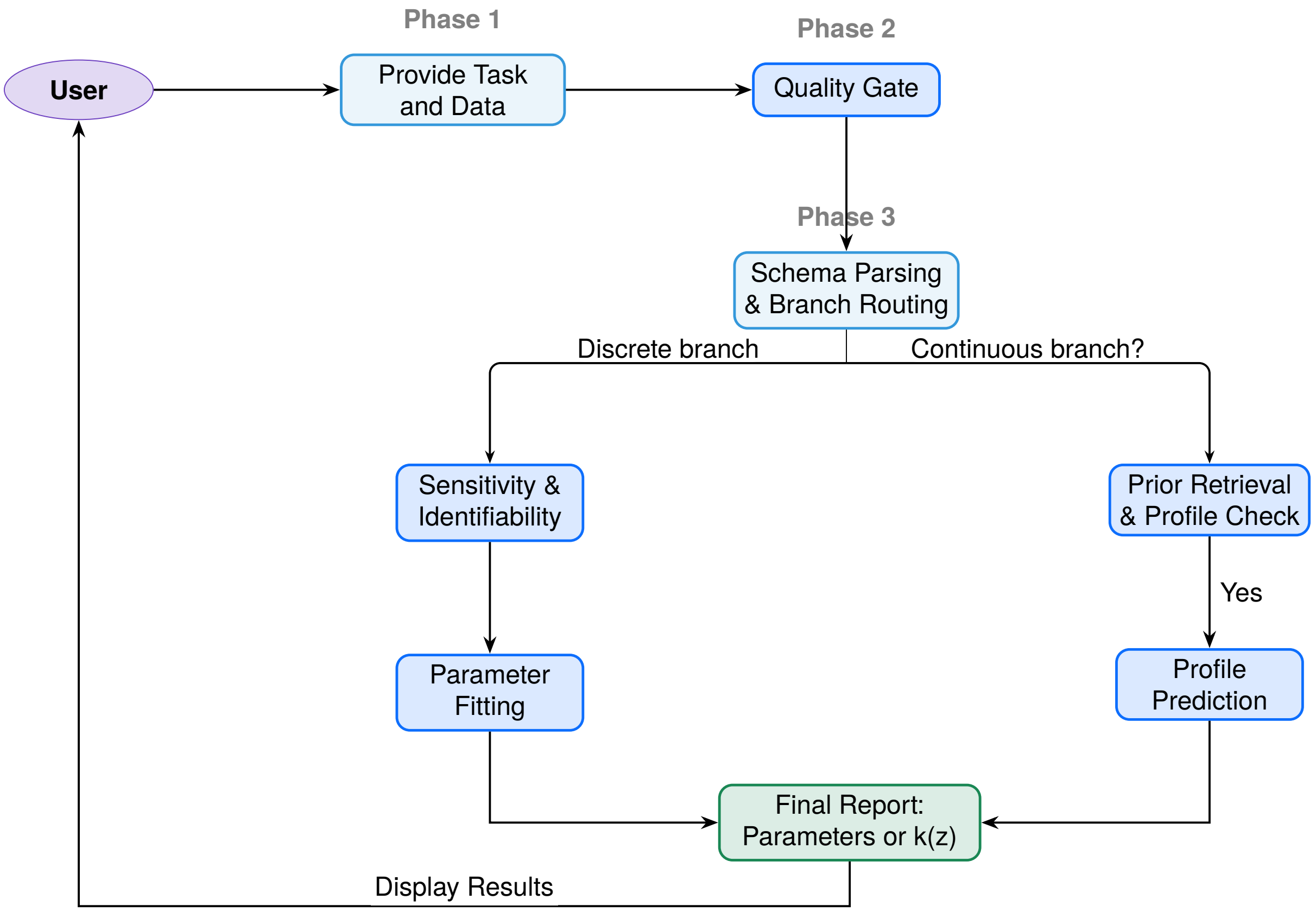

\begin{figure}[H]

  \centering

  \includegraphics[width=0.99\textwidth]{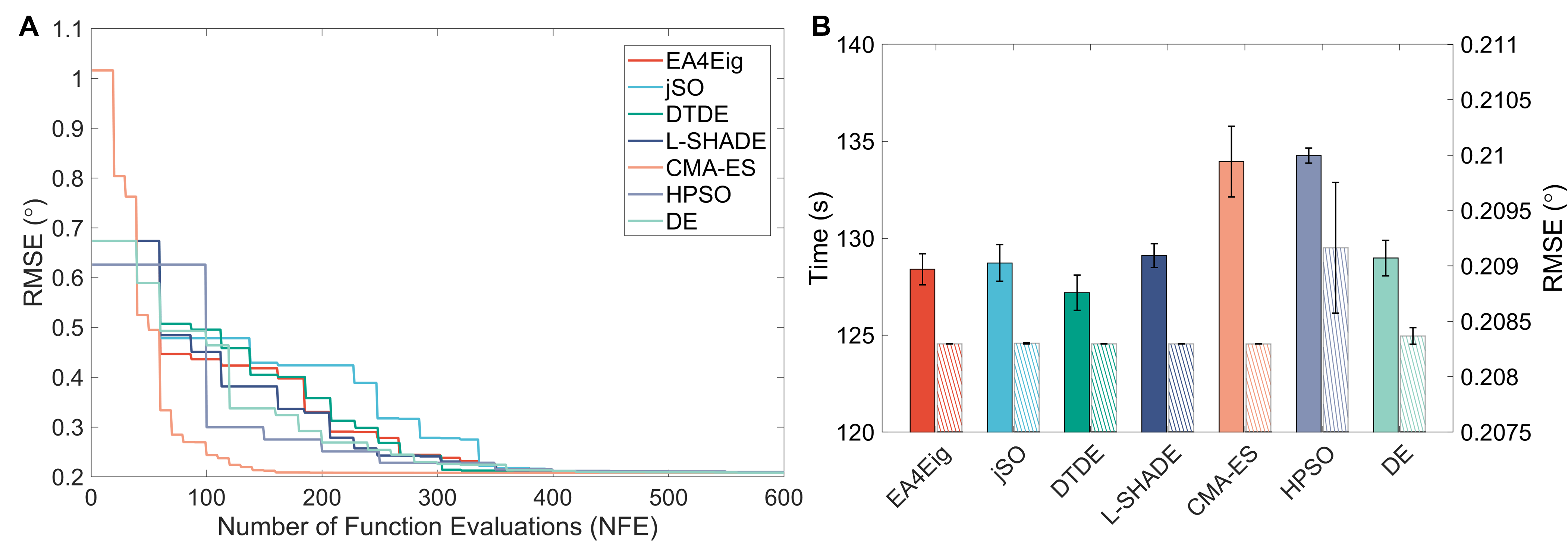}

  \caption{Comparison of seven global optimization algorithms for FDTR parameter extraction. (A) RMSE convergence trajectories (stair-step, reflecting discrete restart/generation updates) as a function of the number of function evaluations (NFE), for all seven optimizers from NFE $=0$ to $600$. Lower curves and earlier descent are better because each NFE corresponds to one full FDTR forward solve. CMA-ES (orange) starts near RMSE $\approx1.02^{\circ}$ and descends in a small number of large steps, crossing below the rest of the field by NFE $\approx60$ and becoming the first method to reach RMSE $\le0.21^{\circ}$, at approximately NFE 140. The remaining six optimizers start from a tighter cluster near RMSE $\approx0.62$--$0.68^{\circ}$ and descend more gradually: jSO (cyan) consistently trails the pack, remaining the highest curve from NFE $\approx100$ to $300$; EA4Eig, DTDE, L-SHADE, HPSO, and DE track each other closely through the middle of the run; and all six converge onto the same terminal RMSE as CMA-ES, $\approx0.2083^{\circ}$, by NFE $\approx400$--$500$, so these runs reach comparable terminal fit quality with different NFE budgets. CMA-ES therefore enters the low-RMSE regime first and reaches RMSE $\le 0.21^\circ$ at the fewest evaluations, giving it the earliest NFE-normalized convergence in this comparison. (B) Dual-axis summary of evolutionary-stage computational cost and evolution-only terminal accuracy across repeated runs. The solid bars referenced to the left axis denote the wall-clock time of the evolutionary search stage, whereas the hatched bars referenced to the right axis denote the terminal RMSE before the post-search \texttt{L-BFGS-B} refinement; the error bars indicate run-to-run variability. DTDE achieves the lowest wall-clock time at comparable terminal RMSE, and is therefore the preferred default deployment optimizer for reduced-space FDTR fitting, whereas CMA-ES reaches the specified RMSE threshold with fewer evaluations in (A).}

  \label{fig:si_optimizer}

\end{figure}

\begin{figure}[H]
  \centering
  \includegraphics[width=0.99\textwidth]{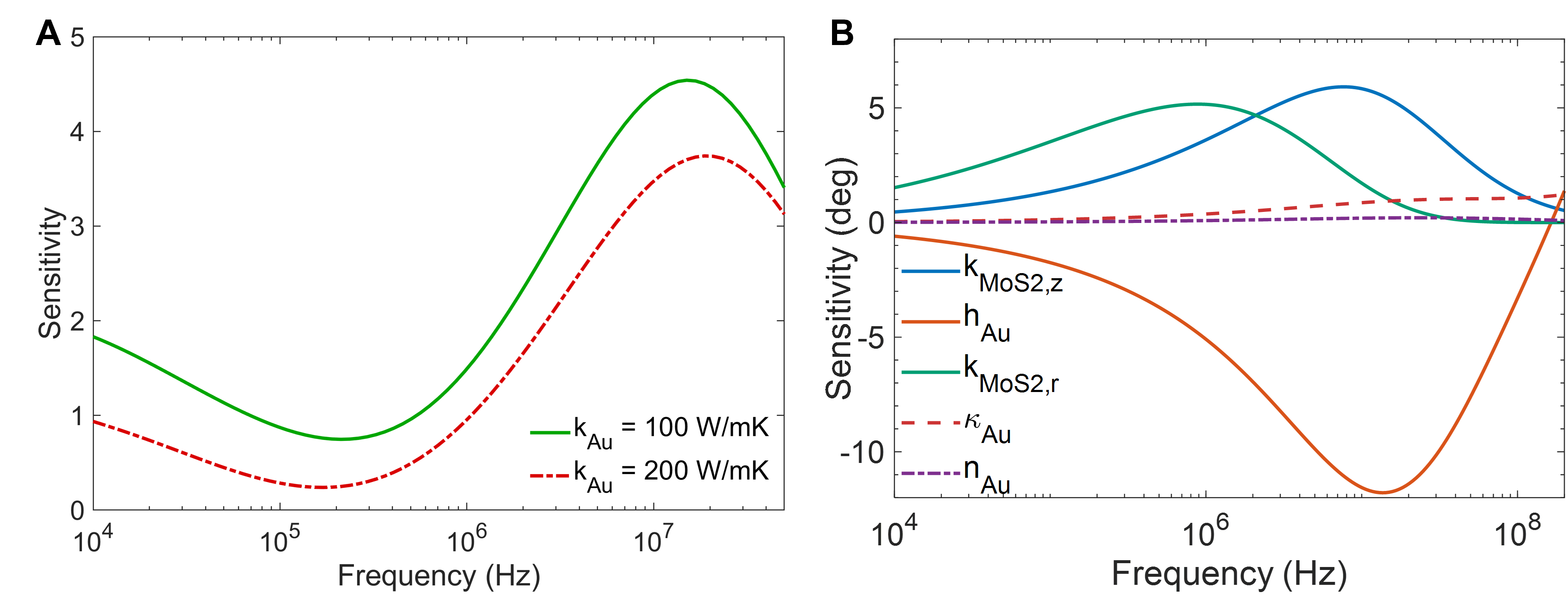}
  \caption{Thermal and optical sensitivity of Au transducers. (A)~Magnitude of the logarithmic phase sensitivity to the SiO\textsubscript{2} thermal conductivity, $|S_{k_{\mathrm{SiO}_2}}|$, for an Au/SiO\textsubscript{2} stack with $k_{\mathrm{Au}}=\SI{100}{\watt\per\metre\per\kelvin}$ (green, solid) and $k_{\mathrm{Au}}=\SI{200}{\watt\per\metre\per\kelvin}$ (red, dash-dot). Reducing the transducer thermal conductivity increases the phase sensitivity to the silica substrate over the plotted frequency range. (B)~Signed logarithmic phase sensitivities $S_p=\partial\phi/\partial\ln p$ (degrees) for an Au/MoS\textsubscript{2} stack with a \SI{100}{\nano\metre} Au transducer. The selected parameters are the cross-plane and in-plane MoS\textsubscript{2} thermal conductivities ($k_{\mathrm{MoS}_2,z}$ and $k_{\mathrm{MoS}_2,r}$), Au thickness $h_{\mathrm{Au}}$, and Au optical constants $n_{\mathrm{Au}}$ and $\kappa_{\mathrm{Au}}$. Here $n_{\mathrm{Au}}$ and $\kappa_{\mathrm{Au}}$ are the refractive index and extinction coefficient, respectively, and are dimensionless. The parameter set is the same as in \Cref{fig:optical}D, which shows the corresponding sensitivities for a \SI{43}{\nano\metre} Au transducer.}
  \label{fig:si_transducer_sensitivity}
\end{figure}

\begin{figure}[H]
  \centering
  \includegraphics[width=0.60\textwidth]{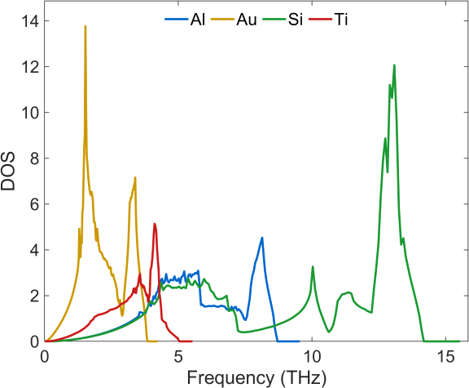}
  \caption{Phonon density of states of Au, Al, Ti, and Si. Phonon DOS (arbitrary units) as a function of phonon frequency (THz) for Au (gold), Al (blue), Ti (red), and Si (green), obtained from the PhononBench online service using its MatterSim machine-learning interatomic potential~\citep{han2025phononbench}. The Au spectrum is concentrated at low frequencies, while the Si spectrum extends to substantially higher frequencies, with high-frequency peaks near \SI{13}{\tera\hertz}.}
  \label{fig:si_phonon_dos}
\end{figure}

\begin{figure}[!htbp]
  \centering
  \includegraphics[width=0.99\textwidth]{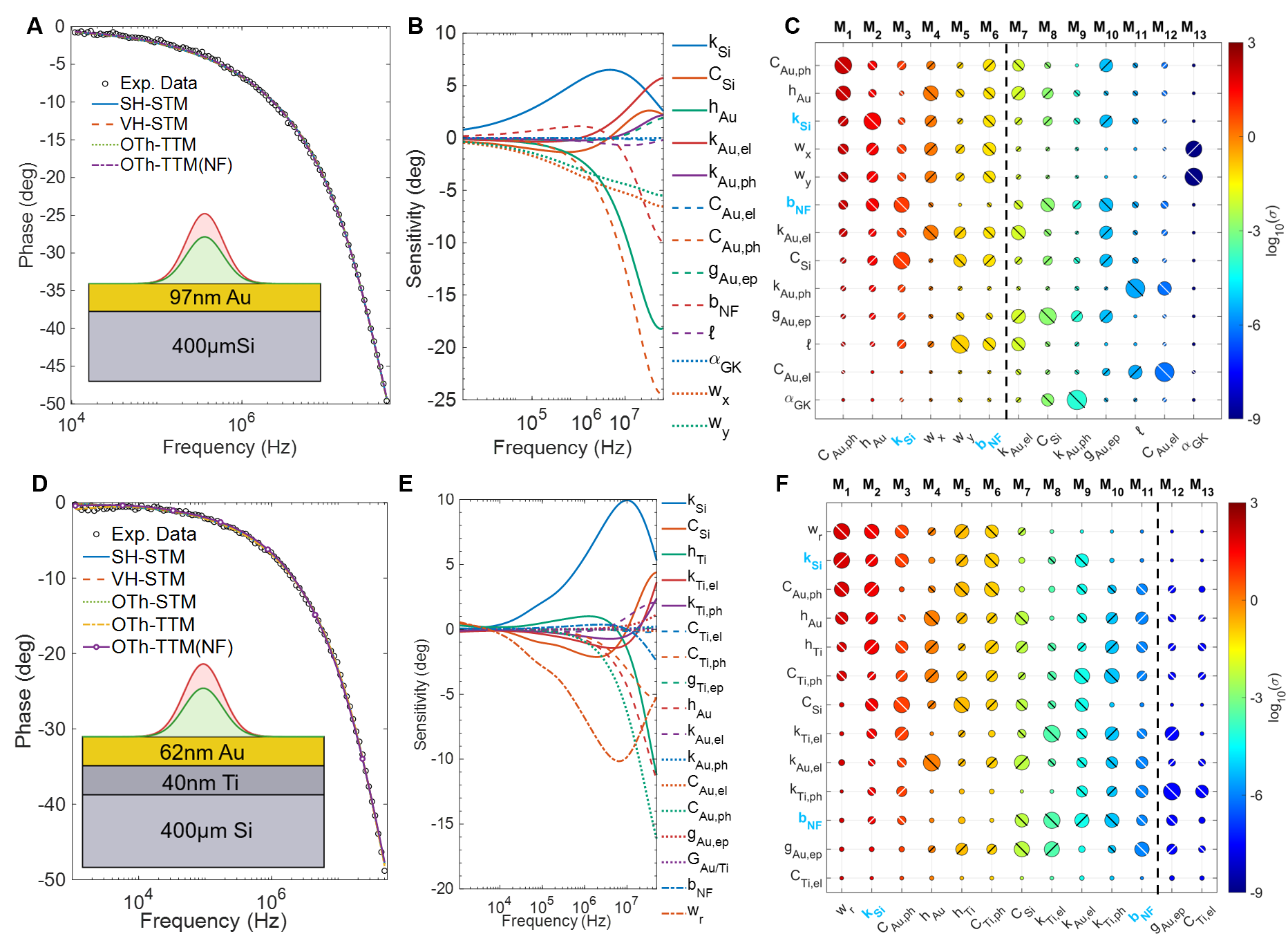}
\caption{Phase fitting, parameter sensitivity, and local SVD analysis for two coaxial ($\Delta = 0$) beam configurations: Au/Si (top row), the coaxial counterpart of Figure~\ref{fig:ausi}D--F, and Au/Ti/Si (bottom row), the small-spot case of Figure~\ref{fig:nonlocal}.
\textbf{Top row (coaxial Au/Si)}: \SI{97}{\nano\metre}~Au / \SI{400}{\micro\metre}~Si with beam radii $w_x = \SI{2.19}{\micro\metre}$ and $w_y = \SI{1.84}{\micro\metre}$, the coaxial partner of the offset measurement in Figure~\ref{fig:ausi}D--F.
(A) Measured phase spectrum (open circles) from \SI{10.9}{\kilo\hertz} to \SI{50}{\mega\hertz}, spanning $-0.7$ to $-49.3$ degrees, with the SH-STM, VH-STM, OTh-TTM and OTh-TTM(NF) fits superimposed. Inset: concentric pump and probe spots and the \SI{97}{\nano\metre} Au / \SI{400}{\micro\metre} Si stack.
(B) Phase sensitivities $S_p = \partial\phi/\partial\ln p$, in degrees, for thirteen parameters, evaluated at the OTh-TTM(NF) operating point ($k_{\mathrm{Si}} = \SI{147.18}{\watt\per\metre\per\kelvin}$, $\bNF = 16.73$, $\ell = \SI{185.0}{\nano\metre}$) by symmetric logarithmic central differences with $\delta = \ln(1.01)$. $S_{k_{\mathrm{Si}}}$ reaches a maximum of $+6.5$ degrees at \SI{5.3}{\mega\hertz}. At \SI{50}{\mega\hertz} the largest sensitivities are $S_{C_{\mathrm{Au,ph}}} = -24.5$, $S_{h_{\mathrm{Au}}} = -18.1$, $S_{\bNF} = -10.0$, $S_{w_x} = -6.5$ and $S_{w_y} = -5.5$ degrees. Root-mean-square sensitivities across the band are $C_{\mathrm{Au,ph}}$ 9.5, $h_{\mathrm{Au}}$ 7.1, $k_{\mathrm{Si}}$ 4.2, $w_x$ 3.7, $w_y$ 3.1, $\bNF$ 3.1, $k_{\mathrm{Au,el}}$ 2.1, $C_{\mathrm{Si}}$ 1.3, $k_{\mathrm{Au,ph}}$ 0.68, $g_{\mathrm{Au,ep}}$ 0.57, $\ell$ 0.35, $C_{\mathrm{Au,el}}$ 0.080 and $\alpha_{\mathrm{GK}}$ 0.016 degrees.
(C) SVD bubble chart of the same Jacobian, with the fit targets $k_{\mathrm{Si}}$ and $\bNF$ highlighted in cyan. The accompanying numerical SVD diagnosis gives the normalized spectrum $\sigma_m/\sigma_1 = 1$, $0.302$, $0.048$, $9.1\times10^{-3}$, $7.6\times10^{-4}$, $4.9\times10^{-4}$, $7.1\times10^{-5}$, $1.1\times10^{-5}$, $6.7\times10^{-7}$, $4.0\times10^{-8}$, $2.1\times10^{-8}$, $4.5\times10^{-9}$ and $4.9\times10^{-12}$; two modes exceed $\sigma_m/\sigma_1 = 0.1$. Leading components are $M_1$: $C_{\mathrm{Au,ph}}$ $+0.71$, $h_{\mathrm{Au}}$ $+0.53$, $w_x$ $+0.25$; $M_2$: $k_{\mathrm{Si}}$ $+0.71$, $\bNF$ $+0.39$, $w_x$ $-0.35$; $M_3$: $C_{\mathrm{Si}}$ $-0.69$, $\bNF$ $-0.57$. The reported terminal mode $M_{13}$ has components $w_y$ $+0.766$ and $w_x$ $-0.643$, indicating a weakly constrained local combination of the two beam radii. The spatial quadrature uses the calibrated directional widths $w_x$ and $w_y$.
\textbf{Bottom row (coaxial Au/Ti/Si)}: \SI{62}{\nano\metre}~Au / \SI{40}{\nano\metre}~Ti / \SI{400}{\micro\metre}~Si with spot radius $w_r = \SI{1.86}{\micro\metre}$, the small-radius geometry of the spot-size-dependent transport case of section~\ref{sec:nonlocal_result}.
(D) Measured phase spectrum (open circles) from \SI{2}{\kilo\hertz} to \SI{50}{\mega\hertz} (fits use the \SI{20}{\kilo\hertz}\text{--}\SI{50}{\mega\hertz} band of Table~\ref{tab:nonlocal_windows}), spanning $\approx-0.5$ to $-49$ degrees, with SH-STM, VH-STM, OTh-STM, OTh-TTM, and OTh-TTM(NF) fits superimposed. Inset: concentric pump and probe beams on the layer stack.
(E) Phase sensitivities $S_p = \partial\phi/\partial\ln p$ ($^{\circ}$) across frequency for the coaxial stack parameters ($k_{\mathrm{Si}}, C_{\mathrm{Si}}, h_{\mathrm{Ti}}, k_{\mathrm{Ti,el}}, k_{\mathrm{Ti,ph}}, C_{\mathrm{Ti,el}}, C_{\mathrm{Ti,ph}}, g_{\mathrm{Ti,ep}}, h_{\mathrm{Au}}, k_{\mathrm{Au,el}}, k_{\mathrm{Au,ph}}, C_{\mathrm{Au,el}}, C_{\mathrm{Au,ph}}, g_{\mathrm{Au,ep}}, G_{\mathrm{Au/Ti}}, \bNF, w_r$), showing $S_{k_{\mathrm{Si}}}$ peaking at $\approx+10.0^{\circ}$ around \SI{2}{\mega\hertz} and high-frequency dominance by $w_r$, $C_{\mathrm{Ti,ph}}$, and $C_{\mathrm{Au,ph}}$.
(F) SVD bubble chart for the coaxial Au/Ti/Si Jacobian. The bottom parameter ranking is ($w_r$, $k_{\mathrm{Si}}$, $C_{\mathrm{Au,ph}}$, $h_{\mathrm{Au}}$, $h_{\mathrm{Ti}}$, $C_{\mathrm{Ti,ph}}$, $C_{\mathrm{Si}}$, $k_{\mathrm{Ti,el}}$, $k_{\mathrm{Au,el}}$, $k_{\mathrm{Ti,ph}}$, $\bNF$, $g_{\mathrm{Au,ep}}$, $C_{\mathrm{Ti,el}}$); the fit targets $k_{\mathrm{Si}}$ and $\bNF$ are highlighted in cyan. The color scale spans $-9$ to $3$, shared with (C) and Figure~\ref{fig:ausi}. The reported fit targets are $k_{\mathrm{Si}}$ and the Ti/Si interface parameter $\bNF$; target admission uses the final intersection list.}
\label{fig:si_spatial_overlap}
\end{figure}

\begin{figure}[!htbp]
  \centering
  \includegraphics[width=0.99\textwidth]{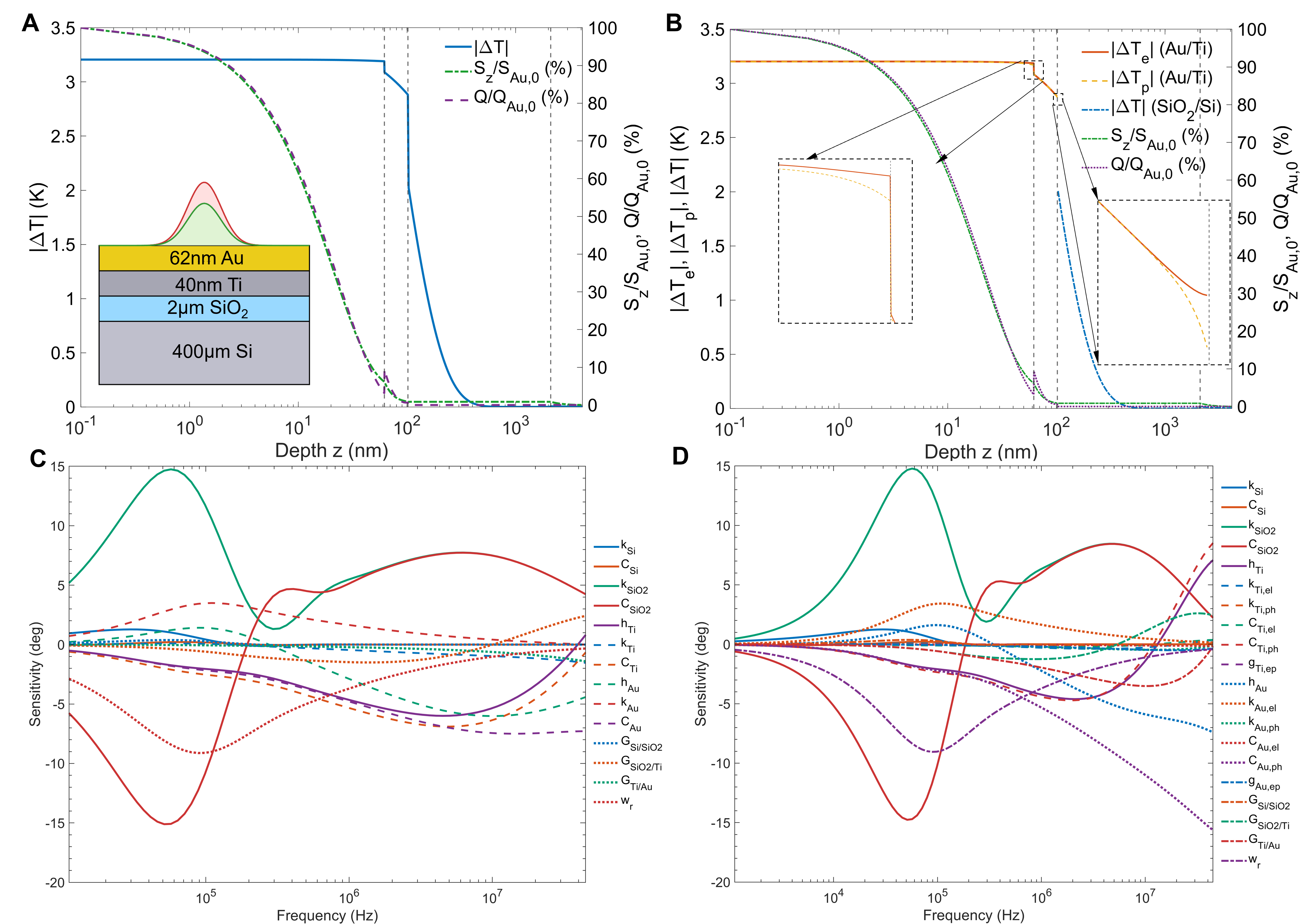}
  \caption{Microscopic thermal response comparison and sensitivity diagnostics for the buried Au/Ti/SiO\textsubscript{2}/Si heterostructure (\SI{62}{\nano\metre} Au / \SI{40}{\nano\metre} Ti / \SI{2}{\micro\metre} SiO\textsubscript{2} / \SI{400}{\micro\metre} Si), evaluated in the small-spot configuration ($w_r=\SI{1.86}{\micro\metre}$). (A) OTh-STM model at a modulation frequency of \SI{50}{\mega\hertz}: depth-resolved lattice temperature amplitude $|\Delta T|$ (blue, left axis) together with the normalized Poynting flux $S_z/S_{\mathrm{Au},0}$ (green dash-dot, right axis, \%) and normalized absorbed power density $Q/Q_{\mathrm{Au},0}$ (purple dashed, right axis, \%), each referenced to its value at the Au surface. $|\Delta T|$ is essentially flat through the Au/Ti bilayer, then decays sharply once $S_z$ and $Q$ themselves collapse toward the SiO\textsubscript{2}/Si region, illustrating that most pump absorption is concentrated within the first $\sim$\SI{100}{\nano\metre}. Vertical dashed lines mark the Au/Ti ($z=\SI{62}{\nano\metre}$), Ti/SiO\textsubscript{2} ($z=\SI{102}{\nano\metre}$), and SiO\textsubscript{2}/Si ($z=\SI{2102}{\nano\metre}$) interfaces. (B) OTh-TTM model at the same frequency: the single lattice curve of (A) splits into separate electron ($|\Delta T_e|$, orange) and phonon ($|\Delta T_p|$, gold dashed) amplitudes in Au/Ti, converging onto the same $S_z$, $Q$, and SiO\textsubscript{2}/Si-region $|\Delta T|$ (blue dotted) curves at depth. The two zoomed insets magnify the near-surface and Au/Ti-interface region, where $|\Delta T_e|$ visibly separates above $|\Delta T_p|$, exposing the electron--phonon temperature divergence that a single-temperature model cannot represent. (C) Phase-sensitivity curves $S_p=\partial\phi/\partial\ln p$ (degrees) for the OTh-STM model across the fourteen candidate parameters $\{k_{\mathrm{Si}}, C_{\mathrm{Si}}, k_{\mathrm{SiO_2}}, C_{\mathrm{SiO_2}}, h_{\mathrm{Ti}}, k_{\mathrm{Ti}}, C_{\mathrm{Ti}}, h_{\mathrm{Au}}, k_{\mathrm{Au}}, C_{\mathrm{Au}}, G_{\mathrm{SiO_2/Si}}, G_{\mathrm{Ti/SiO_2}}, G_{\mathrm{Au/Ti}}, w_r\}$ of Figure~\ref{fig:si_microscopic_2}, plotted from \SI{2e4}{\hertz} to $\sim$\SI{4e7}{\hertz}: $k_{\mathrm{SiO_2}}$ (green) dominates at low-to-mid frequency, peaking near $+15^{\circ}$ around \SI{5e4}{\hertz} before reaching a positive minimum near \SI{3e5}{\hertz} and recovering to $\sim+8^{\circ}$ at the highest frequencies, while $C_{\mathrm{SiO_2}}$ (red) is anti-correlated, reaching a minimum near $-15^{\circ}$ around \SI{1e5}{\hertz} before crossing zero near \SI{2e5}{\hertz} and rising to $\sim+8^{\circ}$ at high frequency; the substrate $k_{\mathrm{Si}}$, $C_{\mathrm{Si}}$, and interface-conductance responses remain smaller than the dominant oxide responses. For this buried-oxide case, the requested target released to fitting is $k_{\mathrm{SiO_2}}$ (Table~\ref{tab:fit_ledger}). (D) Corresponding phase-sensitivity curves for the OTh-TTM model, now over twenty parameters that additionally split $k_{\mathrm{Ti}}$, $C_{\mathrm{Ti}}$, $k_{\mathrm{Au}}$, and $C_{\mathrm{Au}}$ into electron/phonon channels and add the volumetric couplings $g_{\mathrm{Ti,ep}}$ and $g_{\mathrm{Au,ep}}$ (cf.\ Figure~\ref{fig:si_microscopic_3}). $k_{\mathrm{SiO_2}}$ and $C_{\mathrm{SiO_2}}$ retain essentially the same extrema and crossing frequency as in (C), with metal phonon heat-capacity sensitivities reaching magnitudes of approximately $8$--$15^{\circ}$; the corresponding SVD diagnosis is given in Figure~\ref{fig:si_microscopic_3}.}
  \label{fig:si_microscopic_1}
\end{figure}

Figure~\ref{fig:complete-workflow} shows the discrete-layer and continuous-profile routes. Validation of a discrete fit can trigger revision of the affected module, as in the substrate-closure change for Au/Ti/Si.

\clearpage
\section[Table Archive: Back-fitted Au/MoS2 Parameters]{Table Archive: Back-fitted Au/MoS\texorpdfstring{\textsubscript{2}}{2} Parameters}

\label{sec:archive_aumos2}

Table~\ref{tab:archive_aumos2} reports the parameters obtained by fitting the published Au/MoS\textsubscript{2} spectrum of Ref.~\cite{guo2026light} with SH-STM, VH-STM, OTh-STM, and OTh-TTM, as shown in Figure~\ref{fig:optical}B.

\begin{table}[!ht]

\centering

\caption{Parameters recovered by three-parameter fits to the published Au/MoS\textsubscript{2} phase spectrum of Ref.~\cite{guo2026light} using the four PRISM branches. Stack and optical inputs are held at the reported values; uncertainties are those of the present refits. OTh-STM reproduces the literature central values, and the corresponding comparison appears in Figure~\ref{fig:optical}B of the main text.}        

\label{tab:archive_aumos2}

\begin{tabular*}{\textwidth}{@{\extracolsep{\fill}}lccc@{}}

\toprule

Model & $k_{\mathrm{MoS_2},z}$ [W/(m K)] & $k_{\mathrm{MoS_2},r}$ [W/(m K)] & $G$ [MW/(m$^2$ K)] \\

\midrule

SH-STM  & $1.684\pm0.3$ & $75.71\pm5.7$ & $21.93\pm1.4$ \\

VH-STM  & $1.683\pm0.3$ & $77.18\pm5.8$ & $22.93\pm1.6$ \\

OTh-STM & $2.790\pm0.5$ & $84.00\pm6.6$ & $31.00\pm2.5$ \\

OTh-TTM & $2.774\pm0.5$ & $84.19\pm6.6$ & $33.34\pm2.9$ \\

\bottomrule

\end{tabular*}

\end{table}


\section{Supplemental figures}

\label{sec:suppfigs}

For the Au/Ti/Si case, the relevant interpretation is therefore two-step. A broader Ti-to-Si coupling window can increase how efficiently the interface injects energy into Si, but the measured transport closure still depends on which Si branches receive that heat and whether those modes can re-thermalize over the FDTR thermal penetration depth. Because room-temperature Si still carries a substantial fraction of heat in intrinsically long-MFP low-frequency modes, stronger spectral coupling can coexist with the frequency-dependent nonequilibrium signatures that motivate the hydrodynamic-style OTh-TTM(NF) closure in the main text~\citep{regner2013mfpfdtr,inelastic2022sharpinterface,liu2025boundary,beardo2020phonon}.


\begin{figure}[!htbp]
  \centering
  \includegraphics[width=0.99\textwidth]{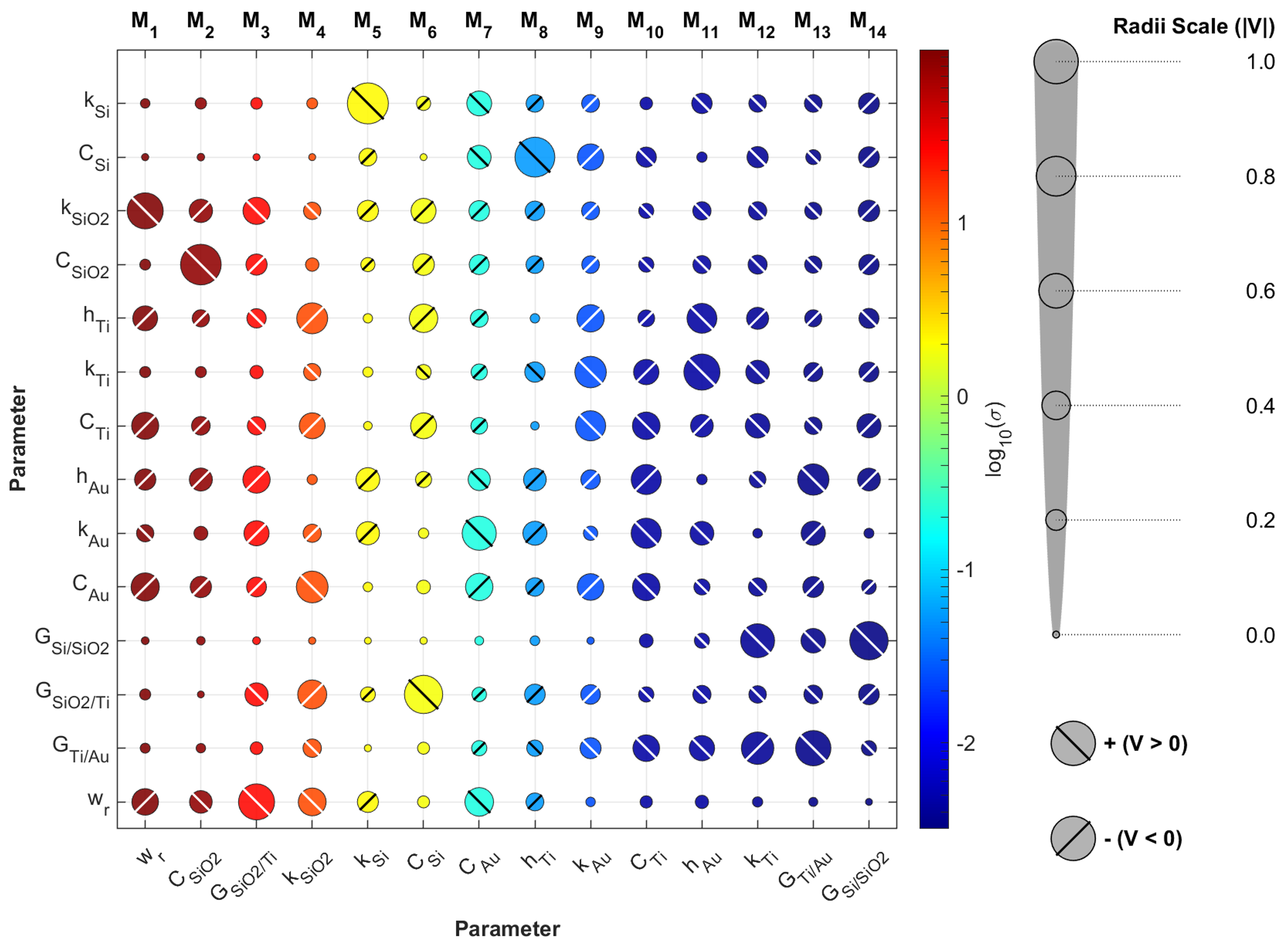}
  \caption{SVD bubble chart for the OTh-STM model of the Au/Ti/SiO\textsubscript{2}/Si stack in the small-spot configuration ($w_r=\SI{1.86}{\micro\metre}$), for the same fourteen parameters as Figure~\ref{fig:si_microscopic_1}C. The bottom parameter ranking is ($w_r$, $C_{\mathrm{SiO_2}}$, $G_{\mathrm{Ti/SiO_2}}$, $k_{\mathrm{SiO_2}}$, $k_{\mathrm{Si}}$, $C_{\mathrm{Si}}$, $C_{\mathrm{Au}}$, $h_{\mathrm{Ti}}$, $k_{\mathrm{Au}}$, $C_{\mathrm{Ti}}$, $h_{\mathrm{Au}}$, $k_{\mathrm{Ti}}$, $G_{\mathrm{Au/Ti}}$, $G_{\mathrm{SiO_2/Si}}$). Modes $M_1$--$M_{14}$ are shown. The leading displayed components include loadings from $w_r$, $C_{\mathrm{SiO_2}}$, $G_{\mathrm{Ti/SiO_2}}$, and $k_{\mathrm{SiO_2}}$. The substrate pair $k_{\mathrm{Si}}$/$C_{\mathrm{Si}}$ also contributes to $M_5$--$M_8$, and several interface and metal-layer parameters have large loadings in weak singular directions. For the requested targets of this buried-oxide case, the final intersection list admits $k_{\mathrm{SiO_2}}$; other stack properties retain the specified input values.}
  \label{fig:si_microscopic_2}
\end{figure}

\begin{figure}[!htbp]
  \centering
  \includegraphics[width=0.99\textwidth]{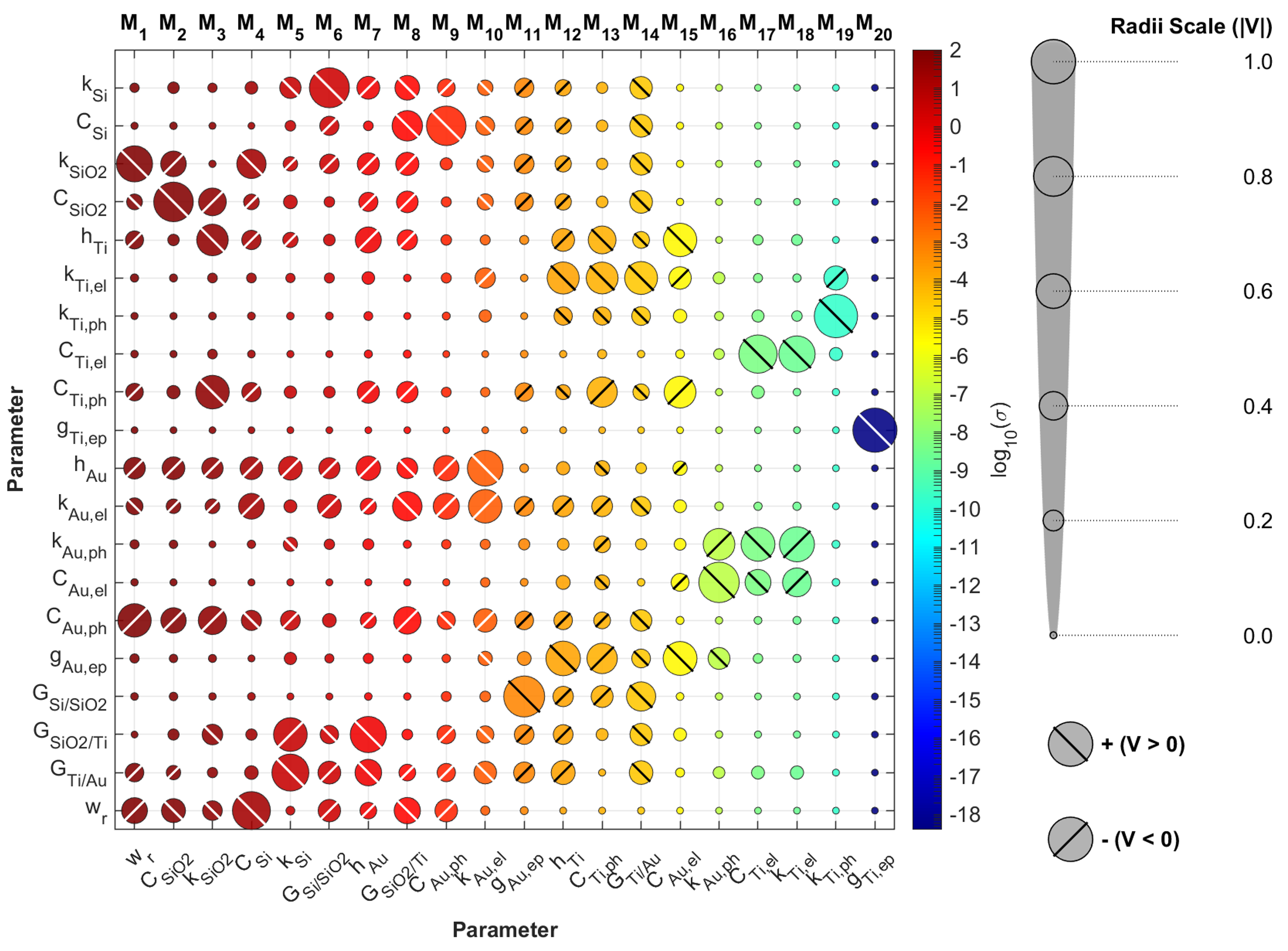}
  \caption{SVD bubble chart for the OTh-TTM model of the Au/Ti/SiO\textsubscript{2}/Si stack in the small-spot configuration ($w_r=\SI{1.86}{\micro\metre}$), with twenty candidate parameters, including the additional electron/phonon material parameters introduced by the two-temperature closure: $k_{\mathrm{Ti,el}}$, $k_{\mathrm{Ti,ph}}$, $C_{\mathrm{Ti,el}}$, $C_{\mathrm{Ti,ph}}$, $g_{\mathrm{Ti,ep}}$, $k_{\mathrm{Au,el}}$, $k_{\mathrm{Au,ph}}$, $C_{\mathrm{Au,el}}$, $C_{\mathrm{Au,ph}}$, and $g_{\mathrm{Au,ep}}$. In the accompanying numerical SVD diagnosis, the leading modes $M_1$--$M_4$ (dark red, $\log_{10}\sigma_m^{\mathrm{raw}}\approx1$--$2$) include loadings from $w_r$, $C_{\mathrm{SiO_2}}$, and $k_{\mathrm{SiO_2}}$, joined here by a large $C_{\mathrm{Au,ph}}$ loading in $M_1$. These are local sensitivity components; final target admission uses the intersection list. Several electron/phonon-channel parameters have their largest loadings in trailing, low-$\sigma$ modes: $C_{\mathrm{Ti,el}}$, $k_{\mathrm{Au,ph}}$, and $C_{\mathrm{Au,el}}$ have large bubbles in the light-green/cyan modes $M_{16}$--$M_{18}$ ($\log_{10}\sigma_m^{\mathrm{raw}}\approx-8$ to $-11$), $k_{\mathrm{Ti,el}}$ and $k_{\mathrm{Ti,ph}}$ dominate the still deeper teal mode $M_{19}$ ($\log_{10}\sigma_m^{\mathrm{raw}}\approx-12$), and the Ti-layer electron--phonon coupling $g_{\mathrm{Ti,ep}}$ loads almost exclusively onto the terminal mode $M_{20}$ at $\log_{10}\sigma_m^{\mathrm{raw}}\approx-18$, at the lower end of this chart's color scale. For the requested buried targets, the final screening retains $k_{\mathrm{SiO_2}}$ under both thermal models (Table~\ref{tab:fit_ledger}).}
  \label{fig:si_microscopic_3}
\end{figure}

\begin{figure}[!htbp]
  \centering
  \includegraphics[width=0.99\textwidth,height=0.66\textheight,keepaspectratio]{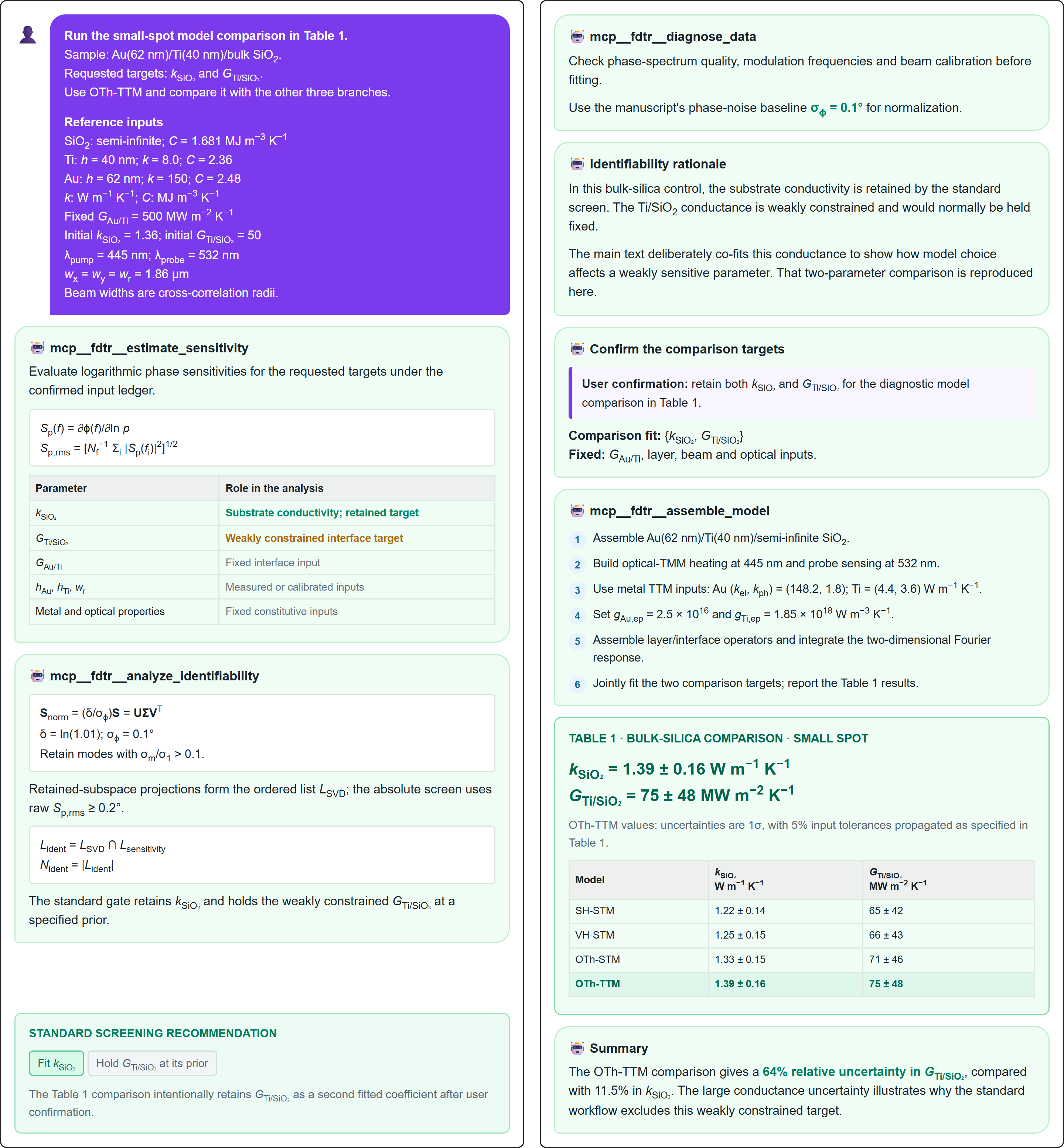}
  \caption{Illustrative PRISM interaction for the two-parameter Au/Ti/SiO$_2$ bulk-silica comparison in \Cref{tab:silica}A. The stack is Au(62\,nm)/Ti(40\,nm) on semi-infinite SiO$_2$, with pump/probe wavelengths of 445/532\,nm and circular cross-correlation radius $w_r=1.86\,\mu$m. Left: the user request, reference inputs, logarithmic phase sensitivity, and the SVD/absolute-sensitivity intersection screen; $N_{\mathrm{ident}}$ denotes the length of the ordered intersection list. Right: spectrum diagnosis, confirmation of the comparison targets, solver assembly, and results reproduced from \Cref{tab:silica}A. The user-confirmed diagnostic co-fit follows the comparison described in section~\ref{sec:optical_result} of the main text. The OTh-TTM values are $k_{\mathrm{SiO_2}}=1.39\pm0.16$\,W\,m$^{-1}$K$^{-1}$ and $G_{\mathrm{Ti/SiO_2}}=75\pm48$\,MW\,m$^{-2}$K$^{-1}$. Uncertainties are one standard deviation under the input-propagation protocol of \Cref{tab:silica}; the conductance relative uncertainty is 64\%, compared with 11.5\% for the conductivity.}

  \label{fig:si_chat_ui}
\end{figure}

\clearpage

\begin{figure}[p]
  \centering
  \includegraphics[width=\textwidth,height=0.46\textheight,keepaspectratio]{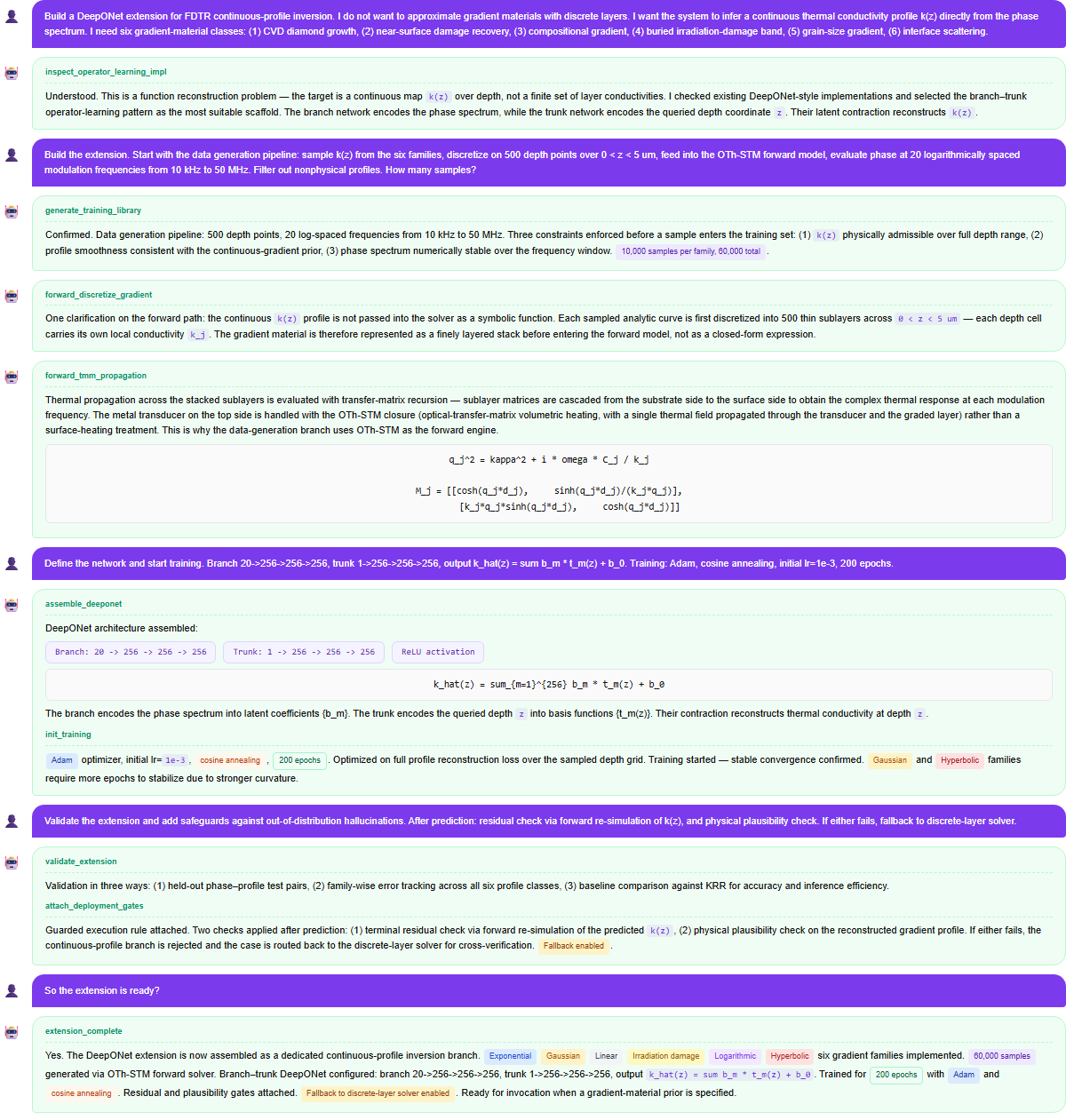}
  \caption{Representative agentic execution trace within PRISM for the DeepONet continuous-profile branch, reproduced as a sequence of user requests (purple) and agent tool calls (green), top to bottom.
    Following the user's request for a continuous $k(z)$ inversion across six gradient-material classes (CVD diamond growth, near-surface damage recovery, compositional gradient, buried irradiation-damage band, grain-size gradient, interface scattering), \texttt{inspect\_operator\_learning\_impl} identifies the task as a function-reconstruction problem and selects the branch--trunk operator-learning scaffold, with the branch network encoding the phase spectrum and the trunk network encoding the queried depth $z$.
    (i)~\emph{Data generation}: \texttt{generate\_training\_library} samples $k(z)$ from the six families, discretizes each profile on 500 depth points over $0<z<\SI{5}{\micro\metre}$, and evaluates the phase at 20 log-spaced modulation frequencies from \SI{10}{\kilo\hertz} to \SI{50}{\mega\hertz}, enforcing physical admissibility, profile regularity, and phase-spectrum numerical stability before accepting a sample, yielding 10{,}000 samples per family (60{,}000 total before the $7{:}2{:}1$ training/validation/test split); \texttt{forward\_discretize\_gradient} clarifies that each continuous profile is itself discretized into 500 thin sublayers before being passed to the forward solver, so the gradient material is represented as a finely layered stack rather than a closed-form expression; \texttt{forward\_tmm\_propagation} then cascades the sublayer transfer matrices $M_j=\left(\begin{smallmatrix}\cosh(q_jd_j) & \sinh(q_jd_j)/(k_jq_j)\\ k_jq_j\sinh(q_jd_j) & \cosh(q_jd_j)\end{smallmatrix}\right)$ (with $q_j^2=\mu_\rho^2+i\omega C_j/k_j$ per sublayer, $k_j$ the sublayer thermal conductivity and $\mu_\rho$ the angular in-plane magnitude $2\pi\sqrt{u^2+v^2}$ in the two-dimensional surface-response Fourier integral). The displayed matrices map each sublayer from top to bottom; the substrate-to-surface recursion described in the trace uses their inverse relations. The graded OTh-STM formulation is described under \protect\hyperlink{par:deeponet_forward}{\emph{Forward model for training-data generation}} (section~\ref{sec:deeponet_training_data}): the optical transfer matrix supplies depth-resolved volumetric heating in the Au transducer, while a single thermal field is propagated through both the transducer and the graded layer.
    (ii)~\emph{Architecture assembly and training}: \texttt{assemble\_deeponet} builds the branch ($20\to256\to256\to256$, ReLU) and trunk ($1\to256\to256\to256$, ReLU) networks with output $\widehat{k}(z)=\sum_{i=1}^{256}b_i\,t_i(z)+b_0$; \texttt{init\_training} configures the Adam optimizer with initial learning rate $10^{-3}$, a decaying learning-rate schedule, and 200 epochs, reporting that the Gaussian and hyperbolic families require more epochs to stabilize owing to their stronger curvature.
    (iii)~\emph{Validation and deployment gating}: \texttt{validate\_extension} checks the trained operator three ways --- held-out phase--profile test pairs, family-wise error tracking across all six classes, and a baseline accuracy/inference-efficiency comparison against kernel ridge regression (KRR; Figure~\ref{fig:deeponet}K) --- while \texttt{attach\_deployment\_gates} enforces two checks after every prediction, a terminal residual check via forward re-simulation of the predicted $k(z)$ and a physical-plausibility check on the reconstructed profile, falling back to the discrete-layer solver if either fails. The trace concludes with \texttt{extension\_complete}, marking completion of the supervised branch extension. Routine invocation uses the profile-prior and deployment checks described above.}
  \label{fig:si_deeponet_log}
\end{figure}
\clearpage

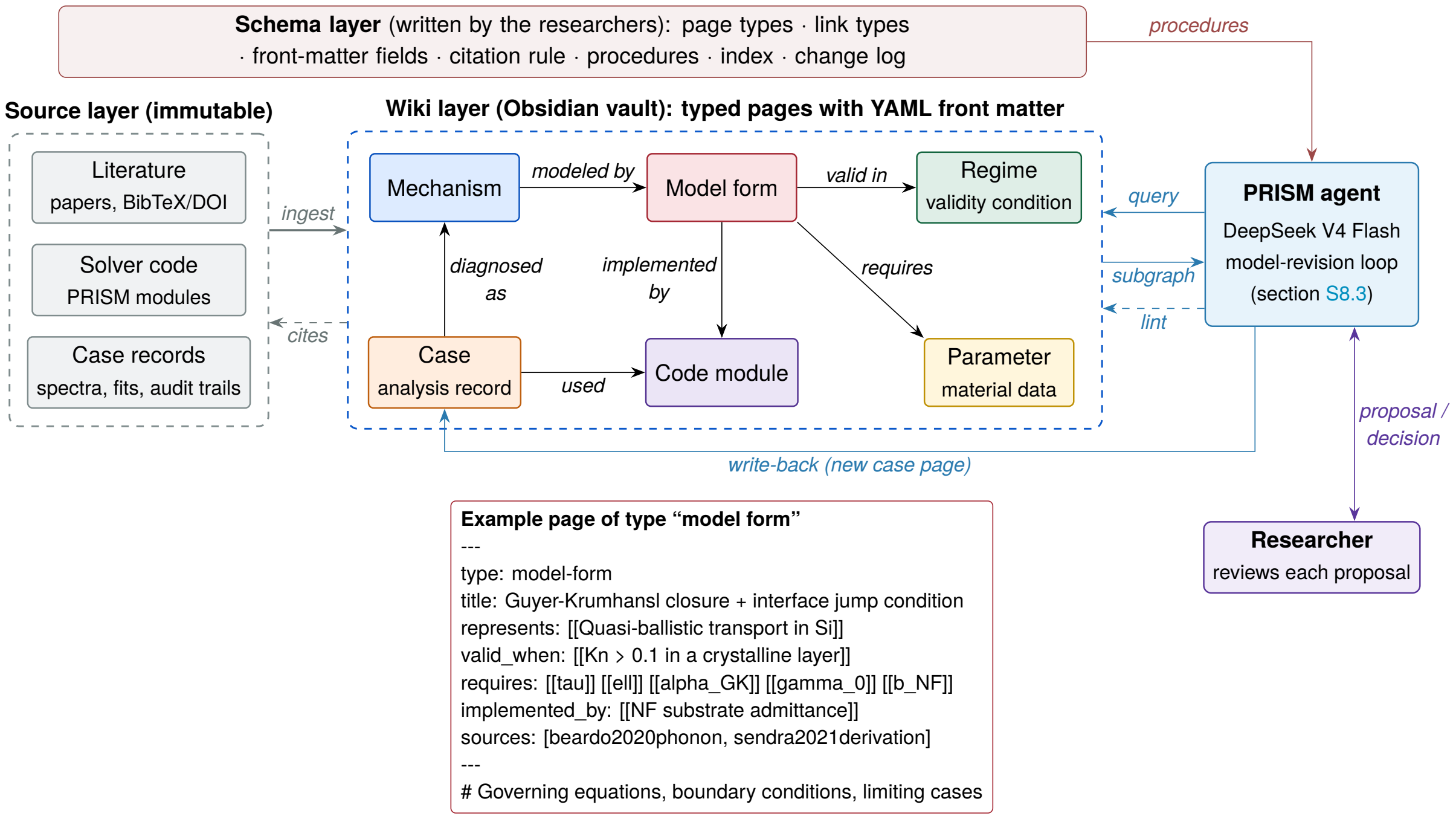
\begin{figure}[H]
\centering
\resizebox{\textwidth}{!}{%
\begin{tikzpicture}[
  font=\prismuifont\footnotesize,
  >={Stealth[length=2mm]},
  page/.style={draw=#1!75!black, fill=#1!14, rounded corners=2pt, minimum width=2.1cm, minimum height=0.95cm, align=center, line width=0.6pt},
  src/.style={draw=stategray!85!black, fill=stategray!14, rounded corners=2pt, minimum width=3.0cm, minimum height=1.0cm, align=center, line width=0.6pt},
  lab/.style={font=\prismuifont\scriptsize\itshape, align=center, inner sep=1pt},
  ttl/.style={font=\prismuifont\footnotesize\bfseries},
  layer/.style={draw=#1, dashed, rounded corners=4pt, line width=0.7pt},
]
\node[src] (lit)   at (1.2,5.4) {Literature\\ \scriptsize papers, BibTeX/DOI};
\node[src] (scode) at (1.2,4.1) {Solver code\\ \scriptsize PRISM modules};
\node[src] (crec)  at (1.2,2.8) {Case records\\ \scriptsize spectra, fits, audit trails};
\node[layer=stategray!85!black, fit=(lit)(scode)(crec), inner sep=7pt] (S) {};
\node[ttl, anchor=south] at (S.north) {Source layer (immutable)};
\node[page=skillblue]  (mech) at (5.5,5.4)  {Mechanism};
\node[page=llmcore]    (mf)   at (9.4,5.4)  {Model form};
\node[page=datagreen]  (reg)  at (13.3,5.4) {Regime\\ \scriptsize validity condition};
\node[page=deeponet]   (case) at (5.5,2.8)  {Case\\ \scriptsize analysis record};
\node[page=userpurple] (cod)  at (9.4,2.8)  {Code module};
\node[page=mcpyellow]  (par)  at (13.3,2.8) {Parameter\\ \scriptsize material data};
\node[layer=skillblue!80!black, fit=(mech)(reg)(case)(par), inner sep=8pt] (W) {};
\node[ttl, anchor=south] at (W.north) {Wiki layer (Obsidian vault): typed pages with YAML front matter};
\draw[->] (mech) -- node[lab, above=1pt]{modeled by} (mf);
\draw[->] (mf)   -- node[lab, above=1pt]{valid in} (reg);
\draw[->] (mf)   -- node[lab, left=1pt]{implemented\\by} (cod);
\draw[->] (mf.south east) -- node[lab, above right=-1pt]{requires} (par.north west);
\draw[->] (case) -- node[lab, right=1pt]{diagnosed\\as} (mech);
\draw[->] (case) -- node[lab, below=1pt]{used} (cod);
\node[draw=sectioncolor, fill=sectioncolor!8, rounded corners=3pt, text width=14.2cm, minimum height=0.8cm, align=center] (schema) at (7.3,7.45)
  {\textbf{Schema layer} (written by the researchers): page types \textperiodcentered{} link types \textperiodcentered{} front-matter fields \textperiodcentered{} citation rule \textperiodcentered{} procedures \textperiodcentered{} index \textperiodcentered{} change log};
\draw[->, thick, stategray!70!black] (S.east |- 0,4.8) -- node[lab, above=1pt]{ingest} (W.west |- 0,4.8);
\draw[->, dashed, stategray!70!black] (W.west |- 0,3.5) -- node[lab, below=1pt]{cites} (S.east |- 0,3.5);
\node[draw=langgraph!80!black, fill=langgraph!12, rounded corners=3pt, minimum width=3.0cm, minimum height=2.3cm, align=center, line width=0.6pt] (agent) at (17.7,4.6)
  {\textbf{PRISM agent}\\[2pt] \scriptsize DeepSeek V4 Flash\\ \scriptsize model-revision loop\\ \scriptsize (section~\ref{sec:si_kg_revision})};
\draw[->, sectioncolor] (schema.east) -| node[lab, pos=0.25, above=1pt]{procedures} (agent.north);
\draw[->, langgraph!80!black] (agent.west |- 0,5.05) -- node[lab, above=1pt]{query} (W.east |- 0,5.05);
\draw[->, langgraph!80!black] (W.east |- 0,4.35) -- node[lab, below=1pt]{subgraph} (agent.west |- 0,4.35);
\draw[->, dashed, langgraph!80!black] (agent.west |- 0,3.7) -- node[lab, below=1pt]{lint} (W.east |- 0,3.7);
\draw[->, langgraph!80!black] ([xshift=-8mm]agent.south) -- ++(0,-1.75) -| node[lab, pos=0.25, below=1pt]{write-back (new case page)} (case.south);
\node[draw=userpurple!80!black, fill=userpurple!10, rounded corners=3pt, minimum width=3.0cm, minimum height=1.0cm, align=center, line width=0.6pt] (res) at (17.7,0.2)
  {\textbf{Researcher}\\ \scriptsize reviews each proposal};
\draw[<->, userpurple!80!black] ([xshift=6mm]agent.south) -- node[lab, right=1pt]{proposal /\\decision} ([xshift=6mm]res.north);
\node[draw=llmcore!75!black, fill=white, rounded corners=2pt, anchor=north, align=left, font=\prismuifont\scriptsize, inner sep=4pt] (card) at (9.4,1.0)
  {\textbf{Example page of type \textquotedblleft model form\textquotedblright}\\
   {-}{-}{-}\\
   type: model-form\\
   title: Guyer-Krumhansl closure + interface jump condition\\
   represents: [[Quasi-ballistic transport in Si]]\\
   valid\_when: [[Kn \textgreater{} 0.1 in a crystalline layer]]\\
   requires: [[tau]] [[ell]] [[alpha\_GK]] [[gamma\_0]] [[b\_NF]]\\
   implemented\_by: [[NF substrate admittance]]\\
   sources: [beardo2020phonon, sendra2021derivation]\\
   {-}{-}{-}\\
   \# Governing equations, boundary conditions, limiting cases};
\end{tikzpicture}%
}
\caption{Structure of the wiki-based knowledge graph used by PRISM (section~\ref{sec:si_knowledge_base}). The source layer holds the literature, the solver code, and the records of past analyses, and is never modified by the agent. The wiki layer consists of Markdown pages in an Obsidian vault; each page has a type, and its YAML front matter declares typed links to other pages (solid arrows) and citations back to the source layer (dashed arrow). The schema layer, written by the researchers, fixes the page types, the link types, the citation rule, and the procedures the agent follows, and keeps an index of all pages and a log of every change. The agent ingests sources into pages, queries the graph with the descriptors of a misfit and receives the matching subgraph with its citations, writes every analysis back as a case page, and lints the wiki for inconsistencies; every proposal it derives from the graph is returned to the researcher for review. The inset shows the front matter of one model-form page.}
\label{fig:si_kg_structure}
\end{figure}
\clearpage

\section{Implementation Notes for the Thermal Solver Stack}
\label{sec:impl_notes}

This section lists the stack inputs, optical and thermal module settings, and non-Fourier constitutive parameters used with the derivations in sections~\ref{sec:tmm} and~\ref{sec:non_fourier_si}.

\subsection{General Stack Case}

The tables list initial guesses and fixed inputs; fitted values for each benchmark are reported in the main text.

\begin{table}[H]

\centering

\caption{Thermal physical properties and default layer thicknesses used in the benchmark stacks. The default thicknesses apply to the Ti-containing stacks; the thicknesses of the Au/SiO$_2$/Si, Au/Si, and Au/MoS$_2$ stacks are given in the main text.}

\label{tab:archive_si_thermal_props}

\small

\begin{tabular*}{\textwidth}{@{\extracolsep{\fill}} l c c c c c}

\toprule

\textbf{Material} & $h$ (nm) & $k$ (W\,m$^{-1}$K$^{-1}$) & $k_{\mathrm{el}}$ (W\,m$^{-1}$K$^{-1}$) & $k_{\mathrm{ph}}$ (W\,m$^{-1}$K$^{-1}$) & $C$ (J\,m$^{-3}$K$^{-1}$) \\

\midrule

Si  & $4\times10^5$$^a$ & 140$^b$  & ---  & ---  & $1.632\times10^6$$^c$ \\

SiO$_2$ & $2\times10^3$$^d$ & 1.36     & ---  & ---  & $1.681\times10^6$ \\

Ti  & 40                & 8.0      & 4.4  & 3.6  & $2.36\times10^6$ \\

Au  & 62                & 150      & 148.2 & 1.8  & $2.48\times10^6$ \\

\bottomrule

\end{tabular*}

\end{table}

{\footnotesize

$^a$\,Si substrate thickness in nm; treated as semi-infinite in the thermal model.

$^b$\,Fitted parameter; listed value is the initial guess.

$^c$\,Room-temperature volumetric heat capacity of crystalline Si at 296.15\,K~\cite{beardo2020phonon}. For the non-Fourier analyses at 311\,K, $C_{\mathrm{Si}}=1.692\times10^6$~J\,m$^{-3}$K$^{-1}$ (Table~\ref{tab:archive_si_nf_params}).

$^d$\,Default SiO$_2$ thickness for the Au/Ti/SiO$_2$/Si and Au/SiO$_2$/Si buried-silica stacks; the Au/Ti/SiO$_2$ control treats SiO$_2$ as the semi-infinite substrate.

{\RaggedRight
\par\noindent The main Au/Ti/SiO$_2$/Si benchmark therefore uses Au(62\,nm)/\allowbreak Ti(40\,nm)/\allowbreak SiO$_2$(2\,$\mu$m)/\allowbreak Si. See the interface-conductance table below for the corresponding interface priors.
\par}

}

\begin{table}[H]

\centering

\caption{Initial interface-conductance inputs ($G$) for the benchmark stacks.}

\label{tab:archive_si_interface_G}

\small

\begin{tabular*}{\textwidth}{@{\extracolsep{\fill}} l c c c}

\toprule

\textbf{Interface} & $G_{\mathrm{ph}}$ (MW\,m$^{-2}$K$^{-1}$) & $G_{\mathrm{el}}$ (MW\,m$^{-2}$K$^{-1}$) & $G_{\mathrm{total}}$ (MW\,m$^{-2}$K$^{-1}$) \\

\midrule

Si/SiO$_2$ & 50 & --- & 50 \\

SiO$_2$/Ti & 50 & --- & 50 \\

Ti/Au & 50 & 450 & 500 \\

\midrule

\multicolumn{4}{l}{\textit{Additional Material Combinations:}} \\

Ti/Si & 100 & --- & 100 \\

Au/Si & 50 & --- & 50 \\

Au/SiO$_2$ & 50 & --- & 50 \\

Au/MoS$_2$ & 31 & --- & 31 \\

\bottomrule

\end{tabular*}

\end{table}

Table~\ref{tab:archive_si_beam_sizes} records the benchmark beam radii and offset settings used in the main text.

\begin{table}[H]

\centering

\caption{Benchmark beam radii and offset settings. For our measurements, the listed radii are the raw $1/e^2$ widths of the pump--probe cross-correlation profile defined in \hyperref[sec:methods]{the materials and methods section} and Equation~\eqref{eq:fourier_combined_kernel}. The slightly elliptical Au/Si spot retains both axes in the two-dimensional Fourier quadrature; $w_r$ denotes a common directional correlation width only for circular spots. The Au/MoS$_2$ entry instead gives the individual pump and probe radii reported in Ref.~\cite{guo2026light}, and uses their combined squared widths in the same general kernel.}

\label{tab:archive_si_beam_sizes}

\small

\begin{tabularx}{\textwidth}{@{} l l >{\raggedright\arraybackslash}X @{}}

\toprule

\textbf{Benchmark} & \textbf{Beam radii} & \textbf{Geometry / offset} \\

\midrule

Au/Si & $w_x=\SI{2.19}{\micro\metre}$, $w_y=\SI{1.84}{\micro\metre}$ & Coaxial ($\Delta=0$) and offset ($\Delta\approx\SI{2}{\micro\metre}$) \\

Au/Ti/SiO$_2$/Si & $w_r=\SI{10.44}{\micro\metre}$; $w_r=\SI{1.86}{\micro\metre}$ & Large- and small-spot fits; target-reduction case of Figures~\ref{fig:si_microscopic_1}--\ref{fig:si_microscopic_3} \\

Au/Ti/SiO$_2$ & $w_r=\SI{1.86}{\micro\metre}$ & Bulk-silica control (Table~\ref{tab:silica}A) \\

Au/SiO$_2$/Si & $w_r=\SI{1.86}{\micro\metre}$; $w_r=\SI{10.44}{\micro\metre}$ & Small- and large-spot control without Ti (Table~\ref{tab:silica}A) \\

Au/Ti/Si & $w_r=\SI{1.86}{\micro\metre}$; $w_r=\SI{10.44}{\micro\metre}$ & Small- and large-spot non-Fourier escalation test \\
Au/MoS$_2$ & $w_r=\SI{2.8}{\micro\metre}$ & Coaxial; spectrum of Ref.~\cite{guo2026light} (Figure~\ref{fig:optical}) \\

\bottomrule

\end{tabularx}

\end{table}

Table~\ref{tab:archive_si_optical_constants} separates the 445~nm inputs for our own experimental cases from the 488~nm inputs of the published Au/MoS$_2$ benchmark.

\begin{table}[H]

\centering

\caption{Optical constants for the optical transfer-matrix calculations near room temperature. The 445~nm pump column (our case) uses Au and Ti data of Johnson and Christy~\cite{Johnson1972Optical,Johnson1974Optical}, the fused-silica dispersion of Malitson~\cite{Malitson1965Interspecimen} with negligible visible absorption, and 300~K crystalline-Si data of Green~\cite{Green2008Self}. Tabulated $n$ and $\kappa$ are linearly interpolated in wavelength, except for SiO$_2$, whose $n$ is evaluated from the Sellmeier formula. For bulk MoS$_2$, the 445~nm value is the ordinary (in-plane) optical component of Ermolaev et al.~\cite{Ermolaev2020Broadband}, appropriate to normal incidence on the basal plane. The 488/532~nm MoS$_2$ inputs come from Ref.~\cite{guo2026light}. The numerical datasets were accessed through the refractiveindex.info database~\cite{Polyanskiy2024Refractiveindex}. The 488~nm column is retained for that literature benchmark; our own experimental cases use 445~nm pumping and 532~nm probing. The 445~nm MoS$_2$ entry is a literature reference value. All $n,\kappa$ values are dimensionless, with $\tilde N=n+i\kappa$.}
\label{tab:archive_si_optical_constants}
\small
\begin{tabular*}{\textwidth}{@{\extracolsep{\fill}} l c c c c c c}
\toprule
& \multicolumn{2}{c}{\shortstack{\textbf{Pump (445 nm)}\\\textbf{our case}}} & \multicolumn{2}{c}{\textbf{Pump (488 nm)}} & \multicolumn{2}{c}{\textbf{Probe (532 nm)}} \\
\cmidrule(lr){2-3} \cmidrule(lr){4-5} \cmidrule(lr){6-7}
\textbf{Material} & $n$ & $\kappa$ & $n$ & $\kappa$ & $n$ & $\kappa$ \\
\midrule
Au & 1.4002 & 1.9238 & 1.09 & 1.75 & 0.44 & 2.27 \\
Ti & 2.2520 & 3.0310 & 2.04 & 2.83 & 2.16 & 2.93 \\
SiO$_2$ & 1.46595 & $\approx0$ & 1.46 & 0 & 1.461 & 0 \\
Si & 4.7515 & 0.09665 & 4.23 & 0.072 & 4.15 & 0.044 \\
MoS$_2$ (bulk) & 4.7258 & 3.3624 & 3.98 & 1.17 & 4.08 & 0.95 \\
\bottomrule
\end{tabular*}

\end{table}

\subsection{Model-Specific Module Settings}

Table~\ref{tab:archive_si_optical_module} records how each solver branch differs in its optical source handling and detection logic.

\begin{table}[H]

\centering

\caption{Optical module and detection logic for the experimental branches. OTh probe readout includes all absorbing layers. In the graded OTh-STM training branch of section~\ref{sec:deeponet_training_data} ($h_{\mathrm{Au}}=\SI{97}{\nano\metre}$), the volumetric probe kernel operates in the shallow-sensing limit and reduces to the surface-temperature readout.}

\label{tab:archive_si_optical_module}

\small

\begin{tabular*}{\textwidth}{@{\extracolsep{\fill}} l l l l}

\toprule

\textbf{Model} & \textbf{Heat Source Type} & \textbf{Depth specification} & \textbf{Detection Mode} \\

\midrule

\textbf{SH-STM} & Surface source & 0 (at boundary) & Surface probe sensitivity \\

\textbf{VH-STM} & Exponential volume source & Fixed $d_0 = \lambda^{\mathrm{pump}} / (4\pi \kappa)$ & Surface probe sensitivity \\

\textbf{OTh-STM} & Volumetric TMM & Layer-resolved $Q_n(z)$ & Volumetric sensing ($C_{f,j}^{(\pm)},g_{f,j}^{(\pm)}$) \\

\textbf{OTh-TTM} & Volumetric TMM & Layer-resolved $Q_n(z)$ & Volumetric sensing ($C_{f,j}^{(\pm)},g_{f,j}^{(\pm)}$) \\

\bottomrule

\end{tabular*}

\end{table}

The OTh-TTM branch additionally requires the explicit electron--phonon coupling parameters listed in Table~\ref{tab:archive_si_ttm_params}.

\begin{table}[H]

\centering

\caption{Thermal Module Settings for the OTh-TTM Model.}

\label{tab:archive_si_ttm_params}

\small

\begin{tabular*}{\textwidth}{@{\extracolsep{\fill}} l l l l}

\toprule

\textbf{Parameter} & \textbf{Unit} & \textbf{Au} & \textbf{Ti} \\

\midrule

$g_{\mathrm{ep}}$ (electron--phonon coupling)  & W\,m$^{-3}$K$^{-1}$ & $2.5\times10^{16}$  & $1.85\times10^{18}$ \\

$k_{\mathrm{el}}$ (electron thermal conductivity) & W\,m$^{-1}$K$^{-1}$ & 148.2 & 4.4 \\

$k_{\mathrm{ph}}$ (phonon thermal conductivity) & W\,m$^{-1}$K$^{-1}$ & 1.8  & 3.6 \\

$C_{\mathrm{el}}$ (electron heat capacity)    & J\,m$^{-3}$K$^{-1}$ & $2.0\times10^4$     & $1.0\times10^5$ \\

$C_{\mathrm{ph}}$ (phonon heat capacity)      & J\,m$^{-3}$K$^{-1}$ & $2.46\times10^6$    & $2.26\times10^6$ \\

Electron fraction $R_{\mathrm{el}}$ & ---  & 0.988 & 0.55 \\

\bottomrule

\end{tabular*}

\end{table}

{\footnotesize

The electron fraction $R_{\mathrm{el}}$ represents the fraction of total thermal conductivity carried by electrons: $k_{\mathrm{el}} = R_{\mathrm{el}}\,k$, $k_{\mathrm{ph}} = (1-R_{\mathrm{el}})\,k$.

For Au: $R_{\mathrm{el}}=0.988$, so that the lattice carries \SI{1.8}{\watt\per\metre\per\kelvin}; the first-principles lattice conductivity of Au is \SI{2}{\watt\per\metre\per\kelvin}~\cite{jain2016thermal}; for Ti: $R_{\mathrm{el}}=0.55$ (d-band transition metal with significant phonon contribution).

}

\subsection{Non-Fourier Phonon Hydrodynamics: OTh-TTM(NF)}

The Au/Ti/Si benchmark uses OTh-TTM(NF) after the diffusive hierarchy fails to reconcile spot-size and frequency-window dependence. This closure incorporates phonon hydrodynamics and a nonequilibrium interface condition, following Beardo \textit{et al.}~\cite{beardo2020phonon}, the TDTR analysis of Xiang \textit{et al.}~\cite{xiang2022time}, and the collective-transport literature of de Tomas \textit{et al.}~\cite{detomas2014collective,detomas2014groupiv}. Table~\ref{tab:archive_si_nf_params} gives the Si bulk and Ti/Si interface inputs for this case.

\begin{table}[H]

\centering

\caption{Bulk Si and reference interface parameters of the OTh-TTM(NF) model, taken directly from the 311\,K first-principles data of Beardo \textit{et al.}~\cite{beardo2020phonon}.}

\label{tab:archive_si_nf_params}

\small

\begin{tabular*}{\textwidth}{@{\extracolsep{\fill}} l l l >{\raggedright\arraybackslash}p{5.2cm}}

\toprule

\textbf{Parameter} & \textbf{Unit} & \textbf{Literature Value$^\dagger$} & \textbf{Physical meaning} \\

\midrule

\multicolumn{4}{l}{\textit{Bulk Si hydrodynamic transport}} \\

$C_{\mathrm{Si}}$      & J\,m$^{-3}$K$^{-1}$ & $1.692\times10^6$ & Volumetric heat capacity \\

$\tau$                 & ps            & 42.0    & Phonon flux relaxation time \\

$\ell$                 & nm            & 185.0   & Nonlocal phonon length scale \\

$\alpha_{\mathrm{GK}}$ & ---           & 1/3     & Nonlocal divergence weight coefficient \\

\midrule

\multicolumn{4}{l}{\textit{Reference interface nonequilibrium parameters}} \\

$\gamma_0$                & MW\,m$^{-2}$K$^{-1}$ & 1068    & Substrate modal energy-transfer coefficient \\

$\gamma_{0\Gamma}$        & MW\,m$^{-2}$K$^{-1}$ & 2237    & Transducer modal energy-transfer coefficient \\

$\beta_{\mathrm{NF}}$     & nm            & $-21.0$ & Interface divergence jump coefficient \\

$\chi_{xx}$               & nm            & $-31.0$ & Cross-plane (interface-normal) gradient correction \\

$\chi_{zz}$               & nm            & $-16.0$ & In-plane (tangential) gradient correction \\

\midrule

\multicolumn{4}{l}{\textit{Experimental conditions}} \\

$T_{\mathrm{amb}}$     & K             & 296.15   & Ambient temperature \\

$T_0$                  & K             & 311      & Representative steady-state temperature \\

$w_r$                  & \textmu m     & 1.86, 10.44    & Small- and large-spot cross-correlation radii ($1/e^2$) \\

\bottomrule

\end{tabular*}

\end{table}

{\footnotesize

$^\dagger$\,First-principles kinetic-collective parameters at 311\,K from Table~I of Beardo \textit{et al.}~\cite{beardo2020phonon}, anchored in the collective-transport formulation of de Tomas \textit{et al.}~\cite{detomas2014collective,detomas2014groupiv}. Note on coordinates: to maintain direct correspondence with Equations~(B7d)--(B7e) of Ref.~\cite{beardo2020phonon}, $\chi_{xx}$ and $\chi_{zz}$ retain the local interface coordinates of section~\ref{sec:non_fourier_si}, where $x$ is normal to the substrate interface (cross-plane) and $y,z$ are tangential (in-plane).

}

\end{document}